%% file: SupGpaper_v1_.tex
\documentclass[10pt, twocolumn, twoside]{IEEEtran}

\usepackage{cite}
\usepackage{enumerate}
\usepackage{graphicx}
\usepackage[cmex10]{amsmath} 
\usepackage{amssymb}
\usepackage{mathtools}
\usepackage{xspace}
\usepackage{subfigure}
\usepackage{algorithm} 
\usepackage{colonequals}
\usepackage[ mathscr]{euscript}
\usepackage{multirow}
\usepackage[table]{xcolor}
\usepackage{siunitx}

\usepackage{epsfig}
\usepackage{epstopdf}

\usepackage{tikz}
\usepackage{pgfplots}

\usetikzlibrary{arrows,shapes,backgrounds,plotmarks,positioning}

\pgfplotsset{compat=newest}                         
\pgfplotsset{plot coordinates/math parser=false}
\newlength\figureheight
\newlength\figurewidth

\newtheorem{theorem}{Theorem}[section]
\newtheorem{lemma}[theorem]{Lemma}
\newtheorem{definition}[theorem]{Definition}
\newtheorem{corollary}[theorem]{Corollary}

\newtheorem{assumption}[theorem]{Assumption}

\newtheorem{example}[theorem]{Example}

\newcommand{\G}{\Omega}                     
\newcommand{\snr}{\gamma}
\newcommand{\Snr}{\boldsymbol{\gamma}}
\newcommand{\SNR}{\Gamma}

\newcommand{\A}{\mathcal{A}}                
\newcommand{\act}{\mathbf{a}}               
\newcommand{\po}{\theta}                    
\newcommand{\Po}{\boldsymbol{\theta}}       
\newcommand{\PoU}{\overline{\Po}^*}         
\newcommand{\PoD}{\underline{\Po}^*}        
\newcommand{\poU}{\overline{\theta}^*}         
\newcommand{\poD}{\underline{\theta}^*}        
\newcommand{\BR}{\boldsymbol{\psi}}
\newcommand{\Br}{\psi}

\newcommand{\PbCostr}{\bar{P_b}}

\newcommand{\Lat}{\mathcal{L}}    
\newcommand{\x}{\mathbf{x}}
\newcommand{\y}{\mathbf{y}}

\newcommand{\figref}[1]{Fig.~\ref{#1}}

\newcommand{\Z}{\mathbb{Z}}
\newcommand{\R}{\mathbb{R}}

\begin{document}

\title{Adaptive Modulation in Network-coded Two-way Relay Channel: A Supermodular Game Approach}

\author{Ni~Ding,~\IEEEmembership{Member,~IEEE}, Parastoo~Sadeghi,~\IEEEmembership{Senior Member,~IEEE}, Rodney~A.~Kennedy,~\IEEEmembership{Fellow,~IEEE}
\thanks{The authors are with the Research School of Engineering, College of Engineering and Computer Science, the Australian National University (ANU), Canberra, ACT 0200, Australia (email: $\{$ni.ding, parastoo.sadeghi, rodney.kennedy$\}$@anu.edu.au).}
}

\markboth{Adaptive Modulation in Network-coded Two-way Relay Channel: A Supermodular Game Approach}%
{Ding, Sadeghi and Kennedy}

\maketitle

\begin{abstract}
We study the adaptive modulation (AM) problem in a network-coded two-way relay channel (NC-TWRC), where each of the two users controls its own bit rate in the $m$-ary quadrature amplitude modulation ($m$-QAM) to minimize the transmission error rate and enhance the spectral efficiency. We show that there exists a strategic complementarity, one user tends to transmit while the other decides to do so in order to enhance the overall spectral efficiency, which is beyond the scope of the conventional single-agent AM scheduling method. We propose a two-player game model parameterized by the signal-to-noise ratios (SNRs) of two user-to-user channels and prove that it is a supermodular game where there always exist the extremal pure strategy Nash equilibria (PSNEs), the largest and smallest PSNEs. We show by simulation results that the extremal PSNEs incur a similar bit error rate (BER) as the conventional single-agent AM scheme, but significantly improve the spectral efficiency in the NC-TWRC system. The study also reveals the Pareto order of the extremal PSNEs: The largest and smallest PSNEs are Pareto worst and best PSNEs, respectively. Finally, we derive the sufficient conditions for the extremal PSNEs to be symmetric and monotonic in channel SNRs. We also discuss how to utilize the symmetry and monotonicity to relieve the complexity in the PSNE learning process.
\end{abstract}

\begin{IEEEkeywords}
adaptive modulation, amplify-and-forward, monotonic comparative statics, physical layer network coding, strategic complementarity, super/submodularity, supermodular game.
\end{IEEEkeywords}

\section{Introduction}

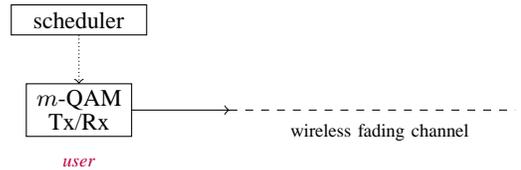
\begin{figure}[tbp]
	\centering
		\centerline{\scalebox{1}{\input{figures/AM.tex}}}
	\caption{Single-user variable-rate adaptive $m$-QAM modulation in wireless transmission \cite{Goldsmith1997,Goldsmith1998,Alouini2000}: the number of bits per QAM symbol of the $m$-QAM transmitter/receiver (Tx/Rx) is controlled by a scheduler.}
	\label{fig:AM}
\end{figure}

The method of adaptive modulation (AM) \cite{Goldsmith1997,Goldsmith1998,Alouini2000,Pursley2000} is to adjust the modulation scheme in response to the time-varying feature of wireless fading channel. Consider \figref{fig:AM} for example. According to the channel state information (CSI), a scheduler increases the transmission rate in $m$-ary quadrature amplitude modulation ($m$-QAM) under favorable channel conditions, e.g., high instantaneous signal-to-noise (SNR) ratio, and reduces it when the channel quality degrades\cite{Goldsmith1997}. The adaptive $m$-QAM modulation system in \figref{fig:AM} poses a single-agent decision-making problem: The scheduler controls the bit rate of $m$-QAM Tx/Rx by considering the SNR of the wireless fading channel. The purpose is to optimize the transmission error rate and spectral efficiency.\footnote{Since there is a tradeoff between transmission error rate and spectral efficiency, the aim of the AM schemes in \cite{Goldsmith1997,Goldsmith1998,Alouini2000,Pursley2000} is to determine the optimal transmission rate by satisfying a certain bit error rate (BER) constraint.}

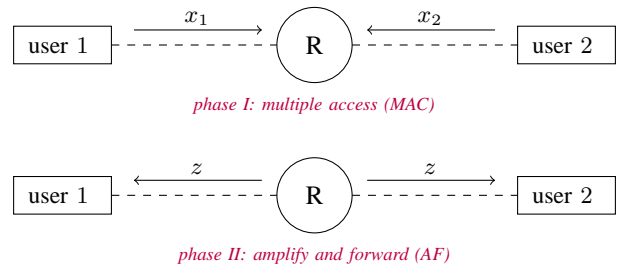
\begin{figure}[tbp]
	\centering
		\centerline{\scalebox{1}{\input{figures/PNC.tex}}}
	\caption{Two-phase physical layer network coding (PNC) scheme using amplify-and-forward (AF) protocol \cite{Popovski2007,Louie2010} in network-coded two-way relay channel (NC-TWRC).}
	\label{fig:PNC}
\end{figure}

Consider the two-way relay network in \figref{fig:PNC}. There are two terminal/end users communicating with each other via a relay, the center node labeled by `R'. The relay exchanges the messages from users $1$ and $2$ by adopting a two-phase amplify-and-forward network coding scheme \cite{Popovski2007,Louie2010}. In phase I, the multiple access (MAC) stage, two users transmit messages $x_1$ and $x_2$ to relay simultaneously. In phase II, the amplify and forward (AF) stage, the relay broadcasts $z$, the superposition of the received signals in phase I. Since a user knows its own message, each user subtracts the self-interference term from $z$ and extracts the message sent by the other. It was shown in \cite{Popovski2007} that this network coding scheme offered a higher spectral efficiency than the conventional one-way relaying method. We call the system in \figref{fig:PNC} network-coded two-way relay channel (NC-TWRC).

By assuming that both terminal users in \figref{fig:PNC} adopt adaptive $m$-QAM modulation mechanism in \figref{fig:AM}, we have an adaptive $m$-QAM NC-TWRC system as shown in \figref{fig:SYSTEM}. The problem left to each scheduler is how to find the best transmission scheme that minimizes the transmission error rate and improves the total spectral efficiency in NC-TWRC. One way to solve the problem is to directly apply the conventional single-agent AM method in \figref{fig:AM} as in \cite{Hwang2011,Yang2012,Ma2011}, e.g., letting the users make decisions separately based on SNRs of their corresponding user-to-user channels without considering the other users decision/action. But, does this method give rise to the optimal AM scheme? We show in the following context that the conventional single-agent AM scheme is not the best choice for NC-TWRC.

\subsection{Motivation} \label{sec:mot}

In any AM method for NC-TWRC, the decisions of both schedulers are made in phase I, the MAC phase. By considering the SNRs of user-to-user channels, the transmission error rate in the system is minimized. But, does the conventional single-agent AM method also optimize the spectral efficiency (the average transmission rate) in NC-TWRC? Before answering this question, it is worth discussing what contributes to the overall spectral efficiency in NC-TWRC. It can be seen in \figref{fig:PNC} that the spectral efficiency is determined by the number of bits per transmission, or broadcast, at the relay in phase II, which is affected by the decisions of both schedulers in phase I. So, a decision made by just considering the user-to-user channels does not necessarily optimize the spectral efficiency.

For example, assume that the SNR of user 1-to-user 2 channel is very low and by following the single-agent AM scheme, scheduler 1 would choose holding (not transmitting) in order to maintain a good average transmission error rate. Assume that scheduler 2 decides to transmit an 8-QAM symbol at this time. Due to scheduler 2's action, if scheduler 1 changes his/her decision to transmitting, the number of bits broadcast by the relay in phase II will be increased from 3 to at least 4. If the gain in spectral efficiency is greater than the loss in transmission error rate, holding is not so good as transmitting for scheduler 1, i.e., the single-agent AM transmission scheme is not optimal in this case. In addition to deciding whether or not to transmit, scheduler 1 should also determine the number of bits in the QAM symbol, e.g., 2, 4 or $8$-QAM, by considering the quantified gains and losses in both transmission error rate and spectral efficiency.

\begin{figure}[tbp]
	\centering
		\centerline{\scalebox{1}{\input{figures/SYSTEM.tex}}}
	\caption{Adaptive $m$-QAM modulation in NC-TWRC: Each scheduler makes its own decision to optimize its transmission error rate and the spectral efficiency in the entire system. }
	\label{fig:SYSTEM}
\end{figure}
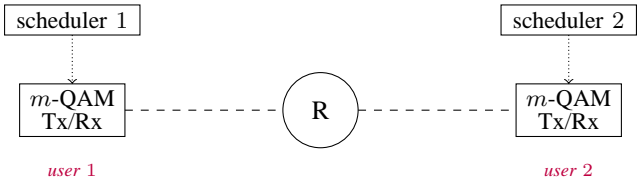

From this example, it can be seen that the spectral efficiency in NC-TWRC depends on the joint decisions of both schedulers. Therefore, to make optimal decisions in the adaptive $m$-QAM NC-TWRC system in \figref{fig:SYSTEM}, a scheduler should consider not only the SNR of the user-to-user channel, but also the decision made by the other scheduler. Then, the AM problem in NC-TWRC is a multi-agent, instead of a single-agent, optimization problem.

\subsection{Main Results}

This paper considers a game theoretic approach, a multi-agent decision-making framework, for the AM problem in the NC-TWRC in \figref{fig:SYSTEM}. Based on the analysis in Section~\ref{sec:mot}, we can see that, for the purpose of enhancing the spectral efficiency, a scheduler tends to transmit by knowing the other decides to do so. This strategic complementarity is naturally captured by the supermodular game theory \cite{Vives1990,Milgrom1990,Topkis2001}, we propose a two-player game model for the AM problem in NC-TWRC and prove the supermodularity of it. We show that the supermodularity of game guarantees the existence of pure strategy Nash equilibria (PSNEs) and the two extremal, the largest and smallest, PSNEs can be determined by a greedy recursive algorithm within a finite number of iterations. Our study also reveals the Pareto order of the extremal PSNEs, where it is shown that the largest and smallest PSNEs are Pareto worst and best PSNEs, respectively. Finally, we derive the sufficient conditions for the extremal PSNEs to be symmetric and monotonic in channel SNRs.

To the best of our knowledge, this is the first time that the AM problem in NC-TWRC is considered and optimized from the perspective of game theory. In this paper, the analysis of the game model and Nash equilibria (NEs) are mainly based on studies on supermodular game theory \cite{Vives1990,Milgrom1990,Topkis2001}. The main results are listed below:

\begin{itemize}
    \item \textit{two-player game formulation}: In order to solve the adaptive $m$-QAM modulation problem in \figref{fig:SYSTEM}, we propose a two-player game model parameterized by the SNRs of two user-to-user channels. The purpose is to determine the joint strategies of schedulers, the bit rates of both $m$-QAM Txs/Rxs, such that the transmission error rate and spectral efficiency in the NC-TWRC are optimized.
    \item \textit{existence and performance of extremal PSNEs}: We prove that PSNEs always exist in this game and the largest and smallest PSNEs can be searched by finite iterations of Cournot tatonnement\cite{Vives2005}. We show that Cournot tatonnement can be implemented online so that the two schedulers gradually update their strategies to and finally reach the extremal PSNEs. We run simulations to show that the extremal PSNEs are superior to the conventional single-agent AM schemes in enhancing the spectral efficiency in the NC-TWRC system.
    \item \textit{Pareto order of extremal PSNEs}: We show that the smallest PSNE Pareto dominates the largest one. Therefore, the smallest PSNE is always preferred to the largest one by both schedulers.
    \item \textit{symmetry and monotonicity of extremal PSNEs}: We derive the sufficient conditions for the extremal PSNEs to be symmetric and nondecreasing in the SNRs of two user-to-user channels. We show that symmetry and monotonicity of PSNEs can be utilized to relieve the computational complexity of the PSNE learning process, e.g., the Cournot tatonnement algorithm.
\end{itemize}

\subsection{Organization}

The rest of the paper is organized as follows. In Section~\ref{sec:System}, we describe the system in \figref{fig:SYSTEM}, clarify the assumptions and state the optimization problem on designing the AM scheme in NC-TWRC. In Section~\ref{sec:SupGMod}, we present a two-player game model formulation for the AM problem in \figref{fig:SYSTEM}. In Section~\ref{sec:ExOrdNE}, we prove the supermodularity of this game, the existence of the extremal PSNEs and their Pareto order. We also run a simulation to show the performance of the extremal PSNEs. In Section~\ref{sec:SymMonoNE}, we study the symmetry and monotonicity of the extremal PSNEs in channel SNRs.

\section{System Model}
\label{sec:System}

In the NC-TWRC system in \figref{fig:SYSTEM}, schedulers 1 and 2 decide the number of bits per QAM symbol of their own $m$-QAM Txs/Rxs. The objective of each scheduler is two-fold: (a) to minimize the BER over the user-to-user channel, and (b) to optimize the spectral efficiency, i.e., maximize the number of bits exchanged between two users. Let $i=\{1,2\}$, we list the assumption in this system as follows.

\begin{assumption} \label{ass:Action}
    Let $a_i\in\A_i=\{0,1,\dotsc,A_m\}$ be the strategy taken by scheduler $i$. $a_i$ denotes the number of bits per QAM symbol except that $a_i=0$ denotes no transmission. A decision that determines the value of $a_i$ is made at each symbol duration by scheduler $i$. 
\end{assumption}

\begin{assumption} \label{ass:AF_PNC}
    We assume that users adopt amplify-and-forward physical layer network coding (AF-PNC) scheme \cite{Nechip2008,Annamal2011,Hwang2011,Louie2010} as follows. In \figref{fig:PNC}, assume that $x_1$ and $x_2$ are two messages sent from user $1$ and user $2$, respectively, in phase I. The received signal at the relay is $z=\sum_{i=1}^{2}\sqrt{P_i}h_ix_i+n_r$. The relay broadcasts an amplified version of $z$ to the end users in phase II. Here, $P_i$ is the symbol power consumed by scheduler $i$ to transmit $x_i$, $h_i$ is the gain of the fading channel from user $i$ to the relay, $n_r$ is the noise at the relay. Scheduler $i$ receives the broadcast signal $y_i=G\sqrt{P_r}h_iz+n_i $ where $P_r$ is the power consumed by the relay to broadcast $z$ and $n_i$ is the noise at user $i$. Since scheduler $i$ knows its own symbol, the self interference term $G\sqrt{P_r}\sqrt{P_i}h_i^{2}x_i$ is subtracted from $y_i$. The remaining signal $$\hat{y}_i=G\sqrt{P_r}\sqrt{P_{-i}}h_ih_{-i}x_{-i}+G\sqrt{P_i}h_in_r+n_i$$ is demodulated. Here, $-i=\{1,2\}\setminus{i}$ denotes the other scheduler/user. Let $\sigma_r^2$, $\sigma_1^2$ and $\sigma_2^2$ be the noise power of $n_r$, $n_1$ and $n_2$, respectively. $G$ is a scaling factor that the relay uses to normalize the average per-symbol power and is given by $G=\sqrt{\frac{1}{P_1|h_1|^2+P_2|h_2|^2+\sigma_r^2}}$ \cite{Nechip2008,Annamal2011,Hwang2011}. Denote $\snr_i$ the SNR of the wireless channel from user $i$ to $-i$ and assume that the user-to-relay channels are reciprocal. The value of $\snr_i$ is \cite{Louie2010}
    \begin{equation}  \label{eq:2TSSNR}
        \snr_{i}=\frac{\frac{P_rP_{i}|h_i|^2|h_{-i}|^2}{\sigma_{-i}^2\sigma_r^2}}{\frac{P_{-i}|h_{-i}|^2}{\sigma_{-i}^2}+\frac{1}{G^2\sigma_r^2}}.
    \end{equation}
\end{assumption}

\begin{assumption} \label{ass:CSI}
    The schedulers and relay transmit/broadcast symbols by unit transmission power, i.e., $P_r=P_1=P_2=1$. Scheduler $i$ obtains the instantaneous values of both $\snr_1$ and $\snr_2$ to assist the decision making. Let $\SNR_i$ be the set that contains all possible values of $\snr_i$ and $\SNR=\SNR_1\times{\SNR_2}$, where $\times$ denotes the Cartesian product. We define an SNR vector $\Snr=(\snr_1,\snr_2)\in\SNR$. Here, $\SNR_i$ could be an infinite or finite set. The game model proposed in this paper can be applied to both cases. But, we just consider the finite case, i.e., the entire SNR range is quantized into finite number of regions so that $\SNR_i$ is a finite set for $i\in\{1,2\}$.
\end{assumption}

It should be noted that there exist channel estimation methods for each scheduler to obtain the instantaneous value of $\snr_1$ and $\snr_2$. For example, by adopting the two-phase training protocol proposed in \cite{Gao2009}, the instantaneous values of $h_1$ and $h_2$ can be estimated. By assuming that both schedulers obtain the values of $\sigma_1^2$, $\sigma_2^2$ and $\sigma_r^2$, $\snr_1$ and $\snr_2$ can be derived by \eqref{eq:2TSSNR}. Since this paper considers game with perfect information, we assume perfect channel estimation. Also, the channel estimation techniques may vary with the network coding (NC) scheme, e.g., if we adopt other NC, the expression of $\snr_{i}$ in \eqref{eq:2TSSNR} would be different. However, it would not affect the main results derived in Sections~\ref{sec:ExOrdNE} and \ref{sec:SymMonoNE}.

\section{Game Formulation}
\label{sec:SupGMod}

As discussed in Section~\ref{sec:mot}, to optimize BER and spectral efficiency in NC-TWRC, a scheduler needs to consider not only the SNR of user-to-user channel but also the other scheduler's decision, i.e., the two users interact with each other via their actions. The AM problem in \figref{fig:SYSTEM} is a multi-agent optimization one. In this section, we propose a two-player game model.

\subsection{Two-player Game model}

Consider modeling the adaptive $m$-QAM modulation problem in \figref{fig:SYSTEM} by a two-player game $\G_{\Snr}=\{N,\SNR,\{\A_i,c_i(\snr_i,\act)\}_{i=1}^{N}\}$.\footnote{The normal form game model is $\G=\{N,\{\A_i,c_i(\act)\}_{i=1}^{N}\}$\cite{Osborne2004}. Here, $\G_{\snr}$ can be considered a normal form game that is parameterized by $\snr$, i.e., for each value of $\snr$, there is a game in normal form. }
In this game,
\begin{itemize}
    \item $N=2$ denotes the number of schedulers/users;
    \item $\act=(a_1,a_2)\in\A$ is a strategy profile, the joint strategy taken by two schedulers, and $\A=\A_1\times\A_2=\{0,1,\dotsc,A_m\}^2$ is the strategy profile set. We also use the notation $\act=(a_i,a_{-i})$ in this paper, where $a_i$  and $a_{-i}$ denotes the strategy of scheduler $i$ and the strategy of the other scheduler, respectively;
    \item $c_i:\SNR_i\times{\A}\mapsto{\R_{+}}$ is the cost function that quantifies the losses associated with transmission error rate and spectrum efficiency. We define $c_i$ as
            \begin{align} \label{eq:Cost}
                c_i(\snr_i,\act)&=c_i(\snr_i,a_i,a_{-i})   \nonumber\\
                              &=wc_e(\snr_i,a_i)+c_t(a_i)+c_r(\act),
            \end{align}
        where $w>0$ is a weight factor. $c_e$, $c_t$ and $c_r$ are defined as follows.
\end{itemize}

Firstly, we define $c_e\colon\SNR_i\times\A_i\mapsto\R_+$ as the cost function that quantifies the loss incurred by the transmission BER.
We consider two expressions of $c_e$. One is
    \begin{equation}  \label{eq:ImmCe1}
        c_e(\snr_i,a_i)=0.2\exp \Big( -\frac{1.5\snr_i}{2^{a_i}-1}  \Big),
    \end{equation}
where $c_e$ denotes the upper BER bound of transmitting $2^{a_i}$-QAM symbol when the SNR of the user-to-user channel is $\snr_i$ \cite{Goldsmith1997,Alouini2000,Foschini1983}. The other is
    \begin{equation} \label{eq:ImmCe2}
        c_e(\snr_i,a_i)=\frac{-\ln(5\PbCostr)(2^{a_{i}}-1)}{1.5\snr_i},
    \end{equation}
where $c_e$ denotes the minimal symbol power required to transmit $2^{a_i}$-QAM symbol with an average BER less than a constraint $\PbCostr\leq{0.2}$. In this paper, we set $\PbCostr=10^{-3}$. Note, defining $c_e$ as in \eqref{eq:ImmCe2} does not necessarily mean that variable symbol power will be consumed to transmit $2^{a_i}$-QAM symbols. As assumed in Assumption~\ref{ass:CSI}, we consider constant transmission power in this paper. Equation~\eqref{eq:ImmCe2} just shows another way of quantifying the transmission error loss. In Section~\ref{sec:SymMonoNE}, we will show the advantage of defining $c_e$ as in \eqref{eq:ImmCe2}: The resulting PSNEs are monotonically increasing in $\Snr$.

Next, we define $c_t\colon\A_i\mapsto\R_+$ as
    $$ c_t(a_i)=\frac{1}{a_{i}+1} $$
so that $c_t$ is inversely proportional to the transmission rate of user $i$. We define $c_r\colon\A_i\mapsto\R_+$ as
    $$ c_r(\act) = \frac{a_{-i}}{a_{i}+1}. $$
Here, $c_r$ models the strategic complementary behaviors of two schedulers when they are trying to enhance the spectral efficiency: In order to minimize $c_r$, if scheduler $-i$ increases $a_{-i}$, scheduler $i$ tends to increase $a_i$. On the other hand, the definition of $c_r$ also promotes equal share of the spectrum of the communication channel between two users. The equilibrium point by considering $c_r$ alone is when users take the same strategy, i.e., $a_i=a_{-i}$.

It can be seen that $c_t(a_i) + c_r(\act) = \frac{a_{-i}+1}{a_{i}+1}$ determines the costs in losing spectral efficiency in NC-TWRC. Therefore, $c_i$ is a weighted-sum cost of transmission error rate and spectral efficiency.\footnote{Since there is naturally a tradeoff between transmission error rate and spectral efficiency, they cannot be optimized at the same time. In the definition of $c_i$ in \eqref{eq:Cost}, we adopted the scalarization approach in the multi-objective optimization \cite{Jaimes2009}: defining the total cost as the weighted-sum of all costs. The purpose is to seek the Pareto optimality between transmission error rate and spectral efficiency. } We show in Section~\ref{sec:comp} that the game model differs from the conventional single-agent AM method in the presence of $c_r$.

\subsection{Nash Equilibria}

Denote a pure strategy $\Po\colon\SNR\mapsto\A$ as $\Po(\Snr)=(\po_1(\Snr),\po_2(\Snr))$ where $\po_i(\Snr)$ determines the strategy of scheduler $i$ by following $\Po$ when the SNR vector $\Snr$ takes certain value. The pure strategy Nash equilibrium (PSNE) is defined as follows.

\begin{definition}[PSNE]
    A pure strategy $\Po^*$ is a Nash equilibrium of $\G_{\Snr}$ if
        $$ c_i(\snr_i,\po_i^*(\Snr),\po_{-i}^*(\Snr)) \leq c_i(\snr_i,\po_i(\Snr),\po_{-i}^*(\Snr)) $$
    for all $i\in\{1,2\}$, $\theta_i$ and $\Snr$ .
\end{definition}

It should be noted that although we show that the schedulers' actions are interacting with each other in Section~\ref{sec:mot}, a PSNE is a function of just $\Snr=(\snr_1,\snr_2)$, i.e., if a PSNE $\Po$ is adopted as the AM scheme in Fig.~\ref{fig:SYSTEM}, each scheduler is only required to know the instantaneous SNRs or two user-to-user channels instead of the other scheduler's strategy.

\subsection{Comparison with Single-agent AM Method} \label{sec:comp}

One can see that if we omit $c_r$ in the cost function $c_i$, the optimization problem reduces to
    \begin{equation} \label{obj:AM}
        \min_{a_i\in\A_i} \{ wc_e(\snr_i,a_i)+c_t(a_i) \}, \quad \forall{i}\in\{1,2\},
    \end{equation}
which is a conventional single-agent AM problem as proposed in \cite{Goldsmith1997,Goldsmith1998,Alouini2000,Pursley2000}. The optimization problem in \eqref{obj:AM} can be solved separately for two terminal users, where the optimal decision of user $i$ is a function of the corresponding user-to-user SNR $\snr_i$. But, this is not the case in the presence of $c_r$ in the cost function $c_i$, since the action should be optimal to not only $\snr_i$ but also $a_{-i}$, the other scheduler's action. This problem cannot be solved separately for the two users. Instead, the interaction of users should be studied in order to seek the optimal or equilibrium point.

It should be pointed out that both the conventional single-agent AM method and the game-theoretical AM approach proposed in this paper require the decision-making at the end users rather than the relay, i.e., in phase II, the relay simply amplifies and forwards whatever it receives from the two end users without any scheduling. The difference lies in what needs to be known to assist the decision-making: In the conventional single-agent AM method, each user only needs to know the SNR of its corresponding user-to-user channel, i.e., user $1$ knows $\snr_1$ and user $2$ knows $\snr_2$; In the game-theoretical AM approach, both users need to know $\snr_1$ and $\snr_2$, the SNRs of two user-to-user channels. However, according to equation~\eqref{eq:2TSSNR}, the values of both $\snr_i$ depend on the estimations of $h_1$ and $h_2$. If we adopt the channel estimation technique proposed in \cite{Gao2009}, the values of $h_1$ and $h_2$ can be estimated at both end users. In this case, the overhead in obtaining the necessary information for adopting an PSNE are the same as in the conventional single-agent AM method.

\section{Existence and Pareto Order of PSNEs}
\label{sec:ExOrdNE}

Pure strategy is always more preferred to randomized one because of its simplicity. But, Narsh equilibrium is not in the form of pure strategy in general. The remaining question is whether one can prove the existence of PSNEs in $\G_{\Snr}$. In this section we show that this question can be easily answered by showing the supermodularity of game $\G_{\Snr}$. The work in this section is based on the studies on supermodular game in \cite{Topkis1979,Milgrom1990,Vives1990}. For the definitions of related concepts, e.g., partially ordered set (poset) and complete lattice, we refer the reader to Appendix~\ref{app:lattice}.

\subsection{Preliminaries}

We first introduce the definitions of super/submodularity and supermodular game (SupG) and some existing results on SupG as follows.

\begin{definition}[Super/submodularity\cite{Topkis2001}] \label{def:SupermodM}
    $f\colon\Z^K\mapsto{\R}$ is supermodular if $f(\x)+f(\y)\leq{f(\x\vee{\y})+f(\x\wedge{\y})}$ for all $\x,\y\in{\Z^K}$. $g\colon\Z^K\mapsto{\R}$ is submodular if $-g$ is supermodular. $\vee$ and $\wedge$ denote componentwise maximization and minimization, respectively.
\end{definition}

\begin{definition}[SupG \cite{Topkis1979,Milgrom1990}] \label{def:SupG}
    $\G_{\Snr}$ is supermodular if $\A_i$ is a complete lattice\footnote{See the definition of complete lattice in Appendix~\ref{app:lattice}.} and $c_i$ is submodular in $\act=(a_1,a_2)$ for all $i$ and $\Snr$.\footnotemark
    \footnotetext{\lq{Supermodular}\rq\ in SupG denotes the supermodularity of the utility function in the game model. Since the cost is the negative of utility, the submodularity of the cost function contributes to the supermodularity of game.}
\end{definition}

Due to Tarski's fixed point theorem \cite{Tarski1955}, PSNEs always exist in SupG \cite{Vives1990}. Besides, a SupG also has the following properties.

\begin{lemma}[Properties of SupG \cite{Topkis1979,Milgrom1990,Vives1990}] \label{lemma:NESupG}
    If $\G_{\Snr}$ is supermodular, the following hold.
    \begin{enumerate}[(a)]
        \item PSNE exists. For each $\Snr$, the set of PSNEs is a complete lattice, i.e., there exist a largest PSNE $\PoU$ and a smallest PSNE $\PoD$.
        \item $\PoU$ and $\PoD$ can be found by \textit{Cournot tatonnement}: Define maximal best response function $\overline{\BR}\colon\SNR\times{\A}\mapsto\A$ with the $i$th tuple being
             $$ \overline{\Br}_i(\Snr,a_{-i})=\max\{\underset{a_i\in\A_i}{\operatorname{argmin}}\ c_i(\snr_i,a_i,a_{-i})\}.$$
            Let $\overline{\Po}^{(0)}(\Snr)=\sup(\A)$ for all $\Snr$.\footnote{$\sup(\A)$ and $\inf(\A)$ denote the componentwise supremum and infimum of $\A$, respectively. Since $\A$ is a finite set, $\sup(\A)$ and $\inf(\A)$ denote the componentwise maximum and minimum of $\A$, respectively.} The sequence $\{\overline{\Po}^{(k)}(\Snr)\}$ generated by iteration
                $$ \overline{\Po}(\Snr) \colonequals \overline{\BR}(\Snr,\overline{\Po}(\Snr)) $$
            converges monotonically downward to $\PoU(\Snr)$ for all $\Snr$. Similarly, Let $\underline{\BR}$ be the minimal best response function and $\underline{\Po}^{(0)}(\Snr)=\inf(\A)$ for all $\Snr$. The sequence $\{\underline{\Po}^{(k)}(\Snr)\}$ generated by iteration
                $$ \underline{\Po}(\Snr) \colonequals \underline{\BR}(\Snr,\underline{\Po}(\Snr)) $$
            converges monotonically upward to $\PoD(\Snr)$ for all $\Snr$.
        \item If $c_i(a_i,a_{-i})$ is nondecreasing in $a_{-i}$ for all $i$, then PSNEs are Pareto ordered such that
                $$ c_i(\snr_i,\PoD(\Snr))\leq c_i(\snr_i,\PoU(\Snr)) $$
              for all $i$ and $\Snr$, i.e., $\PoD$ Pareto dominates $\PoU$. \hfill\IEEEQED
    \end{enumerate}
\end{lemma}

\subsection{Existence and Cournot Tatonnement of Extremal PSNEs}

Based on Lemma~\ref{lemma:NESupG}(a), we prove the existence of extremal PSNEs in $\G_{\Snr}$ as follows.

\begin{theorem}  \label{theo:ExistPSNE}
    $\G_{\Snr}$ is supermodular for all $\Snr$, where PSNEs exist. The largest and smallest PSNEs, $\PoU$ and $\PoD$, can be found by Cournot tatonnement.
\end{theorem}
\begin{IEEEproof}
According to Definition~\ref{def:CompL}, it is straightforward to see that $\A_i=\{0,1,\dotsc,A_m\}$ is a complete lattice. Consider the submodularity of $c_i$ on the strategy profile space. Since
    \begin{align} \label{eq:theo:ExistPSNE}
        &\quad c_i(\snr_i,a_i+1,a_{-i})+c_i(\snr_i,a_i,a_{-i}+1)    \nonumber \\
        &\qquad -c_i(\snr_i,a_i+1,a_{-i}+1)-c_i(\snr_i,a_i,a_{-i})   \nonumber \\
        &=c_r(a_i+1,a_{-i})+c_r(a_i,a_{-i}+1)   \nonumber \\
        &\qquad -c_r(a_i,a_{-i})-c_r(a_i+1,a_{-i}+1)  \nonumber\\
        &=\frac{1}{a_i}-\frac{1}{a_i+1} \geq 0,
    \end{align}
by Definition~\ref{def:SupermodM}, $c_i$ is submodular in $\act=(a_i,a_{-i})$ for all $\Snr$ and $i\in\{1,2\}$.\footnote{In \eqref{eq:theo:ExistPSNE}, we show submodularity for the function defined on 2-dimensional integer space. Based on Definition~\ref{def:SupermodM}, $f:\Z^2\mapsto{\R}$ is submodular, $f(x_1^+,x_2^-)+f(x_1^-,x_2^+)\geq{f(x_1^-,x_2^-)+f(x_1^+,x_2^+)}$ for all $x_1^-,x_1^+,x_2^-,x_2^+\in\Z$ such that $x_1^+ \geq x_1^-$ and $x_2^+ \geq x_2^-$.}
Therefore, according to Definition~\ref{def:SupG}, game $\G_{\Snr}$ is supermodular. By Lemma~\ref{lemma:NESupG}(a) and (b), the largest and smallest PSNEs, $\PoU$ and $\PoD$, exists, and can be found by Cournot tatonnement.
\end{IEEEproof}

\begin{figure}[tbp]
	\centering
		\centerline{\scalebox{0.95}{\input{figures/Cournot.tex}}}
	\caption{The sequences $\{\overline{\po}_1^{(k)}(\Snr)\}$ and $\{\underline{\po}_1^{(k)}(\Snr)\}$ generated by Cournot tatonnement, when $A_m=9$ and the SNR vector (in ratio) is $\Snr=(\snr_1,\snr_2)=(7,8)$. In this case, the sequence $\{\overline{\po}_1^{(k)}(\Snr)\}$ converges to $\poU_1(\Snr)$, the first tuple in the largest PSNE $\PoU(\Snr)$. The sequence $\{\overline{\po}_1^{(k)}(\Snr)\}$ converges to $\poD_1(\Snr)$, the first tuple in the smallest PSNE $\PoD(\Snr)$, at the $3$rd iteration.}
	\label{fig:Cournot}
\end{figure}
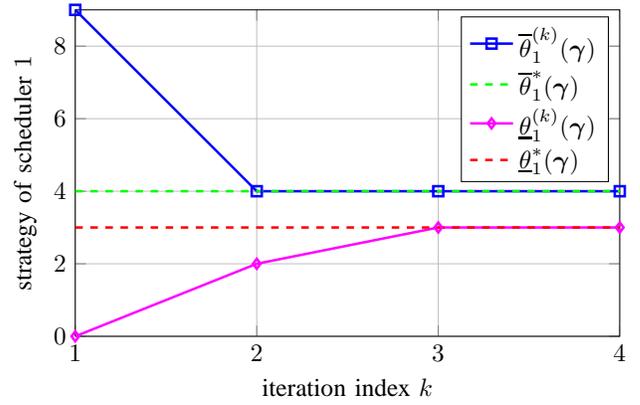


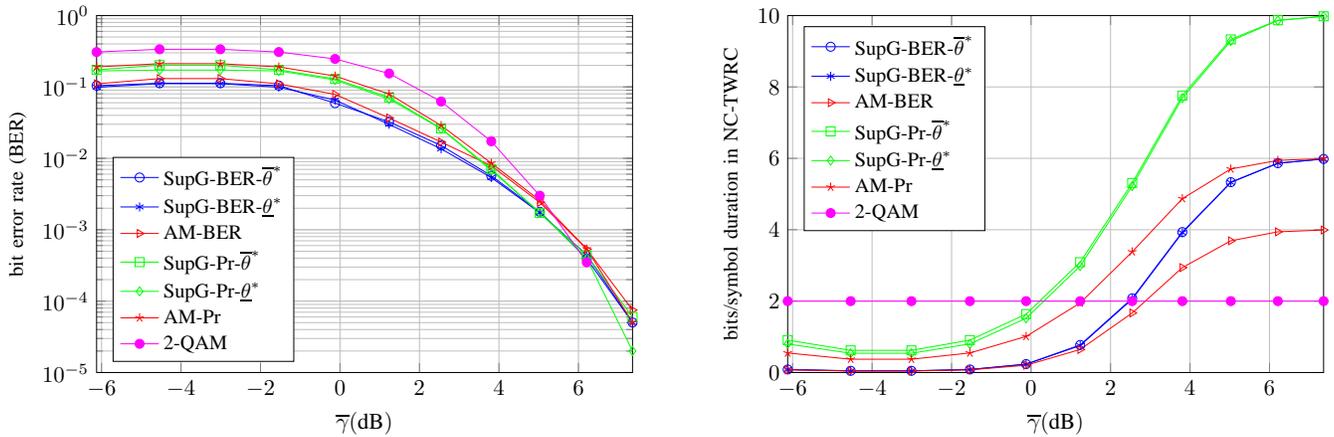
\begin{figure*}[tbp]
	\centering
        \subfigure{\scalebox{0.85}{\input{SIM/BER.tex}}}\qquad\quad
		\subfigure{\scalebox{0.85}{\input{SIM/BitRate.tex}}}
	\caption{BER (left) and spectral efficiency (right) vs. average SNR $\bar{\snr}$ by following the strategies in the experiment in Example~\ref{ex:sim}. The simulation lasts for $10^4$ symbol durations. Spectral efficiency is obtained as the number of bits exchanged at the relay averaged over symbol durations, which also denotes the average number of bits sent by both users. $\bar{\snr}$ denotes the average SNR of both user-to-user channels. We set $A_m=9$. The extremal PSNEs, $\PoU$ and $\PoD$, in the experiment in Example~\ref{ex:sim} are obtained by Cournot tatonnement.}
	\label{fig:BER}
\end{figure*}

\begin{example}
In \figref{fig:Cournot}, we show the examples of the sequences $\{\overline{\po}_1^{(k)}(\Snr)\}$ and $\{\underline{\po}_1^{(k)}(\Snr)\}$. We set $A_m=9$ and the SNR vector (in ratio) as $\Snr=(\snr_1,\snr_2)=(7,8)$. We first start the iteration $\overline{\Po}(\Snr) \colonequals \overline{\BR}(\Snr,\overline{\Po}(\Snr))$ in Cournot tatonnement with $\overline{\Po}^{(0)}=\sup(\A)=(A_m,A_m)=(9,9)$. We plot $\overline{\po}_1^{(k)}(\Snr)$, the first tuple in $\overline{\Po}^{(k)}(\Snr)$, in \figref{fig:Cournot}. It can be seen that the sequence $\{\overline{\po}_1^{(k)}(\Snr)\}$ converges monotonically downward to the largest PSNE $\poU_1(\Snr)$. We then run the iteration $\underline{\Po}(\Snr) \colonequals \underline{\BR}(\Snr,\underline{\Po}(\Snr))$ that starts with $\overline{\Po}^{(0)}=\inf(\A)=(0,0)$. We plot $\underline{\po}_1^{(k)}(\Snr)$ in \figref{fig:Cournot}. It can be seen that the sequence $\{\underline{\po}_1^{(k)}(\Snr)\}$ converges monotonically upward to the smallest PSNE $\{\poD_1(\Snr)\}$.
\end{example}

In fact, \figref{fig:Cournot} presents a method of learning the extremal PSNEs, $\PoU$ and $\PoD$, by Cournot tatonnemen. Assume that $\PoU$ and $\PoD$ are not available to the two users at the beginning. To learn them, or one of them, the only thing to do is to implement the greedy Cournot tatonnemen algorithm. Based on the convergence results in \figref{fig:Cournot}, $\PoU$ and $\PoD$ for each $\Snr$ can be learned within a finite number of iterations. In addition, the users can learn the extremal PSNEs online. Let the two users adopt strategy $\overline{\Po}(\Snr)=\sup(\A)$ initially. In the first symbol duration, let user $i$ take the action $a_i=\max\{{\operatorname{argmin}}_{a_i\in\A_i}\ c_i(\snr_i,a_i,\overline{\theta}_{-i}(\Snr))\}$ based on the channel SNRs $\Snr=(\snr_1,\snr_2)$. Then, at the end of the symbol duration, each user knows the other's action $a_{-i}$ and does the update $\overline{\Po}(\Snr)=(a_i,a_{-i})$.\footnote{The actual implementation may require more signaling information. For example, the user needs to let the other know which action he/she takes for demodulation and/or online implementation of the Cournot tatonnemen algorithm. This problem can be solved by implementing the AM method at a packet level instead of a symbol level so that the overhead information can be included in the packet header. Whether the AM is done on packet or symbol level does not affect the main results in this paper. } In the second symbol duration, repeat the same procedure and so on. By doing so, each user can adapt the strategy $\overline{\Po}(\Snr)$ towards the largest PSNE $\PoU$ for each $\Snr$ gradually. The smallest PSNE can be learned online in the same way.

It should be pointed out that there exist PSNEs other than the two extremal ones. But, we are just interested in $\PoU$ and $\PoD$ since they are the ones that can be found directly by the existing Cournot tatonnemen algorithm. Also, in Section~\ref{sec:ParetoOrder}, the Pareto order of the PSNEs shows that the smallest PSNE $\PoD$ is the best AM scheme since it Pareto dominates all other PSNEs. Therefore, in reality $\PoD$ may be the only one that is of interest.

\subsection{Simulation Results}

To show the performance of the extremal PSNEs, $\PoU$ and $\PoD$, we do the following experiment.
\begin{example} \label{ex:sim}
We run a simulation that lasts $10^4$ symbol durations, where $h_1$ and $h_2$, the gains of two user-to-relay channels, are both Rayleigh distributed. Let $\bar{\snr}_i$ denote the average channel SNR from user $i$ to user $-i$. We set $\bar{\snr}_1=\bar{\snr}_2=\bar{\snr}$ and vary the value of $\bar{\snr}$ from $-6\si{\decibel}$ to $7\si{\decibel}$. Let $\snr_{i\text{min}}$ and $\snr_{i\text{max}}$ denote the maximum and minimum values of $\snr_i$ (in ratio), respectively. We choose $100$ SNR levels in $[\snr_{i\text{min}}, \snr_{i\text{max}}]$ with step size $\frac{\snr_{i\text{max}}-\snr_{i\text{min}}}{100}$ to constitute the finite set $\SNR_i$ for all $i\in\{1,2\}$. The strategies used in simulation are listed below.

\begin{itemize}
    \item We consider two sets of PSNEs based on SupG model with $A_m=9$. The first group contains:
        \begin{itemize}
            \item SupG-BER-$\PoU$ and SupG-BER-$\PoD$: They are the largest and smallest PSNEs when we adopt the SupG model with the cost function $c_e$ being $c_e(\snr_i,a_i)=0.2\exp \big( -\frac{1.5\snr_i}{2^{a_i}-1}\big)$ and weight factor $w$ being $50$.
            \item SupG-Pr-$\PoU$ and SupG-Pr-$\PoD$: They are the largest and smallest PSNEs when we adopt the SupG model with the cost function $c_e$ being $c_e(\snr_i,a_i)=\frac{-\ln(5\PbCostr)(2^{a_{i}}-1)}{1.5\snr_i}$ and weight factor $w$ being $0.05$.
        \end{itemize}
    \item We also consider the conventional single-agent AM methods corresponding to the two sets of PSNEs above.
        \begin{itemize}
            \item AM-BER: In this strategy, user $i$ adopts the strategy $a_i^*(\snr_i)=\arg\min_{a_i}\{wc_e(\snr_i,a_i)+c_t(a_i)\}$ where $c_e(\snr_i,a_i)=0.2\exp \big( -\frac{1.5\snr_i}{2^{a_i}-1}\big)$ and weight factor $w=50$. This is the single-agent AM method that corresponds to SupG-BER-$\PoU$ and SupG-BER-$\PoD$.
            \item AM-Pr: In this strategy, user $i$ adopts the strategy $a_i^*(\snr_i)=\arg\min_{a_i}\{wc_e(\snr_i,a_i)+c_t(a_i)\}$ where $c_e(\snr_i,a_i)=\frac{-\ln(5\PbCostr)(2^{a_{i}}-1)}{1.5\snr_i}$ and weight factor $w=0.05$. This is the single-agent AM method that corresponds to SupG-Pr-$\PoU$ and SupG-Pr-$\PoD$.
        \end{itemize}
    \item We finally try a fixed rate strategy, $2$-QAM (BPSK), i.e., the schedulers always transmit BPSK symbols whatever the channel SNRs $\Snr=(\snr_1,\snr_2)$ are.
\end{itemize}
\end{example}

The simulation results are shown in \figref{fig:BER}. We present the transmission error rate by the BER in the entire NC-TWRC system: the total number of erroneous bits divided by the total number of bits sent by two schedulers. It can be seen that the extremal PSNEs always incur slightly lower transmission error rates than the corresponding conventional single-agent AM methods. We also present the spectral efficiency as the number of bits broadcast by the relay averaged over symbol durations. It can be seen that the extremal PSNEs incur higher spectral efficiencies than the corresponding conventional single-agent AM methods. Both extremal PSNEs and conventional AM schemes offer more flexible transmission control methods than the fixed rate strategy. But, when the user-to-user channel SNR is above $2\si{\decibel}$, the spectral efficiency in NC-TWRC is significantly improved by following the extremal PSNEs of two-player SupG formulation. For example, when $\bar{\snr}=6\si{\decibel}$, following SupG-Pr-$\PoU$ or SupG-Pr-$\PoD$ allows the relay to exchange $4$ more bits on average than AM-Pr.

\begin{figure}[tbp]
	\centering
		\centerline{\scalebox{0.85}{\input{SIM/TransRate_R.tex}}}
	\caption{The broadcast rate, the number of broadcasts averaged over symbol durations, at the relay vs. average SNR $\bar{\snr}$ by following the strategies listed in the experiment in Example~\ref{ex:sim}. The simulation and system settings are the same as in \figref{fig:BER}}
	\label{fig:BitRateR}
\end{figure}
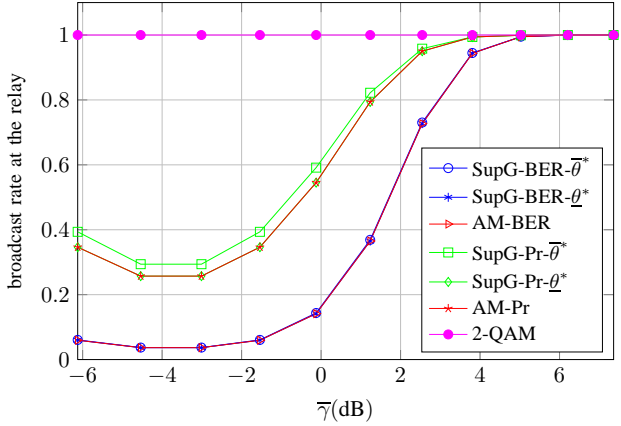

\begin{figure*}[tbp]
	\centering
		\subfigure{\scalebox{0.85}{\input{SIM/MeanCost1.tex}}} \qquad \qquad
        \subfigure{\scalebox{0.85}{\input{SIM/MeanCost2.tex}}}
	\caption{$\bar{c}_1$ and $\bar{c}_2$, the average costs incurred at scheduler $1$ and $2$, respectively, by following the extremal PSNEs in the experiment in Example~\ref{ex:sim}. $\bar{c}_i$ is the costs actually incurred at scheduler $i$ in the simulation averaged over symbol durations. The simulation and system settings are the same as in \figref{fig:BER}. According to Theorem~\ref{theo:OrderPSNE}, the costs incurred by $\PoD$ are always lower than those incurred by $\PoU$. }
	\label{fig:MeanCost}
\end{figure*}

In \figref{fig:BitRateR}, we also show the broadcast rate at the relay in the simulation. The broadcast rate is obtained as the number of broadcasts at the relay averaged over symbol durations. Since the transmission powers are all constant in our system (Assumption~\ref{ass:CSI}), the broadcast rate shown in \figref{fig:BitRateR} also indicates the average power consumption at the relay. It can be seen that the adaptive transmission methods, the conventional single-agent AM scheme and extremal PSNEs based on SupG model, can adjust the transmission rate accordingly to channel quality, i.e., the adaptive transmission methods become more conservative when the channel SNR reduces. It also shows that the smallest PSNEs, SupG-BER-$\PoD$ and SupG-Pr-$\PoD$, incur approximately the same broadcast rate at the relay as the corresponding conventional single-agent AM methods, AM-BER and AM-Pr, respectively. Combined with the results in \figref{fig:BER}, we can see that by following the smallest PSNEs there are more bits delivered at the relay in each broadcast. It means by consuming the same amount of transmission power as in the conventional single-agent AM scheme, we can transmit more symbols in NC-TWRC by following the smallest PSNEs.


\subsection{Pareto Order of Extremal PSNEs} \label{sec:ParetoOrder}

Based on Lemma~\ref{lemma:NESupG}(c), we prove the Pareto order of extremal PSNEs in $\G_{\Snr}$ as follows.

\begin{theorem} \label{theo:OrderPSNE}
     In $\G_{\Snr}$, $\PoD$ Pareto dominates $\PoU$, i.e.,
        $$ c_i(\snr_i,\PoD(\Snr)) \leq c_i(\snr_i,\PoU(\Snr)),$$
    for all $\Snr$ and $i\in\{1,2\}$.
\end{theorem}
\begin{IEEEproof}
It can be seen that $c_r(a_i,a_{-i})=\frac{a_{-i}}{a_{i}+1}$ is nondecreasing in $a_{-i}$. So, $c_i(\Snr,a_i,a_{-i})$ is nondecreasing in $a_{-i}$. According to Lemma~\ref{lemma:NESupG}(c), Theorem~\ref{theo:OrderPSNE} holds.
\end{IEEEproof}

Theorem~\ref{theo:OrderPSNE} in fact describes the preference order of the extremal PSNEs. Since for both schedulers, $\PoD$ incurs less cost than $\PoU$, they always prefer $\PoD$ to $\PoU$, i.e., in reality, both of the schedulers will adopt $\PoD$. In fact, $\PoD$ Pareto dominates not only $\PoU$ but also all other PSNEs. Since $\PoD$ is the smallest PSNE of all PSNEs, if $c_i$ is nondecreasing in $a_{-i}$ in SupG $\G_{\Snr}$, $\PoD$ Pareto dominates all other PSNEs including $\PoU$, i.e., $\PoD$ is the most preferred equilibrium strategy among all PSNEs \cite{Vives1990}.

\begin{example}
To show an example of Theorem~\ref{theo:OrderPSNE}, we also record the mean costs incurred by following four extremal PSNEs adopted in the experiment in Example~\ref{ex:sim}. The results are shown in \figref{fig:MeanCost}, where we can see that the average costs incurred by the smallest PSNEs are always lower than those incurred by the largest ones.
\end{example}

\section{Symmetry and  Monotonicity of Nash Equilibria}
\label{sec:SymMonoNE}

In addition to the well-known results on SupG, we study two more properties of PSNEs in $\G_{\Snr}$: the symmetry and the monotonicity of $\PoU$ and $\PoD$ in SNR vector $\Snr=(\snr_1,\snr_2)$. We discuss how to use these two properties to simplify the PSNE searching process.

If we assume homogenous players: The schedulers not only adopt the same strategy set, i.e., $\A_i=\A_{-i}$,\footnote{As stated in Assumption~\ref{ass:Action}, $\A_1=\A_2=\{0,1,\dotsc,A_m\}$.} but also use the same SNR set, i.e., $\SNR_i=\SNR_{-i}$,\footnote{$\SNR_i=\SNR_{-i}$ means: If $\SNR_1$ and $\SNR_2$ are infinite, the SNR variation range of two user-to-user channels are the same; If $\SNR_1$ and $\SNR_2$ are finite, the quantization levels of the SNR variation range of two user-to-user channels are the same.} we can show that the extremal PSNEs are symmetric in $\Snr$.
\begin{theorem}[Symmetry of PSNEs] \label{theo:SymmetrPSNE}
    In $\G_{\Snr}$, if $\SNR_i=\SNR_{-i}$, $\PoU$ and $\PoD$ are symmetric in $\Snr$, i.e.,
      \begin{equation}
        \begin{aligned}
            \poU_i(\snr_i,\snr_{-i})&=\poU_{-i}(\snr_{-i},\snr_i),  \\
            \poD_i(\snr_i,\snr_{-i})&=\poD_{-i}(\snr_{-i},\snr_i),
        \end{aligned} \nonumber
      \end{equation}
    for all $i\in\{1,2\}$.
\end{theorem}
\begin{IEEEproof}
Since $\SNR_i=\SNR_{-i}$, the cost function is indifferent to the identity of player, i.e., $c_i(\snr_i,a_i,a_{-i})=c_{-i}(\snr_{-i},a_{-i},a_{i})$. Then, $\overline{\BR}$ is also identity-indifferent, i.e.,
    $$ \overline{\Br}_i(\snr_i,\snr_{-i},a_{-i})=\overline{\Br}_{-i}(\snr_{-i},\snr_i,a_i). $$
Since $\PoU(\Snr)$ is a fixed point of the maximal best response function $\overline{\BR}$, we have $\PoU(\Snr)=\overline{\BR}(\Snr,\PoU(\Snr))$. Then,
    \begin{equation}  \label{eq:app:SymL}
        \begin{aligned}
            \poU_i(\snr_i,\snr_{-i})&=\overline{\Br}_i(\snr_i,\snr_{-i},\poU_{-i}(\snr_i,\snr_{-i}))  \\
                                    &=\overline{\Br}_{-i}(\snr_{-i},\snr_i,\poU_i(\snr_{-i},\snr_i)) \\
                                    &=\poU_{-i}(\snr_{-i},\snr_i),
        \end{aligned}
    \end{equation}
In the same way, we can show that the minimal best response function $\underline{\BR}$ is also identity-indifferent and prove $\poD_i(\snr_i,\snr_{-i})=\poD_{-i}(\snr_{-i},\snr_i)$.
\end{IEEEproof}

Based on Theorem~\ref{theo:SymmetrPSNE}, if we know the values of $\PoU$ and $\PoD$ at some $\Snr=(\snr_1,\snr_2)$, then we automatically know the values of $\PoU$ and $\PoD$ at $\Snr=(\snr_2,\snr_1)$. This property can be used to reduce the computational complexity in learning $\PoU$ and $\PoD$ in real applications. For example, by implementing the Cournot tatonnement, we find that $\PoU(5,7)=(1,4)$, then we can directly assign $\PoU(7,5)=(4,1)$ without running Cournot tatonnement again for $\Snr=(7,5)$. By doing so, we just need to know the values of $\PoU$ and $\PoD$ in half of the space of $\SNR$. So, the complexity of Cournot tatonnement is halved.

As explained in Section~\ref{sec:SupGMod}, a PSNE is a function of $\Snr=(\snr_1,\snr_2)$, i.e., the PSNE assigns a strategy for each value of SNR vector. It is then worth discussing how the extremal PSNEs vary with $\Snr$, or whether there is some regularity in the variations of $\PoU$ and $\PoD$. We study the monotonicity of the $\PoU$ and $\PoD$ in $\Snr$ in the following context. We first introduce a result in \textit{monotonic comparative statics}\footnotemark\ as follows.
\footnotetext{In economics, monotone comparative statics denotes the situation that the optimal solution varies monotonically with the system parameters \cite{Milgrom1994}.}

\begin{lemma}[Monotonic PSNEs in SupG \cite{Milgrom1990}] \label{lemma:MonoNE}
    In SupG $\G_{\Snr}$, if $\SNR$ is a poset\footnote{See the definition of poset in Appendix~\ref{app:lattice}} and the cost function $c_i$ is submodular in $(\snr_i,a_i)$ for all $a_{-i}$, the largest and smallest PSNEs, $\PoU$ and $\PoD$, are nondecreasing in $\Snr$.
\end{lemma}

Based on Lemma~\ref{lemma:MonoNE}, we can derive immediately the sufficient condition for the monotonicity of $\PoU$ and $\PoD$ in $\G_{\Snr}$.

\begin{theorem}[Monotonicity of PSNEs in $\Snr$]  \label{theo:MonoPSNE}
    If $c_e$ is submodular in $(\snr_i,a_i)$, $\PoU$ and $\PoD$ are nondecreasing in $\Snr$.
\end{theorem}
\begin{IEEEproof}
    $\SNR=\SNR_1\times{\SNR_2}$ is a poset according to the definition in Appendix~\ref{app:lattice}, and $c_i$ is submodular in $(\snr_i,a_i)$ for all $i$ if $c_e$ is submodular in $(\snr_i,a_i)$. Therefore, Lemma~\ref{lemma:MonoNE} holds for game $\G_{\Snr}$ if $c_e$ is submodular in $(\snr_i,a_i)$. Therefore, $\PoU$ and $\PoD$ are nondecreasing in $\Snr$.
\end{IEEEproof}

The following corollary shows one way to make Theorem~\ref{theo:MonoPSNE} hold by choosing a proper $c_e$ function.

\begin{corollary}  \label{cor:SubCe}
    If $c_e(\snr_i,a_i)=\frac{-\ln(5\PbCostr)(2^{a_{i}}-1)}{1.5\snr_i^{\phantom{g}}}$, Theorem~\ref{theo:MonoPSNE} holds.
\end{corollary}
\begin{IEEEproof}
    When $c_e(\snr_i,a_i)=\frac{-\ln(5\PbCostr)(2^{a_{i}}-1)}{1.5\snr_i^{\phantom{g}}}$, we have
    \begin{equation}
        \begin{aligned}
            &\quad c_e(\snr_i^+,a_i)+ c_e(\snr_i^-,a_i+1)-c_e(\snr_i^-,a_i) \\
            & \qquad -c_e(\snr_i^+,a_i+1)    \\
            &=-\frac{\ln(5\PbCostr)}{1.5} \Big( \frac{2^{a_{i}+1}-1}{\snr_i^{-}}+\frac{2^{a_{i}}-1}{\snr_i^{+}}-\frac{2^{a_{i}}-1}{\snr_i^{-}} \\
            &\qquad -\frac{2^{a_{i}+1}-1}{\snr_i^{+}}  \Big)    \\
            &=-\frac{\ln(5\PbCostr)}{1.5} \Big( \frac{2^{a_i}}{\snr_i^-}-\frac{2^{a_i}}{\snr_i^+} \Big) \geq 0
        \end{aligned} \nonumber
    \end{equation}
   for all $\snr_i^+\geq{\snr_i^-}$. By Definition~\ref{def:SupermodM}, $c_e$ is submodular. Therefore, Theorem~\ref{theo:MonoPSNE} holds.
\end{IEEEproof}

\begin{figure*}[tbp]
	\centering
        \subfigure{\scalebox{1}{\input{Policy/MonoH1.tex}}} \qquad
        \subfigure{\scalebox{1}{\input{Policy/MonoH2.tex}}}
	\caption{$\poU_1$ and $\poU_2$, the equilibrium strategies of schedulers $1$ and $2$, respectively, in the largest PSNE $\PoU$ of game $\G_{\Snr}$. In this game model, we use $c_e(\snr_i,a_i)=\frac{-\ln(5\PbCostr)(2^{a_{i}}-1)}{1.5\snr_i}$ and set $w=0.05$. According to Corollary~\ref{cor:SubCe}, Theorem~\ref{theo:MonoPSNE} holds. Both $\poU_1$ and $\poU_2$ are nondecreasing in $\Snr=(\snr_1,\snr_2)$. Also, since $\A_1=\A_2=\{0,1,\dotsc,9\}$ and $\SNR_1=\SNR_2=\{0.1,1,2,\dotsc,10\}$, due to Theorem~\ref{theo:SymmetrPSNE}, $\poU_i$ is symmetric in $\Snr$, i.e., $\poU_1(\snr_1,\snr_2)=\poU_2(\snr_2,\snr_1)$. }
	\label{fig:MonoH}
\end{figure*}

The monotonicity of $\PoU$ and $\PoD$ can also be used to relieve the complexity of the PSNE learning process. For example, assume Theorem~\ref{theo:MonoPSNE} holds and, by implementing the Cournot tatonnement algorithm, we find that $\PoU(6,7)=(4,6)$ and $\PoD(6,7)=(2,3)$. Due to the monotonicity of $\PoU$ and $\PoD$ in $\Snr$, we know that $\PoU(5,7)\leq(4,6)$ and $\PoD(6,8)\geq(2,3)$. So, when using Cournot tatonnement to find the value of $\PoU$ for $\Snr=(5,7)$, we can start with $\overline{\Po}^{(0)}(5,7)=(4,6)$  instead of $\overline{\Po}^{(0)}(5,7)=\sup(\A)$; when using Cournot tatonnement to find the value of $\PoD$ for $\Snr=(6,8)$, we can start with $\underline{\Po}^{(0)}(6,8)=(2,3)$  instead of $\underline{\Po}^{(0)}(6,8)=\inf(\A)$. It can be shown that a proper implementation of this method will have the size of PSNE searching space decreasing gradually in $\snr$ and therefore the complexity is reduced accordingly. There also exist many other algorithms concerning how to utilize the monotonicity of PSNEs to relive the computational complexity. For example, the idea based on the works in \cite{Krishn2009,Ding2013Mono} is that we just need to know the turning points in the optimal monotonic strategy instead of learning the overview of it, which sometimes can be accomplished by a multivariate optimization algorithm. We do not extend further in this aspect since it is beyond the scope our work and could be one of the research directions in the future.

\begin{figure}[tbp]
	\centering
        \scalebox{1}{\input{Policy/MonoL1.tex}}
	\caption{$\poU_1$, equilibrium strategy of scheduler $1$ in the largest PSNE $\PoU$ of game $\G_{\Snr}$. In this game model, we use $c_e(\snr_i,a_i)=\frac{-\ln(5\PbCostr)(2^{a_{i}}-1)}{1.5\snr_i}$ and set $w=0.05$. According to Corollary~\ref{cor:SubCe}, Theorem~\ref{theo:MonoPSNE} holds. $\poU_1$ is nondecreasing in $\Snr=(\snr_1,\snr_2)$.}
	\label{fig:MonoL}
\end{figure}
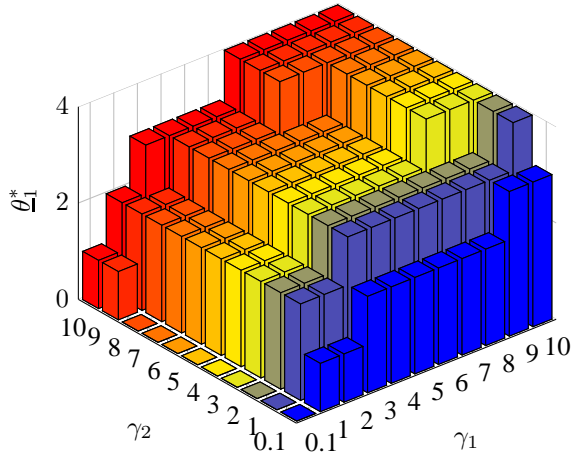

\begin{figure}[tbp]
	\centering
        \scalebox{1}{\input{Policy/NonMono1.tex}}
	\caption{$\poU_1$, equilibrium strategy of scheduler $1$ in the largest PSNE $\PoU$ of game $\G_{\Snr}$. In this game model, we use $c_e(\snr_i,a_i)=0.2\exp \big( -\frac{1.5\snr_i}{2^{a_i}-1}  \big)$ and set $w=50$. Theorem~\ref{theo:MonoPSNE} no longer holds. In this case, $\poU_1$ is not monotonic in $\Snr=(\snr_1,\snr_2)$.}
	\label{fig:NonMono}
\end{figure}
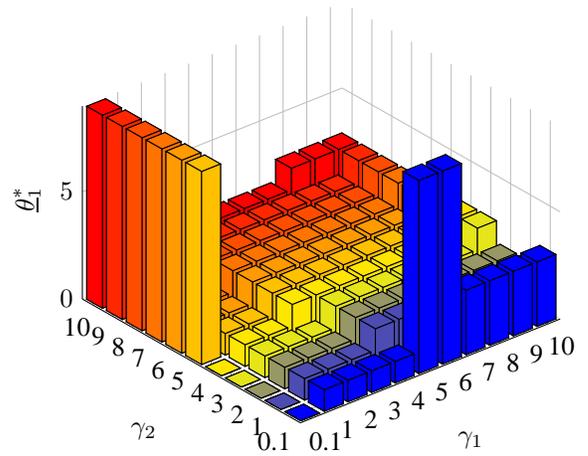

We show examples of Theorem~\ref{theo:SymmetrPSNE} and Theorem~\ref{theo:MonoPSNE} in Figs.~\ref{fig:MonoH}-\ref{fig:NonMono}, where we set $A_m=9$ and the SNR sets (in ratio) as $\SNR_1=\SNR_2=\{0.1,1,2,\dotsc,10\}$.\footnote{Since SNR value is greater than $0$, $\SNR_i$ starts from $0.1$, instead of $0$.} \figref{fig:MonoH} shows the largest PSNE $\PoU$ of $\G_{\Snr}$. In the game model, we use $c_e(\snr_i,a_i)=\frac{-\ln(5\PbCostr)(2^{a_{i}}-1)}{1.5\snr_i}$ and set $w=0.05$. The symmetry and monotonicity of $\poU_1$ and $\poU_2$ in $\Snr$ can be clearly seen from \figref{fig:MonoH}. The monotonicity of the smallest PSNEs $\PoD$ is shown in \figref{fig:MonoL}. Since $\PoD$ is also symmetric in $\Snr$, we just show one tuple of $\PoD$. We then change the $c_e$ in the game model to $c_e(\snr_i,a_i)=0.2\exp \Big( -\frac{1.5\snr_i}{2^{a_i}-1}  \Big)$ and set $w=50$. The purpose of changing the expression of $c_e$ is to breach the condition, the submodularity of $c_e$, so that Theorem~\ref{theo:MonoPSNE} no longer holds. We show $\poU_1$, the first tuple of $\PoU$, in \figref{fig:NonMono}. It can be seen that $\poU_1$ is not nondecreasing in $\Snr$.

\section{Conclusion and Future Work}

In this paper, we proposed a two-player game model for the adaptive $m$-QAM modulation problem in a NC-TWRC where a two-phase AF-PNC scheme was adopted. We proved that this game was supermodular so that the largest and smallest PSNEs always existed. We showed that extremal PSNEs were superior to the conventional AM scheme in enhancing spectral efficiency while providing slightly better BERs. We proved that the smallest PSNE always Pareto dominated the largest one and derived the sufficient conditions for the largest and smallest PSNEs to be symmetric and monotonic in user-to-user channel SNRs.

As part of the conclusion, we point out to two directions of the research work in the future. One is the equilibria searching problem. The Cournot tatonnement presented in this paper finds the PSNEs for each value of $\Snr$. The complexity would be really high if the cardinality of SNR set $\SNR$ is large (an extremal case is when $\SNR$ is infinite set, e.g., a continuum in the real number set). To propose low complexity algorithms, the results in Section~\ref{sec:SymMonoNE} can be utilized. The other is to consider the case of imperfect channel estimation cases, which is a more realistic assumption. In this case, one can model the game as a Bayesian game and study its supermodularity and show the existence and properties of Bayesian equilibria.

\appendices

\section{}
\label{app:lattice}

A set $(\Lat,\geq)$ is a \textit{poset},  partially ordered set, if the binary relation $\geq$ is reflexive, antisymmetric and transitive. If $\Lat$ contains $M$-tuple numerical elements, $\geq$ is componentwise greater than or equal to, i.e., let $\x=(x_1,\dotsc,x_M)\in\Lat$, $\x \geq \y$ if and only if $x_m \geq y_m$ for all $m\in\{1,\dotsc,M\}$.

\begin{definition}[Lattice\cite{Topkis2001}] \label{def:poset}
    A poset $\Lat$ is a lattice if $\x_1\vee{\x_2}\in{\Lat}$ and $\x_1\wedge{\x_2}\in{\Lat}$ for all $\x_1,\x_2\in{\Lat}$, where $\vee$ and $\wedge$ denotes componentwise maximization and minimization.
\end{definition}

\begin{definition}[Complete Lattice\cite{Topkis2001}]  \label{def:CompL}
    A lattic $\Lat$ is complete if $\sup(\tilde\Lat)\in{\Lat}$ and $\inf(\tilde\Lat)\in{\Lat}$ for all nonempty $\tilde{\Lat}\subseteq{\Lat}$. $\sup(\tilde\Lat)$ and $\inf(\tilde\Lat)$ denote the componentwise supremum and infimum of $\tilde\Lat$, respectively.
\end{definition}

\bibliographystyle{ieeetr}
\bibliography{SupGpaper_bib}

\end{document}

%% file: figures/AM.tex
\begin{tikzpicture}

\node at (0.7,1.2) {\small scheduler};
\draw (-0.2,1.4) rectangle (1.6,1);
\draw [->,densely dotted] (0.7,1)--(0.7,0.35);

\draw (0,0.35) rectangle (1.4,-0.35);
\node at (0.7,0.14) {\small $m$-QAM};
\node at (0.7,-0.15) {\small Tx/Rx};

\draw [->] (1.4,0)--(2.7,0);
\draw [dashed] (2.7,0)--(6.5,0);
\node at (4.7,-0.3) {\scriptsize wireless fading channel};

\node at (0.7,-0.7) [color=purple] {\scriptsize \textit{user} };

\end{tikzpicture}

%% file: figures/PNC.tex
\begin{tikzpicture}

\draw (0,0.25) rectangle (1.3,-0.25);
\node at (0.6,0) {\small user $1$};

\draw [dashed] (1.3,0)--(3.5,0);
\draw [->] (1.6,0.2)--(3.3,0.2);
\node at (2.435,0.35) {\small $x_1$};

\draw (4,0) circle (0.5);
\node at (4,0) {R};

\draw [dashed] (4.5,0)--(6.7,0);
\draw [<-] (4.7,0.2)--(6.4,0.2);
\node at (5.535,0.35) {\small $x_2$};

\draw (6.7,0.25) rectangle (8,-0.25);
\node at (7.3,0) {\small user $2$};

\node at (4,-0.8) [color=purple] {\scriptsize \textit{phase I: multiple access (MAC)}};

\draw (0,-1.75) rectangle (1.3,-2.25);
\node at (0.6,-2) {\small user $1$};

\draw [dashed] (1.3,-2)--(3.5,-2);
\draw [<-] (1.6,-1.8)--(3.3,-1.8);
\node at (2.435,-1.65) {\small $z$};

\draw (4,-2) circle (0.5);
\node at (4,-2) {R};

\draw [dashed] (4.5,-2)--(6.7,-2);
\draw [->] (4.7,-1.8)--(6.4,-1.8);
\node at (5.535,-1.65) {\small $z$};

\draw (6.7,-1.75) rectangle (8,-2.25);
\node at (7.3,-2) {\small user $2$};

\node at (4,-2.8) [color=purple] {\scriptsize \textit{phase II: amplify and forward (AF)}};

\end{tikzpicture}

%% file: figures/SYSTEM.tex
\begin{tikzpicture}

\node at (0.7,1.2) {\small scheduler $1$};
\draw (-0.2,1.4) rectangle (1.6,1);
\draw [->,densely dotted] (0.7,1)--(0.7,0.35);

\draw (0,0.35) rectangle (1.4,-0.35);
\node at (0.7,0.14) {\small $m$-QAM};
\node at (0.7,-0.15) {\small Tx/Rx};

\draw [dashed] (1.4,0)--(3.5,0);

\draw (4,0) circle (0.5);
\node at (4,0) {R};

\draw [dashed] (4.5,0)--(6.6,0);

\node at (7.3,1.2) {\small scheduler $2$};
\draw (6.4,1.4) rectangle (8.2,1);
\draw [->,densely dotted] (7.3,1)--(7.3,0.35);

\draw (6.6,0.35) rectangle (8,-0.35);
\node at (7.3,0.14) {\small $m$-QAM};
\node at (7.3,-0.14) {\small Tx/Rx};

\node at (0.7,-0.8) [color=purple] {\scriptsize \textit{user $1$}};
\node at (7.3,-0.8) [color=purple] {\scriptsize \textit{user $2$}};

\end{tikzpicture}

%% file: figures/Cournot.tex
%
%
%
\definecolor{mycolor1}{rgb}{1,0,1}%
\begin{tikzpicture}

\begin{axis}[%
width=3in,
height=1.8in,
scale only axis,
xmin=1,
xmax=4,
xlabel={iteration index $k$},
xtick={1,2,3,4},
ymin=0,
ymax=9,
ylabel={strategy of scheduler $1$},
grid = both,
legend style={draw=black,fill=white,legend cell align=left}
]
\addplot [
color=blue,
solid,
line width=1.0pt,
mark=square,
mark options={solid}
]
table[row sep=crcr]{
1 9\\
2 4\\
3 4\\
4 4\\
};
\addlegendentry{ $\overline{\po}_1^{(k)}(\Snr)$};

\addplot [
color=green,
dashed,
line width=1.0pt
]
table[row sep=crcr]{
1 4\\
2 4\\
3 4\\
4 4\\
};
\addlegendentry{ $\overline{\po}_1^{*}(\Snr)$};

\addplot [
color=mycolor1,
solid,
line width=1.0pt,
mark=diamond,
mark options={solid}
]
table[row sep=crcr]{
1 0\\
2 2\\
3 3\\
4 3\\
};
\addlegendentry{ $\underline{\po}_1^{(k)}(\Snr)$};

\addplot [
color=red,
dashed,
line width=1.0pt
]
table[row sep=crcr]{
1 3\\
2 3\\
3 3\\
4 3\\
};
\addlegendentry{ $\underline{\po}_1^{*}(\Snr)$};

\end{axis}
\end{tikzpicture}%

%% file: SIM/BER.tex
%
%
%
\definecolor{mycolor1}{rgb}{1.00000,0.00000,1.00000}%
\begin{tikzpicture}

\begin{axis}[%
width=3.3in,
height=2.2in,
scale only axis,
xmin=-6.1244260279434,
xmax=7.3610796585662,
xlabel={$\overline{\gamma} (\si{\decibel})$},
ymode=log,
ymin=1e-05,
ymax=1,
yminorticks=true,
ylabel={\small bit error rate (BER)},
grid = both,
legend style={at={(0.03,0.03)},anchor=south west,draw=black,fill=white,legend cell align=left}
]
\addplot [color=blue,solid,mark=o,mark options={solid}]
  table[row sep=crcr]{SIM/BER-1.tsv};
\addlegendentry{\small SupG-BER-$\overline{\mathbf{\theta}}^*$};

\addplot [color=blue,solid,mark=asterisk,mark options={solid}]
  table[row sep=crcr]{SIM/BER-2.tsv};
\addlegendentry{\small SupG-BER-$\underline{\mathbf{\theta}}^*$};

\addplot [color=red,solid,mark=triangle,mark options={solid,,rotate=270}]
  table[row sep=crcr]{SIM/BER-3.tsv};
\addlegendentry{\small AM-BER};

\addplot [color=green,solid,mark=square,mark options={solid}]
  table[row sep=crcr]{SIM/BER-4.tsv};
\addlegendentry{\small SupG-Pr-$\overline{\mathbf{\theta}}^*$};

\addplot [color=green,solid,mark=diamond,mark options={solid}]
  table[row sep=crcr]{SIM/BER-5.tsv};
\addlegendentry{\small SupG-Pr-$\underline{\mathbf{\theta}}^*$};

\addplot [color=red,solid,mark=star,mark options={solid}]
  table[row sep=crcr]{SIM/BER-6.tsv};
\addlegendentry{\small AM-Pr};

\addplot [color=mycolor1,solid,mark=*,mark options={solid}]
  table[row sep=crcr]{SIM/BER-7.tsv};
\addlegendentry{\small 2-QAM};

\end{axis}
\end{tikzpicture}%

%% file: SIM/BitRate.tex
%
%
%
\definecolor{mycolor1}{rgb}{1.00000,0.00000,1.00000}%
\begin{tikzpicture}

\begin{axis}[%
width=3.3in,
height=2.2in,
scale only axis,
xmin=-6.1244260279434,
xmax=7.3610796585662,
xlabel={$\overline{\gamma} (\si{\decibel})$},
ymin=0,
ymax=10,
ylabel={\small bits/symbol duration in NC-TWRC},
grid = both,
legend style={at={(0.03,0.97)},anchor=north west,draw=black,fill=white,legend cell align=left}
]
\addplot [color=blue,solid,mark=o,mark options={solid}]
  table[row sep=crcr]{SIM/BitRate-1.tsv};
\addlegendentry{\small SupG-BER-$\overline{\mathbf{\theta}}^*$};

\addplot [color=blue,solid,mark=asterisk,mark options={solid}]
  table[row sep=crcr]{SIM/BitRate-2.tsv};
\addlegendentry{\small SupG-BER-$\underline{\mathbf{\theta}}^*$};

\addplot [color=red,solid,mark=triangle,mark options={solid,,rotate=270}]
  table[row sep=crcr]{SIM/BitRate-3.tsv};
\addlegendentry{\small AM-BER};

\addplot [color=green,solid,mark=square,mark options={solid}]
  table[row sep=crcr]{SIM/BitRate-4.tsv};
\addlegendentry{\small SupG-Pr-$\overline{\mathbf{\theta}}^*$};

\addplot [color=green,solid,mark=diamond,mark options={solid}]
  table[row sep=crcr]{SIM/BitRate-5.tsv};
\addlegendentry{\small SupG-Pr-$\underline{\mathbf{\theta}}^*$};

\addplot [color=red,solid,mark=star,mark options={solid}]
  table[row sep=crcr]{SIM/BitRate-6.tsv};
\addlegendentry{\small AM-Pr};

\addplot [color=mycolor1,solid,mark=*,mark options={solid}]
  table[row sep=crcr]{SIM/BitRate-7.tsv};
\addlegendentry{\small 2-QAM};

\end{axis}
\end{tikzpicture}%

%% file: SIM/TransRate_R.tex
%
%
%
\definecolor{mycolor1}{rgb}{1.00000,0.00000,1.00000}%
\begin{tikzpicture}

\begin{axis}[%
width=3.3in,
height=2.2in,
scale only axis,
xmin=-6.1244260279434,
xmax=7.3610796585662,
xlabel={$\overline{\gamma} (\si{\decibel})$},
ymin=0,
ymax=1.1,
ylabel={\small broadcast rate at the relay},
grid = both,
legend style={at={(0.98,0.02)},anchor=south east,draw=black,fill=white,legend cell align=left}
]
\addplot [color=blue,solid,mark=o,mark options={solid}]
  table[row sep=crcr]{SIM/TransRate_R-1.tsv};
\addlegendentry{\small SupG-BER-$\overline{\mathbf{\theta}}^*$};

\addplot [color=blue,solid,mark=asterisk,mark options={solid}]
  table[row sep=crcr]{SIM/TransRate_R-2.tsv};
\addlegendentry{\small SupG-BER-$\underline{\mathbf{\theta}}^*$};

\addplot [color=red,solid,mark=triangle,mark options={solid,,rotate=270}]
  table[row sep=crcr]{SIM/TransRate_R-3.tsv};
\addlegendentry{\small AM-BER};

\addplot [color=green,solid,mark=square,mark options={solid}]
  table[row sep=crcr]{SIM/TransRate_R-4.tsv};
\addlegendentry{\small SupG-Pr-$\overline{\mathbf{\theta}}^*$};

\addplot [color=green,solid,mark=diamond,mark options={solid}]
  table[row sep=crcr]{SIM/TransRate_R-5.tsv};
\addlegendentry{\small SupG-Pr-$\underline{\mathbf{\theta}}^*$};

\addplot [color=red,solid,mark=star,mark options={solid}]
  table[row sep=crcr]{SIM/TransRate_R-6.tsv};
\addlegendentry{\small AM-Pr};

\addplot [color=mycolor1,solid,mark=*,mark options={solid}]
  table[row sep=crcr]{SIM/TransRate_R-7.tsv};
\addlegendentry{\small 2-QAM};

\end{axis}
\end{tikzpicture}%

%% file: SIM/MeanCost1.tex
%
%
\begin{tikzpicture}

\begin{axis}[%
width=3.3in,
height=2.2in,
scale only axis,
xmin=-6.1244260279434,
xmax=7.3610796585662,
xlabel={$\overline{\gamma} (\si{\decibel})$},
ymin=1,
ymax=1.8,
ylabel={average cost $\bar{c}_1$},
grid = both,
legend style={at={(0.03,0.55)},anchor=south west,draw=black,fill=white,legend cell align=left}
]
\addplot [
color=blue,
solid,
mark=o,
mark options={solid}
]
table[row sep=crcr]{
-6.1244260279434 1.21589583333333\\
-4.53901891043867 1.15904166666667\\
-3.01029995663981 1.15904166666667\\
-1.53901891043867 1.21589583333333\\
-0.124426027943397 1.35684583333333\\
1.23565137563515 1.592325\\
2.54459536890706 1.66122916666666\\
3.80668951933905 1.46039583333334\\
5.02677206291304 1.17068333333334\\
6.20990250347433 1.0489625\\
7.3610796585662 1.01345\\
};
\addlegendentry{SupG-BER-$\overline{\mathbf{\theta}}^*$};

\addplot [
color=blue,
solid,
mark=asterisk,
mark options={solid}
]
table[row sep=crcr]{
-6.1244260279434 1.20219583333333\\
-4.53901891043867 1.14905833333333\\
-3.01029995663981 1.14905833333333\\
-1.53901891043867 1.20219583333333\\
-0.124426027943397 1.35120833333333\\
1.23565137563515 1.5735375\\
2.54459536890706 1.65629166666666\\
3.80668951933905 1.43960416666667\\
5.02677206291304 1.16837500000001\\
6.20990250347433 1.0532125\\
7.3610796585662 1.00849166666667\\
};
\addlegendentry{SupG-BER-$\underline{\mathbf{\theta}}^*$};

\addplot [
color=green,
solid,
mark=square,
mark options={solid}
]
table[row sep=crcr]{
-6.1244260279434 1.17968765209282\\
-4.53901891043867 1.13473077709755\\
-3.01029995663981 1.13473077709755\\
-1.53901891043867 1.17968765209282\\
-0.124426027943397 1.25803287128334\\
1.23565137563515 1.33650507208409\\
2.54459536890706 1.34864425248528\\
3.80668951933905 1.27836252518606\\
5.02677206291304 1.15383368161818\\
6.20990250347433 1.05352858838933\\
7.3610796585662 1.01332230076996\\
};
\addlegendentry{SupG-Pr-$\overline{\mathbf{\theta}}^*$};

\addplot [
color=green,
solid,
mark=diamond,
mark options={solid}
]
table[row sep=crcr]{
-6.1244260279434 1.14689383649027\\
-4.53901891043867 1.10922769312\\
-3.01029995663981 1.10922769312\\
-1.53901891043867 1.14689383649027\\
-0.124426027943397 1.22385511076921\\
1.23565137563515 1.3111739915308\\
2.54459536890706 1.33296064652895\\
3.80668951933905 1.26930111714826\\
5.02677206291304 1.14893607876827\\
6.20990250347433 1.05294283269526\\
7.3610796585662 1.01237010196645\\
};
\addlegendentry{SupG-Pr-$\underline{\mathbf{\theta}}^*$};

\end{axis}
\end{tikzpicture}%

%% file: SIM/MeanCost2.tex
%
%
\begin{tikzpicture}

\begin{axis}[%
width=3.3in,
height=2.2in,
scale only axis,
xmin=-6.1244260279434,
xmax=7.3610796585662,
xlabel={$\overline{\gamma} (\si{\decibel})$},
ymin=1,
ymax=1.8,
ylabel={average cost $\bar{c}_2$},
grid = both,
legend style={at={(0.03,0.55)},anchor=south west,draw=black,fill=white,legend cell align=left}
]
\addplot [
color=blue,
solid,
mark=o,
mark options={solid}
]
table[row sep=crcr]{
-6.1244260279434 1.24150416666667\\
-4.53901891043867 1.18193333333333\\
-3.01029995663981 1.18193333333333\\
-1.53901891043867 1.24150416666667\\
-0.124426027943397 1.36858333333333\\
1.23565137563515 1.5775125\\
2.54459536890706 1.71286249999999\\
3.80668951933905 1.52873333333334\\
5.02677206291304 1.18991666666667\\
6.20990250347433 1.05130833333333\\
7.3610796585662 1.00678333333333\\
};
\addlegendentry{SupG-BER-$\overline{\mathbf{\theta}}^*$};

\addplot [
color=blue,
solid,
mark=asterisk,
mark options={solid}
]
table[row sep=crcr]{
-6.1244260279434 1.2153125\\
-4.53901891043867 1.18000833333333\\
-3.01029995663981 1.18100833333333\\
-1.53901891043867 1.2353125\\
-0.124426027943397 1.35491666666667\\
1.23565137563515 1.56134166666666\\
2.54459536890706 1.71172916666666\\
3.80668951933905 1.5022875\\
5.02677206291304 1.18434166666668\\
6.20990250347433 1.0466\\
7.3610796585662 1.00565833333333\\
};
\addlegendentry{SupG-BER-$\underline{\mathbf{\theta}}^*$};

\addplot [
color=green,
solid,
mark=square,
mark options={solid}
]
table[row sep=crcr]{
-6.1244260279434 1.17968765209282\\
-4.53901891043867 1.13473077709755\\
-3.01029995663981 1.13473077709755\\
-1.53901891043867 1.17968765209282\\
-0.124426027943397 1.25803287128334\\
1.23565137563515 1.33650507208409\\
2.54459536890706 1.34864425248528\\
3.80668951933905 1.27836252518606\\
5.02677206291304 1.15383368161818\\
6.20990250347433 1.05352858838933\\
7.3610796585662 1.01332230076996\\
};
\addlegendentry{SupG-Pr-$\overline{\mathbf{\theta}}^*$};

\addplot [
color=green,
solid,
mark=diamond,
mark options={solid}
]
table[row sep=crcr]{
-6.1244260279434 1.15689383649027\\
-4.53901891043867 1.12922769311999\\
-3.01029995663981 1.12922769311999\\
-1.53901891043867 1.15689383649027\\
-0.124426027943397 1.24385511076921\\
1.23565137563515 1.3111739915308\\
2.54459536890706 1.33296064652895\\
3.80668951933905 1.26930111714826\\
5.02677206291304 1.14893607876827\\
6.20990250347433 1.05294283269526\\
7.3610796585662 1.01237010196645\\
};
\addlegendentry{SupG-Pr-$\underline{\mathbf{\theta}}^*$};

\end{axis}
\end{tikzpicture}%

%% file: Policy/MonoH1.tex
%
%
\begin{tikzpicture}

\begin{axis}[%
width=2.5in,
height=2.2in,
unbounded coords=jump,
view={-130}{40},
scale only axis,
xmin=0.5,
xmax=11.5,
xtick={1,2,3,4,5,6,7,8,9,10,11},
xticklabels={0.1,1,2,3,4,5,6,7,8,9,10},
xlabel={$\gamma_2$},
xmajorgrids,
y dir=reverse,
ymin=0.5,
ymax=11.5,
ytick={1,2,3,4,5,6,7,8,9,10,11},
yticklabels={0.1,1,2,3,4,5,6,7,8,9,10},
ylabel={$\gamma_1$},
ymajorgrids,
zmin=0,
zmax=4,
zlabel={$\overline{\theta}_1^*$},
zmajorgrids,
name=plot1,
axis x line*=bottom,
axis y line*=left,
axis z line*=left
]

\addplot3[%
surf,
shader=flat,
draw=black,
point meta=explicit,
mesh/rows=4]
table[row sep=crcr,header=false,meta index=3] {
NaN NaN NaN 11\\
10.6 0.6 0 11\\
10.6 1.4 0 11\\
NaN NaN NaN 11\\
NaN NaN NaN 11\\
NaN NaN NaN 11\\
NaN NaN NaN 11\\
10.6 1.6 0 11\\
10.6 2.4 0 11\\
NaN NaN NaN 11\\
NaN NaN NaN 11\\
NaN NaN NaN 11\\
NaN NaN NaN 11\\
10.6 2.6 0 11\\
10.6 3.4 0 11\\
NaN NaN NaN 11\\
NaN NaN NaN 11\\
NaN NaN NaN 11\\
NaN NaN NaN 11\\
10.6 3.6 0 11\\
10.6 4.4 0 11\\
NaN NaN NaN 11\\
NaN NaN NaN 11\\
NaN NaN NaN 11\\
NaN NaN NaN 11\\
10.6 4.6 0 11\\
10.6 5.4 0 11\\
NaN NaN NaN 11\\
NaN NaN NaN 11\\
NaN NaN NaN 11\\
NaN NaN NaN 11\\
10.6 5.6 0 11\\
10.6 6.4 0 11\\
NaN NaN NaN 11\\
NaN NaN NaN 11\\
NaN NaN NaN 11\\
NaN NaN NaN 11\\
10.6 6.6 0 11\\
10.6 7.4 0 11\\
NaN NaN NaN 11\\
NaN NaN NaN 11\\
NaN NaN NaN 11\\
NaN NaN NaN 11\\
10.6 7.6 0 11\\
10.6 8.4 0 11\\
NaN NaN NaN 11\\
NaN NaN NaN 11\\
NaN NaN NaN 11\\
NaN NaN NaN 11\\
10.6 8.6 0 11\\
10.6 9.4 0 11\\
NaN NaN NaN 11\\
NaN NaN NaN 11\\
NaN NaN NaN 11\\
NaN NaN NaN 11\\
10.6 9.6 0 11\\
10.6 10.4 0 11\\
NaN NaN NaN 11\\
NaN NaN NaN 11\\
NaN NaN NaN 11\\
NaN NaN NaN 11\\
10.6 10.6 0 11\\
10.6 11.4 0 11\\
NaN NaN NaN 11\\
NaN NaN NaN 11\\
NaN NaN NaN 11\\
10.6 0.6 0 11\\
10.6 0.6 1 11\\
10.6 1.4 1 11\\
10.6 1.4 0 11\\
10.6 0.6 0 11\\
NaN NaN NaN 11\\
10.6 1.6 0 11\\
10.6 1.6 2 11\\
10.6 2.4 2 11\\
10.6 2.4 0 11\\
10.6 1.6 0 11\\
NaN NaN NaN 11\\
10.6 2.6 0 11\\
10.6 2.6 3 11\\
10.6 3.4 3 11\\
10.6 3.4 0 11\\
10.6 2.6 0 11\\
NaN NaN NaN 11\\
10.6 3.6 0 11\\
10.6 3.6 3 11\\
10.6 4.4 3 11\\
10.6 4.4 0 11\\
10.6 3.6 0 11\\
NaN NaN NaN 11\\
10.6 4.6 0 11\\
10.6 4.6 3 11\\
10.6 5.4 3 11\\
10.6 5.4 0 11\\
10.6 4.6 0 11\\
NaN NaN NaN 11\\
10.6 5.6 0 11\\
10.6 5.6 3 11\\
10.6 6.4 3 11\\
10.6 6.4 0 11\\
10.6 5.6 0 11\\
NaN NaN NaN 11\\
10.6 6.6 0 11\\
10.6 6.6 4 11\\
10.6 7.4 4 11\\
10.6 7.4 0 11\\
10.6 6.6 0 11\\
NaN NaN NaN 11\\
10.6 7.6 0 11\\
10.6 7.6 4 11\\
10.6 8.4 4 11\\
10.6 8.4 0 11\\
10.6 7.6 0 11\\
NaN NaN NaN 11\\
10.6 8.6 0 11\\
10.6 8.6 4 11\\
10.6 9.4 4 11\\
10.6 9.4 0 11\\
10.6 8.6 0 11\\
NaN NaN NaN 11\\
10.6 9.6 0 11\\
10.6 9.6 4 11\\
10.6 10.4 4 11\\
10.6 10.4 0 11\\
10.6 9.6 0 11\\
NaN NaN NaN 11\\
10.6 10.6 0 11\\
10.6 10.6 4 11\\
10.6 11.4 4 11\\
10.6 11.4 0 11\\
10.6 10.6 0 11\\
NaN NaN NaN 11\\
11.4 0.6 0 11\\
11.4 0.6 1 11\\
11.4 1.4 1 11\\
11.4 1.4 0 11\\
11.4 0.6 0 11\\
NaN NaN NaN 11\\
11.4 1.6 0 11\\
11.4 1.6 2 11\\
11.4 2.4 2 11\\
11.4 2.4 0 11\\
11.4 1.6 0 11\\
NaN NaN NaN 11\\
11.4 2.6 0 11\\
11.4 2.6 3 11\\
11.4 3.4 3 11\\
11.4 3.4 0 11\\
11.4 2.6 0 11\\
NaN NaN NaN 11\\
11.4 3.6 0 11\\
11.4 3.6 3 11\\
11.4 4.4 3 11\\
11.4 4.4 0 11\\
11.4 3.6 0 11\\
NaN NaN NaN 11\\
11.4 4.6 0 11\\
11.4 4.6 3 11\\
11.4 5.4 3 11\\
11.4 5.4 0 11\\
11.4 4.6 0 11\\
NaN NaN NaN 11\\
11.4 5.6 0 11\\
11.4 5.6 3 11\\
11.4 6.4 3 11\\
11.4 6.4 0 11\\
11.4 5.6 0 11\\
NaN NaN NaN 11\\
11.4 6.6 0 11\\
11.4 6.6 4 11\\
11.4 7.4 4 11\\
11.4 7.4 0 11\\
11.4 6.6 0 11\\
NaN NaN NaN 11\\
11.4 7.6 0 11\\
11.4 7.6 4 11\\
11.4 8.4 4 11\\
11.4 8.4 0 11\\
11.4 7.6 0 11\\
NaN NaN NaN 11\\
11.4 8.6 0 11\\
11.4 8.6 4 11\\
11.4 9.4 4 11\\
11.4 9.4 0 11\\
11.4 8.6 0 11\\
NaN NaN NaN 11\\
11.4 9.6 0 11\\
11.4 9.6 4 11\\
11.4 10.4 4 11\\
11.4 10.4 0 11\\
11.4 9.6 0 11\\
NaN NaN NaN 11\\
11.4 10.6 0 11\\
11.4 10.6 4 11\\
11.4 11.4 4 11\\
11.4 11.4 0 11\\
11.4 10.6 0 11\\
NaN NaN NaN 11\\
NaN NaN NaN 11\\
11.4 0.6 0 11\\
11.4 1.4 0 11\\
NaN NaN NaN 11\\
NaN NaN NaN 11\\
NaN NaN NaN 11\\
NaN NaN NaN 11\\
11.4 1.6 0 11\\
11.4 2.4 0 11\\
NaN NaN NaN 11\\
NaN NaN NaN 11\\
NaN NaN NaN 11\\
NaN NaN NaN 11\\
11.4 2.6 0 11\\
11.4 3.4 0 11\\
NaN NaN NaN 11\\
NaN NaN NaN 11\\
NaN NaN NaN 11\\
NaN NaN NaN 11\\
11.4 3.6 0 11\\
11.4 4.4 0 11\\
NaN NaN NaN 11\\
NaN NaN NaN 11\\
NaN NaN NaN 11\\
NaN NaN NaN 11\\
11.4 4.6 0 11\\
11.4 5.4 0 11\\
NaN NaN NaN 11\\
NaN NaN NaN 11\\
NaN NaN NaN 11\\
NaN NaN NaN 11\\
11.4 5.6 0 11\\
11.4 6.4 0 11\\
NaN NaN NaN 11\\
NaN NaN NaN 11\\
NaN NaN NaN 11\\
NaN NaN NaN 11\\
11.4 6.6 0 11\\
11.4 7.4 0 11\\
NaN NaN NaN 11\\
NaN NaN NaN 11\\
NaN NaN NaN 11\\
NaN NaN NaN 11\\
11.4 7.6 0 11\\
11.4 8.4 0 11\\
NaN NaN NaN 11\\
NaN NaN NaN 11\\
NaN NaN NaN 11\\
NaN NaN NaN 11\\
11.4 8.6 0 11\\
11.4 9.4 0 11\\
NaN NaN NaN 11\\
NaN NaN NaN 11\\
NaN NaN NaN 11\\
NaN NaN NaN 11\\
11.4 9.6 0 11\\
11.4 10.4 0 11\\
NaN NaN NaN 11\\
NaN NaN NaN 11\\
NaN NaN NaN 11\\
NaN NaN NaN 11\\
11.4 10.6 0 11\\
11.4 11.4 0 11\\
NaN NaN NaN 11\\
NaN NaN NaN 11\\
NaN NaN NaN 11\\
};

\addplot3[%
surf,
shader=flat,
draw=black,
point meta=explicit,
mesh/rows=4]
table[row sep=crcr,header=false,meta index=3] {
NaN NaN NaN 10\\
9.6 0.6 0 10\\
9.6 1.4 0 10\\
NaN NaN NaN 10\\
NaN NaN NaN 10\\
NaN NaN NaN 10\\
NaN NaN NaN 10\\
9.6 1.6 0 10\\
9.6 2.4 0 10\\
NaN NaN NaN 10\\
NaN NaN NaN 10\\
NaN NaN NaN 10\\
NaN NaN NaN 10\\
9.6 2.6 0 10\\
9.6 3.4 0 10\\
NaN NaN NaN 10\\
NaN NaN NaN 10\\
NaN NaN NaN 10\\
NaN NaN NaN 10\\
9.6 3.6 0 10\\
9.6 4.4 0 10\\
NaN NaN NaN 10\\
NaN NaN NaN 10\\
NaN NaN NaN 10\\
NaN NaN NaN 10\\
9.6 4.6 0 10\\
9.6 5.4 0 10\\
NaN NaN NaN 10\\
NaN NaN NaN 10\\
NaN NaN NaN 10\\
NaN NaN NaN 10\\
9.6 5.6 0 10\\
9.6 6.4 0 10\\
NaN NaN NaN 10\\
NaN NaN NaN 10\\
NaN NaN NaN 10\\
NaN NaN NaN 10\\
9.6 6.6 0 10\\
9.6 7.4 0 10\\
NaN NaN NaN 10\\
NaN NaN NaN 10\\
NaN NaN NaN 10\\
NaN NaN NaN 10\\
9.6 7.6 0 10\\
9.6 8.4 0 10\\
NaN NaN NaN 10\\
NaN NaN NaN 10\\
NaN NaN NaN 10\\
NaN NaN NaN 10\\
9.6 8.6 0 10\\
9.6 9.4 0 10\\
NaN NaN NaN 10\\
NaN NaN NaN 10\\
NaN NaN NaN 10\\
NaN NaN NaN 10\\
9.6 9.6 0 10\\
9.6 10.4 0 10\\
NaN NaN NaN 10\\
NaN NaN NaN 10\\
NaN NaN NaN 10\\
NaN NaN NaN 10\\
9.6 10.6 0 10\\
9.6 11.4 0 10\\
NaN NaN NaN 10\\
NaN NaN NaN 10\\
NaN NaN NaN 10\\
9.6 0.6 0 10\\
9.6 0.6 1 10\\
9.6 1.4 1 10\\
9.6 1.4 0 10\\
9.6 0.6 0 10\\
NaN NaN NaN 10\\
9.6 1.6 0 10\\
9.6 1.6 2 10\\
9.6 2.4 2 10\\
9.6 2.4 0 10\\
9.6 1.6 0 10\\
NaN NaN NaN 10\\
9.6 2.6 0 10\\
9.6 2.6 3 10\\
9.6 3.4 3 10\\
9.6 3.4 0 10\\
9.6 2.6 0 10\\
NaN NaN NaN 10\\
9.6 3.6 0 10\\
9.6 3.6 3 10\\
9.6 4.4 3 10\\
9.6 4.4 0 10\\
9.6 3.6 0 10\\
NaN NaN NaN 10\\
9.6 4.6 0 10\\
9.6 4.6 3 10\\
9.6 5.4 3 10\\
9.6 5.4 0 10\\
9.6 4.6 0 10\\
NaN NaN NaN 10\\
9.6 5.6 0 10\\
9.6 5.6 3 10\\
9.6 6.4 3 10\\
9.6 6.4 0 10\\
9.6 5.6 0 10\\
NaN NaN NaN 10\\
9.6 6.6 0 10\\
9.6 6.6 4 10\\
9.6 7.4 4 10\\
9.6 7.4 0 10\\
9.6 6.6 0 10\\
NaN NaN NaN 10\\
9.6 7.6 0 10\\
9.6 7.6 4 10\\
9.6 8.4 4 10\\
9.6 8.4 0 10\\
9.6 7.6 0 10\\
NaN NaN NaN 10\\
9.6 8.6 0 10\\
9.6 8.6 4 10\\
9.6 9.4 4 10\\
9.6 9.4 0 10\\
9.6 8.6 0 10\\
NaN NaN NaN 10\\
9.6 9.6 0 10\\
9.6 9.6 4 10\\
9.6 10.4 4 10\\
9.6 10.4 0 10\\
9.6 9.6 0 10\\
NaN NaN NaN 10\\
9.6 10.6 0 10\\
9.6 10.6 4 10\\
9.6 11.4 4 10\\
9.6 11.4 0 10\\
9.6 10.6 0 10\\
NaN NaN NaN 10\\
10.4 0.6 0 10\\
10.4 0.6 1 10\\
10.4 1.4 1 10\\
10.4 1.4 0 10\\
10.4 0.6 0 10\\
NaN NaN NaN 10\\
10.4 1.6 0 10\\
10.4 1.6 2 10\\
10.4 2.4 2 10\\
10.4 2.4 0 10\\
10.4 1.6 0 10\\
NaN NaN NaN 10\\
10.4 2.6 0 10\\
10.4 2.6 3 10\\
10.4 3.4 3 10\\
10.4 3.4 0 10\\
10.4 2.6 0 10\\
NaN NaN NaN 10\\
10.4 3.6 0 10\\
10.4 3.6 3 10\\
10.4 4.4 3 10\\
10.4 4.4 0 10\\
10.4 3.6 0 10\\
NaN NaN NaN 10\\
10.4 4.6 0 10\\
10.4 4.6 3 10\\
10.4 5.4 3 10\\
10.4 5.4 0 10\\
10.4 4.6 0 10\\
NaN NaN NaN 10\\
10.4 5.6 0 10\\
10.4 5.6 3 10\\
10.4 6.4 3 10\\
10.4 6.4 0 10\\
10.4 5.6 0 10\\
NaN NaN NaN 10\\
10.4 6.6 0 10\\
10.4 6.6 4 10\\
10.4 7.4 4 10\\
10.4 7.4 0 10\\
10.4 6.6 0 10\\
NaN NaN NaN 10\\
10.4 7.6 0 10\\
10.4 7.6 4 10\\
10.4 8.4 4 10\\
10.4 8.4 0 10\\
10.4 7.6 0 10\\
NaN NaN NaN 10\\
10.4 8.6 0 10\\
10.4 8.6 4 10\\
10.4 9.4 4 10\\
10.4 9.4 0 10\\
10.4 8.6 0 10\\
NaN NaN NaN 10\\
10.4 9.6 0 10\\
10.4 9.6 4 10\\
10.4 10.4 4 10\\
10.4 10.4 0 10\\
10.4 9.6 0 10\\
NaN NaN NaN 10\\
10.4 10.6 0 10\\
10.4 10.6 4 10\\
10.4 11.4 4 10\\
10.4 11.4 0 10\\
10.4 10.6 0 10\\
NaN NaN NaN 10\\
NaN NaN NaN 10\\
10.4 0.6 0 10\\
10.4 1.4 0 10\\
NaN NaN NaN 10\\
NaN NaN NaN 10\\
NaN NaN NaN 10\\
NaN NaN NaN 10\\
10.4 1.6 0 10\\
10.4 2.4 0 10\\
NaN NaN NaN 10\\
NaN NaN NaN 10\\
NaN NaN NaN 10\\
NaN NaN NaN 10\\
10.4 2.6 0 10\\
10.4 3.4 0 10\\
NaN NaN NaN 10\\
NaN NaN NaN 10\\
NaN NaN NaN 10\\
NaN NaN NaN 10\\
10.4 3.6 0 10\\
10.4 4.4 0 10\\
NaN NaN NaN 10\\
NaN NaN NaN 10\\
NaN NaN NaN 10\\
NaN NaN NaN 10\\
10.4 4.6 0 10\\
10.4 5.4 0 10\\
NaN NaN NaN 10\\
NaN NaN NaN 10\\
NaN NaN NaN 10\\
NaN NaN NaN 10\\
10.4 5.6 0 10\\
10.4 6.4 0 10\\
NaN NaN NaN 10\\
NaN NaN NaN 10\\
NaN NaN NaN 10\\
NaN NaN NaN 10\\
10.4 6.6 0 10\\
10.4 7.4 0 10\\
NaN NaN NaN 10\\
NaN NaN NaN 10\\
NaN NaN NaN 10\\
NaN NaN NaN 10\\
10.4 7.6 0 10\\
10.4 8.4 0 10\\
NaN NaN NaN 10\\
NaN NaN NaN 10\\
NaN NaN NaN 10\\
NaN NaN NaN 10\\
10.4 8.6 0 10\\
10.4 9.4 0 10\\
NaN NaN NaN 10\\
NaN NaN NaN 10\\
NaN NaN NaN 10\\
NaN NaN NaN 10\\
10.4 9.6 0 10\\
10.4 10.4 0 10\\
NaN NaN NaN 10\\
NaN NaN NaN 10\\
NaN NaN NaN 10\\
NaN NaN NaN 10\\
10.4 10.6 0 10\\
10.4 11.4 0 10\\
NaN NaN NaN 10\\
NaN NaN NaN 10\\
NaN NaN NaN 10\\
};

\addplot3[%
surf,
shader=flat,
draw=black,
point meta=explicit,
mesh/rows=4]
table[row sep=crcr,header=false,meta index=3] {
NaN NaN NaN 9\\
8.6 0.6 0 9\\
8.6 1.4 0 9\\
NaN NaN NaN 9\\
NaN NaN NaN 9\\
NaN NaN NaN 9\\
NaN NaN NaN 9\\
8.6 1.6 0 9\\
8.6 2.4 0 9\\
NaN NaN NaN 9\\
NaN NaN NaN 9\\
NaN NaN NaN 9\\
NaN NaN NaN 9\\
8.6 2.6 0 9\\
8.6 3.4 0 9\\
NaN NaN NaN 9\\
NaN NaN NaN 9\\
NaN NaN NaN 9\\
NaN NaN NaN 9\\
8.6 3.6 0 9\\
8.6 4.4 0 9\\
NaN NaN NaN 9\\
NaN NaN NaN 9\\
NaN NaN NaN 9\\
NaN NaN NaN 9\\
8.6 4.6 0 9\\
8.6 5.4 0 9\\
NaN NaN NaN 9\\
NaN NaN NaN 9\\
NaN NaN NaN 9\\
NaN NaN NaN 9\\
8.6 5.6 0 9\\
8.6 6.4 0 9\\
NaN NaN NaN 9\\
NaN NaN NaN 9\\
NaN NaN NaN 9\\
NaN NaN NaN 9\\
8.6 6.6 0 9\\
8.6 7.4 0 9\\
NaN NaN NaN 9\\
NaN NaN NaN 9\\
NaN NaN NaN 9\\
NaN NaN NaN 9\\
8.6 7.6 0 9\\
8.6 8.4 0 9\\
NaN NaN NaN 9\\
NaN NaN NaN 9\\
NaN NaN NaN 9\\
NaN NaN NaN 9\\
8.6 8.6 0 9\\
8.6 9.4 0 9\\
NaN NaN NaN 9\\
NaN NaN NaN 9\\
NaN NaN NaN 9\\
NaN NaN NaN 9\\
8.6 9.6 0 9\\
8.6 10.4 0 9\\
NaN NaN NaN 9\\
NaN NaN NaN 9\\
NaN NaN NaN 9\\
NaN NaN NaN 9\\
8.6 10.6 0 9\\
8.6 11.4 0 9\\
NaN NaN NaN 9\\
NaN NaN NaN 9\\
NaN NaN NaN 9\\
8.6 0.6 0 9\\
8.6 0.6 1 9\\
8.6 1.4 1 9\\
8.6 1.4 0 9\\
8.6 0.6 0 9\\
NaN NaN NaN 9\\
8.6 1.6 0 9\\
8.6 1.6 2 9\\
8.6 2.4 2 9\\
8.6 2.4 0 9\\
8.6 1.6 0 9\\
NaN NaN NaN 9\\
8.6 2.6 0 9\\
8.6 2.6 3 9\\
8.6 3.4 3 9\\
8.6 3.4 0 9\\
8.6 2.6 0 9\\
NaN NaN NaN 9\\
8.6 3.6 0 9\\
8.6 3.6 3 9\\
8.6 4.4 3 9\\
8.6 4.4 0 9\\
8.6 3.6 0 9\\
NaN NaN NaN 9\\
8.6 4.6 0 9\\
8.6 4.6 3 9\\
8.6 5.4 3 9\\
8.6 5.4 0 9\\
8.6 4.6 0 9\\
NaN NaN NaN 9\\
8.6 5.6 0 9\\
8.6 5.6 3 9\\
8.6 6.4 3 9\\
8.6 6.4 0 9\\
8.6 5.6 0 9\\
NaN NaN NaN 9\\
8.6 6.6 0 9\\
8.6 6.6 4 9\\
8.6 7.4 4 9\\
8.6 7.4 0 9\\
8.6 6.6 0 9\\
NaN NaN NaN 9\\
8.6 7.6 0 9\\
8.6 7.6 4 9\\
8.6 8.4 4 9\\
8.6 8.4 0 9\\
8.6 7.6 0 9\\
NaN NaN NaN 9\\
8.6 8.6 0 9\\
8.6 8.6 4 9\\
8.6 9.4 4 9\\
8.6 9.4 0 9\\
8.6 8.6 0 9\\
NaN NaN NaN 9\\
8.6 9.6 0 9\\
8.6 9.6 4 9\\
8.6 10.4 4 9\\
8.6 10.4 0 9\\
8.6 9.6 0 9\\
NaN NaN NaN 9\\
8.6 10.6 0 9\\
8.6 10.6 4 9\\
8.6 11.4 4 9\\
8.6 11.4 0 9\\
8.6 10.6 0 9\\
NaN NaN NaN 9\\
9.4 0.6 0 9\\
9.4 0.6 1 9\\
9.4 1.4 1 9\\
9.4 1.4 0 9\\
9.4 0.6 0 9\\
NaN NaN NaN 9\\
9.4 1.6 0 9\\
9.4 1.6 2 9\\
9.4 2.4 2 9\\
9.4 2.4 0 9\\
9.4 1.6 0 9\\
NaN NaN NaN 9\\
9.4 2.6 0 9\\
9.4 2.6 3 9\\
9.4 3.4 3 9\\
9.4 3.4 0 9\\
9.4 2.6 0 9\\
NaN NaN NaN 9\\
9.4 3.6 0 9\\
9.4 3.6 3 9\\
9.4 4.4 3 9\\
9.4 4.4 0 9\\
9.4 3.6 0 9\\
NaN NaN NaN 9\\
9.4 4.6 0 9\\
9.4 4.6 3 9\\
9.4 5.4 3 9\\
9.4 5.4 0 9\\
9.4 4.6 0 9\\
NaN NaN NaN 9\\
9.4 5.6 0 9\\
9.4 5.6 3 9\\
9.4 6.4 3 9\\
9.4 6.4 0 9\\
9.4 5.6 0 9\\
NaN NaN NaN 9\\
9.4 6.6 0 9\\
9.4 6.6 4 9\\
9.4 7.4 4 9\\
9.4 7.4 0 9\\
9.4 6.6 0 9\\
NaN NaN NaN 9\\
9.4 7.6 0 9\\
9.4 7.6 4 9\\
9.4 8.4 4 9\\
9.4 8.4 0 9\\
9.4 7.6 0 9\\
NaN NaN NaN 9\\
9.4 8.6 0 9\\
9.4 8.6 4 9\\
9.4 9.4 4 9\\
9.4 9.4 0 9\\
9.4 8.6 0 9\\
NaN NaN NaN 9\\
9.4 9.6 0 9\\
9.4 9.6 4 9\\
9.4 10.4 4 9\\
9.4 10.4 0 9\\
9.4 9.6 0 9\\
NaN NaN NaN 9\\
9.4 10.6 0 9\\
9.4 10.6 4 9\\
9.4 11.4 4 9\\
9.4 11.4 0 9\\
9.4 10.6 0 9\\
NaN NaN NaN 9\\
NaN NaN NaN 9\\
9.4 0.6 0 9\\
9.4 1.4 0 9\\
NaN NaN NaN 9\\
NaN NaN NaN 9\\
NaN NaN NaN 9\\
NaN NaN NaN 9\\
9.4 1.6 0 9\\
9.4 2.4 0 9\\
NaN NaN NaN 9\\
NaN NaN NaN 9\\
NaN NaN NaN 9\\
NaN NaN NaN 9\\
9.4 2.6 0 9\\
9.4 3.4 0 9\\
NaN NaN NaN 9\\
NaN NaN NaN 9\\
NaN NaN NaN 9\\
NaN NaN NaN 9\\
9.4 3.6 0 9\\
9.4 4.4 0 9\\
NaN NaN NaN 9\\
NaN NaN NaN 9\\
NaN NaN NaN 9\\
NaN NaN NaN 9\\
9.4 4.6 0 9\\
9.4 5.4 0 9\\
NaN NaN NaN 9\\
NaN NaN NaN 9\\
NaN NaN NaN 9\\
NaN NaN NaN 9\\
9.4 5.6 0 9\\
9.4 6.4 0 9\\
NaN NaN NaN 9\\
NaN NaN NaN 9\\
NaN NaN NaN 9\\
NaN NaN NaN 9\\
9.4 6.6 0 9\\
9.4 7.4 0 9\\
NaN NaN NaN 9\\
NaN NaN NaN 9\\
NaN NaN NaN 9\\
NaN NaN NaN 9\\
9.4 7.6 0 9\\
9.4 8.4 0 9\\
NaN NaN NaN 9\\
NaN NaN NaN 9\\
NaN NaN NaN 9\\
NaN NaN NaN 9\\
9.4 8.6 0 9\\
9.4 9.4 0 9\\
NaN NaN NaN 9\\
NaN NaN NaN 9\\
NaN NaN NaN 9\\
NaN NaN NaN 9\\
9.4 9.6 0 9\\
9.4 10.4 0 9\\
NaN NaN NaN 9\\
NaN NaN NaN 9\\
NaN NaN NaN 9\\
NaN NaN NaN 9\\
9.4 10.6 0 9\\
9.4 11.4 0 9\\
NaN NaN NaN 9\\
NaN NaN NaN 9\\
NaN NaN NaN 9\\
};

\addplot3[%
surf,
shader=flat,
draw=black,
point meta=explicit,
mesh/rows=4]
table[row sep=crcr,header=false,meta index=3] {
NaN NaN NaN 8\\
7.6 0.6 0 8\\
7.6 1.4 0 8\\
NaN NaN NaN 8\\
NaN NaN NaN 8\\
NaN NaN NaN 8\\
NaN NaN NaN 8\\
7.6 1.6 0 8\\
7.6 2.4 0 8\\
NaN NaN NaN 8\\
NaN NaN NaN 8\\
NaN NaN NaN 8\\
NaN NaN NaN 8\\
7.6 2.6 0 8\\
7.6 3.4 0 8\\
NaN NaN NaN 8\\
NaN NaN NaN 8\\
NaN NaN NaN 8\\
NaN NaN NaN 8\\
7.6 3.6 0 8\\
7.6 4.4 0 8\\
NaN NaN NaN 8\\
NaN NaN NaN 8\\
NaN NaN NaN 8\\
NaN NaN NaN 8\\
7.6 4.6 0 8\\
7.6 5.4 0 8\\
NaN NaN NaN 8\\
NaN NaN NaN 8\\
NaN NaN NaN 8\\
NaN NaN NaN 8\\
7.6 5.6 0 8\\
7.6 6.4 0 8\\
NaN NaN NaN 8\\
NaN NaN NaN 8\\
NaN NaN NaN 8\\
NaN NaN NaN 8\\
7.6 6.6 0 8\\
7.6 7.4 0 8\\
NaN NaN NaN 8\\
NaN NaN NaN 8\\
NaN NaN NaN 8\\
NaN NaN NaN 8\\
7.6 7.6 0 8\\
7.6 8.4 0 8\\
NaN NaN NaN 8\\
NaN NaN NaN 8\\
NaN NaN NaN 8\\
NaN NaN NaN 8\\
7.6 8.6 0 8\\
7.6 9.4 0 8\\
NaN NaN NaN 8\\
NaN NaN NaN 8\\
NaN NaN NaN 8\\
NaN NaN NaN 8\\
7.6 9.6 0 8\\
7.6 10.4 0 8\\
NaN NaN NaN 8\\
NaN NaN NaN 8\\
NaN NaN NaN 8\\
NaN NaN NaN 8\\
7.6 10.6 0 8\\
7.6 11.4 0 8\\
NaN NaN NaN 8\\
NaN NaN NaN 8\\
NaN NaN NaN 8\\
7.6 0.6 0 8\\
7.6 0.6 1 8\\
7.6 1.4 1 8\\
7.6 1.4 0 8\\
7.6 0.6 0 8\\
NaN NaN NaN 8\\
7.6 1.6 0 8\\
7.6 1.6 2 8\\
7.6 2.4 2 8\\
7.6 2.4 0 8\\
7.6 1.6 0 8\\
NaN NaN NaN 8\\
7.6 2.6 0 8\\
7.6 2.6 2 8\\
7.6 3.4 2 8\\
7.6 3.4 0 8\\
7.6 2.6 0 8\\
NaN NaN NaN 8\\
7.6 3.6 0 8\\
7.6 3.6 3 8\\
7.6 4.4 3 8\\
7.6 4.4 0 8\\
7.6 3.6 0 8\\
NaN NaN NaN 8\\
7.6 4.6 0 8\\
7.6 4.6 3 8\\
7.6 5.4 3 8\\
7.6 5.4 0 8\\
7.6 4.6 0 8\\
NaN NaN NaN 8\\
7.6 5.6 0 8\\
7.6 5.6 3 8\\
7.6 6.4 3 8\\
7.6 6.4 0 8\\
7.6 5.6 0 8\\
NaN NaN NaN 8\\
7.6 6.6 0 8\\
7.6 6.6 4 8\\
7.6 7.4 4 8\\
7.6 7.4 0 8\\
7.6 6.6 0 8\\
NaN NaN NaN 8\\
7.6 7.6 0 8\\
7.6 7.6 4 8\\
7.6 8.4 4 8\\
7.6 8.4 0 8\\
7.6 7.6 0 8\\
NaN NaN NaN 8\\
7.6 8.6 0 8\\
7.6 8.6 4 8\\
7.6 9.4 4 8\\
7.6 9.4 0 8\\
7.6 8.6 0 8\\
NaN NaN NaN 8\\
7.6 9.6 0 8\\
7.6 9.6 4 8\\
7.6 10.4 4 8\\
7.6 10.4 0 8\\
7.6 9.6 0 8\\
NaN NaN NaN 8\\
7.6 10.6 0 8\\
7.6 10.6 4 8\\
7.6 11.4 4 8\\
7.6 11.4 0 8\\
7.6 10.6 0 8\\
NaN NaN NaN 8\\
8.4 0.6 0 8\\
8.4 0.6 1 8\\
8.4 1.4 1 8\\
8.4 1.4 0 8\\
8.4 0.6 0 8\\
NaN NaN NaN 8\\
8.4 1.6 0 8\\
8.4 1.6 2 8\\
8.4 2.4 2 8\\
8.4 2.4 0 8\\
8.4 1.6 0 8\\
NaN NaN NaN 8\\
8.4 2.6 0 8\\
8.4 2.6 2 8\\
8.4 3.4 2 8\\
8.4 3.4 0 8\\
8.4 2.6 0 8\\
NaN NaN NaN 8\\
8.4 3.6 0 8\\
8.4 3.6 3 8\\
8.4 4.4 3 8\\
8.4 4.4 0 8\\
8.4 3.6 0 8\\
NaN NaN NaN 8\\
8.4 4.6 0 8\\
8.4 4.6 3 8\\
8.4 5.4 3 8\\
8.4 5.4 0 8\\
8.4 4.6 0 8\\
NaN NaN NaN 8\\
8.4 5.6 0 8\\
8.4 5.6 3 8\\
8.4 6.4 3 8\\
8.4 6.4 0 8\\
8.4 5.6 0 8\\
NaN NaN NaN 8\\
8.4 6.6 0 8\\
8.4 6.6 4 8\\
8.4 7.4 4 8\\
8.4 7.4 0 8\\
8.4 6.6 0 8\\
NaN NaN NaN 8\\
8.4 7.6 0 8\\
8.4 7.6 4 8\\
8.4 8.4 4 8\\
8.4 8.4 0 8\\
8.4 7.6 0 8\\
NaN NaN NaN 8\\
8.4 8.6 0 8\\
8.4 8.6 4 8\\
8.4 9.4 4 8\\
8.4 9.4 0 8\\
8.4 8.6 0 8\\
NaN NaN NaN 8\\
8.4 9.6 0 8\\
8.4 9.6 4 8\\
8.4 10.4 4 8\\
8.4 10.4 0 8\\
8.4 9.6 0 8\\
NaN NaN NaN 8\\
8.4 10.6 0 8\\
8.4 10.6 4 8\\
8.4 11.4 4 8\\
8.4 11.4 0 8\\
8.4 10.6 0 8\\
NaN NaN NaN 8\\
NaN NaN NaN 8\\
8.4 0.6 0 8\\
8.4 1.4 0 8\\
NaN NaN NaN 8\\
NaN NaN NaN 8\\
NaN NaN NaN 8\\
NaN NaN NaN 8\\
8.4 1.6 0 8\\
8.4 2.4 0 8\\
NaN NaN NaN 8\\
NaN NaN NaN 8\\
NaN NaN NaN 8\\
NaN NaN NaN 8\\
8.4 2.6 0 8\\
8.4 3.4 0 8\\
NaN NaN NaN 8\\
NaN NaN NaN 8\\
NaN NaN NaN 8\\
NaN NaN NaN 8\\
8.4 3.6 0 8\\
8.4 4.4 0 8\\
NaN NaN NaN 8\\
NaN NaN NaN 8\\
NaN NaN NaN 8\\
NaN NaN NaN 8\\
8.4 4.6 0 8\\
8.4 5.4 0 8\\
NaN NaN NaN 8\\
NaN NaN NaN 8\\
NaN NaN NaN 8\\
NaN NaN NaN 8\\
8.4 5.6 0 8\\
8.4 6.4 0 8\\
NaN NaN NaN 8\\
NaN NaN NaN 8\\
NaN NaN NaN 8\\
NaN NaN NaN 8\\
8.4 6.6 0 8\\
8.4 7.4 0 8\\
NaN NaN NaN 8\\
NaN NaN NaN 8\\
NaN NaN NaN 8\\
NaN NaN NaN 8\\
8.4 7.6 0 8\\
8.4 8.4 0 8\\
NaN NaN NaN 8\\
NaN NaN NaN 8\\
NaN NaN NaN 8\\
NaN NaN NaN 8\\
8.4 8.6 0 8\\
8.4 9.4 0 8\\
NaN NaN NaN 8\\
NaN NaN NaN 8\\
NaN NaN NaN 8\\
NaN NaN NaN 8\\
8.4 9.6 0 8\\
8.4 10.4 0 8\\
NaN NaN NaN 8\\
NaN NaN NaN 8\\
NaN NaN NaN 8\\
NaN NaN NaN 8\\
8.4 10.6 0 8\\
8.4 11.4 0 8\\
NaN NaN NaN 8\\
NaN NaN NaN 8\\
NaN NaN NaN 8\\
};

\addplot3[%
surf,
shader=flat,
draw=black,
point meta=explicit,
mesh/rows=4]
table[row sep=crcr,header=false,meta index=3] {
NaN NaN NaN 7\\
6.6 0.6 0 7\\
6.6 1.4 0 7\\
NaN NaN NaN 7\\
NaN NaN NaN 7\\
NaN NaN NaN 7\\
NaN NaN NaN 7\\
6.6 1.6 0 7\\
6.6 2.4 0 7\\
NaN NaN NaN 7\\
NaN NaN NaN 7\\
NaN NaN NaN 7\\
NaN NaN NaN 7\\
6.6 2.6 0 7\\
6.6 3.4 0 7\\
NaN NaN NaN 7\\
NaN NaN NaN 7\\
NaN NaN NaN 7\\
NaN NaN NaN 7\\
6.6 3.6 0 7\\
6.6 4.4 0 7\\
NaN NaN NaN 7\\
NaN NaN NaN 7\\
NaN NaN NaN 7\\
NaN NaN NaN 7\\
6.6 4.6 0 7\\
6.6 5.4 0 7\\
NaN NaN NaN 7\\
NaN NaN NaN 7\\
NaN NaN NaN 7\\
NaN NaN NaN 7\\
6.6 5.6 0 7\\
6.6 6.4 0 7\\
NaN NaN NaN 7\\
NaN NaN NaN 7\\
NaN NaN NaN 7\\
NaN NaN NaN 7\\
6.6 6.6 0 7\\
6.6 7.4 0 7\\
NaN NaN NaN 7\\
NaN NaN NaN 7\\
NaN NaN NaN 7\\
NaN NaN NaN 7\\
6.6 7.6 0 7\\
6.6 8.4 0 7\\
NaN NaN NaN 7\\
NaN NaN NaN 7\\
NaN NaN NaN 7\\
NaN NaN NaN 7\\
6.6 8.6 0 7\\
6.6 9.4 0 7\\
NaN NaN NaN 7\\
NaN NaN NaN 7\\
NaN NaN NaN 7\\
NaN NaN NaN 7\\
6.6 9.6 0 7\\
6.6 10.4 0 7\\
NaN NaN NaN 7\\
NaN NaN NaN 7\\
NaN NaN NaN 7\\
NaN NaN NaN 7\\
6.6 10.6 0 7\\
6.6 11.4 0 7\\
NaN NaN NaN 7\\
NaN NaN NaN 7\\
NaN NaN NaN 7\\
6.6 0.6 0 7\\
6.6 0.6 1 7\\
6.6 1.4 1 7\\
6.6 1.4 0 7\\
6.6 0.6 0 7\\
NaN NaN NaN 7\\
6.6 1.6 0 7\\
6.6 1.6 2 7\\
6.6 2.4 2 7\\
6.6 2.4 0 7\\
6.6 1.6 0 7\\
NaN NaN NaN 7\\
6.6 2.6 0 7\\
6.6 2.6 2 7\\
6.6 3.4 2 7\\
6.6 3.4 0 7\\
6.6 2.6 0 7\\
NaN NaN NaN 7\\
6.6 3.6 0 7\\
6.6 3.6 3 7\\
6.6 4.4 3 7\\
6.6 4.4 0 7\\
6.6 3.6 0 7\\
NaN NaN NaN 7\\
6.6 4.6 0 7\\
6.6 4.6 3 7\\
6.6 5.4 3 7\\
6.6 5.4 0 7\\
6.6 4.6 0 7\\
NaN NaN NaN 7\\
6.6 5.6 0 7\\
6.6 5.6 3 7\\
6.6 6.4 3 7\\
6.6 6.4 0 7\\
6.6 5.6 0 7\\
NaN NaN NaN 7\\
6.6 6.6 0 7\\
6.6 6.6 4 7\\
6.6 7.4 4 7\\
6.6 7.4 0 7\\
6.6 6.6 0 7\\
NaN NaN NaN 7\\
6.6 7.6 0 7\\
6.6 7.6 4 7\\
6.6 8.4 4 7\\
6.6 8.4 0 7\\
6.6 7.6 0 7\\
NaN NaN NaN 7\\
6.6 8.6 0 7\\
6.6 8.6 4 7\\
6.6 9.4 4 7\\
6.6 9.4 0 7\\
6.6 8.6 0 7\\
NaN NaN NaN 7\\
6.6 9.6 0 7\\
6.6 9.6 4 7\\
6.6 10.4 4 7\\
6.6 10.4 0 7\\
6.6 9.6 0 7\\
NaN NaN NaN 7\\
6.6 10.6 0 7\\
6.6 10.6 4 7\\
6.6 11.4 4 7\\
6.6 11.4 0 7\\
6.6 10.6 0 7\\
NaN NaN NaN 7\\
7.4 0.6 0 7\\
7.4 0.6 1 7\\
7.4 1.4 1 7\\
7.4 1.4 0 7\\
7.4 0.6 0 7\\
NaN NaN NaN 7\\
7.4 1.6 0 7\\
7.4 1.6 2 7\\
7.4 2.4 2 7\\
7.4 2.4 0 7\\
7.4 1.6 0 7\\
NaN NaN NaN 7\\
7.4 2.6 0 7\\
7.4 2.6 2 7\\
7.4 3.4 2 7\\
7.4 3.4 0 7\\
7.4 2.6 0 7\\
NaN NaN NaN 7\\
7.4 3.6 0 7\\
7.4 3.6 3 7\\
7.4 4.4 3 7\\
7.4 4.4 0 7\\
7.4 3.6 0 7\\
NaN NaN NaN 7\\
7.4 4.6 0 7\\
7.4 4.6 3 7\\
7.4 5.4 3 7\\
7.4 5.4 0 7\\
7.4 4.6 0 7\\
NaN NaN NaN 7\\
7.4 5.6 0 7\\
7.4 5.6 3 7\\
7.4 6.4 3 7\\
7.4 6.4 0 7\\
7.4 5.6 0 7\\
NaN NaN NaN 7\\
7.4 6.6 0 7\\
7.4 6.6 4 7\\
7.4 7.4 4 7\\
7.4 7.4 0 7\\
7.4 6.6 0 7\\
NaN NaN NaN 7\\
7.4 7.6 0 7\\
7.4 7.6 4 7\\
7.4 8.4 4 7\\
7.4 8.4 0 7\\
7.4 7.6 0 7\\
NaN NaN NaN 7\\
7.4 8.6 0 7\\
7.4 8.6 4 7\\
7.4 9.4 4 7\\
7.4 9.4 0 7\\
7.4 8.6 0 7\\
NaN NaN NaN 7\\
7.4 9.6 0 7\\
7.4 9.6 4 7\\
7.4 10.4 4 7\\
7.4 10.4 0 7\\
7.4 9.6 0 7\\
NaN NaN NaN 7\\
7.4 10.6 0 7\\
7.4 10.6 4 7\\
7.4 11.4 4 7\\
7.4 11.4 0 7\\
7.4 10.6 0 7\\
NaN NaN NaN 7\\
NaN NaN NaN 7\\
7.4 0.6 0 7\\
7.4 1.4 0 7\\
NaN NaN NaN 7\\
NaN NaN NaN 7\\
NaN NaN NaN 7\\
NaN NaN NaN 7\\
7.4 1.6 0 7\\
7.4 2.4 0 7\\
NaN NaN NaN 7\\
NaN NaN NaN 7\\
NaN NaN NaN 7\\
NaN NaN NaN 7\\
7.4 2.6 0 7\\
7.4 3.4 0 7\\
NaN NaN NaN 7\\
NaN NaN NaN 7\\
NaN NaN NaN 7\\
NaN NaN NaN 7\\
7.4 3.6 0 7\\
7.4 4.4 0 7\\
NaN NaN NaN 7\\
NaN NaN NaN 7\\
NaN NaN NaN 7\\
NaN NaN NaN 7\\
7.4 4.6 0 7\\
7.4 5.4 0 7\\
NaN NaN NaN 7\\
NaN NaN NaN 7\\
NaN NaN NaN 7\\
NaN NaN NaN 7\\
7.4 5.6 0 7\\
7.4 6.4 0 7\\
NaN NaN NaN 7\\
NaN NaN NaN 7\\
NaN NaN NaN 7\\
NaN NaN NaN 7\\
7.4 6.6 0 7\\
7.4 7.4 0 7\\
NaN NaN NaN 7\\
NaN NaN NaN 7\\
NaN NaN NaN 7\\
NaN NaN NaN 7\\
7.4 7.6 0 7\\
7.4 8.4 0 7\\
NaN NaN NaN 7\\
NaN NaN NaN 7\\
NaN NaN NaN 7\\
NaN NaN NaN 7\\
7.4 8.6 0 7\\
7.4 9.4 0 7\\
NaN NaN NaN 7\\
NaN NaN NaN 7\\
NaN NaN NaN 7\\
NaN NaN NaN 7\\
7.4 9.6 0 7\\
7.4 10.4 0 7\\
NaN NaN NaN 7\\
NaN NaN NaN 7\\
NaN NaN NaN 7\\
NaN NaN NaN 7\\
7.4 10.6 0 7\\
7.4 11.4 0 7\\
NaN NaN NaN 7\\
NaN NaN NaN 7\\
NaN NaN NaN 7\\
};

\addplot3[%
surf,
shader=flat,
draw=black,
point meta=explicit,
mesh/rows=4]
table[row sep=crcr,header=false,meta index=3] {
NaN NaN NaN 6\\
5.6 0.6 0 6\\
5.6 1.4 0 6\\
NaN NaN NaN 6\\
NaN NaN NaN 6\\
NaN NaN NaN 6\\
NaN NaN NaN 6\\
5.6 1.6 0 6\\
5.6 2.4 0 6\\
NaN NaN NaN 6\\
NaN NaN NaN 6\\
NaN NaN NaN 6\\
NaN NaN NaN 6\\
5.6 2.6 0 6\\
5.6 3.4 0 6\\
NaN NaN NaN 6\\
NaN NaN NaN 6\\
NaN NaN NaN 6\\
NaN NaN NaN 6\\
5.6 3.6 0 6\\
5.6 4.4 0 6\\
NaN NaN NaN 6\\
NaN NaN NaN 6\\
NaN NaN NaN 6\\
NaN NaN NaN 6\\
5.6 4.6 0 6\\
5.6 5.4 0 6\\
NaN NaN NaN 6\\
NaN NaN NaN 6\\
NaN NaN NaN 6\\
NaN NaN NaN 6\\
5.6 5.6 0 6\\
5.6 6.4 0 6\\
NaN NaN NaN 6\\
NaN NaN NaN 6\\
NaN NaN NaN 6\\
NaN NaN NaN 6\\
5.6 6.6 0 6\\
5.6 7.4 0 6\\
NaN NaN NaN 6\\
NaN NaN NaN 6\\
NaN NaN NaN 6\\
NaN NaN NaN 6\\
5.6 7.6 0 6\\
5.6 8.4 0 6\\
NaN NaN NaN 6\\
NaN NaN NaN 6\\
NaN NaN NaN 6\\
NaN NaN NaN 6\\
5.6 8.6 0 6\\
5.6 9.4 0 6\\
NaN NaN NaN 6\\
NaN NaN NaN 6\\
NaN NaN NaN 6\\
NaN NaN NaN 6\\
5.6 9.6 0 6\\
5.6 10.4 0 6\\
NaN NaN NaN 6\\
NaN NaN NaN 6\\
NaN NaN NaN 6\\
NaN NaN NaN 6\\
5.6 10.6 0 6\\
5.6 11.4 0 6\\
NaN NaN NaN 6\\
NaN NaN NaN 6\\
NaN NaN NaN 6\\
5.6 0.6 0 6\\
5.6 0.6 1 6\\
5.6 1.4 1 6\\
5.6 1.4 0 6\\
5.6 0.6 0 6\\
NaN NaN NaN 6\\
5.6 1.6 0 6\\
5.6 1.6 2 6\\
5.6 2.4 2 6\\
5.6 2.4 0 6\\
5.6 1.6 0 6\\
NaN NaN NaN 6\\
5.6 2.6 0 6\\
5.6 2.6 2 6\\
5.6 3.4 2 6\\
5.6 3.4 0 6\\
5.6 2.6 0 6\\
NaN NaN NaN 6\\
5.6 3.6 0 6\\
5.6 3.6 3 6\\
5.6 4.4 3 6\\
5.6 4.4 0 6\\
5.6 3.6 0 6\\
NaN NaN NaN 6\\
5.6 4.6 0 6\\
5.6 4.6 3 6\\
5.6 5.4 3 6\\
5.6 5.4 0 6\\
5.6 4.6 0 6\\
NaN NaN NaN 6\\
5.6 5.6 0 6\\
5.6 5.6 3 6\\
5.6 6.4 3 6\\
5.6 6.4 0 6\\
5.6 5.6 0 6\\
NaN NaN NaN 6\\
5.6 6.6 0 6\\
5.6 6.6 3 6\\
5.6 7.4 3 6\\
5.6 7.4 0 6\\
5.6 6.6 0 6\\
NaN NaN NaN 6\\
5.6 7.6 0 6\\
5.6 7.6 3 6\\
5.6 8.4 3 6\\
5.6 8.4 0 6\\
5.6 7.6 0 6\\
NaN NaN NaN 6\\
5.6 8.6 0 6\\
5.6 8.6 4 6\\
5.6 9.4 4 6\\
5.6 9.4 0 6\\
5.6 8.6 0 6\\
NaN NaN NaN 6\\
5.6 9.6 0 6\\
5.6 9.6 4 6\\
5.6 10.4 4 6\\
5.6 10.4 0 6\\
5.6 9.6 0 6\\
NaN NaN NaN 6\\
5.6 10.6 0 6\\
5.6 10.6 4 6\\
5.6 11.4 4 6\\
5.6 11.4 0 6\\
5.6 10.6 0 6\\
NaN NaN NaN 6\\
6.4 0.6 0 6\\
6.4 0.6 1 6\\
6.4 1.4 1 6\\
6.4 1.4 0 6\\
6.4 0.6 0 6\\
NaN NaN NaN 6\\
6.4 1.6 0 6\\
6.4 1.6 2 6\\
6.4 2.4 2 6\\
6.4 2.4 0 6\\
6.4 1.6 0 6\\
NaN NaN NaN 6\\
6.4 2.6 0 6\\
6.4 2.6 2 6\\
6.4 3.4 2 6\\
6.4 3.4 0 6\\
6.4 2.6 0 6\\
NaN NaN NaN 6\\
6.4 3.6 0 6\\
6.4 3.6 3 6\\
6.4 4.4 3 6\\
6.4 4.4 0 6\\
6.4 3.6 0 6\\
NaN NaN NaN 6\\
6.4 4.6 0 6\\
6.4 4.6 3 6\\
6.4 5.4 3 6\\
6.4 5.4 0 6\\
6.4 4.6 0 6\\
NaN NaN NaN 6\\
6.4 5.6 0 6\\
6.4 5.6 3 6\\
6.4 6.4 3 6\\
6.4 6.4 0 6\\
6.4 5.6 0 6\\
NaN NaN NaN 6\\
6.4 6.6 0 6\\
6.4 6.6 3 6\\
6.4 7.4 3 6\\
6.4 7.4 0 6\\
6.4 6.6 0 6\\
NaN NaN NaN 6\\
6.4 7.6 0 6\\
6.4 7.6 3 6\\
6.4 8.4 3 6\\
6.4 8.4 0 6\\
6.4 7.6 0 6\\
NaN NaN NaN 6\\
6.4 8.6 0 6\\
6.4 8.6 4 6\\
6.4 9.4 4 6\\
6.4 9.4 0 6\\
6.4 8.6 0 6\\
NaN NaN NaN 6\\
6.4 9.6 0 6\\
6.4 9.6 4 6\\
6.4 10.4 4 6\\
6.4 10.4 0 6\\
6.4 9.6 0 6\\
NaN NaN NaN 6\\
6.4 10.6 0 6\\
6.4 10.6 4 6\\
6.4 11.4 4 6\\
6.4 11.4 0 6\\
6.4 10.6 0 6\\
NaN NaN NaN 6\\
NaN NaN NaN 6\\
6.4 0.6 0 6\\
6.4 1.4 0 6\\
NaN NaN NaN 6\\
NaN NaN NaN 6\\
NaN NaN NaN 6\\
NaN NaN NaN 6\\
6.4 1.6 0 6\\
6.4 2.4 0 6\\
NaN NaN NaN 6\\
NaN NaN NaN 6\\
NaN NaN NaN 6\\
NaN NaN NaN 6\\
6.4 2.6 0 6\\
6.4 3.4 0 6\\
NaN NaN NaN 6\\
NaN NaN NaN 6\\
NaN NaN NaN 6\\
NaN NaN NaN 6\\
6.4 3.6 0 6\\
6.4 4.4 0 6\\
NaN NaN NaN 6\\
NaN NaN NaN 6\\
NaN NaN NaN 6\\
NaN NaN NaN 6\\
6.4 4.6 0 6\\
6.4 5.4 0 6\\
NaN NaN NaN 6\\
NaN NaN NaN 6\\
NaN NaN NaN 6\\
NaN NaN NaN 6\\
6.4 5.6 0 6\\
6.4 6.4 0 6\\
NaN NaN NaN 6\\
NaN NaN NaN 6\\
NaN NaN NaN 6\\
NaN NaN NaN 6\\
6.4 6.6 0 6\\
6.4 7.4 0 6\\
NaN NaN NaN 6\\
NaN NaN NaN 6\\
NaN NaN NaN 6\\
NaN NaN NaN 6\\
6.4 7.6 0 6\\
6.4 8.4 0 6\\
NaN NaN NaN 6\\
NaN NaN NaN 6\\
NaN NaN NaN 6\\
NaN NaN NaN 6\\
6.4 8.6 0 6\\
6.4 9.4 0 6\\
NaN NaN NaN 6\\
NaN NaN NaN 6\\
NaN NaN NaN 6\\
NaN NaN NaN 6\\
6.4 9.6 0 6\\
6.4 10.4 0 6\\
NaN NaN NaN 6\\
NaN NaN NaN 6\\
NaN NaN NaN 6\\
NaN NaN NaN 6\\
6.4 10.6 0 6\\
6.4 11.4 0 6\\
NaN NaN NaN 6\\
NaN NaN NaN 6\\
NaN NaN NaN 6\\
};

\addplot3[%
surf,
shader=flat,
draw=black,
point meta=explicit,
mesh/rows=4]
table[row sep=crcr,header=false,meta index=3] {
NaN NaN NaN 5\\
4.6 0.6 0 5\\
4.6 1.4 0 5\\
NaN NaN NaN 5\\
NaN NaN NaN 5\\
NaN NaN NaN 5\\
NaN NaN NaN 5\\
4.6 1.6 0 5\\
4.6 2.4 0 5\\
NaN NaN NaN 5\\
NaN NaN NaN 5\\
NaN NaN NaN 5\\
NaN NaN NaN 5\\
4.6 2.6 0 5\\
4.6 3.4 0 5\\
NaN NaN NaN 5\\
NaN NaN NaN 5\\
NaN NaN NaN 5\\
NaN NaN NaN 5\\
4.6 3.6 0 5\\
4.6 4.4 0 5\\
NaN NaN NaN 5\\
NaN NaN NaN 5\\
NaN NaN NaN 5\\
NaN NaN NaN 5\\
4.6 4.6 0 5\\
4.6 5.4 0 5\\
NaN NaN NaN 5\\
NaN NaN NaN 5\\
NaN NaN NaN 5\\
NaN NaN NaN 5\\
4.6 5.6 0 5\\
4.6 6.4 0 5\\
NaN NaN NaN 5\\
NaN NaN NaN 5\\
NaN NaN NaN 5\\
NaN NaN NaN 5\\
4.6 6.6 0 5\\
4.6 7.4 0 5\\
NaN NaN NaN 5\\
NaN NaN NaN 5\\
NaN NaN NaN 5\\
NaN NaN NaN 5\\
4.6 7.6 0 5\\
4.6 8.4 0 5\\
NaN NaN NaN 5\\
NaN NaN NaN 5\\
NaN NaN NaN 5\\
NaN NaN NaN 5\\
4.6 8.6 0 5\\
4.6 9.4 0 5\\
NaN NaN NaN 5\\
NaN NaN NaN 5\\
NaN NaN NaN 5\\
NaN NaN NaN 5\\
4.6 9.6 0 5\\
4.6 10.4 0 5\\
NaN NaN NaN 5\\
NaN NaN NaN 5\\
NaN NaN NaN 5\\
NaN NaN NaN 5\\
4.6 10.6 0 5\\
4.6 11.4 0 5\\
NaN NaN NaN 5\\
NaN NaN NaN 5\\
NaN NaN NaN 5\\
4.6 0.6 0 5\\
4.6 0.6 0 5\\
4.6 1.4 0 5\\
4.6 1.4 0 5\\
4.6 0.6 0 5\\
NaN NaN NaN 5\\
4.6 1.6 0 5\\
4.6 1.6 2 5\\
4.6 2.4 2 5\\
4.6 2.4 0 5\\
4.6 1.6 0 5\\
NaN NaN NaN 5\\
4.6 2.6 0 5\\
4.6 2.6 2 5\\
4.6 3.4 2 5\\
4.6 3.4 0 5\\
4.6 2.6 0 5\\
NaN NaN NaN 5\\
4.6 3.6 0 5\\
4.6 3.6 3 5\\
4.6 4.4 3 5\\
4.6 4.4 0 5\\
4.6 3.6 0 5\\
NaN NaN NaN 5\\
4.6 4.6 0 5\\
4.6 4.6 3 5\\
4.6 5.4 3 5\\
4.6 5.4 0 5\\
4.6 4.6 0 5\\
NaN NaN NaN 5\\
4.6 5.6 0 5\\
4.6 5.6 3 5\\
4.6 6.4 3 5\\
4.6 6.4 0 5\\
4.6 5.6 0 5\\
NaN NaN NaN 5\\
4.6 6.6 0 5\\
4.6 6.6 3 5\\
4.6 7.4 3 5\\
4.6 7.4 0 5\\
4.6 6.6 0 5\\
NaN NaN NaN 5\\
4.6 7.6 0 5\\
4.6 7.6 3 5\\
4.6 8.4 3 5\\
4.6 8.4 0 5\\
4.6 7.6 0 5\\
NaN NaN NaN 5\\
4.6 8.6 0 5\\
4.6 8.6 4 5\\
4.6 9.4 4 5\\
4.6 9.4 0 5\\
4.6 8.6 0 5\\
NaN NaN NaN 5\\
4.6 9.6 0 5\\
4.6 9.6 4 5\\
4.6 10.4 4 5\\
4.6 10.4 0 5\\
4.6 9.6 0 5\\
NaN NaN NaN 5\\
4.6 10.6 0 5\\
4.6 10.6 4 5\\
4.6 11.4 4 5\\
4.6 11.4 0 5\\
4.6 10.6 0 5\\
NaN NaN NaN 5\\
5.4 0.6 0 5\\
5.4 0.6 0 5\\
5.4 1.4 0 5\\
5.4 1.4 0 5\\
5.4 0.6 0 5\\
NaN NaN NaN 5\\
5.4 1.6 0 5\\
5.4 1.6 2 5\\
5.4 2.4 2 5\\
5.4 2.4 0 5\\
5.4 1.6 0 5\\
NaN NaN NaN 5\\
5.4 2.6 0 5\\
5.4 2.6 2 5\\
5.4 3.4 2 5\\
5.4 3.4 0 5\\
5.4 2.6 0 5\\
NaN NaN NaN 5\\
5.4 3.6 0 5\\
5.4 3.6 3 5\\
5.4 4.4 3 5\\
5.4 4.4 0 5\\
5.4 3.6 0 5\\
NaN NaN NaN 5\\
5.4 4.6 0 5\\
5.4 4.6 3 5\\
5.4 5.4 3 5\\
5.4 5.4 0 5\\
5.4 4.6 0 5\\
NaN NaN NaN 5\\
5.4 5.6 0 5\\
5.4 5.6 3 5\\
5.4 6.4 3 5\\
5.4 6.4 0 5\\
5.4 5.6 0 5\\
NaN NaN NaN 5\\
5.4 6.6 0 5\\
5.4 6.6 3 5\\
5.4 7.4 3 5\\
5.4 7.4 0 5\\
5.4 6.6 0 5\\
NaN NaN NaN 5\\
5.4 7.6 0 5\\
5.4 7.6 3 5\\
5.4 8.4 3 5\\
5.4 8.4 0 5\\
5.4 7.6 0 5\\
NaN NaN NaN 5\\
5.4 8.6 0 5\\
5.4 8.6 4 5\\
5.4 9.4 4 5\\
5.4 9.4 0 5\\
5.4 8.6 0 5\\
NaN NaN NaN 5\\
5.4 9.6 0 5\\
5.4 9.6 4 5\\
5.4 10.4 4 5\\
5.4 10.4 0 5\\
5.4 9.6 0 5\\
NaN NaN NaN 5\\
5.4 10.6 0 5\\
5.4 10.6 4 5\\
5.4 11.4 4 5\\
5.4 11.4 0 5\\
5.4 10.6 0 5\\
NaN NaN NaN 5\\
NaN NaN NaN 5\\
5.4 0.6 0 5\\
5.4 1.4 0 5\\
NaN NaN NaN 5\\
NaN NaN NaN 5\\
NaN NaN NaN 5\\
NaN NaN NaN 5\\
5.4 1.6 0 5\\
5.4 2.4 0 5\\
NaN NaN NaN 5\\
NaN NaN NaN 5\\
NaN NaN NaN 5\\
NaN NaN NaN 5\\
5.4 2.6 0 5\\
5.4 3.4 0 5\\
NaN NaN NaN 5\\
NaN NaN NaN 5\\
NaN NaN NaN 5\\
NaN NaN NaN 5\\
5.4 3.6 0 5\\
5.4 4.4 0 5\\
NaN NaN NaN 5\\
NaN NaN NaN 5\\
NaN NaN NaN 5\\
NaN NaN NaN 5\\
5.4 4.6 0 5\\
5.4 5.4 0 5\\
NaN NaN NaN 5\\
NaN NaN NaN 5\\
NaN NaN NaN 5\\
NaN NaN NaN 5\\
5.4 5.6 0 5\\
5.4 6.4 0 5\\
NaN NaN NaN 5\\
NaN NaN NaN 5\\
NaN NaN NaN 5\\
NaN NaN NaN 5\\
5.4 6.6 0 5\\
5.4 7.4 0 5\\
NaN NaN NaN 5\\
NaN NaN NaN 5\\
NaN NaN NaN 5\\
NaN NaN NaN 5\\
5.4 7.6 0 5\\
5.4 8.4 0 5\\
NaN NaN NaN 5\\
NaN NaN NaN 5\\
NaN NaN NaN 5\\
NaN NaN NaN 5\\
5.4 8.6 0 5\\
5.4 9.4 0 5\\
NaN NaN NaN 5\\
NaN NaN NaN 5\\
NaN NaN NaN 5\\
NaN NaN NaN 5\\
5.4 9.6 0 5\\
5.4 10.4 0 5\\
NaN NaN NaN 5\\
NaN NaN NaN 5\\
NaN NaN NaN 5\\
NaN NaN NaN 5\\
5.4 10.6 0 5\\
5.4 11.4 0 5\\
NaN NaN NaN 5\\
NaN NaN NaN 5\\
NaN NaN NaN 5\\
};

\addplot3[%
surf,
shader=flat,
draw=black,
point meta=explicit,
mesh/rows=4]
table[row sep=crcr,header=false,meta index=3] {
NaN NaN NaN 4\\
3.6 0.6 0 4\\
3.6 1.4 0 4\\
NaN NaN NaN 4\\
NaN NaN NaN 4\\
NaN NaN NaN 4\\
NaN NaN NaN 4\\
3.6 1.6 0 4\\
3.6 2.4 0 4\\
NaN NaN NaN 4\\
NaN NaN NaN 4\\
NaN NaN NaN 4\\
NaN NaN NaN 4\\
3.6 2.6 0 4\\
3.6 3.4 0 4\\
NaN NaN NaN 4\\
NaN NaN NaN 4\\
NaN NaN NaN 4\\
NaN NaN NaN 4\\
3.6 3.6 0 4\\
3.6 4.4 0 4\\
NaN NaN NaN 4\\
NaN NaN NaN 4\\
NaN NaN NaN 4\\
NaN NaN NaN 4\\
3.6 4.6 0 4\\
3.6 5.4 0 4\\
NaN NaN NaN 4\\
NaN NaN NaN 4\\
NaN NaN NaN 4\\
NaN NaN NaN 4\\
3.6 5.6 0 4\\
3.6 6.4 0 4\\
NaN NaN NaN 4\\
NaN NaN NaN 4\\
NaN NaN NaN 4\\
NaN NaN NaN 4\\
3.6 6.6 0 4\\
3.6 7.4 0 4\\
NaN NaN NaN 4\\
NaN NaN NaN 4\\
NaN NaN NaN 4\\
NaN NaN NaN 4\\
3.6 7.6 0 4\\
3.6 8.4 0 4\\
NaN NaN NaN 4\\
NaN NaN NaN 4\\
NaN NaN NaN 4\\
NaN NaN NaN 4\\
3.6 8.6 0 4\\
3.6 9.4 0 4\\
NaN NaN NaN 4\\
NaN NaN NaN 4\\
NaN NaN NaN 4\\
NaN NaN NaN 4\\
3.6 9.6 0 4\\
3.6 10.4 0 4\\
NaN NaN NaN 4\\
NaN NaN NaN 4\\
NaN NaN NaN 4\\
NaN NaN NaN 4\\
3.6 10.6 0 4\\
3.6 11.4 0 4\\
NaN NaN NaN 4\\
NaN NaN NaN 4\\
NaN NaN NaN 4\\
3.6 0.6 0 4\\
3.6 0.6 0 4\\
3.6 1.4 0 4\\
3.6 1.4 0 4\\
3.6 0.6 0 4\\
NaN NaN NaN 4\\
3.6 1.6 0 4\\
3.6 1.6 2 4\\
3.6 2.4 2 4\\
3.6 2.4 0 4\\
3.6 1.6 0 4\\
NaN NaN NaN 4\\
3.6 2.6 0 4\\
3.6 2.6 2 4\\
3.6 3.4 2 4\\
3.6 3.4 0 4\\
3.6 2.6 0 4\\
NaN NaN NaN 4\\
3.6 3.6 0 4\\
3.6 3.6 3 4\\
3.6 4.4 3 4\\
3.6 4.4 0 4\\
3.6 3.6 0 4\\
NaN NaN NaN 4\\
3.6 4.6 0 4\\
3.6 4.6 3 4\\
3.6 5.4 3 4\\
3.6 5.4 0 4\\
3.6 4.6 0 4\\
NaN NaN NaN 4\\
3.6 5.6 0 4\\
3.6 5.6 3 4\\
3.6 6.4 3 4\\
3.6 6.4 0 4\\
3.6 5.6 0 4\\
NaN NaN NaN 4\\
3.6 6.6 0 4\\
3.6 6.6 3 4\\
3.6 7.4 3 4\\
3.6 7.4 0 4\\
3.6 6.6 0 4\\
NaN NaN NaN 4\\
3.6 7.6 0 4\\
3.6 7.6 3 4\\
3.6 8.4 3 4\\
3.6 8.4 0 4\\
3.6 7.6 0 4\\
NaN NaN NaN 4\\
3.6 8.6 0 4\\
3.6 8.6 4 4\\
3.6 9.4 4 4\\
3.6 9.4 0 4\\
3.6 8.6 0 4\\
NaN NaN NaN 4\\
3.6 9.6 0 4\\
3.6 9.6 4 4\\
3.6 10.4 4 4\\
3.6 10.4 0 4\\
3.6 9.6 0 4\\
NaN NaN NaN 4\\
3.6 10.6 0 4\\
3.6 10.6 4 4\\
3.6 11.4 4 4\\
3.6 11.4 0 4\\
3.6 10.6 0 4\\
NaN NaN NaN 4\\
4.4 0.6 0 4\\
4.4 0.6 0 4\\
4.4 1.4 0 4\\
4.4 1.4 0 4\\
4.4 0.6 0 4\\
NaN NaN NaN 4\\
4.4 1.6 0 4\\
4.4 1.6 2 4\\
4.4 2.4 2 4\\
4.4 2.4 0 4\\
4.4 1.6 0 4\\
NaN NaN NaN 4\\
4.4 2.6 0 4\\
4.4 2.6 2 4\\
4.4 3.4 2 4\\
4.4 3.4 0 4\\
4.4 2.6 0 4\\
NaN NaN NaN 4\\
4.4 3.6 0 4\\
4.4 3.6 3 4\\
4.4 4.4 3 4\\
4.4 4.4 0 4\\
4.4 3.6 0 4\\
NaN NaN NaN 4\\
4.4 4.6 0 4\\
4.4 4.6 3 4\\
4.4 5.4 3 4\\
4.4 5.4 0 4\\
4.4 4.6 0 4\\
NaN NaN NaN 4\\
4.4 5.6 0 4\\
4.4 5.6 3 4\\
4.4 6.4 3 4\\
4.4 6.4 0 4\\
4.4 5.6 0 4\\
NaN NaN NaN 4\\
4.4 6.6 0 4\\
4.4 6.6 3 4\\
4.4 7.4 3 4\\
4.4 7.4 0 4\\
4.4 6.6 0 4\\
NaN NaN NaN 4\\
4.4 7.6 0 4\\
4.4 7.6 3 4\\
4.4 8.4 3 4\\
4.4 8.4 0 4\\
4.4 7.6 0 4\\
NaN NaN NaN 4\\
4.4 8.6 0 4\\
4.4 8.6 4 4\\
4.4 9.4 4 4\\
4.4 9.4 0 4\\
4.4 8.6 0 4\\
NaN NaN NaN 4\\
4.4 9.6 0 4\\
4.4 9.6 4 4\\
4.4 10.4 4 4\\
4.4 10.4 0 4\\
4.4 9.6 0 4\\
NaN NaN NaN 4\\
4.4 10.6 0 4\\
4.4 10.6 4 4\\
4.4 11.4 4 4\\
4.4 11.4 0 4\\
4.4 10.6 0 4\\
NaN NaN NaN 4\\
NaN NaN NaN 4\\
4.4 0.6 0 4\\
4.4 1.4 0 4\\
NaN NaN NaN 4\\
NaN NaN NaN 4\\
NaN NaN NaN 4\\
NaN NaN NaN 4\\
4.4 1.6 0 4\\
4.4 2.4 0 4\\
NaN NaN NaN 4\\
NaN NaN NaN 4\\
NaN NaN NaN 4\\
NaN NaN NaN 4\\
4.4 2.6 0 4\\
4.4 3.4 0 4\\
NaN NaN NaN 4\\
NaN NaN NaN 4\\
NaN NaN NaN 4\\
NaN NaN NaN 4\\
4.4 3.6 0 4\\
4.4 4.4 0 4\\
NaN NaN NaN 4\\
NaN NaN NaN 4\\
NaN NaN NaN 4\\
NaN NaN NaN 4\\
4.4 4.6 0 4\\
4.4 5.4 0 4\\
NaN NaN NaN 4\\
NaN NaN NaN 4\\
NaN NaN NaN 4\\
NaN NaN NaN 4\\
4.4 5.6 0 4\\
4.4 6.4 0 4\\
NaN NaN NaN 4\\
NaN NaN NaN 4\\
NaN NaN NaN 4\\
NaN NaN NaN 4\\
4.4 6.6 0 4\\
4.4 7.4 0 4\\
NaN NaN NaN 4\\
NaN NaN NaN 4\\
NaN NaN NaN 4\\
NaN NaN NaN 4\\
4.4 7.6 0 4\\
4.4 8.4 0 4\\
NaN NaN NaN 4\\
NaN NaN NaN 4\\
NaN NaN NaN 4\\
NaN NaN NaN 4\\
4.4 8.6 0 4\\
4.4 9.4 0 4\\
NaN NaN NaN 4\\
NaN NaN NaN 4\\
NaN NaN NaN 4\\
NaN NaN NaN 4\\
4.4 9.6 0 4\\
4.4 10.4 0 4\\
NaN NaN NaN 4\\
NaN NaN NaN 4\\
NaN NaN NaN 4\\
NaN NaN NaN 4\\
4.4 10.6 0 4\\
4.4 11.4 0 4\\
NaN NaN NaN 4\\
NaN NaN NaN 4\\
NaN NaN NaN 4\\
};

\addplot3[%
surf,
shader=flat,
draw=black,
point meta=explicit,
mesh/rows=4]
table[row sep=crcr,header=false,meta index=3] {
NaN NaN NaN 3\\
2.6 0.6 0 3\\
2.6 1.4 0 3\\
NaN NaN NaN 3\\
NaN NaN NaN 3\\
NaN NaN NaN 3\\
NaN NaN NaN 3\\
2.6 1.6 0 3\\
2.6 2.4 0 3\\
NaN NaN NaN 3\\
NaN NaN NaN 3\\
NaN NaN NaN 3\\
NaN NaN NaN 3\\
2.6 2.6 0 3\\
2.6 3.4 0 3\\
NaN NaN NaN 3\\
NaN NaN NaN 3\\
NaN NaN NaN 3\\
NaN NaN NaN 3\\
2.6 3.6 0 3\\
2.6 4.4 0 3\\
NaN NaN NaN 3\\
NaN NaN NaN 3\\
NaN NaN NaN 3\\
NaN NaN NaN 3\\
2.6 4.6 0 3\\
2.6 5.4 0 3\\
NaN NaN NaN 3\\
NaN NaN NaN 3\\
NaN NaN NaN 3\\
NaN NaN NaN 3\\
2.6 5.6 0 3\\
2.6 6.4 0 3\\
NaN NaN NaN 3\\
NaN NaN NaN 3\\
NaN NaN NaN 3\\
NaN NaN NaN 3\\
2.6 6.6 0 3\\
2.6 7.4 0 3\\
NaN NaN NaN 3\\
NaN NaN NaN 3\\
NaN NaN NaN 3\\
NaN NaN NaN 3\\
2.6 7.6 0 3\\
2.6 8.4 0 3\\
NaN NaN NaN 3\\
NaN NaN NaN 3\\
NaN NaN NaN 3\\
NaN NaN NaN 3\\
2.6 8.6 0 3\\
2.6 9.4 0 3\\
NaN NaN NaN 3\\
NaN NaN NaN 3\\
NaN NaN NaN 3\\
NaN NaN NaN 3\\
2.6 9.6 0 3\\
2.6 10.4 0 3\\
NaN NaN NaN 3\\
NaN NaN NaN 3\\
NaN NaN NaN 3\\
NaN NaN NaN 3\\
2.6 10.6 0 3\\
2.6 11.4 0 3\\
NaN NaN NaN 3\\
NaN NaN NaN 3\\
NaN NaN NaN 3\\
2.6 0.6 0 3\\
2.6 0.6 0 3\\
2.6 1.4 0 3\\
2.6 1.4 0 3\\
2.6 0.6 0 3\\
NaN NaN NaN 3\\
2.6 1.6 0 3\\
2.6 1.6 2 3\\
2.6 2.4 2 3\\
2.6 2.4 0 3\\
2.6 1.6 0 3\\
NaN NaN NaN 3\\
2.6 2.6 0 3\\
2.6 2.6 2 3\\
2.6 3.4 2 3\\
2.6 3.4 0 3\\
2.6 2.6 0 3\\
NaN NaN NaN 3\\
2.6 3.6 0 3\\
2.6 3.6 3 3\\
2.6 4.4 3 3\\
2.6 4.4 0 3\\
2.6 3.6 0 3\\
NaN NaN NaN 3\\
2.6 4.6 0 3\\
2.6 4.6 3 3\\
2.6 5.4 3 3\\
2.6 5.4 0 3\\
2.6 4.6 0 3\\
NaN NaN NaN 3\\
2.6 5.6 0 3\\
2.6 5.6 3 3\\
2.6 6.4 3 3\\
2.6 6.4 0 3\\
2.6 5.6 0 3\\
NaN NaN NaN 3\\
2.6 6.6 0 3\\
2.6 6.6 3 3\\
2.6 7.4 3 3\\
2.6 7.4 0 3\\
2.6 6.6 0 3\\
NaN NaN NaN 3\\
2.6 7.6 0 3\\
2.6 7.6 3 3\\
2.6 8.4 3 3\\
2.6 8.4 0 3\\
2.6 7.6 0 3\\
NaN NaN NaN 3\\
2.6 8.6 0 3\\
2.6 8.6 4 3\\
2.6 9.4 4 3\\
2.6 9.4 0 3\\
2.6 8.6 0 3\\
NaN NaN NaN 3\\
2.6 9.6 0 3\\
2.6 9.6 4 3\\
2.6 10.4 4 3\\
2.6 10.4 0 3\\
2.6 9.6 0 3\\
NaN NaN NaN 3\\
2.6 10.6 0 3\\
2.6 10.6 4 3\\
2.6 11.4 4 3\\
2.6 11.4 0 3\\
2.6 10.6 0 3\\
NaN NaN NaN 3\\
3.4 0.6 0 3\\
3.4 0.6 0 3\\
3.4 1.4 0 3\\
3.4 1.4 0 3\\
3.4 0.6 0 3\\
NaN NaN NaN 3\\
3.4 1.6 0 3\\
3.4 1.6 2 3\\
3.4 2.4 2 3\\
3.4 2.4 0 3\\
3.4 1.6 0 3\\
NaN NaN NaN 3\\
3.4 2.6 0 3\\
3.4 2.6 2 3\\
3.4 3.4 2 3\\
3.4 3.4 0 3\\
3.4 2.6 0 3\\
NaN NaN NaN 3\\
3.4 3.6 0 3\\
3.4 3.6 3 3\\
3.4 4.4 3 3\\
3.4 4.4 0 3\\
3.4 3.6 0 3\\
NaN NaN NaN 3\\
3.4 4.6 0 3\\
3.4 4.6 3 3\\
3.4 5.4 3 3\\
3.4 5.4 0 3\\
3.4 4.6 0 3\\
NaN NaN NaN 3\\
3.4 5.6 0 3\\
3.4 5.6 3 3\\
3.4 6.4 3 3\\
3.4 6.4 0 3\\
3.4 5.6 0 3\\
NaN NaN NaN 3\\
3.4 6.6 0 3\\
3.4 6.6 3 3\\
3.4 7.4 3 3\\
3.4 7.4 0 3\\
3.4 6.6 0 3\\
NaN NaN NaN 3\\
3.4 7.6 0 3\\
3.4 7.6 3 3\\
3.4 8.4 3 3\\
3.4 8.4 0 3\\
3.4 7.6 0 3\\
NaN NaN NaN 3\\
3.4 8.6 0 3\\
3.4 8.6 4 3\\
3.4 9.4 4 3\\
3.4 9.4 0 3\\
3.4 8.6 0 3\\
NaN NaN NaN 3\\
3.4 9.6 0 3\\
3.4 9.6 4 3\\
3.4 10.4 4 3\\
3.4 10.4 0 3\\
3.4 9.6 0 3\\
NaN NaN NaN 3\\
3.4 10.6 0 3\\
3.4 10.6 4 3\\
3.4 11.4 4 3\\
3.4 11.4 0 3\\
3.4 10.6 0 3\\
NaN NaN NaN 3\\
NaN NaN NaN 3\\
3.4 0.6 0 3\\
3.4 1.4 0 3\\
NaN NaN NaN 3\\
NaN NaN NaN 3\\
NaN NaN NaN 3\\
NaN NaN NaN 3\\
3.4 1.6 0 3\\
3.4 2.4 0 3\\
NaN NaN NaN 3\\
NaN NaN NaN 3\\
NaN NaN NaN 3\\
NaN NaN NaN 3\\
3.4 2.6 0 3\\
3.4 3.4 0 3\\
NaN NaN NaN 3\\
NaN NaN NaN 3\\
NaN NaN NaN 3\\
NaN NaN NaN 3\\
3.4 3.6 0 3\\
3.4 4.4 0 3\\
NaN NaN NaN 3\\
NaN NaN NaN 3\\
NaN NaN NaN 3\\
NaN NaN NaN 3\\
3.4 4.6 0 3\\
3.4 5.4 0 3\\
NaN NaN NaN 3\\
NaN NaN NaN 3\\
NaN NaN NaN 3\\
NaN NaN NaN 3\\
3.4 5.6 0 3\\
3.4 6.4 0 3\\
NaN NaN NaN 3\\
NaN NaN NaN 3\\
NaN NaN NaN 3\\
NaN NaN NaN 3\\
3.4 6.6 0 3\\
3.4 7.4 0 3\\
NaN NaN NaN 3\\
NaN NaN NaN 3\\
NaN NaN NaN 3\\
NaN NaN NaN 3\\
3.4 7.6 0 3\\
3.4 8.4 0 3\\
NaN NaN NaN 3\\
NaN NaN NaN 3\\
NaN NaN NaN 3\\
NaN NaN NaN 3\\
3.4 8.6 0 3\\
3.4 9.4 0 3\\
NaN NaN NaN 3\\
NaN NaN NaN 3\\
NaN NaN NaN 3\\
NaN NaN NaN 3\\
3.4 9.6 0 3\\
3.4 10.4 0 3\\
NaN NaN NaN 3\\
NaN NaN NaN 3\\
NaN NaN NaN 3\\
NaN NaN NaN 3\\
3.4 10.6 0 3\\
3.4 11.4 0 3\\
NaN NaN NaN 3\\
NaN NaN NaN 3\\
NaN NaN NaN 3\\
};

\addplot3[%
surf,
shader=flat,
draw=black,
point meta=explicit,
mesh/rows=4]
table[row sep=crcr,header=false,meta index=3] {
NaN NaN NaN 2\\
1.6 0.6 0 2\\
1.6 1.4 0 2\\
NaN NaN NaN 2\\
NaN NaN NaN 2\\
NaN NaN NaN 2\\
NaN NaN NaN 2\\
1.6 1.6 0 2\\
1.6 2.4 0 2\\
NaN NaN NaN 2\\
NaN NaN NaN 2\\
NaN NaN NaN 2\\
NaN NaN NaN 2\\
1.6 2.6 0 2\\
1.6 3.4 0 2\\
NaN NaN NaN 2\\
NaN NaN NaN 2\\
NaN NaN NaN 2\\
NaN NaN NaN 2\\
1.6 3.6 0 2\\
1.6 4.4 0 2\\
NaN NaN NaN 2\\
NaN NaN NaN 2\\
NaN NaN NaN 2\\
NaN NaN NaN 2\\
1.6 4.6 0 2\\
1.6 5.4 0 2\\
NaN NaN NaN 2\\
NaN NaN NaN 2\\
NaN NaN NaN 2\\
NaN NaN NaN 2\\
1.6 5.6 0 2\\
1.6 6.4 0 2\\
NaN NaN NaN 2\\
NaN NaN NaN 2\\
NaN NaN NaN 2\\
NaN NaN NaN 2\\
1.6 6.6 0 2\\
1.6 7.4 0 2\\
NaN NaN NaN 2\\
NaN NaN NaN 2\\
NaN NaN NaN 2\\
NaN NaN NaN 2\\
1.6 7.6 0 2\\
1.6 8.4 0 2\\
NaN NaN NaN 2\\
NaN NaN NaN 2\\
NaN NaN NaN 2\\
NaN NaN NaN 2\\
1.6 8.6 0 2\\
1.6 9.4 0 2\\
NaN NaN NaN 2\\
NaN NaN NaN 2\\
NaN NaN NaN 2\\
NaN NaN NaN 2\\
1.6 9.6 0 2\\
1.6 10.4 0 2\\
NaN NaN NaN 2\\
NaN NaN NaN 2\\
NaN NaN NaN 2\\
NaN NaN NaN 2\\
1.6 10.6 0 2\\
1.6 11.4 0 2\\
NaN NaN NaN 2\\
NaN NaN NaN 2\\
NaN NaN NaN 2\\
1.6 0.6 0 2\\
1.6 0.6 0 2\\
1.6 1.4 0 2\\
1.6 1.4 0 2\\
1.6 0.6 0 2\\
NaN NaN NaN 2\\
1.6 1.6 0 2\\
1.6 1.6 2 2\\
1.6 2.4 2 2\\
1.6 2.4 0 2\\
1.6 1.6 0 2\\
NaN NaN NaN 2\\
1.6 2.6 0 2\\
1.6 2.6 2 2\\
1.6 3.4 2 2\\
1.6 3.4 0 2\\
1.6 2.6 0 2\\
NaN NaN NaN 2\\
1.6 3.6 0 2\\
1.6 3.6 3 2\\
1.6 4.4 3 2\\
1.6 4.4 0 2\\
1.6 3.6 0 2\\
NaN NaN NaN 2\\
1.6 4.6 0 2\\
1.6 4.6 3 2\\
1.6 5.4 3 2\\
1.6 5.4 0 2\\
1.6 4.6 0 2\\
NaN NaN NaN 2\\
1.6 5.6 0 2\\
1.6 5.6 3 2\\
1.6 6.4 3 2\\
1.6 6.4 0 2\\
1.6 5.6 0 2\\
NaN NaN NaN 2\\
1.6 6.6 0 2\\
1.6 6.6 3 2\\
1.6 7.4 3 2\\
1.6 7.4 0 2\\
1.6 6.6 0 2\\
NaN NaN NaN 2\\
1.6 7.6 0 2\\
1.6 7.6 3 2\\
1.6 8.4 3 2\\
1.6 8.4 0 2\\
1.6 7.6 0 2\\
NaN NaN NaN 2\\
1.6 8.6 0 2\\
1.6 8.6 3 2\\
1.6 9.4 3 2\\
1.6 9.4 0 2\\
1.6 8.6 0 2\\
NaN NaN NaN 2\\
1.6 9.6 0 2\\
1.6 9.6 3 2\\
1.6 10.4 3 2\\
1.6 10.4 0 2\\
1.6 9.6 0 2\\
NaN NaN NaN 2\\
1.6 10.6 0 2\\
1.6 10.6 4 2\\
1.6 11.4 4 2\\
1.6 11.4 0 2\\
1.6 10.6 0 2\\
NaN NaN NaN 2\\
2.4 0.6 0 2\\
2.4 0.6 0 2\\
2.4 1.4 0 2\\
2.4 1.4 0 2\\
2.4 0.6 0 2\\
NaN NaN NaN 2\\
2.4 1.6 0 2\\
2.4 1.6 2 2\\
2.4 2.4 2 2\\
2.4 2.4 0 2\\
2.4 1.6 0 2\\
NaN NaN NaN 2\\
2.4 2.6 0 2\\
2.4 2.6 2 2\\
2.4 3.4 2 2\\
2.4 3.4 0 2\\
2.4 2.6 0 2\\
NaN NaN NaN 2\\
2.4 3.6 0 2\\
2.4 3.6 3 2\\
2.4 4.4 3 2\\
2.4 4.4 0 2\\
2.4 3.6 0 2\\
NaN NaN NaN 2\\
2.4 4.6 0 2\\
2.4 4.6 3 2\\
2.4 5.4 3 2\\
2.4 5.4 0 2\\
2.4 4.6 0 2\\
NaN NaN NaN 2\\
2.4 5.6 0 2\\
2.4 5.6 3 2\\
2.4 6.4 3 2\\
2.4 6.4 0 2\\
2.4 5.6 0 2\\
NaN NaN NaN 2\\
2.4 6.6 0 2\\
2.4 6.6 3 2\\
2.4 7.4 3 2\\
2.4 7.4 0 2\\
2.4 6.6 0 2\\
NaN NaN NaN 2\\
2.4 7.6 0 2\\
2.4 7.6 3 2\\
2.4 8.4 3 2\\
2.4 8.4 0 2\\
2.4 7.6 0 2\\
NaN NaN NaN 2\\
2.4 8.6 0 2\\
2.4 8.6 3 2\\
2.4 9.4 3 2\\
2.4 9.4 0 2\\
2.4 8.6 0 2\\
NaN NaN NaN 2\\
2.4 9.6 0 2\\
2.4 9.6 3 2\\
2.4 10.4 3 2\\
2.4 10.4 0 2\\
2.4 9.6 0 2\\
NaN NaN NaN 2\\
2.4 10.6 0 2\\
2.4 10.6 4 2\\
2.4 11.4 4 2\\
2.4 11.4 0 2\\
2.4 10.6 0 2\\
NaN NaN NaN 2\\
NaN NaN NaN 2\\
2.4 0.6 0 2\\
2.4 1.4 0 2\\
NaN NaN NaN 2\\
NaN NaN NaN 2\\
NaN NaN NaN 2\\
NaN NaN NaN 2\\
2.4 1.6 0 2\\
2.4 2.4 0 2\\
NaN NaN NaN 2\\
NaN NaN NaN 2\\
NaN NaN NaN 2\\
NaN NaN NaN 2\\
2.4 2.6 0 2\\
2.4 3.4 0 2\\
NaN NaN NaN 2\\
NaN NaN NaN 2\\
NaN NaN NaN 2\\
NaN NaN NaN 2\\
2.4 3.6 0 2\\
2.4 4.4 0 2\\
NaN NaN NaN 2\\
NaN NaN NaN 2\\
NaN NaN NaN 2\\
NaN NaN NaN 2\\
2.4 4.6 0 2\\
2.4 5.4 0 2\\
NaN NaN NaN 2\\
NaN NaN NaN 2\\
NaN NaN NaN 2\\
NaN NaN NaN 2\\
2.4 5.6 0 2\\
2.4 6.4 0 2\\
NaN NaN NaN 2\\
NaN NaN NaN 2\\
NaN NaN NaN 2\\
NaN NaN NaN 2\\
2.4 6.6 0 2\\
2.4 7.4 0 2\\
NaN NaN NaN 2\\
NaN NaN NaN 2\\
NaN NaN NaN 2\\
NaN NaN NaN 2\\
2.4 7.6 0 2\\
2.4 8.4 0 2\\
NaN NaN NaN 2\\
NaN NaN NaN 2\\
NaN NaN NaN 2\\
NaN NaN NaN 2\\
2.4 8.6 0 2\\
2.4 9.4 0 2\\
NaN NaN NaN 2\\
NaN NaN NaN 2\\
NaN NaN NaN 2\\
NaN NaN NaN 2\\
2.4 9.6 0 2\\
2.4 10.4 0 2\\
NaN NaN NaN 2\\
NaN NaN NaN 2\\
NaN NaN NaN 2\\
NaN NaN NaN 2\\
2.4 10.6 0 2\\
2.4 11.4 0 2\\
NaN NaN NaN 2\\
NaN NaN NaN 2\\
NaN NaN NaN 2\\
};

\addplot3[%
surf,
shader=flat,
draw=black,
point meta=explicit,
mesh/rows=4]
table[row sep=crcr,header=false,meta index=3] {
NaN NaN NaN 1\\
0.6 0.6 0 1\\
0.6 1.4 0 1\\
NaN NaN NaN 1\\
NaN NaN NaN 1\\
NaN NaN NaN 1\\
NaN NaN NaN 1\\
0.6 1.6 0 1\\
0.6 2.4 0 1\\
NaN NaN NaN 1\\
NaN NaN NaN 1\\
NaN NaN NaN 1\\
NaN NaN NaN 1\\
0.6 2.6 0 1\\
0.6 3.4 0 1\\
NaN NaN NaN 1\\
NaN NaN NaN 1\\
NaN NaN NaN 1\\
NaN NaN NaN 1\\
0.6 3.6 0 1\\
0.6 4.4 0 1\\
NaN NaN NaN 1\\
NaN NaN NaN 1\\
NaN NaN NaN 1\\
NaN NaN NaN 1\\
0.6 4.6 0 1\\
0.6 5.4 0 1\\
NaN NaN NaN 1\\
NaN NaN NaN 1\\
NaN NaN NaN 1\\
NaN NaN NaN 1\\
0.6 5.6 0 1\\
0.6 6.4 0 1\\
NaN NaN NaN 1\\
NaN NaN NaN 1\\
NaN NaN NaN 1\\
NaN NaN NaN 1\\
0.6 6.6 0 1\\
0.6 7.4 0 1\\
NaN NaN NaN 1\\
NaN NaN NaN 1\\
NaN NaN NaN 1\\
NaN NaN NaN 1\\
0.6 7.6 0 1\\
0.6 8.4 0 1\\
NaN NaN NaN 1\\
NaN NaN NaN 1\\
NaN NaN NaN 1\\
NaN NaN NaN 1\\
0.6 8.6 0 1\\
0.6 9.4 0 1\\
NaN NaN NaN 1\\
NaN NaN NaN 1\\
NaN NaN NaN 1\\
NaN NaN NaN 1\\
0.6 9.6 0 1\\
0.6 10.4 0 1\\
NaN NaN NaN 1\\
NaN NaN NaN 1\\
NaN NaN NaN 1\\
NaN NaN NaN 1\\
0.6 10.6 0 1\\
0.6 11.4 0 1\\
NaN NaN NaN 1\\
NaN NaN NaN 1\\
NaN NaN NaN 1\\
0.6 0.6 0 1\\
0.6 0.6 0 1\\
0.6 1.4 0 1\\
0.6 1.4 0 1\\
0.6 0.6 0 1\\
NaN NaN NaN 1\\
0.6 1.6 0 1\\
0.6 1.6 1 1\\
0.6 2.4 1 1\\
0.6 2.4 0 1\\
0.6 1.6 0 1\\
NaN NaN NaN 1\\
0.6 2.6 0 1\\
0.6 2.6 1 1\\
0.6 3.4 1 1\\
0.6 3.4 0 1\\
0.6 2.6 0 1\\
NaN NaN NaN 1\\
0.6 3.6 0 1\\
0.6 3.6 2 1\\
0.6 4.4 2 1\\
0.6 4.4 0 1\\
0.6 3.6 0 1\\
NaN NaN NaN 1\\
0.6 4.6 0 1\\
0.6 4.6 2 1\\
0.6 5.4 2 1\\
0.6 5.4 0 1\\
0.6 4.6 0 1\\
NaN NaN NaN 1\\
0.6 5.6 0 1\\
0.6 5.6 3 1\\
0.6 6.4 3 1\\
0.6 6.4 0 1\\
0.6 5.6 0 1\\
NaN NaN NaN 1\\
0.6 6.6 0 1\\
0.6 6.6 3 1\\
0.6 7.4 3 1\\
0.6 7.4 0 1\\
0.6 6.6 0 1\\
NaN NaN NaN 1\\
0.6 7.6 0 1\\
0.6 7.6 3 1\\
0.6 8.4 3 1\\
0.6 8.4 0 1\\
0.6 7.6 0 1\\
NaN NaN NaN 1\\
0.6 8.6 0 1\\
0.6 8.6 3 1\\
0.6 9.4 3 1\\
0.6 9.4 0 1\\
0.6 8.6 0 1\\
NaN NaN NaN 1\\
0.6 9.6 0 1\\
0.6 9.6 3 1\\
0.6 10.4 3 1\\
0.6 10.4 0 1\\
0.6 9.6 0 1\\
NaN NaN NaN 1\\
0.6 10.6 0 1\\
0.6 10.6 3 1\\
0.6 11.4 3 1\\
0.6 11.4 0 1\\
0.6 10.6 0 1\\
NaN NaN NaN 1\\
1.4 0.6 0 1\\
1.4 0.6 0 1\\
1.4 1.4 0 1\\
1.4 1.4 0 1\\
1.4 0.6 0 1\\
NaN NaN NaN 1\\
1.4 1.6 0 1\\
1.4 1.6 1 1\\
1.4 2.4 1 1\\
1.4 2.4 0 1\\
1.4 1.6 0 1\\
NaN NaN NaN 1\\
1.4 2.6 0 1\\
1.4 2.6 1 1\\
1.4 3.4 1 1\\
1.4 3.4 0 1\\
1.4 2.6 0 1\\
NaN NaN NaN 1\\
1.4 3.6 0 1\\
1.4 3.6 2 1\\
1.4 4.4 2 1\\
1.4 4.4 0 1\\
1.4 3.6 0 1\\
NaN NaN NaN 1\\
1.4 4.6 0 1\\
1.4 4.6 2 1\\
1.4 5.4 2 1\\
1.4 5.4 0 1\\
1.4 4.6 0 1\\
NaN NaN NaN 1\\
1.4 5.6 0 1\\
1.4 5.6 3 1\\
1.4 6.4 3 1\\
1.4 6.4 0 1\\
1.4 5.6 0 1\\
NaN NaN NaN 1\\
1.4 6.6 0 1\\
1.4 6.6 3 1\\
1.4 7.4 3 1\\
1.4 7.4 0 1\\
1.4 6.6 0 1\\
NaN NaN NaN 1\\
1.4 7.6 0 1\\
1.4 7.6 3 1\\
1.4 8.4 3 1\\
1.4 8.4 0 1\\
1.4 7.6 0 1\\
NaN NaN NaN 1\\
1.4 8.6 0 1\\
1.4 8.6 3 1\\
1.4 9.4 3 1\\
1.4 9.4 0 1\\
1.4 8.6 0 1\\
NaN NaN NaN 1\\
1.4 9.6 0 1\\
1.4 9.6 3 1\\
1.4 10.4 3 1\\
1.4 10.4 0 1\\
1.4 9.6 0 1\\
NaN NaN NaN 1\\
1.4 10.6 0 1\\
1.4 10.6 3 1\\
1.4 11.4 3 1\\
1.4 11.4 0 1\\
1.4 10.6 0 1\\
NaN NaN NaN 1\\
NaN NaN NaN 1\\
1.4 0.6 0 1\\
1.4 1.4 0 1\\
NaN NaN NaN 1\\
NaN NaN NaN 1\\
NaN NaN NaN 1\\
NaN NaN NaN 1\\
1.4 1.6 0 1\\
1.4 2.4 0 1\\
NaN NaN NaN 1\\
NaN NaN NaN 1\\
NaN NaN NaN 1\\
NaN NaN NaN 1\\
1.4 2.6 0 1\\
1.4 3.4 0 1\\
NaN NaN NaN 1\\
NaN NaN NaN 1\\
NaN NaN NaN 1\\
NaN NaN NaN 1\\
1.4 3.6 0 1\\
1.4 4.4 0 1\\
NaN NaN NaN 1\\
NaN NaN NaN 1\\
NaN NaN NaN 1\\
NaN NaN NaN 1\\
1.4 4.6 0 1\\
1.4 5.4 0 1\\
NaN NaN NaN 1\\
NaN NaN NaN 1\\
NaN NaN NaN 1\\
NaN NaN NaN 1\\
1.4 5.6 0 1\\
1.4 6.4 0 1\\
NaN NaN NaN 1\\
NaN NaN NaN 1\\
NaN NaN NaN 1\\
NaN NaN NaN 1\\
1.4 6.6 0 1\\
1.4 7.4 0 1\\
NaN NaN NaN 1\\
NaN NaN NaN 1\\
NaN NaN NaN 1\\
NaN NaN NaN 1\\
1.4 7.6 0 1\\
1.4 8.4 0 1\\
NaN NaN NaN 1\\
NaN NaN NaN 1\\
NaN NaN NaN 1\\
NaN NaN NaN 1\\
1.4 8.6 0 1\\
1.4 9.4 0 1\\
NaN NaN NaN 1\\
NaN NaN NaN 1\\
NaN NaN NaN 1\\
NaN NaN NaN 1\\
1.4 9.6 0 1\\
1.4 10.4 0 1\\
NaN NaN NaN 1\\
NaN NaN NaN 1\\
NaN NaN NaN 1\\
NaN NaN NaN 1\\
1.4 10.6 0 1\\
1.4 11.4 0 1\\
NaN NaN NaN 1\\
NaN NaN NaN 1\\
NaN NaN NaN 1\\
};
\end{axis}
\end{tikzpicture}%

%% file: Policy/MonoH2.tex
%
%
\begin{tikzpicture}
\begin{axis}[%
width=2.5in,
height=2.2in,
unbounded coords=jump,
view={-130}{40},
scale only axis,
xmin=0.5,
xmax=11.5,
xtick={1,2,3,4,5,6,7,8,9,10,11},
xticklabels={0.1,1,2,3,4,5,6,7,8,9,10},
xlabel={$\gamma_2$},
xmajorgrids,
y dir=reverse,
ymin=0.5,
ymax=11.5,
ytick={1,2,3,4,5,6,7,8,9,10,11},
yticklabels={0.1,1,2,3,4,5,6,7,8,9,10},
ylabel={$\gamma_1$},
ymajorgrids,
zmin=0,
zmax=4,
zlabel={$\overline{\theta}_2^*$},
zmajorgrids,
at=(plot1.right of south east),
anchor=left of south west,
axis x line*=bottom,
axis y line*=left,
axis z line*=left
]

\addplot3[%
surf,
shader=flat,
draw=black,
point meta=explicit,
mesh/rows=4]
table[row sep=crcr,header=false,meta index=3] {
NaN NaN NaN 11\\
10.6 0.6 0 11\\
10.6 1.4 0 11\\
NaN NaN NaN 11\\
NaN NaN NaN 11\\
NaN NaN NaN 11\\
NaN NaN NaN 11\\
10.6 1.6 0 11\\
10.6 2.4 0 11\\
NaN NaN NaN 11\\
NaN NaN NaN 11\\
NaN NaN NaN 11\\
NaN NaN NaN 11\\
10.6 2.6 0 11\\
10.6 3.4 0 11\\
NaN NaN NaN 11\\
NaN NaN NaN 11\\
NaN NaN NaN 11\\
NaN NaN NaN 11\\
10.6 3.6 0 11\\
10.6 4.4 0 11\\
NaN NaN NaN 11\\
NaN NaN NaN 11\\
NaN NaN NaN 11\\
NaN NaN NaN 11\\
10.6 4.6 0 11\\
10.6 5.4 0 11\\
NaN NaN NaN 11\\
NaN NaN NaN 11\\
NaN NaN NaN 11\\
NaN NaN NaN 11\\
10.6 5.6 0 11\\
10.6 6.4 0 11\\
NaN NaN NaN 11\\
NaN NaN NaN 11\\
NaN NaN NaN 11\\
NaN NaN NaN 11\\
10.6 6.6 0 11\\
10.6 7.4 0 11\\
NaN NaN NaN 11\\
NaN NaN NaN 11\\
NaN NaN NaN 11\\
NaN NaN NaN 11\\
10.6 7.6 0 11\\
10.6 8.4 0 11\\
NaN NaN NaN 11\\
NaN NaN NaN 11\\
NaN NaN NaN 11\\
NaN NaN NaN 11\\
10.6 8.6 0 11\\
10.6 9.4 0 11\\
NaN NaN NaN 11\\
NaN NaN NaN 11\\
NaN NaN NaN 11\\
NaN NaN NaN 11\\
10.6 9.6 0 11\\
10.6 10.4 0 11\\
NaN NaN NaN 11\\
NaN NaN NaN 11\\
NaN NaN NaN 11\\
NaN NaN NaN 11\\
10.6 10.6 0 11\\
10.6 11.4 0 11\\
NaN NaN NaN 11\\
NaN NaN NaN 11\\
NaN NaN NaN 11\\
10.6 0.6 0 11\\
10.6 0.6 3 11\\
10.6 1.4 3 11\\
10.6 1.4 0 11\\
10.6 0.6 0 11\\
NaN NaN NaN 11\\
10.6 1.6 0 11\\
10.6 1.6 4 11\\
10.6 2.4 4 11\\
10.6 2.4 0 11\\
10.6 1.6 0 11\\
NaN NaN NaN 11\\
10.6 2.6 0 11\\
10.6 2.6 4 11\\
10.6 3.4 4 11\\
10.6 3.4 0 11\\
10.6 2.6 0 11\\
NaN NaN NaN 11\\
10.6 3.6 0 11\\
10.6 3.6 4 11\\
10.6 4.4 4 11\\
10.6 4.4 0 11\\
10.6 3.6 0 11\\
NaN NaN NaN 11\\
10.6 4.6 0 11\\
10.6 4.6 4 11\\
10.6 5.4 4 11\\
10.6 5.4 0 11\\
10.6 4.6 0 11\\
NaN NaN NaN 11\\
10.6 5.6 0 11\\
10.6 5.6 4 11\\
10.6 6.4 4 11\\
10.6 6.4 0 11\\
10.6 5.6 0 11\\
NaN NaN NaN 11\\
10.6 6.6 0 11\\
10.6 6.6 4 11\\
10.6 7.4 4 11\\
10.6 7.4 0 11\\
10.6 6.6 0 11\\
NaN NaN NaN 11\\
10.6 7.6 0 11\\
10.6 7.6 4 11\\
10.6 8.4 4 11\\
10.6 8.4 0 11\\
10.6 7.6 0 11\\
NaN NaN NaN 11\\
10.6 8.6 0 11\\
10.6 8.6 4 11\\
10.6 9.4 4 11\\
10.6 9.4 0 11\\
10.6 8.6 0 11\\
NaN NaN NaN 11\\
10.6 9.6 0 11\\
10.6 9.6 4 11\\
10.6 10.4 4 11\\
10.6 10.4 0 11\\
10.6 9.6 0 11\\
NaN NaN NaN 11\\
10.6 10.6 0 11\\
10.6 10.6 4 11\\
10.6 11.4 4 11\\
10.6 11.4 0 11\\
10.6 10.6 0 11\\
NaN NaN NaN 11\\
11.4 0.6 0 11\\
11.4 0.6 3 11\\
11.4 1.4 3 11\\
11.4 1.4 0 11\\
11.4 0.6 0 11\\
NaN NaN NaN 11\\
11.4 1.6 0 11\\
11.4 1.6 4 11\\
11.4 2.4 4 11\\
11.4 2.4 0 11\\
11.4 1.6 0 11\\
NaN NaN NaN 11\\
11.4 2.6 0 11\\
11.4 2.6 4 11\\
11.4 3.4 4 11\\
11.4 3.4 0 11\\
11.4 2.6 0 11\\
NaN NaN NaN 11\\
11.4 3.6 0 11\\
11.4 3.6 4 11\\
11.4 4.4 4 11\\
11.4 4.4 0 11\\
11.4 3.6 0 11\\
NaN NaN NaN 11\\
11.4 4.6 0 11\\
11.4 4.6 4 11\\
11.4 5.4 4 11\\
11.4 5.4 0 11\\
11.4 4.6 0 11\\
NaN NaN NaN 11\\
11.4 5.6 0 11\\
11.4 5.6 4 11\\
11.4 6.4 4 11\\
11.4 6.4 0 11\\
11.4 5.6 0 11\\
NaN NaN NaN 11\\
11.4 6.6 0 11\\
11.4 6.6 4 11\\
11.4 7.4 4 11\\
11.4 7.4 0 11\\
11.4 6.6 0 11\\
NaN NaN NaN 11\\
11.4 7.6 0 11\\
11.4 7.6 4 11\\
11.4 8.4 4 11\\
11.4 8.4 0 11\\
11.4 7.6 0 11\\
NaN NaN NaN 11\\
11.4 8.6 0 11\\
11.4 8.6 4 11\\
11.4 9.4 4 11\\
11.4 9.4 0 11\\
11.4 8.6 0 11\\
NaN NaN NaN 11\\
11.4 9.6 0 11\\
11.4 9.6 4 11\\
11.4 10.4 4 11\\
11.4 10.4 0 11\\
11.4 9.6 0 11\\
NaN NaN NaN 11\\
11.4 10.6 0 11\\
11.4 10.6 4 11\\
11.4 11.4 4 11\\
11.4 11.4 0 11\\
11.4 10.6 0 11\\
NaN NaN NaN 11\\
NaN NaN NaN 11\\
11.4 0.6 0 11\\
11.4 1.4 0 11\\
NaN NaN NaN 11\\
NaN NaN NaN 11\\
NaN NaN NaN 11\\
NaN NaN NaN 11\\
11.4 1.6 0 11\\
11.4 2.4 0 11\\
NaN NaN NaN 11\\
NaN NaN NaN 11\\
NaN NaN NaN 11\\
NaN NaN NaN 11\\
11.4 2.6 0 11\\
11.4 3.4 0 11\\
NaN NaN NaN 11\\
NaN NaN NaN 11\\
NaN NaN NaN 11\\
NaN NaN NaN 11\\
11.4 3.6 0 11\\
11.4 4.4 0 11\\
NaN NaN NaN 11\\
NaN NaN NaN 11\\
NaN NaN NaN 11\\
NaN NaN NaN 11\\
11.4 4.6 0 11\\
11.4 5.4 0 11\\
NaN NaN NaN 11\\
NaN NaN NaN 11\\
NaN NaN NaN 11\\
NaN NaN NaN 11\\
11.4 5.6 0 11\\
11.4 6.4 0 11\\
NaN NaN NaN 11\\
NaN NaN NaN 11\\
NaN NaN NaN 11\\
NaN NaN NaN 11\\
11.4 6.6 0 11\\
11.4 7.4 0 11\\
NaN NaN NaN 11\\
NaN NaN NaN 11\\
NaN NaN NaN 11\\
NaN NaN NaN 11\\
11.4 7.6 0 11\\
11.4 8.4 0 11\\
NaN NaN NaN 11\\
NaN NaN NaN 11\\
NaN NaN NaN 11\\
NaN NaN NaN 11\\
11.4 8.6 0 11\\
11.4 9.4 0 11\\
NaN NaN NaN 11\\
NaN NaN NaN 11\\
NaN NaN NaN 11\\
NaN NaN NaN 11\\
11.4 9.6 0 11\\
11.4 10.4 0 11\\
NaN NaN NaN 11\\
NaN NaN NaN 11\\
NaN NaN NaN 11\\
NaN NaN NaN 11\\
11.4 10.6 0 11\\
11.4 11.4 0 11\\
NaN NaN NaN 11\\
NaN NaN NaN 11\\
NaN NaN NaN 11\\
};

\addplot3[%
surf,
shader=flat,
draw=black,
point meta=explicit,
mesh/rows=4]
table[row sep=crcr,header=false,meta index=3] {
NaN NaN NaN 10\\
9.6 0.6 0 10\\
9.6 1.4 0 10\\
NaN NaN NaN 10\\
NaN NaN NaN 10\\
NaN NaN NaN 10\\
NaN NaN NaN 10\\
9.6 1.6 0 10\\
9.6 2.4 0 10\\
NaN NaN NaN 10\\
NaN NaN NaN 10\\
NaN NaN NaN 10\\
NaN NaN NaN 10\\
9.6 2.6 0 10\\
9.6 3.4 0 10\\
NaN NaN NaN 10\\
NaN NaN NaN 10\\
NaN NaN NaN 10\\
NaN NaN NaN 10\\
9.6 3.6 0 10\\
9.6 4.4 0 10\\
NaN NaN NaN 10\\
NaN NaN NaN 10\\
NaN NaN NaN 10\\
NaN NaN NaN 10\\
9.6 4.6 0 10\\
9.6 5.4 0 10\\
NaN NaN NaN 10\\
NaN NaN NaN 10\\
NaN NaN NaN 10\\
NaN NaN NaN 10\\
9.6 5.6 0 10\\
9.6 6.4 0 10\\
NaN NaN NaN 10\\
NaN NaN NaN 10\\
NaN NaN NaN 10\\
NaN NaN NaN 10\\
9.6 6.6 0 10\\
9.6 7.4 0 10\\
NaN NaN NaN 10\\
NaN NaN NaN 10\\
NaN NaN NaN 10\\
NaN NaN NaN 10\\
9.6 7.6 0 10\\
9.6 8.4 0 10\\
NaN NaN NaN 10\\
NaN NaN NaN 10\\
NaN NaN NaN 10\\
NaN NaN NaN 10\\
9.6 8.6 0 10\\
9.6 9.4 0 10\\
NaN NaN NaN 10\\
NaN NaN NaN 10\\
NaN NaN NaN 10\\
NaN NaN NaN 10\\
9.6 9.6 0 10\\
9.6 10.4 0 10\\
NaN NaN NaN 10\\
NaN NaN NaN 10\\
NaN NaN NaN 10\\
NaN NaN NaN 10\\
9.6 10.6 0 10\\
9.6 11.4 0 10\\
NaN NaN NaN 10\\
NaN NaN NaN 10\\
NaN NaN NaN 10\\
9.6 0.6 0 10\\
9.6 0.6 3 10\\
9.6 1.4 3 10\\
9.6 1.4 0 10\\
9.6 0.6 0 10\\
NaN NaN NaN 10\\
9.6 1.6 0 10\\
9.6 1.6 3 10\\
9.6 2.4 3 10\\
9.6 2.4 0 10\\
9.6 1.6 0 10\\
NaN NaN NaN 10\\
9.6 2.6 0 10\\
9.6 2.6 4 10\\
9.6 3.4 4 10\\
9.6 3.4 0 10\\
9.6 2.6 0 10\\
NaN NaN NaN 10\\
9.6 3.6 0 10\\
9.6 3.6 4 10\\
9.6 4.4 4 10\\
9.6 4.4 0 10\\
9.6 3.6 0 10\\
NaN NaN NaN 10\\
9.6 4.6 0 10\\
9.6 4.6 4 10\\
9.6 5.4 4 10\\
9.6 5.4 0 10\\
9.6 4.6 0 10\\
NaN NaN NaN 10\\
9.6 5.6 0 10\\
9.6 5.6 4 10\\
9.6 6.4 4 10\\
9.6 6.4 0 10\\
9.6 5.6 0 10\\
NaN NaN NaN 10\\
9.6 6.6 0 10\\
9.6 6.6 4 10\\
9.6 7.4 4 10\\
9.6 7.4 0 10\\
9.6 6.6 0 10\\
NaN NaN NaN 10\\
9.6 7.6 0 10\\
9.6 7.6 4 10\\
9.6 8.4 4 10\\
9.6 8.4 0 10\\
9.6 7.6 0 10\\
NaN NaN NaN 10\\
9.6 8.6 0 10\\
9.6 8.6 4 10\\
9.6 9.4 4 10\\
9.6 9.4 0 10\\
9.6 8.6 0 10\\
NaN NaN NaN 10\\
9.6 9.6 0 10\\
9.6 9.6 4 10\\
9.6 10.4 4 10\\
9.6 10.4 0 10\\
9.6 9.6 0 10\\
NaN NaN NaN 10\\
9.6 10.6 0 10\\
9.6 10.6 4 10\\
9.6 11.4 4 10\\
9.6 11.4 0 10\\
9.6 10.6 0 10\\
NaN NaN NaN 10\\
10.4 0.6 0 10\\
10.4 0.6 3 10\\
10.4 1.4 3 10\\
10.4 1.4 0 10\\
10.4 0.6 0 10\\
NaN NaN NaN 10\\
10.4 1.6 0 10\\
10.4 1.6 3 10\\
10.4 2.4 3 10\\
10.4 2.4 0 10\\
10.4 1.6 0 10\\
NaN NaN NaN 10\\
10.4 2.6 0 10\\
10.4 2.6 4 10\\
10.4 3.4 4 10\\
10.4 3.4 0 10\\
10.4 2.6 0 10\\
NaN NaN NaN 10\\
10.4 3.6 0 10\\
10.4 3.6 4 10\\
10.4 4.4 4 10\\
10.4 4.4 0 10\\
10.4 3.6 0 10\\
NaN NaN NaN 10\\
10.4 4.6 0 10\\
10.4 4.6 4 10\\
10.4 5.4 4 10\\
10.4 5.4 0 10\\
10.4 4.6 0 10\\
NaN NaN NaN 10\\
10.4 5.6 0 10\\
10.4 5.6 4 10\\
10.4 6.4 4 10\\
10.4 6.4 0 10\\
10.4 5.6 0 10\\
NaN NaN NaN 10\\
10.4 6.6 0 10\\
10.4 6.6 4 10\\
10.4 7.4 4 10\\
10.4 7.4 0 10\\
10.4 6.6 0 10\\
NaN NaN NaN 10\\
10.4 7.6 0 10\\
10.4 7.6 4 10\\
10.4 8.4 4 10\\
10.4 8.4 0 10\\
10.4 7.6 0 10\\
NaN NaN NaN 10\\
10.4 8.6 0 10\\
10.4 8.6 4 10\\
10.4 9.4 4 10\\
10.4 9.4 0 10\\
10.4 8.6 0 10\\
NaN NaN NaN 10\\
10.4 9.6 0 10\\
10.4 9.6 4 10\\
10.4 10.4 4 10\\
10.4 10.4 0 10\\
10.4 9.6 0 10\\
NaN NaN NaN 10\\
10.4 10.6 0 10\\
10.4 10.6 4 10\\
10.4 11.4 4 10\\
10.4 11.4 0 10\\
10.4 10.6 0 10\\
NaN NaN NaN 10\\
NaN NaN NaN 10\\
10.4 0.6 0 10\\
10.4 1.4 0 10\\
NaN NaN NaN 10\\
NaN NaN NaN 10\\
NaN NaN NaN 10\\
NaN NaN NaN 10\\
10.4 1.6 0 10\\
10.4 2.4 0 10\\
NaN NaN NaN 10\\
NaN NaN NaN 10\\
NaN NaN NaN 10\\
NaN NaN NaN 10\\
10.4 2.6 0 10\\
10.4 3.4 0 10\\
NaN NaN NaN 10\\
NaN NaN NaN 10\\
NaN NaN NaN 10\\
NaN NaN NaN 10\\
10.4 3.6 0 10\\
10.4 4.4 0 10\\
NaN NaN NaN 10\\
NaN NaN NaN 10\\
NaN NaN NaN 10\\
NaN NaN NaN 10\\
10.4 4.6 0 10\\
10.4 5.4 0 10\\
NaN NaN NaN 10\\
NaN NaN NaN 10\\
NaN NaN NaN 10\\
NaN NaN NaN 10\\
10.4 5.6 0 10\\
10.4 6.4 0 10\\
NaN NaN NaN 10\\
NaN NaN NaN 10\\
NaN NaN NaN 10\\
NaN NaN NaN 10\\
10.4 6.6 0 10\\
10.4 7.4 0 10\\
NaN NaN NaN 10\\
NaN NaN NaN 10\\
NaN NaN NaN 10\\
NaN NaN NaN 10\\
10.4 7.6 0 10\\
10.4 8.4 0 10\\
NaN NaN NaN 10\\
NaN NaN NaN 10\\
NaN NaN NaN 10\\
NaN NaN NaN 10\\
10.4 8.6 0 10\\
10.4 9.4 0 10\\
NaN NaN NaN 10\\
NaN NaN NaN 10\\
NaN NaN NaN 10\\
NaN NaN NaN 10\\
10.4 9.6 0 10\\
10.4 10.4 0 10\\
NaN NaN NaN 10\\
NaN NaN NaN 10\\
NaN NaN NaN 10\\
NaN NaN NaN 10\\
10.4 10.6 0 10\\
10.4 11.4 0 10\\
NaN NaN NaN 10\\
NaN NaN NaN 10\\
NaN NaN NaN 10\\
};

\addplot3[%
surf,
shader=flat,
draw=black,
point meta=explicit,
mesh/rows=4]
table[row sep=crcr,header=false,meta index=3] {
NaN NaN NaN 9\\
8.6 0.6 0 9\\
8.6 1.4 0 9\\
NaN NaN NaN 9\\
NaN NaN NaN 9\\
NaN NaN NaN 9\\
NaN NaN NaN 9\\
8.6 1.6 0 9\\
8.6 2.4 0 9\\
NaN NaN NaN 9\\
NaN NaN NaN 9\\
NaN NaN NaN 9\\
NaN NaN NaN 9\\
8.6 2.6 0 9\\
8.6 3.4 0 9\\
NaN NaN NaN 9\\
NaN NaN NaN 9\\
NaN NaN NaN 9\\
NaN NaN NaN 9\\
8.6 3.6 0 9\\
8.6 4.4 0 9\\
NaN NaN NaN 9\\
NaN NaN NaN 9\\
NaN NaN NaN 9\\
NaN NaN NaN 9\\
8.6 4.6 0 9\\
8.6 5.4 0 9\\
NaN NaN NaN 9\\
NaN NaN NaN 9\\
NaN NaN NaN 9\\
NaN NaN NaN 9\\
8.6 5.6 0 9\\
8.6 6.4 0 9\\
NaN NaN NaN 9\\
NaN NaN NaN 9\\
NaN NaN NaN 9\\
NaN NaN NaN 9\\
8.6 6.6 0 9\\
8.6 7.4 0 9\\
NaN NaN NaN 9\\
NaN NaN NaN 9\\
NaN NaN NaN 9\\
NaN NaN NaN 9\\
8.6 7.6 0 9\\
8.6 8.4 0 9\\
NaN NaN NaN 9\\
NaN NaN NaN 9\\
NaN NaN NaN 9\\
NaN NaN NaN 9\\
8.6 8.6 0 9\\
8.6 9.4 0 9\\
NaN NaN NaN 9\\
NaN NaN NaN 9\\
NaN NaN NaN 9\\
NaN NaN NaN 9\\
8.6 9.6 0 9\\
8.6 10.4 0 9\\
NaN NaN NaN 9\\
NaN NaN NaN 9\\
NaN NaN NaN 9\\
NaN NaN NaN 9\\
8.6 10.6 0 9\\
8.6 11.4 0 9\\
NaN NaN NaN 9\\
NaN NaN NaN 9\\
NaN NaN NaN 9\\
8.6 0.6 0 9\\
8.6 0.6 3 9\\
8.6 1.4 3 9\\
8.6 1.4 0 9\\
8.6 0.6 0 9\\
NaN NaN NaN 9\\
8.6 1.6 0 9\\
8.6 1.6 3 9\\
8.6 2.4 3 9\\
8.6 2.4 0 9\\
8.6 1.6 0 9\\
NaN NaN NaN 9\\
8.6 2.6 0 9\\
8.6 2.6 4 9\\
8.6 3.4 4 9\\
8.6 3.4 0 9\\
8.6 2.6 0 9\\
NaN NaN NaN 9\\
8.6 3.6 0 9\\
8.6 3.6 4 9\\
8.6 4.4 4 9\\
8.6 4.4 0 9\\
8.6 3.6 0 9\\
NaN NaN NaN 9\\
8.6 4.6 0 9\\
8.6 4.6 4 9\\
8.6 5.4 4 9\\
8.6 5.4 0 9\\
8.6 4.6 0 9\\
NaN NaN NaN 9\\
8.6 5.6 0 9\\
8.6 5.6 4 9\\
8.6 6.4 4 9\\
8.6 6.4 0 9\\
8.6 5.6 0 9\\
NaN NaN NaN 9\\
8.6 6.6 0 9\\
8.6 6.6 4 9\\
8.6 7.4 4 9\\
8.6 7.4 0 9\\
8.6 6.6 0 9\\
NaN NaN NaN 9\\
8.6 7.6 0 9\\
8.6 7.6 4 9\\
8.6 8.4 4 9\\
8.6 8.4 0 9\\
8.6 7.6 0 9\\
NaN NaN NaN 9\\
8.6 8.6 0 9\\
8.6 8.6 4 9\\
8.6 9.4 4 9\\
8.6 9.4 0 9\\
8.6 8.6 0 9\\
NaN NaN NaN 9\\
8.6 9.6 0 9\\
8.6 9.6 4 9\\
8.6 10.4 4 9\\
8.6 10.4 0 9\\
8.6 9.6 0 9\\
NaN NaN NaN 9\\
8.6 10.6 0 9\\
8.6 10.6 4 9\\
8.6 11.4 4 9\\
8.6 11.4 0 9\\
8.6 10.6 0 9\\
NaN NaN NaN 9\\
9.4 0.6 0 9\\
9.4 0.6 3 9\\
9.4 1.4 3 9\\
9.4 1.4 0 9\\
9.4 0.6 0 9\\
NaN NaN NaN 9\\
9.4 1.6 0 9\\
9.4 1.6 3 9\\
9.4 2.4 3 9\\
9.4 2.4 0 9\\
9.4 1.6 0 9\\
NaN NaN NaN 9\\
9.4 2.6 0 9\\
9.4 2.6 4 9\\
9.4 3.4 4 9\\
9.4 3.4 0 9\\
9.4 2.6 0 9\\
NaN NaN NaN 9\\
9.4 3.6 0 9\\
9.4 3.6 4 9\\
9.4 4.4 4 9\\
9.4 4.4 0 9\\
9.4 3.6 0 9\\
NaN NaN NaN 9\\
9.4 4.6 0 9\\
9.4 4.6 4 9\\
9.4 5.4 4 9\\
9.4 5.4 0 9\\
9.4 4.6 0 9\\
NaN NaN NaN 9\\
9.4 5.6 0 9\\
9.4 5.6 4 9\\
9.4 6.4 4 9\\
9.4 6.4 0 9\\
9.4 5.6 0 9\\
NaN NaN NaN 9\\
9.4 6.6 0 9\\
9.4 6.6 4 9\\
9.4 7.4 4 9\\
9.4 7.4 0 9\\
9.4 6.6 0 9\\
NaN NaN NaN 9\\
9.4 7.6 0 9\\
9.4 7.6 4 9\\
9.4 8.4 4 9\\
9.4 8.4 0 9\\
9.4 7.6 0 9\\
NaN NaN NaN 9\\
9.4 8.6 0 9\\
9.4 8.6 4 9\\
9.4 9.4 4 9\\
9.4 9.4 0 9\\
9.4 8.6 0 9\\
NaN NaN NaN 9\\
9.4 9.6 0 9\\
9.4 9.6 4 9\\
9.4 10.4 4 9\\
9.4 10.4 0 9\\
9.4 9.6 0 9\\
NaN NaN NaN 9\\
9.4 10.6 0 9\\
9.4 10.6 4 9\\
9.4 11.4 4 9\\
9.4 11.4 0 9\\
9.4 10.6 0 9\\
NaN NaN NaN 9\\
NaN NaN NaN 9\\
9.4 0.6 0 9\\
9.4 1.4 0 9\\
NaN NaN NaN 9\\
NaN NaN NaN 9\\
NaN NaN NaN 9\\
NaN NaN NaN 9\\
9.4 1.6 0 9\\
9.4 2.4 0 9\\
NaN NaN NaN 9\\
NaN NaN NaN 9\\
NaN NaN NaN 9\\
NaN NaN NaN 9\\
9.4 2.6 0 9\\
9.4 3.4 0 9\\
NaN NaN NaN 9\\
NaN NaN NaN 9\\
NaN NaN NaN 9\\
NaN NaN NaN 9\\
9.4 3.6 0 9\\
9.4 4.4 0 9\\
NaN NaN NaN 9\\
NaN NaN NaN 9\\
NaN NaN NaN 9\\
NaN NaN NaN 9\\
9.4 4.6 0 9\\
9.4 5.4 0 9\\
NaN NaN NaN 9\\
NaN NaN NaN 9\\
NaN NaN NaN 9\\
NaN NaN NaN 9\\
9.4 5.6 0 9\\
9.4 6.4 0 9\\
NaN NaN NaN 9\\
NaN NaN NaN 9\\
NaN NaN NaN 9\\
NaN NaN NaN 9\\
9.4 6.6 0 9\\
9.4 7.4 0 9\\
NaN NaN NaN 9\\
NaN NaN NaN 9\\
NaN NaN NaN 9\\
NaN NaN NaN 9\\
9.4 7.6 0 9\\
9.4 8.4 0 9\\
NaN NaN NaN 9\\
NaN NaN NaN 9\\
NaN NaN NaN 9\\
NaN NaN NaN 9\\
9.4 8.6 0 9\\
9.4 9.4 0 9\\
NaN NaN NaN 9\\
NaN NaN NaN 9\\
NaN NaN NaN 9\\
NaN NaN NaN 9\\
9.4 9.6 0 9\\
9.4 10.4 0 9\\
NaN NaN NaN 9\\
NaN NaN NaN 9\\
NaN NaN NaN 9\\
NaN NaN NaN 9\\
9.4 10.6 0 9\\
9.4 11.4 0 9\\
NaN NaN NaN 9\\
NaN NaN NaN 9\\
NaN NaN NaN 9\\
};

\addplot3[%
surf,
shader=flat,
draw=black,
point meta=explicit,
mesh/rows=4]
table[row sep=crcr,header=false,meta index=3] {
NaN NaN NaN 8\\
7.6 0.6 0 8\\
7.6 1.4 0 8\\
NaN NaN NaN 8\\
NaN NaN NaN 8\\
NaN NaN NaN 8\\
NaN NaN NaN 8\\
7.6 1.6 0 8\\
7.6 2.4 0 8\\
NaN NaN NaN 8\\
NaN NaN NaN 8\\
NaN NaN NaN 8\\
NaN NaN NaN 8\\
7.6 2.6 0 8\\
7.6 3.4 0 8\\
NaN NaN NaN 8\\
NaN NaN NaN 8\\
NaN NaN NaN 8\\
NaN NaN NaN 8\\
7.6 3.6 0 8\\
7.6 4.4 0 8\\
NaN NaN NaN 8\\
NaN NaN NaN 8\\
NaN NaN NaN 8\\
NaN NaN NaN 8\\
7.6 4.6 0 8\\
7.6 5.4 0 8\\
NaN NaN NaN 8\\
NaN NaN NaN 8\\
NaN NaN NaN 8\\
NaN NaN NaN 8\\
7.6 5.6 0 8\\
7.6 6.4 0 8\\
NaN NaN NaN 8\\
NaN NaN NaN 8\\
NaN NaN NaN 8\\
NaN NaN NaN 8\\
7.6 6.6 0 8\\
7.6 7.4 0 8\\
NaN NaN NaN 8\\
NaN NaN NaN 8\\
NaN NaN NaN 8\\
NaN NaN NaN 8\\
7.6 7.6 0 8\\
7.6 8.4 0 8\\
NaN NaN NaN 8\\
NaN NaN NaN 8\\
NaN NaN NaN 8\\
NaN NaN NaN 8\\
7.6 8.6 0 8\\
7.6 9.4 0 8\\
NaN NaN NaN 8\\
NaN NaN NaN 8\\
NaN NaN NaN 8\\
NaN NaN NaN 8\\
7.6 9.6 0 8\\
7.6 10.4 0 8\\
NaN NaN NaN 8\\
NaN NaN NaN 8\\
NaN NaN NaN 8\\
NaN NaN NaN 8\\
7.6 10.6 0 8\\
7.6 11.4 0 8\\
NaN NaN NaN 8\\
NaN NaN NaN 8\\
NaN NaN NaN 8\\
7.6 0.6 0 8\\
7.6 0.6 3 8\\
7.6 1.4 3 8\\
7.6 1.4 0 8\\
7.6 0.6 0 8\\
NaN NaN NaN 8\\
7.6 1.6 0 8\\
7.6 1.6 3 8\\
7.6 2.4 3 8\\
7.6 2.4 0 8\\
7.6 1.6 0 8\\
NaN NaN NaN 8\\
7.6 2.6 0 8\\
7.6 2.6 3 8\\
7.6 3.4 3 8\\
7.6 3.4 0 8\\
7.6 2.6 0 8\\
NaN NaN NaN 8\\
7.6 3.6 0 8\\
7.6 3.6 3 8\\
7.6 4.4 3 8\\
7.6 4.4 0 8\\
7.6 3.6 0 8\\
NaN NaN NaN 8\\
7.6 4.6 0 8\\
7.6 4.6 3 8\\
7.6 5.4 3 8\\
7.6 5.4 0 8\\
7.6 4.6 0 8\\
NaN NaN NaN 8\\
7.6 5.6 0 8\\
7.6 5.6 3 8\\
7.6 6.4 3 8\\
7.6 6.4 0 8\\
7.6 5.6 0 8\\
NaN NaN NaN 8\\
7.6 6.6 0 8\\
7.6 6.6 4 8\\
7.6 7.4 4 8\\
7.6 7.4 0 8\\
7.6 6.6 0 8\\
NaN NaN NaN 8\\
7.6 7.6 0 8\\
7.6 7.6 4 8\\
7.6 8.4 4 8\\
7.6 8.4 0 8\\
7.6 7.6 0 8\\
NaN NaN NaN 8\\
7.6 8.6 0 8\\
7.6 8.6 4 8\\
7.6 9.4 4 8\\
7.6 9.4 0 8\\
7.6 8.6 0 8\\
NaN NaN NaN 8\\
7.6 9.6 0 8\\
7.6 9.6 4 8\\
7.6 10.4 4 8\\
7.6 10.4 0 8\\
7.6 9.6 0 8\\
NaN NaN NaN 8\\
7.6 10.6 0 8\\
7.6 10.6 4 8\\
7.6 11.4 4 8\\
7.6 11.4 0 8\\
7.6 10.6 0 8\\
NaN NaN NaN 8\\
8.4 0.6 0 8\\
8.4 0.6 3 8\\
8.4 1.4 3 8\\
8.4 1.4 0 8\\
8.4 0.6 0 8\\
NaN NaN NaN 8\\
8.4 1.6 0 8\\
8.4 1.6 3 8\\
8.4 2.4 3 8\\
8.4 2.4 0 8\\
8.4 1.6 0 8\\
NaN NaN NaN 8\\
8.4 2.6 0 8\\
8.4 2.6 3 8\\
8.4 3.4 3 8\\
8.4 3.4 0 8\\
8.4 2.6 0 8\\
NaN NaN NaN 8\\
8.4 3.6 0 8\\
8.4 3.6 3 8\\
8.4 4.4 3 8\\
8.4 4.4 0 8\\
8.4 3.6 0 8\\
NaN NaN NaN 8\\
8.4 4.6 0 8\\
8.4 4.6 3 8\\
8.4 5.4 3 8\\
8.4 5.4 0 8\\
8.4 4.6 0 8\\
NaN NaN NaN 8\\
8.4 5.6 0 8\\
8.4 5.6 3 8\\
8.4 6.4 3 8\\
8.4 6.4 0 8\\
8.4 5.6 0 8\\
NaN NaN NaN 8\\
8.4 6.6 0 8\\
8.4 6.6 4 8\\
8.4 7.4 4 8\\
8.4 7.4 0 8\\
8.4 6.6 0 8\\
NaN NaN NaN 8\\
8.4 7.6 0 8\\
8.4 7.6 4 8\\
8.4 8.4 4 8\\
8.4 8.4 0 8\\
8.4 7.6 0 8\\
NaN NaN NaN 8\\
8.4 8.6 0 8\\
8.4 8.6 4 8\\
8.4 9.4 4 8\\
8.4 9.4 0 8\\
8.4 8.6 0 8\\
NaN NaN NaN 8\\
8.4 9.6 0 8\\
8.4 9.6 4 8\\
8.4 10.4 4 8\\
8.4 10.4 0 8\\
8.4 9.6 0 8\\
NaN NaN NaN 8\\
8.4 10.6 0 8\\
8.4 10.6 4 8\\
8.4 11.4 4 8\\
8.4 11.4 0 8\\
8.4 10.6 0 8\\
NaN NaN NaN 8\\
NaN NaN NaN 8\\
8.4 0.6 0 8\\
8.4 1.4 0 8\\
NaN NaN NaN 8\\
NaN NaN NaN 8\\
NaN NaN NaN 8\\
NaN NaN NaN 8\\
8.4 1.6 0 8\\
8.4 2.4 0 8\\
NaN NaN NaN 8\\
NaN NaN NaN 8\\
NaN NaN NaN 8\\
NaN NaN NaN 8\\
8.4 2.6 0 8\\
8.4 3.4 0 8\\
NaN NaN NaN 8\\
NaN NaN NaN 8\\
NaN NaN NaN 8\\
NaN NaN NaN 8\\
8.4 3.6 0 8\\
8.4 4.4 0 8\\
NaN NaN NaN 8\\
NaN NaN NaN 8\\
NaN NaN NaN 8\\
NaN NaN NaN 8\\
8.4 4.6 0 8\\
8.4 5.4 0 8\\
NaN NaN NaN 8\\
NaN NaN NaN 8\\
NaN NaN NaN 8\\
NaN NaN NaN 8\\
8.4 5.6 0 8\\
8.4 6.4 0 8\\
NaN NaN NaN 8\\
NaN NaN NaN 8\\
NaN NaN NaN 8\\
NaN NaN NaN 8\\
8.4 6.6 0 8\\
8.4 7.4 0 8\\
NaN NaN NaN 8\\
NaN NaN NaN 8\\
NaN NaN NaN 8\\
NaN NaN NaN 8\\
8.4 7.6 0 8\\
8.4 8.4 0 8\\
NaN NaN NaN 8\\
NaN NaN NaN 8\\
NaN NaN NaN 8\\
NaN NaN NaN 8\\
8.4 8.6 0 8\\
8.4 9.4 0 8\\
NaN NaN NaN 8\\
NaN NaN NaN 8\\
NaN NaN NaN 8\\
NaN NaN NaN 8\\
8.4 9.6 0 8\\
8.4 10.4 0 8\\
NaN NaN NaN 8\\
NaN NaN NaN 8\\
NaN NaN NaN 8\\
NaN NaN NaN 8\\
8.4 10.6 0 8\\
8.4 11.4 0 8\\
NaN NaN NaN 8\\
NaN NaN NaN 8\\
NaN NaN NaN 8\\
};

\addplot3[%
surf,
shader=flat,
draw=black,
point meta=explicit,
mesh/rows=4]
table[row sep=crcr,header=false,meta index=3] {
NaN NaN NaN 7\\
6.6 0.6 0 7\\
6.6 1.4 0 7\\
NaN NaN NaN 7\\
NaN NaN NaN 7\\
NaN NaN NaN 7\\
NaN NaN NaN 7\\
6.6 1.6 0 7\\
6.6 2.4 0 7\\
NaN NaN NaN 7\\
NaN NaN NaN 7\\
NaN NaN NaN 7\\
NaN NaN NaN 7\\
6.6 2.6 0 7\\
6.6 3.4 0 7\\
NaN NaN NaN 7\\
NaN NaN NaN 7\\
NaN NaN NaN 7\\
NaN NaN NaN 7\\
6.6 3.6 0 7\\
6.6 4.4 0 7\\
NaN NaN NaN 7\\
NaN NaN NaN 7\\
NaN NaN NaN 7\\
NaN NaN NaN 7\\
6.6 4.6 0 7\\
6.6 5.4 0 7\\
NaN NaN NaN 7\\
NaN NaN NaN 7\\
NaN NaN NaN 7\\
NaN NaN NaN 7\\
6.6 5.6 0 7\\
6.6 6.4 0 7\\
NaN NaN NaN 7\\
NaN NaN NaN 7\\
NaN NaN NaN 7\\
NaN NaN NaN 7\\
6.6 6.6 0 7\\
6.6 7.4 0 7\\
NaN NaN NaN 7\\
NaN NaN NaN 7\\
NaN NaN NaN 7\\
NaN NaN NaN 7\\
6.6 7.6 0 7\\
6.6 8.4 0 7\\
NaN NaN NaN 7\\
NaN NaN NaN 7\\
NaN NaN NaN 7\\
NaN NaN NaN 7\\
6.6 8.6 0 7\\
6.6 9.4 0 7\\
NaN NaN NaN 7\\
NaN NaN NaN 7\\
NaN NaN NaN 7\\
NaN NaN NaN 7\\
6.6 9.6 0 7\\
6.6 10.4 0 7\\
NaN NaN NaN 7\\
NaN NaN NaN 7\\
NaN NaN NaN 7\\
NaN NaN NaN 7\\
6.6 10.6 0 7\\
6.6 11.4 0 7\\
NaN NaN NaN 7\\
NaN NaN NaN 7\\
NaN NaN NaN 7\\
6.6 0.6 0 7\\
6.6 0.6 3 7\\
6.6 1.4 3 7\\
6.6 1.4 0 7\\
6.6 0.6 0 7\\
NaN NaN NaN 7\\
6.6 1.6 0 7\\
6.6 1.6 3 7\\
6.6 2.4 3 7\\
6.6 2.4 0 7\\
6.6 1.6 0 7\\
NaN NaN NaN 7\\
6.6 2.6 0 7\\
6.6 2.6 3 7\\
6.6 3.4 3 7\\
6.6 3.4 0 7\\
6.6 2.6 0 7\\
NaN NaN NaN 7\\
6.6 3.6 0 7\\
6.6 3.6 3 7\\
6.6 4.4 3 7\\
6.6 4.4 0 7\\
6.6 3.6 0 7\\
NaN NaN NaN 7\\
6.6 4.6 0 7\\
6.6 4.6 3 7\\
6.6 5.4 3 7\\
6.6 5.4 0 7\\
6.6 4.6 0 7\\
NaN NaN NaN 7\\
6.6 5.6 0 7\\
6.6 5.6 3 7\\
6.6 6.4 3 7\\
6.6 6.4 0 7\\
6.6 5.6 0 7\\
NaN NaN NaN 7\\
6.6 6.6 0 7\\
6.6 6.6 4 7\\
6.6 7.4 4 7\\
6.6 7.4 0 7\\
6.6 6.6 0 7\\
NaN NaN NaN 7\\
6.6 7.6 0 7\\
6.6 7.6 4 7\\
6.6 8.4 4 7\\
6.6 8.4 0 7\\
6.6 7.6 0 7\\
NaN NaN NaN 7\\
6.6 8.6 0 7\\
6.6 8.6 4 7\\
6.6 9.4 4 7\\
6.6 9.4 0 7\\
6.6 8.6 0 7\\
NaN NaN NaN 7\\
6.6 9.6 0 7\\
6.6 9.6 4 7\\
6.6 10.4 4 7\\
6.6 10.4 0 7\\
6.6 9.6 0 7\\
NaN NaN NaN 7\\
6.6 10.6 0 7\\
6.6 10.6 4 7\\
6.6 11.4 4 7\\
6.6 11.4 0 7\\
6.6 10.6 0 7\\
NaN NaN NaN 7\\
7.4 0.6 0 7\\
7.4 0.6 3 7\\
7.4 1.4 3 7\\
7.4 1.4 0 7\\
7.4 0.6 0 7\\
NaN NaN NaN 7\\
7.4 1.6 0 7\\
7.4 1.6 3 7\\
7.4 2.4 3 7\\
7.4 2.4 0 7\\
7.4 1.6 0 7\\
NaN NaN NaN 7\\
7.4 2.6 0 7\\
7.4 2.6 3 7\\
7.4 3.4 3 7\\
7.4 3.4 0 7\\
7.4 2.6 0 7\\
NaN NaN NaN 7\\
7.4 3.6 0 7\\
7.4 3.6 3 7\\
7.4 4.4 3 7\\
7.4 4.4 0 7\\
7.4 3.6 0 7\\
NaN NaN NaN 7\\
7.4 4.6 0 7\\
7.4 4.6 3 7\\
7.4 5.4 3 7\\
7.4 5.4 0 7\\
7.4 4.6 0 7\\
NaN NaN NaN 7\\
7.4 5.6 0 7\\
7.4 5.6 3 7\\
7.4 6.4 3 7\\
7.4 6.4 0 7\\
7.4 5.6 0 7\\
NaN NaN NaN 7\\
7.4 6.6 0 7\\
7.4 6.6 4 7\\
7.4 7.4 4 7\\
7.4 7.4 0 7\\
7.4 6.6 0 7\\
NaN NaN NaN 7\\
7.4 7.6 0 7\\
7.4 7.6 4 7\\
7.4 8.4 4 7\\
7.4 8.4 0 7\\
7.4 7.6 0 7\\
NaN NaN NaN 7\\
7.4 8.6 0 7\\
7.4 8.6 4 7\\
7.4 9.4 4 7\\
7.4 9.4 0 7\\
7.4 8.6 0 7\\
NaN NaN NaN 7\\
7.4 9.6 0 7\\
7.4 9.6 4 7\\
7.4 10.4 4 7\\
7.4 10.4 0 7\\
7.4 9.6 0 7\\
NaN NaN NaN 7\\
7.4 10.6 0 7\\
7.4 10.6 4 7\\
7.4 11.4 4 7\\
7.4 11.4 0 7\\
7.4 10.6 0 7\\
NaN NaN NaN 7\\
NaN NaN NaN 7\\
7.4 0.6 0 7\\
7.4 1.4 0 7\\
NaN NaN NaN 7\\
NaN NaN NaN 7\\
NaN NaN NaN 7\\
NaN NaN NaN 7\\
7.4 1.6 0 7\\
7.4 2.4 0 7\\
NaN NaN NaN 7\\
NaN NaN NaN 7\\
NaN NaN NaN 7\\
NaN NaN NaN 7\\
7.4 2.6 0 7\\
7.4 3.4 0 7\\
NaN NaN NaN 7\\
NaN NaN NaN 7\\
NaN NaN NaN 7\\
NaN NaN NaN 7\\
7.4 3.6 0 7\\
7.4 4.4 0 7\\
NaN NaN NaN 7\\
NaN NaN NaN 7\\
NaN NaN NaN 7\\
NaN NaN NaN 7\\
7.4 4.6 0 7\\
7.4 5.4 0 7\\
NaN NaN NaN 7\\
NaN NaN NaN 7\\
NaN NaN NaN 7\\
NaN NaN NaN 7\\
7.4 5.6 0 7\\
7.4 6.4 0 7\\
NaN NaN NaN 7\\
NaN NaN NaN 7\\
NaN NaN NaN 7\\
NaN NaN NaN 7\\
7.4 6.6 0 7\\
7.4 7.4 0 7\\
NaN NaN NaN 7\\
NaN NaN NaN 7\\
NaN NaN NaN 7\\
NaN NaN NaN 7\\
7.4 7.6 0 7\\
7.4 8.4 0 7\\
NaN NaN NaN 7\\
NaN NaN NaN 7\\
NaN NaN NaN 7\\
NaN NaN NaN 7\\
7.4 8.6 0 7\\
7.4 9.4 0 7\\
NaN NaN NaN 7\\
NaN NaN NaN 7\\
NaN NaN NaN 7\\
NaN NaN NaN 7\\
7.4 9.6 0 7\\
7.4 10.4 0 7\\
NaN NaN NaN 7\\
NaN NaN NaN 7\\
NaN NaN NaN 7\\
NaN NaN NaN 7\\
7.4 10.6 0 7\\
7.4 11.4 0 7\\
NaN NaN NaN 7\\
NaN NaN NaN 7\\
NaN NaN NaN 7\\
};

\addplot3[%
surf,
shader=flat,
draw=black,
point meta=explicit,
mesh/rows=4]
table[row sep=crcr,header=false,meta index=3] {
NaN NaN NaN 6\\
5.6 0.6 0 6\\
5.6 1.4 0 6\\
NaN NaN NaN 6\\
NaN NaN NaN 6\\
NaN NaN NaN 6\\
NaN NaN NaN 6\\
5.6 1.6 0 6\\
5.6 2.4 0 6\\
NaN NaN NaN 6\\
NaN NaN NaN 6\\
NaN NaN NaN 6\\
NaN NaN NaN 6\\
5.6 2.6 0 6\\
5.6 3.4 0 6\\
NaN NaN NaN 6\\
NaN NaN NaN 6\\
NaN NaN NaN 6\\
NaN NaN NaN 6\\
5.6 3.6 0 6\\
5.6 4.4 0 6\\
NaN NaN NaN 6\\
NaN NaN NaN 6\\
NaN NaN NaN 6\\
NaN NaN NaN 6\\
5.6 4.6 0 6\\
5.6 5.4 0 6\\
NaN NaN NaN 6\\
NaN NaN NaN 6\\
NaN NaN NaN 6\\
NaN NaN NaN 6\\
5.6 5.6 0 6\\
5.6 6.4 0 6\\
NaN NaN NaN 6\\
NaN NaN NaN 6\\
NaN NaN NaN 6\\
NaN NaN NaN 6\\
5.6 6.6 0 6\\
5.6 7.4 0 6\\
NaN NaN NaN 6\\
NaN NaN NaN 6\\
NaN NaN NaN 6\\
NaN NaN NaN 6\\
5.6 7.6 0 6\\
5.6 8.4 0 6\\
NaN NaN NaN 6\\
NaN NaN NaN 6\\
NaN NaN NaN 6\\
NaN NaN NaN 6\\
5.6 8.6 0 6\\
5.6 9.4 0 6\\
NaN NaN NaN 6\\
NaN NaN NaN 6\\
NaN NaN NaN 6\\
NaN NaN NaN 6\\
5.6 9.6 0 6\\
5.6 10.4 0 6\\
NaN NaN NaN 6\\
NaN NaN NaN 6\\
NaN NaN NaN 6\\
NaN NaN NaN 6\\
5.6 10.6 0 6\\
5.6 11.4 0 6\\
NaN NaN NaN 6\\
NaN NaN NaN 6\\
NaN NaN NaN 6\\
5.6 0.6 0 6\\
5.6 0.6 3 6\\
5.6 1.4 3 6\\
5.6 1.4 0 6\\
5.6 0.6 0 6\\
NaN NaN NaN 6\\
5.6 1.6 0 6\\
5.6 1.6 3 6\\
5.6 2.4 3 6\\
5.6 2.4 0 6\\
5.6 1.6 0 6\\
NaN NaN NaN 6\\
5.6 2.6 0 6\\
5.6 2.6 3 6\\
5.6 3.4 3 6\\
5.6 3.4 0 6\\
5.6 2.6 0 6\\
NaN NaN NaN 6\\
5.6 3.6 0 6\\
5.6 3.6 3 6\\
5.6 4.4 3 6\\
5.6 4.4 0 6\\
5.6 3.6 0 6\\
NaN NaN NaN 6\\
5.6 4.6 0 6\\
5.6 4.6 3 6\\
5.6 5.4 3 6\\
5.6 5.4 0 6\\
5.6 4.6 0 6\\
NaN NaN NaN 6\\
5.6 5.6 0 6\\
5.6 5.6 3 6\\
5.6 6.4 3 6\\
5.6 6.4 0 6\\
5.6 5.6 0 6\\
NaN NaN NaN 6\\
5.6 6.6 0 6\\
5.6 6.6 3 6\\
5.6 7.4 3 6\\
5.6 7.4 0 6\\
5.6 6.6 0 6\\
NaN NaN NaN 6\\
5.6 7.6 0 6\\
5.6 7.6 3 6\\
5.6 8.4 3 6\\
5.6 8.4 0 6\\
5.6 7.6 0 6\\
NaN NaN NaN 6\\
5.6 8.6 0 6\\
5.6 8.6 3 6\\
5.6 9.4 3 6\\
5.6 9.4 0 6\\
5.6 8.6 0 6\\
NaN NaN NaN 6\\
5.6 9.6 0 6\\
5.6 9.6 3 6\\
5.6 10.4 3 6\\
5.6 10.4 0 6\\
5.6 9.6 0 6\\
NaN NaN NaN 6\\
5.6 10.6 0 6\\
5.6 10.6 3 6\\
5.6 11.4 3 6\\
5.6 11.4 0 6\\
5.6 10.6 0 6\\
NaN NaN NaN 6\\
6.4 0.6 0 6\\
6.4 0.6 3 6\\
6.4 1.4 3 6\\
6.4 1.4 0 6\\
6.4 0.6 0 6\\
NaN NaN NaN 6\\
6.4 1.6 0 6\\
6.4 1.6 3 6\\
6.4 2.4 3 6\\
6.4 2.4 0 6\\
6.4 1.6 0 6\\
NaN NaN NaN 6\\
6.4 2.6 0 6\\
6.4 2.6 3 6\\
6.4 3.4 3 6\\
6.4 3.4 0 6\\
6.4 2.6 0 6\\
NaN NaN NaN 6\\
6.4 3.6 0 6\\
6.4 3.6 3 6\\
6.4 4.4 3 6\\
6.4 4.4 0 6\\
6.4 3.6 0 6\\
NaN NaN NaN 6\\
6.4 4.6 0 6\\
6.4 4.6 3 6\\
6.4 5.4 3 6\\
6.4 5.4 0 6\\
6.4 4.6 0 6\\
NaN NaN NaN 6\\
6.4 5.6 0 6\\
6.4 5.6 3 6\\
6.4 6.4 3 6\\
6.4 6.4 0 6\\
6.4 5.6 0 6\\
NaN NaN NaN 6\\
6.4 6.6 0 6\\
6.4 6.6 3 6\\
6.4 7.4 3 6\\
6.4 7.4 0 6\\
6.4 6.6 0 6\\
NaN NaN NaN 6\\
6.4 7.6 0 6\\
6.4 7.6 3 6\\
6.4 8.4 3 6\\
6.4 8.4 0 6\\
6.4 7.6 0 6\\
NaN NaN NaN 6\\
6.4 8.6 0 6\\
6.4 8.6 3 6\\
6.4 9.4 3 6\\
6.4 9.4 0 6\\
6.4 8.6 0 6\\
NaN NaN NaN 6\\
6.4 9.6 0 6\\
6.4 9.6 3 6\\
6.4 10.4 3 6\\
6.4 10.4 0 6\\
6.4 9.6 0 6\\
NaN NaN NaN 6\\
6.4 10.6 0 6\\
6.4 10.6 3 6\\
6.4 11.4 3 6\\
6.4 11.4 0 6\\
6.4 10.6 0 6\\
NaN NaN NaN 6\\
NaN NaN NaN 6\\
6.4 0.6 0 6\\
6.4 1.4 0 6\\
NaN NaN NaN 6\\
NaN NaN NaN 6\\
NaN NaN NaN 6\\
NaN NaN NaN 6\\
6.4 1.6 0 6\\
6.4 2.4 0 6\\
NaN NaN NaN 6\\
NaN NaN NaN 6\\
NaN NaN NaN 6\\
NaN NaN NaN 6\\
6.4 2.6 0 6\\
6.4 3.4 0 6\\
NaN NaN NaN 6\\
NaN NaN NaN 6\\
NaN NaN NaN 6\\
NaN NaN NaN 6\\
6.4 3.6 0 6\\
6.4 4.4 0 6\\
NaN NaN NaN 6\\
NaN NaN NaN 6\\
NaN NaN NaN 6\\
NaN NaN NaN 6\\
6.4 4.6 0 6\\
6.4 5.4 0 6\\
NaN NaN NaN 6\\
NaN NaN NaN 6\\
NaN NaN NaN 6\\
NaN NaN NaN 6\\
6.4 5.6 0 6\\
6.4 6.4 0 6\\
NaN NaN NaN 6\\
NaN NaN NaN 6\\
NaN NaN NaN 6\\
NaN NaN NaN 6\\
6.4 6.6 0 6\\
6.4 7.4 0 6\\
NaN NaN NaN 6\\
NaN NaN NaN 6\\
NaN NaN NaN 6\\
NaN NaN NaN 6\\
6.4 7.6 0 6\\
6.4 8.4 0 6\\
NaN NaN NaN 6\\
NaN NaN NaN 6\\
NaN NaN NaN 6\\
NaN NaN NaN 6\\
6.4 8.6 0 6\\
6.4 9.4 0 6\\
NaN NaN NaN 6\\
NaN NaN NaN 6\\
NaN NaN NaN 6\\
NaN NaN NaN 6\\
6.4 9.6 0 6\\
6.4 10.4 0 6\\
NaN NaN NaN 6\\
NaN NaN NaN 6\\
NaN NaN NaN 6\\
NaN NaN NaN 6\\
6.4 10.6 0 6\\
6.4 11.4 0 6\\
NaN NaN NaN 6\\
NaN NaN NaN 6\\
NaN NaN NaN 6\\
};

\addplot3[%
surf,
shader=flat,
draw=black,
point meta=explicit,
mesh/rows=4]
table[row sep=crcr,header=false,meta index=3] {
NaN NaN NaN 5\\
4.6 0.6 0 5\\
4.6 1.4 0 5\\
NaN NaN NaN 5\\
NaN NaN NaN 5\\
NaN NaN NaN 5\\
NaN NaN NaN 5\\
4.6 1.6 0 5\\
4.6 2.4 0 5\\
NaN NaN NaN 5\\
NaN NaN NaN 5\\
NaN NaN NaN 5\\
NaN NaN NaN 5\\
4.6 2.6 0 5\\
4.6 3.4 0 5\\
NaN NaN NaN 5\\
NaN NaN NaN 5\\
NaN NaN NaN 5\\
NaN NaN NaN 5\\
4.6 3.6 0 5\\
4.6 4.4 0 5\\
NaN NaN NaN 5\\
NaN NaN NaN 5\\
NaN NaN NaN 5\\
NaN NaN NaN 5\\
4.6 4.6 0 5\\
4.6 5.4 0 5\\
NaN NaN NaN 5\\
NaN NaN NaN 5\\
NaN NaN NaN 5\\
NaN NaN NaN 5\\
4.6 5.6 0 5\\
4.6 6.4 0 5\\
NaN NaN NaN 5\\
NaN NaN NaN 5\\
NaN NaN NaN 5\\
NaN NaN NaN 5\\
4.6 6.6 0 5\\
4.6 7.4 0 5\\
NaN NaN NaN 5\\
NaN NaN NaN 5\\
NaN NaN NaN 5\\
NaN NaN NaN 5\\
4.6 7.6 0 5\\
4.6 8.4 0 5\\
NaN NaN NaN 5\\
NaN NaN NaN 5\\
NaN NaN NaN 5\\
NaN NaN NaN 5\\
4.6 8.6 0 5\\
4.6 9.4 0 5\\
NaN NaN NaN 5\\
NaN NaN NaN 5\\
NaN NaN NaN 5\\
NaN NaN NaN 5\\
4.6 9.6 0 5\\
4.6 10.4 0 5\\
NaN NaN NaN 5\\
NaN NaN NaN 5\\
NaN NaN NaN 5\\
NaN NaN NaN 5\\
4.6 10.6 0 5\\
4.6 11.4 0 5\\
NaN NaN NaN 5\\
NaN NaN NaN 5\\
NaN NaN NaN 5\\
4.6 0.6 0 5\\
4.6 0.6 2 5\\
4.6 1.4 2 5\\
4.6 1.4 0 5\\
4.6 0.6 0 5\\
NaN NaN NaN 5\\
4.6 1.6 0 5\\
4.6 1.6 3 5\\
4.6 2.4 3 5\\
4.6 2.4 0 5\\
4.6 1.6 0 5\\
NaN NaN NaN 5\\
4.6 2.6 0 5\\
4.6 2.6 3 5\\
4.6 3.4 3 5\\
4.6 3.4 0 5\\
4.6 2.6 0 5\\
NaN NaN NaN 5\\
4.6 3.6 0 5\\
4.6 3.6 3 5\\
4.6 4.4 3 5\\
4.6 4.4 0 5\\
4.6 3.6 0 5\\
NaN NaN NaN 5\\
4.6 4.6 0 5\\
4.6 4.6 3 5\\
4.6 5.4 3 5\\
4.6 5.4 0 5\\
4.6 4.6 0 5\\
NaN NaN NaN 5\\
4.6 5.6 0 5\\
4.6 5.6 3 5\\
4.6 6.4 3 5\\
4.6 6.4 0 5\\
4.6 5.6 0 5\\
NaN NaN NaN 5\\
4.6 6.6 0 5\\
4.6 6.6 3 5\\
4.6 7.4 3 5\\
4.6 7.4 0 5\\
4.6 6.6 0 5\\
NaN NaN NaN 5\\
4.6 7.6 0 5\\
4.6 7.6 3 5\\
4.6 8.4 3 5\\
4.6 8.4 0 5\\
4.6 7.6 0 5\\
NaN NaN NaN 5\\
4.6 8.6 0 5\\
4.6 8.6 3 5\\
4.6 9.4 3 5\\
4.6 9.4 0 5\\
4.6 8.6 0 5\\
NaN NaN NaN 5\\
4.6 9.6 0 5\\
4.6 9.6 3 5\\
4.6 10.4 3 5\\
4.6 10.4 0 5\\
4.6 9.6 0 5\\
NaN NaN NaN 5\\
4.6 10.6 0 5\\
4.6 10.6 3 5\\
4.6 11.4 3 5\\
4.6 11.4 0 5\\
4.6 10.6 0 5\\
NaN NaN NaN 5\\
5.4 0.6 0 5\\
5.4 0.6 2 5\\
5.4 1.4 2 5\\
5.4 1.4 0 5\\
5.4 0.6 0 5\\
NaN NaN NaN 5\\
5.4 1.6 0 5\\
5.4 1.6 3 5\\
5.4 2.4 3 5\\
5.4 2.4 0 5\\
5.4 1.6 0 5\\
NaN NaN NaN 5\\
5.4 2.6 0 5\\
5.4 2.6 3 5\\
5.4 3.4 3 5\\
5.4 3.4 0 5\\
5.4 2.6 0 5\\
NaN NaN NaN 5\\
5.4 3.6 0 5\\
5.4 3.6 3 5\\
5.4 4.4 3 5\\
5.4 4.4 0 5\\
5.4 3.6 0 5\\
NaN NaN NaN 5\\
5.4 4.6 0 5\\
5.4 4.6 3 5\\
5.4 5.4 3 5\\
5.4 5.4 0 5\\
5.4 4.6 0 5\\
NaN NaN NaN 5\\
5.4 5.6 0 5\\
5.4 5.6 3 5\\
5.4 6.4 3 5\\
5.4 6.4 0 5\\
5.4 5.6 0 5\\
NaN NaN NaN 5\\
5.4 6.6 0 5\\
5.4 6.6 3 5\\
5.4 7.4 3 5\\
5.4 7.4 0 5\\
5.4 6.6 0 5\\
NaN NaN NaN 5\\
5.4 7.6 0 5\\
5.4 7.6 3 5\\
5.4 8.4 3 5\\
5.4 8.4 0 5\\
5.4 7.6 0 5\\
NaN NaN NaN 5\\
5.4 8.6 0 5\\
5.4 8.6 3 5\\
5.4 9.4 3 5\\
5.4 9.4 0 5\\
5.4 8.6 0 5\\
NaN NaN NaN 5\\
5.4 9.6 0 5\\
5.4 9.6 3 5\\
5.4 10.4 3 5\\
5.4 10.4 0 5\\
5.4 9.6 0 5\\
NaN NaN NaN 5\\
5.4 10.6 0 5\\
5.4 10.6 3 5\\
5.4 11.4 3 5\\
5.4 11.4 0 5\\
5.4 10.6 0 5\\
NaN NaN NaN 5\\
NaN NaN NaN 5\\
5.4 0.6 0 5\\
5.4 1.4 0 5\\
NaN NaN NaN 5\\
NaN NaN NaN 5\\
NaN NaN NaN 5\\
NaN NaN NaN 5\\
5.4 1.6 0 5\\
5.4 2.4 0 5\\
NaN NaN NaN 5\\
NaN NaN NaN 5\\
NaN NaN NaN 5\\
NaN NaN NaN 5\\
5.4 2.6 0 5\\
5.4 3.4 0 5\\
NaN NaN NaN 5\\
NaN NaN NaN 5\\
NaN NaN NaN 5\\
NaN NaN NaN 5\\
5.4 3.6 0 5\\
5.4 4.4 0 5\\
NaN NaN NaN 5\\
NaN NaN NaN 5\\
NaN NaN NaN 5\\
NaN NaN NaN 5\\
5.4 4.6 0 5\\
5.4 5.4 0 5\\
NaN NaN NaN 5\\
NaN NaN NaN 5\\
NaN NaN NaN 5\\
NaN NaN NaN 5\\
5.4 5.6 0 5\\
5.4 6.4 0 5\\
NaN NaN NaN 5\\
NaN NaN NaN 5\\
NaN NaN NaN 5\\
NaN NaN NaN 5\\
5.4 6.6 0 5\\
5.4 7.4 0 5\\
NaN NaN NaN 5\\
NaN NaN NaN 5\\
NaN NaN NaN 5\\
NaN NaN NaN 5\\
5.4 7.6 0 5\\
5.4 8.4 0 5\\
NaN NaN NaN 5\\
NaN NaN NaN 5\\
NaN NaN NaN 5\\
NaN NaN NaN 5\\
5.4 8.6 0 5\\
5.4 9.4 0 5\\
NaN NaN NaN 5\\
NaN NaN NaN 5\\
NaN NaN NaN 5\\
NaN NaN NaN 5\\
5.4 9.6 0 5\\
5.4 10.4 0 5\\
NaN NaN NaN 5\\
NaN NaN NaN 5\\
NaN NaN NaN 5\\
NaN NaN NaN 5\\
5.4 10.6 0 5\\
5.4 11.4 0 5\\
NaN NaN NaN 5\\
NaN NaN NaN 5\\
NaN NaN NaN 5\\
};

\addplot3[%
surf,
shader=flat,
draw=black,
point meta=explicit,
mesh/rows=4]
table[row sep=crcr,header=false,meta index=3] {
NaN NaN NaN 4\\
3.6 0.6 0 4\\
3.6 1.4 0 4\\
NaN NaN NaN 4\\
NaN NaN NaN 4\\
NaN NaN NaN 4\\
NaN NaN NaN 4\\
3.6 1.6 0 4\\
3.6 2.4 0 4\\
NaN NaN NaN 4\\
NaN NaN NaN 4\\
NaN NaN NaN 4\\
NaN NaN NaN 4\\
3.6 2.6 0 4\\
3.6 3.4 0 4\\
NaN NaN NaN 4\\
NaN NaN NaN 4\\
NaN NaN NaN 4\\
NaN NaN NaN 4\\
3.6 3.6 0 4\\
3.6 4.4 0 4\\
NaN NaN NaN 4\\
NaN NaN NaN 4\\
NaN NaN NaN 4\\
NaN NaN NaN 4\\
3.6 4.6 0 4\\
3.6 5.4 0 4\\
NaN NaN NaN 4\\
NaN NaN NaN 4\\
NaN NaN NaN 4\\
NaN NaN NaN 4\\
3.6 5.6 0 4\\
3.6 6.4 0 4\\
NaN NaN NaN 4\\
NaN NaN NaN 4\\
NaN NaN NaN 4\\
NaN NaN NaN 4\\
3.6 6.6 0 4\\
3.6 7.4 0 4\\
NaN NaN NaN 4\\
NaN NaN NaN 4\\
NaN NaN NaN 4\\
NaN NaN NaN 4\\
3.6 7.6 0 4\\
3.6 8.4 0 4\\
NaN NaN NaN 4\\
NaN NaN NaN 4\\
NaN NaN NaN 4\\
NaN NaN NaN 4\\
3.6 8.6 0 4\\
3.6 9.4 0 4\\
NaN NaN NaN 4\\
NaN NaN NaN 4\\
NaN NaN NaN 4\\
NaN NaN NaN 4\\
3.6 9.6 0 4\\
3.6 10.4 0 4\\
NaN NaN NaN 4\\
NaN NaN NaN 4\\
NaN NaN NaN 4\\
NaN NaN NaN 4\\
3.6 10.6 0 4\\
3.6 11.4 0 4\\
NaN NaN NaN 4\\
NaN NaN NaN 4\\
NaN NaN NaN 4\\
3.6 0.6 0 4\\
3.6 0.6 2 4\\
3.6 1.4 2 4\\
3.6 1.4 0 4\\
3.6 0.6 0 4\\
NaN NaN NaN 4\\
3.6 1.6 0 4\\
3.6 1.6 3 4\\
3.6 2.4 3 4\\
3.6 2.4 0 4\\
3.6 1.6 0 4\\
NaN NaN NaN 4\\
3.6 2.6 0 4\\
3.6 2.6 3 4\\
3.6 3.4 3 4\\
3.6 3.4 0 4\\
3.6 2.6 0 4\\
NaN NaN NaN 4\\
3.6 3.6 0 4\\
3.6 3.6 3 4\\
3.6 4.4 3 4\\
3.6 4.4 0 4\\
3.6 3.6 0 4\\
NaN NaN NaN 4\\
3.6 4.6 0 4\\
3.6 4.6 3 4\\
3.6 5.4 3 4\\
3.6 5.4 0 4\\
3.6 4.6 0 4\\
NaN NaN NaN 4\\
3.6 5.6 0 4\\
3.6 5.6 3 4\\
3.6 6.4 3 4\\
3.6 6.4 0 4\\
3.6 5.6 0 4\\
NaN NaN NaN 4\\
3.6 6.6 0 4\\
3.6 6.6 3 4\\
3.6 7.4 3 4\\
3.6 7.4 0 4\\
3.6 6.6 0 4\\
NaN NaN NaN 4\\
3.6 7.6 0 4\\
3.6 7.6 3 4\\
3.6 8.4 3 4\\
3.6 8.4 0 4\\
3.6 7.6 0 4\\
NaN NaN NaN 4\\
3.6 8.6 0 4\\
3.6 8.6 3 4\\
3.6 9.4 3 4\\
3.6 9.4 0 4\\
3.6 8.6 0 4\\
NaN NaN NaN 4\\
3.6 9.6 0 4\\
3.6 9.6 3 4\\
3.6 10.4 3 4\\
3.6 10.4 0 4\\
3.6 9.6 0 4\\
NaN NaN NaN 4\\
3.6 10.6 0 4\\
3.6 10.6 3 4\\
3.6 11.4 3 4\\
3.6 11.4 0 4\\
3.6 10.6 0 4\\
NaN NaN NaN 4\\
4.4 0.6 0 4\\
4.4 0.6 2 4\\
4.4 1.4 2 4\\
4.4 1.4 0 4\\
4.4 0.6 0 4\\
NaN NaN NaN 4\\
4.4 1.6 0 4\\
4.4 1.6 3 4\\
4.4 2.4 3 4\\
4.4 2.4 0 4\\
4.4 1.6 0 4\\
NaN NaN NaN 4\\
4.4 2.6 0 4\\
4.4 2.6 3 4\\
4.4 3.4 3 4\\
4.4 3.4 0 4\\
4.4 2.6 0 4\\
NaN NaN NaN 4\\
4.4 3.6 0 4\\
4.4 3.6 3 4\\
4.4 4.4 3 4\\
4.4 4.4 0 4\\
4.4 3.6 0 4\\
NaN NaN NaN 4\\
4.4 4.6 0 4\\
4.4 4.6 3 4\\
4.4 5.4 3 4\\
4.4 5.4 0 4\\
4.4 4.6 0 4\\
NaN NaN NaN 4\\
4.4 5.6 0 4\\
4.4 5.6 3 4\\
4.4 6.4 3 4\\
4.4 6.4 0 4\\
4.4 5.6 0 4\\
NaN NaN NaN 4\\
4.4 6.6 0 4\\
4.4 6.6 3 4\\
4.4 7.4 3 4\\
4.4 7.4 0 4\\
4.4 6.6 0 4\\
NaN NaN NaN 4\\
4.4 7.6 0 4\\
4.4 7.6 3 4\\
4.4 8.4 3 4\\
4.4 8.4 0 4\\
4.4 7.6 0 4\\
NaN NaN NaN 4\\
4.4 8.6 0 4\\
4.4 8.6 3 4\\
4.4 9.4 3 4\\
4.4 9.4 0 4\\
4.4 8.6 0 4\\
NaN NaN NaN 4\\
4.4 9.6 0 4\\
4.4 9.6 3 4\\
4.4 10.4 3 4\\
4.4 10.4 0 4\\
4.4 9.6 0 4\\
NaN NaN NaN 4\\
4.4 10.6 0 4\\
4.4 10.6 3 4\\
4.4 11.4 3 4\\
4.4 11.4 0 4\\
4.4 10.6 0 4\\
NaN NaN NaN 4\\
NaN NaN NaN 4\\
4.4 0.6 0 4\\
4.4 1.4 0 4\\
NaN NaN NaN 4\\
NaN NaN NaN 4\\
NaN NaN NaN 4\\
NaN NaN NaN 4\\
4.4 1.6 0 4\\
4.4 2.4 0 4\\
NaN NaN NaN 4\\
NaN NaN NaN 4\\
NaN NaN NaN 4\\
NaN NaN NaN 4\\
4.4 2.6 0 4\\
4.4 3.4 0 4\\
NaN NaN NaN 4\\
NaN NaN NaN 4\\
NaN NaN NaN 4\\
NaN NaN NaN 4\\
4.4 3.6 0 4\\
4.4 4.4 0 4\\
NaN NaN NaN 4\\
NaN NaN NaN 4\\
NaN NaN NaN 4\\
NaN NaN NaN 4\\
4.4 4.6 0 4\\
4.4 5.4 0 4\\
NaN NaN NaN 4\\
NaN NaN NaN 4\\
NaN NaN NaN 4\\
NaN NaN NaN 4\\
4.4 5.6 0 4\\
4.4 6.4 0 4\\
NaN NaN NaN 4\\
NaN NaN NaN 4\\
NaN NaN NaN 4\\
NaN NaN NaN 4\\
4.4 6.6 0 4\\
4.4 7.4 0 4\\
NaN NaN NaN 4\\
NaN NaN NaN 4\\
NaN NaN NaN 4\\
NaN NaN NaN 4\\
4.4 7.6 0 4\\
4.4 8.4 0 4\\
NaN NaN NaN 4\\
NaN NaN NaN 4\\
NaN NaN NaN 4\\
NaN NaN NaN 4\\
4.4 8.6 0 4\\
4.4 9.4 0 4\\
NaN NaN NaN 4\\
NaN NaN NaN 4\\
NaN NaN NaN 4\\
NaN NaN NaN 4\\
4.4 9.6 0 4\\
4.4 10.4 0 4\\
NaN NaN NaN 4\\
NaN NaN NaN 4\\
NaN NaN NaN 4\\
NaN NaN NaN 4\\
4.4 10.6 0 4\\
4.4 11.4 0 4\\
NaN NaN NaN 4\\
NaN NaN NaN 4\\
NaN NaN NaN 4\\
};

\addplot3[%
surf,
shader=flat,
draw=black,
point meta=explicit,
mesh/rows=4]
table[row sep=crcr,header=false,meta index=3] {
NaN NaN NaN 3\\
2.6 0.6 0 3\\
2.6 1.4 0 3\\
NaN NaN NaN 3\\
NaN NaN NaN 3\\
NaN NaN NaN 3\\
NaN NaN NaN 3\\
2.6 1.6 0 3\\
2.6 2.4 0 3\\
NaN NaN NaN 3\\
NaN NaN NaN 3\\
NaN NaN NaN 3\\
NaN NaN NaN 3\\
2.6 2.6 0 3\\
2.6 3.4 0 3\\
NaN NaN NaN 3\\
NaN NaN NaN 3\\
NaN NaN NaN 3\\
NaN NaN NaN 3\\
2.6 3.6 0 3\\
2.6 4.4 0 3\\
NaN NaN NaN 3\\
NaN NaN NaN 3\\
NaN NaN NaN 3\\
NaN NaN NaN 3\\
2.6 4.6 0 3\\
2.6 5.4 0 3\\
NaN NaN NaN 3\\
NaN NaN NaN 3\\
NaN NaN NaN 3\\
NaN NaN NaN 3\\
2.6 5.6 0 3\\
2.6 6.4 0 3\\
NaN NaN NaN 3\\
NaN NaN NaN 3\\
NaN NaN NaN 3\\
NaN NaN NaN 3\\
2.6 6.6 0 3\\
2.6 7.4 0 3\\
NaN NaN NaN 3\\
NaN NaN NaN 3\\
NaN NaN NaN 3\\
NaN NaN NaN 3\\
2.6 7.6 0 3\\
2.6 8.4 0 3\\
NaN NaN NaN 3\\
NaN NaN NaN 3\\
NaN NaN NaN 3\\
NaN NaN NaN 3\\
2.6 8.6 0 3\\
2.6 9.4 0 3\\
NaN NaN NaN 3\\
NaN NaN NaN 3\\
NaN NaN NaN 3\\
NaN NaN NaN 3\\
2.6 9.6 0 3\\
2.6 10.4 0 3\\
NaN NaN NaN 3\\
NaN NaN NaN 3\\
NaN NaN NaN 3\\
NaN NaN NaN 3\\
2.6 10.6 0 3\\
2.6 11.4 0 3\\
NaN NaN NaN 3\\
NaN NaN NaN 3\\
NaN NaN NaN 3\\
2.6 0.6 0 3\\
2.6 0.6 1 3\\
2.6 1.4 1 3\\
2.6 1.4 0 3\\
2.6 0.6 0 3\\
NaN NaN NaN 3\\
2.6 1.6 0 3\\
2.6 1.6 2 3\\
2.6 2.4 2 3\\
2.6 2.4 0 3\\
2.6 1.6 0 3\\
NaN NaN NaN 3\\
2.6 2.6 0 3\\
2.6 2.6 2 3\\
2.6 3.4 2 3\\
2.6 3.4 0 3\\
2.6 2.6 0 3\\
NaN NaN NaN 3\\
2.6 3.6 0 3\\
2.6 3.6 2 3\\
2.6 4.4 2 3\\
2.6 4.4 0 3\\
2.6 3.6 0 3\\
NaN NaN NaN 3\\
2.6 4.6 0 3\\
2.6 4.6 2 3\\
2.6 5.4 2 3\\
2.6 5.4 0 3\\
2.6 4.6 0 3\\
NaN NaN NaN 3\\
2.6 5.6 0 3\\
2.6 5.6 2 3\\
2.6 6.4 2 3\\
2.6 6.4 0 3\\
2.6 5.6 0 3\\
NaN NaN NaN 3\\
2.6 6.6 0 3\\
2.6 6.6 2 3\\
2.6 7.4 2 3\\
2.6 7.4 0 3\\
2.6 6.6 0 3\\
NaN NaN NaN 3\\
2.6 7.6 0 3\\
2.6 7.6 2 3\\
2.6 8.4 2 3\\
2.6 8.4 0 3\\
2.6 7.6 0 3\\
NaN NaN NaN 3\\
2.6 8.6 0 3\\
2.6 8.6 3 3\\
2.6 9.4 3 3\\
2.6 9.4 0 3\\
2.6 8.6 0 3\\
NaN NaN NaN 3\\
2.6 9.6 0 3\\
2.6 9.6 3 3\\
2.6 10.4 3 3\\
2.6 10.4 0 3\\
2.6 9.6 0 3\\
NaN NaN NaN 3\\
2.6 10.6 0 3\\
2.6 10.6 3 3\\
2.6 11.4 3 3\\
2.6 11.4 0 3\\
2.6 10.6 0 3\\
NaN NaN NaN 3\\
3.4 0.6 0 3\\
3.4 0.6 1 3\\
3.4 1.4 1 3\\
3.4 1.4 0 3\\
3.4 0.6 0 3\\
NaN NaN NaN 3\\
3.4 1.6 0 3\\
3.4 1.6 2 3\\
3.4 2.4 2 3\\
3.4 2.4 0 3\\
3.4 1.6 0 3\\
NaN NaN NaN 3\\
3.4 2.6 0 3\\
3.4 2.6 2 3\\
3.4 3.4 2 3\\
3.4 3.4 0 3\\
3.4 2.6 0 3\\
NaN NaN NaN 3\\
3.4 3.6 0 3\\
3.4 3.6 2 3\\
3.4 4.4 2 3\\
3.4 4.4 0 3\\
3.4 3.6 0 3\\
NaN NaN NaN 3\\
3.4 4.6 0 3\\
3.4 4.6 2 3\\
3.4 5.4 2 3\\
3.4 5.4 0 3\\
3.4 4.6 0 3\\
NaN NaN NaN 3\\
3.4 5.6 0 3\\
3.4 5.6 2 3\\
3.4 6.4 2 3\\
3.4 6.4 0 3\\
3.4 5.6 0 3\\
NaN NaN NaN 3\\
3.4 6.6 0 3\\
3.4 6.6 2 3\\
3.4 7.4 2 3\\
3.4 7.4 0 3\\
3.4 6.6 0 3\\
NaN NaN NaN 3\\
3.4 7.6 0 3\\
3.4 7.6 2 3\\
3.4 8.4 2 3\\
3.4 8.4 0 3\\
3.4 7.6 0 3\\
NaN NaN NaN 3\\
3.4 8.6 0 3\\
3.4 8.6 3 3\\
3.4 9.4 3 3\\
3.4 9.4 0 3\\
3.4 8.6 0 3\\
NaN NaN NaN 3\\
3.4 9.6 0 3\\
3.4 9.6 3 3\\
3.4 10.4 3 3\\
3.4 10.4 0 3\\
3.4 9.6 0 3\\
NaN NaN NaN 3\\
3.4 10.6 0 3\\
3.4 10.6 3 3\\
3.4 11.4 3 3\\
3.4 11.4 0 3\\
3.4 10.6 0 3\\
NaN NaN NaN 3\\
NaN NaN NaN 3\\
3.4 0.6 0 3\\
3.4 1.4 0 3\\
NaN NaN NaN 3\\
NaN NaN NaN 3\\
NaN NaN NaN 3\\
NaN NaN NaN 3\\
3.4 1.6 0 3\\
3.4 2.4 0 3\\
NaN NaN NaN 3\\
NaN NaN NaN 3\\
NaN NaN NaN 3\\
NaN NaN NaN 3\\
3.4 2.6 0 3\\
3.4 3.4 0 3\\
NaN NaN NaN 3\\
NaN NaN NaN 3\\
NaN NaN NaN 3\\
NaN NaN NaN 3\\
3.4 3.6 0 3\\
3.4 4.4 0 3\\
NaN NaN NaN 3\\
NaN NaN NaN 3\\
NaN NaN NaN 3\\
NaN NaN NaN 3\\
3.4 4.6 0 3\\
3.4 5.4 0 3\\
NaN NaN NaN 3\\
NaN NaN NaN 3\\
NaN NaN NaN 3\\
NaN NaN NaN 3\\
3.4 5.6 0 3\\
3.4 6.4 0 3\\
NaN NaN NaN 3\\
NaN NaN NaN 3\\
NaN NaN NaN 3\\
NaN NaN NaN 3\\
3.4 6.6 0 3\\
3.4 7.4 0 3\\
NaN NaN NaN 3\\
NaN NaN NaN 3\\
NaN NaN NaN 3\\
NaN NaN NaN 3\\
3.4 7.6 0 3\\
3.4 8.4 0 3\\
NaN NaN NaN 3\\
NaN NaN NaN 3\\
NaN NaN NaN 3\\
NaN NaN NaN 3\\
3.4 8.6 0 3\\
3.4 9.4 0 3\\
NaN NaN NaN 3\\
NaN NaN NaN 3\\
NaN NaN NaN 3\\
NaN NaN NaN 3\\
3.4 9.6 0 3\\
3.4 10.4 0 3\\
NaN NaN NaN 3\\
NaN NaN NaN 3\\
NaN NaN NaN 3\\
NaN NaN NaN 3\\
3.4 10.6 0 3\\
3.4 11.4 0 3\\
NaN NaN NaN 3\\
NaN NaN NaN 3\\
NaN NaN NaN 3\\
};

\addplot3[%
surf,
shader=flat,
draw=black,
point meta=explicit,
mesh/rows=4]
table[row sep=crcr,header=false,meta index=3] {
NaN NaN NaN 2\\
1.6 0.6 0 2\\
1.6 1.4 0 2\\
NaN NaN NaN 2\\
NaN NaN NaN 2\\
NaN NaN NaN 2\\
NaN NaN NaN 2\\
1.6 1.6 0 2\\
1.6 2.4 0 2\\
NaN NaN NaN 2\\
NaN NaN NaN 2\\
NaN NaN NaN 2\\
NaN NaN NaN 2\\
1.6 2.6 0 2\\
1.6 3.4 0 2\\
NaN NaN NaN 2\\
NaN NaN NaN 2\\
NaN NaN NaN 2\\
NaN NaN NaN 2\\
1.6 3.6 0 2\\
1.6 4.4 0 2\\
NaN NaN NaN 2\\
NaN NaN NaN 2\\
NaN NaN NaN 2\\
NaN NaN NaN 2\\
1.6 4.6 0 2\\
1.6 5.4 0 2\\
NaN NaN NaN 2\\
NaN NaN NaN 2\\
NaN NaN NaN 2\\
NaN NaN NaN 2\\
1.6 5.6 0 2\\
1.6 6.4 0 2\\
NaN NaN NaN 2\\
NaN NaN NaN 2\\
NaN NaN NaN 2\\
NaN NaN NaN 2\\
1.6 6.6 0 2\\
1.6 7.4 0 2\\
NaN NaN NaN 2\\
NaN NaN NaN 2\\
NaN NaN NaN 2\\
NaN NaN NaN 2\\
1.6 7.6 0 2\\
1.6 8.4 0 2\\
NaN NaN NaN 2\\
NaN NaN NaN 2\\
NaN NaN NaN 2\\
NaN NaN NaN 2\\
1.6 8.6 0 2\\
1.6 9.4 0 2\\
NaN NaN NaN 2\\
NaN NaN NaN 2\\
NaN NaN NaN 2\\
NaN NaN NaN 2\\
1.6 9.6 0 2\\
1.6 10.4 0 2\\
NaN NaN NaN 2\\
NaN NaN NaN 2\\
NaN NaN NaN 2\\
NaN NaN NaN 2\\
1.6 10.6 0 2\\
1.6 11.4 0 2\\
NaN NaN NaN 2\\
NaN NaN NaN 2\\
NaN NaN NaN 2\\
1.6 0.6 0 2\\
1.6 0.6 1 2\\
1.6 1.4 1 2\\
1.6 1.4 0 2\\
1.6 0.6 0 2\\
NaN NaN NaN 2\\
1.6 1.6 0 2\\
1.6 1.6 2 2\\
1.6 2.4 2 2\\
1.6 2.4 0 2\\
1.6 1.6 0 2\\
NaN NaN NaN 2\\
1.6 2.6 0 2\\
1.6 2.6 2 2\\
1.6 3.4 2 2\\
1.6 3.4 0 2\\
1.6 2.6 0 2\\
NaN NaN NaN 2\\
1.6 3.6 0 2\\
1.6 3.6 2 2\\
1.6 4.4 2 2\\
1.6 4.4 0 2\\
1.6 3.6 0 2\\
NaN NaN NaN 2\\
1.6 4.6 0 2\\
1.6 4.6 2 2\\
1.6 5.4 2 2\\
1.6 5.4 0 2\\
1.6 4.6 0 2\\
NaN NaN NaN 2\\
1.6 5.6 0 2\\
1.6 5.6 2 2\\
1.6 6.4 2 2\\
1.6 6.4 0 2\\
1.6 5.6 0 2\\
NaN NaN NaN 2\\
1.6 6.6 0 2\\
1.6 6.6 2 2\\
1.6 7.4 2 2\\
1.6 7.4 0 2\\
1.6 6.6 0 2\\
NaN NaN NaN 2\\
1.6 7.6 0 2\\
1.6 7.6 2 2\\
1.6 8.4 2 2\\
1.6 8.4 0 2\\
1.6 7.6 0 2\\
NaN NaN NaN 2\\
1.6 8.6 0 2\\
1.6 8.6 2 2\\
1.6 9.4 2 2\\
1.6 9.4 0 2\\
1.6 8.6 0 2\\
NaN NaN NaN 2\\
1.6 9.6 0 2\\
1.6 9.6 2 2\\
1.6 10.4 2 2\\
1.6 10.4 0 2\\
1.6 9.6 0 2\\
NaN NaN NaN 2\\
1.6 10.6 0 2\\
1.6 10.6 2 2\\
1.6 11.4 2 2\\
1.6 11.4 0 2\\
1.6 10.6 0 2\\
NaN NaN NaN 2\\
2.4 0.6 0 2\\
2.4 0.6 1 2\\
2.4 1.4 1 2\\
2.4 1.4 0 2\\
2.4 0.6 0 2\\
NaN NaN NaN 2\\
2.4 1.6 0 2\\
2.4 1.6 2 2\\
2.4 2.4 2 2\\
2.4 2.4 0 2\\
2.4 1.6 0 2\\
NaN NaN NaN 2\\
2.4 2.6 0 2\\
2.4 2.6 2 2\\
2.4 3.4 2 2\\
2.4 3.4 0 2\\
2.4 2.6 0 2\\
NaN NaN NaN 2\\
2.4 3.6 0 2\\
2.4 3.6 2 2\\
2.4 4.4 2 2\\
2.4 4.4 0 2\\
2.4 3.6 0 2\\
NaN NaN NaN 2\\
2.4 4.6 0 2\\
2.4 4.6 2 2\\
2.4 5.4 2 2\\
2.4 5.4 0 2\\
2.4 4.6 0 2\\
NaN NaN NaN 2\\
2.4 5.6 0 2\\
2.4 5.6 2 2\\
2.4 6.4 2 2\\
2.4 6.4 0 2\\
2.4 5.6 0 2\\
NaN NaN NaN 2\\
2.4 6.6 0 2\\
2.4 6.6 2 2\\
2.4 7.4 2 2\\
2.4 7.4 0 2\\
2.4 6.6 0 2\\
NaN NaN NaN 2\\
2.4 7.6 0 2\\
2.4 7.6 2 2\\
2.4 8.4 2 2\\
2.4 8.4 0 2\\
2.4 7.6 0 2\\
NaN NaN NaN 2\\
2.4 8.6 0 2\\
2.4 8.6 2 2\\
2.4 9.4 2 2\\
2.4 9.4 0 2\\
2.4 8.6 0 2\\
NaN NaN NaN 2\\
2.4 9.6 0 2\\
2.4 9.6 2 2\\
2.4 10.4 2 2\\
2.4 10.4 0 2\\
2.4 9.6 0 2\\
NaN NaN NaN 2\\
2.4 10.6 0 2\\
2.4 10.6 2 2\\
2.4 11.4 2 2\\
2.4 11.4 0 2\\
2.4 10.6 0 2\\
NaN NaN NaN 2\\
NaN NaN NaN 2\\
2.4 0.6 0 2\\
2.4 1.4 0 2\\
NaN NaN NaN 2\\
NaN NaN NaN 2\\
NaN NaN NaN 2\\
NaN NaN NaN 2\\
2.4 1.6 0 2\\
2.4 2.4 0 2\\
NaN NaN NaN 2\\
NaN NaN NaN 2\\
NaN NaN NaN 2\\
NaN NaN NaN 2\\
2.4 2.6 0 2\\
2.4 3.4 0 2\\
NaN NaN NaN 2\\
NaN NaN NaN 2\\
NaN NaN NaN 2\\
NaN NaN NaN 2\\
2.4 3.6 0 2\\
2.4 4.4 0 2\\
NaN NaN NaN 2\\
NaN NaN NaN 2\\
NaN NaN NaN 2\\
NaN NaN NaN 2\\
2.4 4.6 0 2\\
2.4 5.4 0 2\\
NaN NaN NaN 2\\
NaN NaN NaN 2\\
NaN NaN NaN 2\\
NaN NaN NaN 2\\
2.4 5.6 0 2\\
2.4 6.4 0 2\\
NaN NaN NaN 2\\
NaN NaN NaN 2\\
NaN NaN NaN 2\\
NaN NaN NaN 2\\
2.4 6.6 0 2\\
2.4 7.4 0 2\\
NaN NaN NaN 2\\
NaN NaN NaN 2\\
NaN NaN NaN 2\\
NaN NaN NaN 2\\
2.4 7.6 0 2\\
2.4 8.4 0 2\\
NaN NaN NaN 2\\
NaN NaN NaN 2\\
NaN NaN NaN 2\\
NaN NaN NaN 2\\
2.4 8.6 0 2\\
2.4 9.4 0 2\\
NaN NaN NaN 2\\
NaN NaN NaN 2\\
NaN NaN NaN 2\\
NaN NaN NaN 2\\
2.4 9.6 0 2\\
2.4 10.4 0 2\\
NaN NaN NaN 2\\
NaN NaN NaN 2\\
NaN NaN NaN 2\\
NaN NaN NaN 2\\
2.4 10.6 0 2\\
2.4 11.4 0 2\\
NaN NaN NaN 2\\
NaN NaN NaN 2\\
NaN NaN NaN 2\\
};


\addplot3[%
surf,
shader=flat,
draw=black,
point meta=explicit,
mesh/rows=4]
table[row sep=crcr,header=false,meta index=3] {
NaN NaN NaN 1\\
0.6 0.6 0 1\\
0.6 1.4 0 1\\
NaN NaN NaN 1\\
NaN NaN NaN 1\\
NaN NaN NaN 1\\
NaN NaN NaN 1\\
0.6 1.6 0 1\\
0.6 2.4 0 1\\
NaN NaN NaN 1\\
NaN NaN NaN 1\\
NaN NaN NaN 1\\
NaN NaN NaN 1\\
0.6 2.6 0 1\\
0.6 3.4 0 1\\
NaN NaN NaN 1\\
NaN NaN NaN 1\\
NaN NaN NaN 1\\
NaN NaN NaN 1\\
0.6 3.6 0 1\\
0.6 4.4 0 1\\
NaN NaN NaN 1\\
NaN NaN NaN 1\\
NaN NaN NaN 1\\
NaN NaN NaN 1\\
0.6 4.6 0 1\\
0.6 5.4 0 1\\
NaN NaN NaN 1\\
NaN NaN NaN 1\\
NaN NaN NaN 1\\
NaN NaN NaN 1\\
0.6 5.6 0 1\\
0.6 6.4 0 1\\
NaN NaN NaN 1\\
NaN NaN NaN 1\\
NaN NaN NaN 1\\
NaN NaN NaN 1\\
0.6 6.6 0 1\\
0.6 7.4 0 1\\
NaN NaN NaN 1\\
NaN NaN NaN 1\\
NaN NaN NaN 1\\
NaN NaN NaN 1\\
0.6 7.6 0 1\\
0.6 8.4 0 1\\
NaN NaN NaN 1\\
NaN NaN NaN 1\\
NaN NaN NaN 1\\
NaN NaN NaN 1\\
0.6 8.6 0 1\\
0.6 9.4 0 1\\
NaN NaN NaN 1\\
NaN NaN NaN 1\\
NaN NaN NaN 1\\
NaN NaN NaN 1\\
0.6 9.6 0 1\\
0.6 10.4 0 1\\
NaN NaN NaN 1\\
NaN NaN NaN 1\\
NaN NaN NaN 1\\
NaN NaN NaN 1\\
0.6 10.6 0 1\\
0.6 11.4 0 1\\
NaN NaN NaN 1\\
NaN NaN NaN 1\\
NaN NaN NaN 1\\
0.6 0.6 0 1\\
0.6 0.6 0 1\\
0.6 1.4 0 1\\
0.6 1.4 0 1\\
0.6 0.6 0 1\\
NaN NaN NaN 1\\
0.6 1.6 0 1\\
0.6 1.6 0 1\\
0.6 2.4 0 1\\
0.6 2.4 0 1\\
0.6 1.6 0 1\\
NaN NaN NaN 1\\
0.6 2.6 0 1\\
0.6 2.6 0 1\\
0.6 3.4 0 1\\
0.6 3.4 0 1\\
0.6 2.6 0 1\\
NaN NaN NaN 1\\
0.6 3.6 0 1\\
0.6 3.6 0 1\\
0.6 4.4 0 1\\
0.6 4.4 0 1\\
0.6 3.6 0 1\\
NaN NaN NaN 1\\
0.6 4.6 0 1\\
0.6 4.6 0 1\\
0.6 5.4 0 1\\
0.6 5.4 0 1\\
0.6 4.6 0 1\\
NaN NaN NaN 1\\
0.6 5.6 0 1\\
0.6 5.6 1 1\\
0.6 6.4 1 1\\
0.6 6.4 0 1\\
0.6 5.6 0 1\\
NaN NaN NaN 1\\
0.6 6.6 0 1\\
0.6 6.6 1 1\\
0.6 7.4 1 1\\
0.6 7.4 0 1\\
0.6 6.6 0 1\\
NaN NaN NaN 1\\
0.6 7.6 0 1\\
0.6 7.6 1 1\\
0.6 8.4 1 1\\
0.6 8.4 0 1\\
0.6 7.6 0 1\\
NaN NaN NaN 1\\
0.6 8.6 0 1\\
0.6 8.6 1 1\\
0.6 9.4 1 1\\
0.6 9.4 0 1\\
0.6 8.6 0 1\\
NaN NaN NaN 1\\
0.6 9.6 0 1\\
0.6 9.6 1 1\\
0.6 10.4 1 1\\
0.6 10.4 0 1\\
0.6 9.6 0 1\\
NaN NaN NaN 1\\
0.6 10.6 0 1\\
0.6 10.6 1 1\\
0.6 11.4 1 1\\
0.6 11.4 0 1\\
0.6 10.6 0 1\\
NaN NaN NaN 1\\
1.4 0.6 0 1\\
1.4 0.6 0 1\\
1.4 1.4 0 1\\
1.4 1.4 0 1\\
1.4 0.6 0 1\\
NaN NaN NaN 1\\
1.4 1.6 0 1\\
1.4 1.6 0 1\\
1.4 2.4 0 1\\
1.4 2.4 0 1\\
1.4 1.6 0 1\\
NaN NaN NaN 1\\
1.4 2.6 0 1\\
1.4 2.6 0 1\\
1.4 3.4 0 1\\
1.4 3.4 0 1\\
1.4 2.6 0 1\\
NaN NaN NaN 1\\
1.4 3.6 0 1\\
1.4 3.6 0 1\\
1.4 4.4 0 1\\
1.4 4.4 0 1\\
1.4 3.6 0 1\\
NaN NaN NaN 1\\
1.4 4.6 0 1\\
1.4 4.6 0 1\\
1.4 5.4 0 1\\
1.4 5.4 0 1\\
1.4 4.6 0 1\\
NaN NaN NaN 1\\
1.4 5.6 0 1\\
1.4 5.6 1 1\\
1.4 6.4 1 1\\
1.4 6.4 0 1\\
1.4 5.6 0 1\\
NaN NaN NaN 1\\
1.4 6.6 0 1\\
1.4 6.6 1 1\\
1.4 7.4 1 1\\
1.4 7.4 0 1\\
1.4 6.6 0 1\\
NaN NaN NaN 1\\
1.4 7.6 0 1\\
1.4 7.6 1 1\\
1.4 8.4 1 1\\
1.4 8.4 0 1\\
1.4 7.6 0 1\\
NaN NaN NaN 1\\
1.4 8.6 0 1\\
1.4 8.6 1 1\\
1.4 9.4 1 1\\
1.4 9.4 0 1\\
1.4 8.6 0 1\\
NaN NaN NaN 1\\
1.4 9.6 0 1\\
1.4 9.6 1 1\\
1.4 10.4 1 1\\
1.4 10.4 0 1\\
1.4 9.6 0 1\\
NaN NaN NaN 1\\
1.4 10.6 0 1\\
1.4 10.6 1 1\\
1.4 11.4 1 1\\
1.4 11.4 0 1\\
1.4 10.6 0 1\\
NaN NaN NaN 1\\
NaN NaN NaN 1\\
1.4 0.6 0 1\\
1.4 1.4 0 1\\
NaN NaN NaN 1\\
NaN NaN NaN 1\\
NaN NaN NaN 1\\
NaN NaN NaN 1\\
1.4 1.6 0 1\\
1.4 2.4 0 1\\
NaN NaN NaN 1\\
NaN NaN NaN 1\\
NaN NaN NaN 1\\
NaN NaN NaN 1\\
1.4 2.6 0 1\\
1.4 3.4 0 1\\
NaN NaN NaN 1\\
NaN NaN NaN 1\\
NaN NaN NaN 1\\
NaN NaN NaN 1\\
1.4 3.6 0 1\\
1.4 4.4 0 1\\
NaN NaN NaN 1\\
NaN NaN NaN 1\\
NaN NaN NaN 1\\
NaN NaN NaN 1\\
1.4 4.6 0 1\\
1.4 5.4 0 1\\
NaN NaN NaN 1\\
NaN NaN NaN 1\\
NaN NaN NaN 1\\
NaN NaN NaN 1\\
1.4 5.6 0 1\\
1.4 6.4 0 1\\
NaN NaN NaN 1\\
NaN NaN NaN 1\\
NaN NaN NaN 1\\
NaN NaN NaN 1\\
1.4 6.6 0 1\\
1.4 7.4 0 1\\
NaN NaN NaN 1\\
NaN NaN NaN 1\\
NaN NaN NaN 1\\
NaN NaN NaN 1\\
1.4 7.6 0 1\\
1.4 8.4 0 1\\
NaN NaN NaN 1\\
NaN NaN NaN 1\\
NaN NaN NaN 1\\
NaN NaN NaN 1\\
1.4 8.6 0 1\\
1.4 9.4 0 1\\
NaN NaN NaN 1\\
NaN NaN NaN 1\\
NaN NaN NaN 1\\
NaN NaN NaN 1\\
1.4 9.6 0 1\\
1.4 10.4 0 1\\
NaN NaN NaN 1\\
NaN NaN NaN 1\\
NaN NaN NaN 1\\
NaN NaN NaN 1\\
1.4 10.6 0 1\\
1.4 11.4 0 1\\
NaN NaN NaN 1\\
NaN NaN NaN 1\\
NaN NaN NaN 1\\
};
\end{axis}
\end{tikzpicture}%

%% file: Policy/MonoL1.tex
%
%
\begin{tikzpicture}

\begin{axis}[%
width=2.5in,
height=2.2in,
unbounded coords=jump,
view={-130}{40},
scale only axis,
xmin=0.5,
xmax=11.5,
xtick={1,2,3,4,5,6,7,8,9,10,11},
xticklabels={0.1,1,2,3,4,5,6,7,8,9,10},
xlabel={$\gamma_2$},
xmajorgrids,
y dir=reverse,
ymin=0.5,
ymax=11.5,
ytick={1,2,3,4,5,6,7,8,9,10,11},
yticklabels={0.1,1,2,3,4,5,6,7,8,9,10},
ylabel={$\gamma_1$},
ymajorgrids,
zmin=0,
zmax=4,
zlabel={$\underline{\theta}_1^*$},
zmajorgrids,
name=plot1,
axis x line*=bottom,
axis y line*=left,
axis z line*=left
]

\addplot3[%
surf,
shader=flat,
draw=black,
point meta=explicit,
mesh/rows=4]
table[row sep=crcr,header=false,meta index=3] {
NaN NaN NaN 11\\
10.6 0.6 0 11\\
10.6 1.4 0 11\\
NaN NaN NaN 11\\
NaN NaN NaN 11\\
NaN NaN NaN 11\\
NaN NaN NaN 11\\
10.6 1.6 0 11\\
10.6 2.4 0 11\\
NaN NaN NaN 11\\
NaN NaN NaN 11\\
NaN NaN NaN 11\\
NaN NaN NaN 11\\
10.6 2.6 0 11\\
10.6 3.4 0 11\\
NaN NaN NaN 11\\
NaN NaN NaN 11\\
NaN NaN NaN 11\\
NaN NaN NaN 11\\
10.6 3.6 0 11\\
10.6 4.4 0 11\\
NaN NaN NaN 11\\
NaN NaN NaN 11\\
NaN NaN NaN 11\\
NaN NaN NaN 11\\
10.6 4.6 0 11\\
10.6 5.4 0 11\\
NaN NaN NaN 11\\
NaN NaN NaN 11\\
NaN NaN NaN 11\\
NaN NaN NaN 11\\
10.6 5.6 0 11\\
10.6 6.4 0 11\\
NaN NaN NaN 11\\
NaN NaN NaN 11\\
NaN NaN NaN 11\\
NaN NaN NaN 11\\
10.6 6.6 0 11\\
10.6 7.4 0 11\\
NaN NaN NaN 11\\
NaN NaN NaN 11\\
NaN NaN NaN 11\\
NaN NaN NaN 11\\
10.6 7.6 0 11\\
10.6 8.4 0 11\\
NaN NaN NaN 11\\
NaN NaN NaN 11\\
NaN NaN NaN 11\\
NaN NaN NaN 11\\
10.6 8.6 0 11\\
10.6 9.4 0 11\\
NaN NaN NaN 11\\
NaN NaN NaN 11\\
NaN NaN NaN 11\\
NaN NaN NaN 11\\
10.6 9.6 0 11\\
10.6 10.4 0 11\\
NaN NaN NaN 11\\
NaN NaN NaN 11\\
NaN NaN NaN 11\\
NaN NaN NaN 11\\
10.6 10.6 0 11\\
10.6 11.4 0 11\\
NaN NaN NaN 11\\
NaN NaN NaN 11\\
NaN NaN NaN 11\\
10.6 0.6 0 11\\
10.6 0.6 1 11\\
10.6 1.4 1 11\\
10.6 1.4 0 11\\
10.6 0.6 0 11\\
NaN NaN NaN 11\\
10.6 1.6 0 11\\
10.6 1.6 2 11\\
10.6 2.4 2 11\\
10.6 2.4 0 11\\
10.6 1.6 0 11\\
NaN NaN NaN 11\\
10.6 2.6 0 11\\
10.6 2.6 3 11\\
10.6 3.4 3 11\\
10.6 3.4 0 11\\
10.6 2.6 0 11\\
NaN NaN NaN 11\\
10.6 3.6 0 11\\
10.6 3.6 3 11\\
10.6 4.4 3 11\\
10.6 4.4 0 11\\
10.6 3.6 0 11\\
NaN NaN NaN 11\\
10.6 4.6 0 11\\
10.6 4.6 3 11\\
10.6 5.4 3 11\\
10.6 5.4 0 11\\
10.6 4.6 0 11\\
NaN NaN NaN 11\\
10.6 5.6 0 11\\
10.6 5.6 3 11\\
10.6 6.4 3 11\\
10.6 6.4 0 11\\
10.6 5.6 0 11\\
NaN NaN NaN 11\\
10.6 6.6 0 11\\
10.6 6.6 4 11\\
10.6 7.4 4 11\\
10.6 7.4 0 11\\
10.6 6.6 0 11\\
NaN NaN NaN 11\\
10.6 7.6 0 11\\
10.6 7.6 4 11\\
10.6 8.4 4 11\\
10.6 8.4 0 11\\
10.6 7.6 0 11\\
NaN NaN NaN 11\\
10.6 8.6 0 11\\
10.6 8.6 4 11\\
10.6 9.4 4 11\\
10.6 9.4 0 11\\
10.6 8.6 0 11\\
NaN NaN NaN 11\\
10.6 9.6 0 11\\
10.6 9.6 4 11\\
10.6 10.4 4 11\\
10.6 10.4 0 11\\
10.6 9.6 0 11\\
NaN NaN NaN 11\\
10.6 10.6 0 11\\
10.6 10.6 4 11\\
10.6 11.4 4 11\\
10.6 11.4 0 11\\
10.6 10.6 0 11\\
NaN NaN NaN 11\\
11.4 0.6 0 11\\
11.4 0.6 1 11\\
11.4 1.4 1 11\\
11.4 1.4 0 11\\
11.4 0.6 0 11\\
NaN NaN NaN 11\\
11.4 1.6 0 11\\
11.4 1.6 2 11\\
11.4 2.4 2 11\\
11.4 2.4 0 11\\
11.4 1.6 0 11\\
NaN NaN NaN 11\\
11.4 2.6 0 11\\
11.4 2.6 3 11\\
11.4 3.4 3 11\\
11.4 3.4 0 11\\
11.4 2.6 0 11\\
NaN NaN NaN 11\\
11.4 3.6 0 11\\
11.4 3.6 3 11\\
11.4 4.4 3 11\\
11.4 4.4 0 11\\
11.4 3.6 0 11\\
NaN NaN NaN 11\\
11.4 4.6 0 11\\
11.4 4.6 3 11\\
11.4 5.4 3 11\\
11.4 5.4 0 11\\
11.4 4.6 0 11\\
NaN NaN NaN 11\\
11.4 5.6 0 11\\
11.4 5.6 3 11\\
11.4 6.4 3 11\\
11.4 6.4 0 11\\
11.4 5.6 0 11\\
NaN NaN NaN 11\\
11.4 6.6 0 11\\
11.4 6.6 4 11\\
11.4 7.4 4 11\\
11.4 7.4 0 11\\
11.4 6.6 0 11\\
NaN NaN NaN 11\\
11.4 7.6 0 11\\
11.4 7.6 4 11\\
11.4 8.4 4 11\\
11.4 8.4 0 11\\
11.4 7.6 0 11\\
NaN NaN NaN 11\\
11.4 8.6 0 11\\
11.4 8.6 4 11\\
11.4 9.4 4 11\\
11.4 9.4 0 11\\
11.4 8.6 0 11\\
NaN NaN NaN 11\\
11.4 9.6 0 11\\
11.4 9.6 4 11\\
11.4 10.4 4 11\\
11.4 10.4 0 11\\
11.4 9.6 0 11\\
NaN NaN NaN 11\\
11.4 10.6 0 11\\
11.4 10.6 4 11\\
11.4 11.4 4 11\\
11.4 11.4 0 11\\
11.4 10.6 0 11\\
NaN NaN NaN 11\\
NaN NaN NaN 11\\
11.4 0.6 0 11\\
11.4 1.4 0 11\\
NaN NaN NaN 11\\
NaN NaN NaN 11\\
NaN NaN NaN 11\\
NaN NaN NaN 11\\
11.4 1.6 0 11\\
11.4 2.4 0 11\\
NaN NaN NaN 11\\
NaN NaN NaN 11\\
NaN NaN NaN 11\\
NaN NaN NaN 11\\
11.4 2.6 0 11\\
11.4 3.4 0 11\\
NaN NaN NaN 11\\
NaN NaN NaN 11\\
NaN NaN NaN 11\\
NaN NaN NaN 11\\
11.4 3.6 0 11\\
11.4 4.4 0 11\\
NaN NaN NaN 11\\
NaN NaN NaN 11\\
NaN NaN NaN 11\\
NaN NaN NaN 11\\
11.4 4.6 0 11\\
11.4 5.4 0 11\\
NaN NaN NaN 11\\
NaN NaN NaN 11\\
NaN NaN NaN 11\\
NaN NaN NaN 11\\
11.4 5.6 0 11\\
11.4 6.4 0 11\\
NaN NaN NaN 11\\
NaN NaN NaN 11\\
NaN NaN NaN 11\\
NaN NaN NaN 11\\
11.4 6.6 0 11\\
11.4 7.4 0 11\\
NaN NaN NaN 11\\
NaN NaN NaN 11\\
NaN NaN NaN 11\\
NaN NaN NaN 11\\
11.4 7.6 0 11\\
11.4 8.4 0 11\\
NaN NaN NaN 11\\
NaN NaN NaN 11\\
NaN NaN NaN 11\\
NaN NaN NaN 11\\
11.4 8.6 0 11\\
11.4 9.4 0 11\\
NaN NaN NaN 11\\
NaN NaN NaN 11\\
NaN NaN NaN 11\\
NaN NaN NaN 11\\
11.4 9.6 0 11\\
11.4 10.4 0 11\\
NaN NaN NaN 11\\
NaN NaN NaN 11\\
NaN NaN NaN 11\\
NaN NaN NaN 11\\
11.4 10.6 0 11\\
11.4 11.4 0 11\\
NaN NaN NaN 11\\
NaN NaN NaN 11\\
NaN NaN NaN 11\\
};

\addplot3[%
surf,
shader=flat,
draw=black,
point meta=explicit,
mesh/rows=4]
table[row sep=crcr,header=false,meta index=3] {
NaN NaN NaN 10\\
9.6 0.6 0 10\\
9.6 1.4 0 10\\
NaN NaN NaN 10\\
NaN NaN NaN 10\\
NaN NaN NaN 10\\
NaN NaN NaN 10\\
9.6 1.6 0 10\\
9.6 2.4 0 10\\
NaN NaN NaN 10\\
NaN NaN NaN 10\\
NaN NaN NaN 10\\
NaN NaN NaN 10\\
9.6 2.6 0 10\\
9.6 3.4 0 10\\
NaN NaN NaN 10\\
NaN NaN NaN 10\\
NaN NaN NaN 10\\
NaN NaN NaN 10\\
9.6 3.6 0 10\\
9.6 4.4 0 10\\
NaN NaN NaN 10\\
NaN NaN NaN 10\\
NaN NaN NaN 10\\
NaN NaN NaN 10\\
9.6 4.6 0 10\\
9.6 5.4 0 10\\
NaN NaN NaN 10\\
NaN NaN NaN 10\\
NaN NaN NaN 10\\
NaN NaN NaN 10\\
9.6 5.6 0 10\\
9.6 6.4 0 10\\
NaN NaN NaN 10\\
NaN NaN NaN 10\\
NaN NaN NaN 10\\
NaN NaN NaN 10\\
9.6 6.6 0 10\\
9.6 7.4 0 10\\
NaN NaN NaN 10\\
NaN NaN NaN 10\\
NaN NaN NaN 10\\
NaN NaN NaN 10\\
9.6 7.6 0 10\\
9.6 8.4 0 10\\
NaN NaN NaN 10\\
NaN NaN NaN 10\\
NaN NaN NaN 10\\
NaN NaN NaN 10\\
9.6 8.6 0 10\\
9.6 9.4 0 10\\
NaN NaN NaN 10\\
NaN NaN NaN 10\\
NaN NaN NaN 10\\
NaN NaN NaN 10\\
9.6 9.6 0 10\\
9.6 10.4 0 10\\
NaN NaN NaN 10\\
NaN NaN NaN 10\\
NaN NaN NaN 10\\
NaN NaN NaN 10\\
9.6 10.6 0 10\\
9.6 11.4 0 10\\
NaN NaN NaN 10\\
NaN NaN NaN 10\\
NaN NaN NaN 10\\
9.6 0.6 0 10\\
9.6 0.6 1 10\\
9.6 1.4 1 10\\
9.6 1.4 0 10\\
9.6 0.6 0 10\\
NaN NaN NaN 10\\
9.6 1.6 0 10\\
9.6 1.6 2 10\\
9.6 2.4 2 10\\
9.6 2.4 0 10\\
9.6 1.6 0 10\\
NaN NaN NaN 10\\
9.6 2.6 0 10\\
9.6 2.6 2 10\\
9.6 3.4 2 10\\
9.6 3.4 0 10\\
9.6 2.6 0 10\\
NaN NaN NaN 10\\
9.6 3.6 0 10\\
9.6 3.6 3 10\\
9.6 4.4 3 10\\
9.6 4.4 0 10\\
9.6 3.6 0 10\\
NaN NaN NaN 10\\
9.6 4.6 0 10\\
9.6 4.6 3 10\\
9.6 5.4 3 10\\
9.6 5.4 0 10\\
9.6 4.6 0 10\\
NaN NaN NaN 10\\
9.6 5.6 0 10\\
9.6 5.6 3 10\\
9.6 6.4 3 10\\
9.6 6.4 0 10\\
9.6 5.6 0 10\\
NaN NaN NaN 10\\
9.6 6.6 0 10\\
9.6 6.6 4 10\\
9.6 7.4 4 10\\
9.6 7.4 0 10\\
9.6 6.6 0 10\\
NaN NaN NaN 10\\
9.6 7.6 0 10\\
9.6 7.6 4 10\\
9.6 8.4 4 10\\
9.6 8.4 0 10\\
9.6 7.6 0 10\\
NaN NaN NaN 10\\
9.6 8.6 0 10\\
9.6 8.6 4 10\\
9.6 9.4 4 10\\
9.6 9.4 0 10\\
9.6 8.6 0 10\\
NaN NaN NaN 10\\
9.6 9.6 0 10\\
9.6 9.6 4 10\\
9.6 10.4 4 10\\
9.6 10.4 0 10\\
9.6 9.6 0 10\\
NaN NaN NaN 10\\
9.6 10.6 0 10\\
9.6 10.6 4 10\\
9.6 11.4 4 10\\
9.6 11.4 0 10\\
9.6 10.6 0 10\\
NaN NaN NaN 10\\
10.4 0.6 0 10\\
10.4 0.6 1 10\\
10.4 1.4 1 10\\
10.4 1.4 0 10\\
10.4 0.6 0 10\\
NaN NaN NaN 10\\
10.4 1.6 0 10\\
10.4 1.6 2 10\\
10.4 2.4 2 10\\
10.4 2.4 0 10\\
10.4 1.6 0 10\\
NaN NaN NaN 10\\
10.4 2.6 0 10\\
10.4 2.6 2 10\\
10.4 3.4 2 10\\
10.4 3.4 0 10\\
10.4 2.6 0 10\\
NaN NaN NaN 10\\
10.4 3.6 0 10\\
10.4 3.6 3 10\\
10.4 4.4 3 10\\
10.4 4.4 0 10\\
10.4 3.6 0 10\\
NaN NaN NaN 10\\
10.4 4.6 0 10\\
10.4 4.6 3 10\\
10.4 5.4 3 10\\
10.4 5.4 0 10\\
10.4 4.6 0 10\\
NaN NaN NaN 10\\
10.4 5.6 0 10\\
10.4 5.6 3 10\\
10.4 6.4 3 10\\
10.4 6.4 0 10\\
10.4 5.6 0 10\\
NaN NaN NaN 10\\
10.4 6.6 0 10\\
10.4 6.6 4 10\\
10.4 7.4 4 10\\
10.4 7.4 0 10\\
10.4 6.6 0 10\\
NaN NaN NaN 10\\
10.4 7.6 0 10\\
10.4 7.6 4 10\\
10.4 8.4 4 10\\
10.4 8.4 0 10\\
10.4 7.6 0 10\\
NaN NaN NaN 10\\
10.4 8.6 0 10\\
10.4 8.6 4 10\\
10.4 9.4 4 10\\
10.4 9.4 0 10\\
10.4 8.6 0 10\\
NaN NaN NaN 10\\
10.4 9.6 0 10\\
10.4 9.6 4 10\\
10.4 10.4 4 10\\
10.4 10.4 0 10\\
10.4 9.6 0 10\\
NaN NaN NaN 10\\
10.4 10.6 0 10\\
10.4 10.6 4 10\\
10.4 11.4 4 10\\
10.4 11.4 0 10\\
10.4 10.6 0 10\\
NaN NaN NaN 10\\
NaN NaN NaN 10\\
10.4 0.6 0 10\\
10.4 1.4 0 10\\
NaN NaN NaN 10\\
NaN NaN NaN 10\\
NaN NaN NaN 10\\
NaN NaN NaN 10\\
10.4 1.6 0 10\\
10.4 2.4 0 10\\
NaN NaN NaN 10\\
NaN NaN NaN 10\\
NaN NaN NaN 10\\
NaN NaN NaN 10\\
10.4 2.6 0 10\\
10.4 3.4 0 10\\
NaN NaN NaN 10\\
NaN NaN NaN 10\\
NaN NaN NaN 10\\
NaN NaN NaN 10\\
10.4 3.6 0 10\\
10.4 4.4 0 10\\
NaN NaN NaN 10\\
NaN NaN NaN 10\\
NaN NaN NaN 10\\
NaN NaN NaN 10\\
10.4 4.6 0 10\\
10.4 5.4 0 10\\
NaN NaN NaN 10\\
NaN NaN NaN 10\\
NaN NaN NaN 10\\
NaN NaN NaN 10\\
10.4 5.6 0 10\\
10.4 6.4 0 10\\
NaN NaN NaN 10\\
NaN NaN NaN 10\\
NaN NaN NaN 10\\
NaN NaN NaN 10\\
10.4 6.6 0 10\\
10.4 7.4 0 10\\
NaN NaN NaN 10\\
NaN NaN NaN 10\\
NaN NaN NaN 10\\
NaN NaN NaN 10\\
10.4 7.6 0 10\\
10.4 8.4 0 10\\
NaN NaN NaN 10\\
NaN NaN NaN 10\\
NaN NaN NaN 10\\
NaN NaN NaN 10\\
10.4 8.6 0 10\\
10.4 9.4 0 10\\
NaN NaN NaN 10\\
NaN NaN NaN 10\\
NaN NaN NaN 10\\
NaN NaN NaN 10\\
10.4 9.6 0 10\\
10.4 10.4 0 10\\
NaN NaN NaN 10\\
NaN NaN NaN 10\\
NaN NaN NaN 10\\
NaN NaN NaN 10\\
10.4 10.6 0 10\\
10.4 11.4 0 10\\
NaN NaN NaN 10\\
NaN NaN NaN 10\\
NaN NaN NaN 10\\
};

\addplot3[%
surf,
shader=flat,
draw=black,
point meta=explicit,
mesh/rows=4]
table[row sep=crcr,header=false,meta index=3] {
NaN NaN NaN 9\\
8.6 0.6 0 9\\
8.6 1.4 0 9\\
NaN NaN NaN 9\\
NaN NaN NaN 9\\
NaN NaN NaN 9\\
NaN NaN NaN 9\\
8.6 1.6 0 9\\
8.6 2.4 0 9\\
NaN NaN NaN 9\\
NaN NaN NaN 9\\
NaN NaN NaN 9\\
NaN NaN NaN 9\\
8.6 2.6 0 9\\
8.6 3.4 0 9\\
NaN NaN NaN 9\\
NaN NaN NaN 9\\
NaN NaN NaN 9\\
NaN NaN NaN 9\\
8.6 3.6 0 9\\
8.6 4.4 0 9\\
NaN NaN NaN 9\\
NaN NaN NaN 9\\
NaN NaN NaN 9\\
NaN NaN NaN 9\\
8.6 4.6 0 9\\
8.6 5.4 0 9\\
NaN NaN NaN 9\\
NaN NaN NaN 9\\
NaN NaN NaN 9\\
NaN NaN NaN 9\\
8.6 5.6 0 9\\
8.6 6.4 0 9\\
NaN NaN NaN 9\\
NaN NaN NaN 9\\
NaN NaN NaN 9\\
NaN NaN NaN 9\\
8.6 6.6 0 9\\
8.6 7.4 0 9\\
NaN NaN NaN 9\\
NaN NaN NaN 9\\
NaN NaN NaN 9\\
NaN NaN NaN 9\\
8.6 7.6 0 9\\
8.6 8.4 0 9\\
NaN NaN NaN 9\\
NaN NaN NaN 9\\
NaN NaN NaN 9\\
NaN NaN NaN 9\\
8.6 8.6 0 9\\
8.6 9.4 0 9\\
NaN NaN NaN 9\\
NaN NaN NaN 9\\
NaN NaN NaN 9\\
NaN NaN NaN 9\\
8.6 9.6 0 9\\
8.6 10.4 0 9\\
NaN NaN NaN 9\\
NaN NaN NaN 9\\
NaN NaN NaN 9\\
NaN NaN NaN 9\\
8.6 10.6 0 9\\
8.6 11.4 0 9\\
NaN NaN NaN 9\\
NaN NaN NaN 9\\
NaN NaN NaN 9\\
8.6 0.6 0 9\\
8.6 0.6 0 9\\
8.6 1.4 0 9\\
8.6 1.4 0 9\\
8.6 0.6 0 9\\
NaN NaN NaN 9\\
8.6 1.6 0 9\\
8.6 1.6 2 9\\
8.6 2.4 2 9\\
8.6 2.4 0 9\\
8.6 1.6 0 9\\
NaN NaN NaN 9\\
8.6 2.6 0 9\\
8.6 2.6 2 9\\
8.6 3.4 2 9\\
8.6 3.4 0 9\\
8.6 2.6 0 9\\
NaN NaN NaN 9\\
8.6 3.6 0 9\\
8.6 3.6 3 9\\
8.6 4.4 3 9\\
8.6 4.4 0 9\\
8.6 3.6 0 9\\
NaN NaN NaN 9\\
8.6 4.6 0 9\\
8.6 4.6 3 9\\
8.6 5.4 3 9\\
8.6 5.4 0 9\\
8.6 4.6 0 9\\
NaN NaN NaN 9\\
8.6 5.6 0 9\\
8.6 5.6 3 9\\
8.6 6.4 3 9\\
8.6 6.4 0 9\\
8.6 5.6 0 9\\
NaN NaN NaN 9\\
8.6 6.6 0 9\\
8.6 6.6 4 9\\
8.6 7.4 4 9\\
8.6 7.4 0 9\\
8.6 6.6 0 9\\
NaN NaN NaN 9\\
8.6 7.6 0 9\\
8.6 7.6 4 9\\
8.6 8.4 4 9\\
8.6 8.4 0 9\\
8.6 7.6 0 9\\
NaN NaN NaN 9\\
8.6 8.6 0 9\\
8.6 8.6 4 9\\
8.6 9.4 4 9\\
8.6 9.4 0 9\\
8.6 8.6 0 9\\
NaN NaN NaN 9\\
8.6 9.6 0 9\\
8.6 9.6 4 9\\
8.6 10.4 4 9\\
8.6 10.4 0 9\\
8.6 9.6 0 9\\
NaN NaN NaN 9\\
8.6 10.6 0 9\\
8.6 10.6 4 9\\
8.6 11.4 4 9\\
8.6 11.4 0 9\\
8.6 10.6 0 9\\
NaN NaN NaN 9\\
9.4 0.6 0 9\\
9.4 0.6 0 9\\
9.4 1.4 0 9\\
9.4 1.4 0 9\\
9.4 0.6 0 9\\
NaN NaN NaN 9\\
9.4 1.6 0 9\\
9.4 1.6 2 9\\
9.4 2.4 2 9\\
9.4 2.4 0 9\\
9.4 1.6 0 9\\
NaN NaN NaN 9\\
9.4 2.6 0 9\\
9.4 2.6 2 9\\
9.4 3.4 2 9\\
9.4 3.4 0 9\\
9.4 2.6 0 9\\
NaN NaN NaN 9\\
9.4 3.6 0 9\\
9.4 3.6 3 9\\
9.4 4.4 3 9\\
9.4 4.4 0 9\\
9.4 3.6 0 9\\
NaN NaN NaN 9\\
9.4 4.6 0 9\\
9.4 4.6 3 9\\
9.4 5.4 3 9\\
9.4 5.4 0 9\\
9.4 4.6 0 9\\
NaN NaN NaN 9\\
9.4 5.6 0 9\\
9.4 5.6 3 9\\
9.4 6.4 3 9\\
9.4 6.4 0 9\\
9.4 5.6 0 9\\
NaN NaN NaN 9\\
9.4 6.6 0 9\\
9.4 6.6 4 9\\
9.4 7.4 4 9\\
9.4 7.4 0 9\\
9.4 6.6 0 9\\
NaN NaN NaN 9\\
9.4 7.6 0 9\\
9.4 7.6 4 9\\
9.4 8.4 4 9\\
9.4 8.4 0 9\\
9.4 7.6 0 9\\
NaN NaN NaN 9\\
9.4 8.6 0 9\\
9.4 8.6 4 9\\
9.4 9.4 4 9\\
9.4 9.4 0 9\\
9.4 8.6 0 9\\
NaN NaN NaN 9\\
9.4 9.6 0 9\\
9.4 9.6 4 9\\
9.4 10.4 4 9\\
9.4 10.4 0 9\\
9.4 9.6 0 9\\
NaN NaN NaN 9\\
9.4 10.6 0 9\\
9.4 10.6 4 9\\
9.4 11.4 4 9\\
9.4 11.4 0 9\\
9.4 10.6 0 9\\
NaN NaN NaN 9\\
NaN NaN NaN 9\\
9.4 0.6 0 9\\
9.4 1.4 0 9\\
NaN NaN NaN 9\\
NaN NaN NaN 9\\
NaN NaN NaN 9\\
NaN NaN NaN 9\\
9.4 1.6 0 9\\
9.4 2.4 0 9\\
NaN NaN NaN 9\\
NaN NaN NaN 9\\
NaN NaN NaN 9\\
NaN NaN NaN 9\\
9.4 2.6 0 9\\
9.4 3.4 0 9\\
NaN NaN NaN 9\\
NaN NaN NaN 9\\
NaN NaN NaN 9\\
NaN NaN NaN 9\\
9.4 3.6 0 9\\
9.4 4.4 0 9\\
NaN NaN NaN 9\\
NaN NaN NaN 9\\
NaN NaN NaN 9\\
NaN NaN NaN 9\\
9.4 4.6 0 9\\
9.4 5.4 0 9\\
NaN NaN NaN 9\\
NaN NaN NaN 9\\
NaN NaN NaN 9\\
NaN NaN NaN 9\\
9.4 5.6 0 9\\
9.4 6.4 0 9\\
NaN NaN NaN 9\\
NaN NaN NaN 9\\
NaN NaN NaN 9\\
NaN NaN NaN 9\\
9.4 6.6 0 9\\
9.4 7.4 0 9\\
NaN NaN NaN 9\\
NaN NaN NaN 9\\
NaN NaN NaN 9\\
NaN NaN NaN 9\\
9.4 7.6 0 9\\
9.4 8.4 0 9\\
NaN NaN NaN 9\\
NaN NaN NaN 9\\
NaN NaN NaN 9\\
NaN NaN NaN 9\\
9.4 8.6 0 9\\
9.4 9.4 0 9\\
NaN NaN NaN 9\\
NaN NaN NaN 9\\
NaN NaN NaN 9\\
NaN NaN NaN 9\\
9.4 9.6 0 9\\
9.4 10.4 0 9\\
NaN NaN NaN 9\\
NaN NaN NaN 9\\
NaN NaN NaN 9\\
NaN NaN NaN 9\\
9.4 10.6 0 9\\
9.4 11.4 0 9\\
NaN NaN NaN 9\\
NaN NaN NaN 9\\
NaN NaN NaN 9\\
};

\addplot3[%
surf,
shader=flat,
draw=black,
point meta=explicit,
mesh/rows=4]
table[row sep=crcr,header=false,meta index=3] {
NaN NaN NaN 8\\
7.6 0.6 0 8\\
7.6 1.4 0 8\\
NaN NaN NaN 8\\
NaN NaN NaN 8\\
NaN NaN NaN 8\\
NaN NaN NaN 8\\
7.6 1.6 0 8\\
7.6 2.4 0 8\\
NaN NaN NaN 8\\
NaN NaN NaN 8\\
NaN NaN NaN 8\\
NaN NaN NaN 8\\
7.6 2.6 0 8\\
7.6 3.4 0 8\\
NaN NaN NaN 8\\
NaN NaN NaN 8\\
NaN NaN NaN 8\\
NaN NaN NaN 8\\
7.6 3.6 0 8\\
7.6 4.4 0 8\\
NaN NaN NaN 8\\
NaN NaN NaN 8\\
NaN NaN NaN 8\\
NaN NaN NaN 8\\
7.6 4.6 0 8\\
7.6 5.4 0 8\\
NaN NaN NaN 8\\
NaN NaN NaN 8\\
NaN NaN NaN 8\\
NaN NaN NaN 8\\
7.6 5.6 0 8\\
7.6 6.4 0 8\\
NaN NaN NaN 8\\
NaN NaN NaN 8\\
NaN NaN NaN 8\\
NaN NaN NaN 8\\
7.6 6.6 0 8\\
7.6 7.4 0 8\\
NaN NaN NaN 8\\
NaN NaN NaN 8\\
NaN NaN NaN 8\\
NaN NaN NaN 8\\
7.6 7.6 0 8\\
7.6 8.4 0 8\\
NaN NaN NaN 8\\
NaN NaN NaN 8\\
NaN NaN NaN 8\\
NaN NaN NaN 8\\
7.6 8.6 0 8\\
7.6 9.4 0 8\\
NaN NaN NaN 8\\
NaN NaN NaN 8\\
NaN NaN NaN 8\\
NaN NaN NaN 8\\
7.6 9.6 0 8\\
7.6 10.4 0 8\\
NaN NaN NaN 8\\
NaN NaN NaN 8\\
NaN NaN NaN 8\\
NaN NaN NaN 8\\
7.6 10.6 0 8\\
7.6 11.4 0 8\\
NaN NaN NaN 8\\
NaN NaN NaN 8\\
NaN NaN NaN 8\\
7.6 0.6 0 8\\
7.6 0.6 0 8\\
7.6 1.4 0 8\\
7.6 1.4 0 8\\
7.6 0.6 0 8\\
NaN NaN NaN 8\\
7.6 1.6 0 8\\
7.6 1.6 2 8\\
7.6 2.4 2 8\\
7.6 2.4 0 8\\
7.6 1.6 0 8\\
NaN NaN NaN 8\\
7.6 2.6 0 8\\
7.6 2.6 2 8\\
7.6 3.4 2 8\\
7.6 3.4 0 8\\
7.6 2.6 0 8\\
NaN NaN NaN 8\\
7.6 3.6 0 8\\
7.6 3.6 3 8\\
7.6 4.4 3 8\\
7.6 4.4 0 8\\
7.6 3.6 0 8\\
NaN NaN NaN 8\\
7.6 4.6 0 8\\
7.6 4.6 3 8\\
7.6 5.4 3 8\\
7.6 5.4 0 8\\
7.6 4.6 0 8\\
NaN NaN NaN 8\\
7.6 5.6 0 8\\
7.6 5.6 3 8\\
7.6 6.4 3 8\\
7.6 6.4 0 8\\
7.6 5.6 0 8\\
NaN NaN NaN 8\\
7.6 6.6 0 8\\
7.6 6.6 3 8\\
7.6 7.4 3 8\\
7.6 7.4 0 8\\
7.6 6.6 0 8\\
NaN NaN NaN 8\\
7.6 7.6 0 8\\
7.6 7.6 3 8\\
7.6 8.4 3 8\\
7.6 8.4 0 8\\
7.6 7.6 0 8\\
NaN NaN NaN 8\\
7.6 8.6 0 8\\
7.6 8.6 4 8\\
7.6 9.4 4 8\\
7.6 9.4 0 8\\
7.6 8.6 0 8\\
NaN NaN NaN 8\\
7.6 9.6 0 8\\
7.6 9.6 4 8\\
7.6 10.4 4 8\\
7.6 10.4 0 8\\
7.6 9.6 0 8\\
NaN NaN NaN 8\\
7.6 10.6 0 8\\
7.6 10.6 4 8\\
7.6 11.4 4 8\\
7.6 11.4 0 8\\
7.6 10.6 0 8\\
NaN NaN NaN 8\\
8.4 0.6 0 8\\
8.4 0.6 0 8\\
8.4 1.4 0 8\\
8.4 1.4 0 8\\
8.4 0.6 0 8\\
NaN NaN NaN 8\\
8.4 1.6 0 8\\
8.4 1.6 2 8\\
8.4 2.4 2 8\\
8.4 2.4 0 8\\
8.4 1.6 0 8\\
NaN NaN NaN 8\\
8.4 2.6 0 8\\
8.4 2.6 2 8\\
8.4 3.4 2 8\\
8.4 3.4 0 8\\
8.4 2.6 0 8\\
NaN NaN NaN 8\\
8.4 3.6 0 8\\
8.4 3.6 3 8\\
8.4 4.4 3 8\\
8.4 4.4 0 8\\
8.4 3.6 0 8\\
NaN NaN NaN 8\\
8.4 4.6 0 8\\
8.4 4.6 3 8\\
8.4 5.4 3 8\\
8.4 5.4 0 8\\
8.4 4.6 0 8\\
NaN NaN NaN 8\\
8.4 5.6 0 8\\
8.4 5.6 3 8\\
8.4 6.4 3 8\\
8.4 6.4 0 8\\
8.4 5.6 0 8\\
NaN NaN NaN 8\\
8.4 6.6 0 8\\
8.4 6.6 3 8\\
8.4 7.4 3 8\\
8.4 7.4 0 8\\
8.4 6.6 0 8\\
NaN NaN NaN 8\\
8.4 7.6 0 8\\
8.4 7.6 3 8\\
8.4 8.4 3 8\\
8.4 8.4 0 8\\
8.4 7.6 0 8\\
NaN NaN NaN 8\\
8.4 8.6 0 8\\
8.4 8.6 4 8\\
8.4 9.4 4 8\\
8.4 9.4 0 8\\
8.4 8.6 0 8\\
NaN NaN NaN 8\\
8.4 9.6 0 8\\
8.4 9.6 4 8\\
8.4 10.4 4 8\\
8.4 10.4 0 8\\
8.4 9.6 0 8\\
NaN NaN NaN 8\\
8.4 10.6 0 8\\
8.4 10.6 4 8\\
8.4 11.4 4 8\\
8.4 11.4 0 8\\
8.4 10.6 0 8\\
NaN NaN NaN 8\\
NaN NaN NaN 8\\
8.4 0.6 0 8\\
8.4 1.4 0 8\\
NaN NaN NaN 8\\
NaN NaN NaN 8\\
NaN NaN NaN 8\\
NaN NaN NaN 8\\
8.4 1.6 0 8\\
8.4 2.4 0 8\\
NaN NaN NaN 8\\
NaN NaN NaN 8\\
NaN NaN NaN 8\\
NaN NaN NaN 8\\
8.4 2.6 0 8\\
8.4 3.4 0 8\\
NaN NaN NaN 8\\
NaN NaN NaN 8\\
NaN NaN NaN 8\\
NaN NaN NaN 8\\
8.4 3.6 0 8\\
8.4 4.4 0 8\\
NaN NaN NaN 8\\
NaN NaN NaN 8\\
NaN NaN NaN 8\\
NaN NaN NaN 8\\
8.4 4.6 0 8\\
8.4 5.4 0 8\\
NaN NaN NaN 8\\
NaN NaN NaN 8\\
NaN NaN NaN 8\\
NaN NaN NaN 8\\
8.4 5.6 0 8\\
8.4 6.4 0 8\\
NaN NaN NaN 8\\
NaN NaN NaN 8\\
NaN NaN NaN 8\\
NaN NaN NaN 8\\
8.4 6.6 0 8\\
8.4 7.4 0 8\\
NaN NaN NaN 8\\
NaN NaN NaN 8\\
NaN NaN NaN 8\\
NaN NaN NaN 8\\
8.4 7.6 0 8\\
8.4 8.4 0 8\\
NaN NaN NaN 8\\
NaN NaN NaN 8\\
NaN NaN NaN 8\\
NaN NaN NaN 8\\
8.4 8.6 0 8\\
8.4 9.4 0 8\\
NaN NaN NaN 8\\
NaN NaN NaN 8\\
NaN NaN NaN 8\\
NaN NaN NaN 8\\
8.4 9.6 0 8\\
8.4 10.4 0 8\\
NaN NaN NaN 8\\
NaN NaN NaN 8\\
NaN NaN NaN 8\\
NaN NaN NaN 8\\
8.4 10.6 0 8\\
8.4 11.4 0 8\\
NaN NaN NaN 8\\
NaN NaN NaN 8\\
NaN NaN NaN 8\\
};

\addplot3[%
surf,
shader=flat,
draw=black,
point meta=explicit,
mesh/rows=4]
table[row sep=crcr,header=false,meta index=3] {
NaN NaN NaN 7\\
6.6 0.6 0 7\\
6.6 1.4 0 7\\
NaN NaN NaN 7\\
NaN NaN NaN 7\\
NaN NaN NaN 7\\
NaN NaN NaN 7\\
6.6 1.6 0 7\\
6.6 2.4 0 7\\
NaN NaN NaN 7\\
NaN NaN NaN 7\\
NaN NaN NaN 7\\
NaN NaN NaN 7\\
6.6 2.6 0 7\\
6.6 3.4 0 7\\
NaN NaN NaN 7\\
NaN NaN NaN 7\\
NaN NaN NaN 7\\
NaN NaN NaN 7\\
6.6 3.6 0 7\\
6.6 4.4 0 7\\
NaN NaN NaN 7\\
NaN NaN NaN 7\\
NaN NaN NaN 7\\
NaN NaN NaN 7\\
6.6 4.6 0 7\\
6.6 5.4 0 7\\
NaN NaN NaN 7\\
NaN NaN NaN 7\\
NaN NaN NaN 7\\
NaN NaN NaN 7\\
6.6 5.6 0 7\\
6.6 6.4 0 7\\
NaN NaN NaN 7\\
NaN NaN NaN 7\\
NaN NaN NaN 7\\
NaN NaN NaN 7\\
6.6 6.6 0 7\\
6.6 7.4 0 7\\
NaN NaN NaN 7\\
NaN NaN NaN 7\\
NaN NaN NaN 7\\
NaN NaN NaN 7\\
6.6 7.6 0 7\\
6.6 8.4 0 7\\
NaN NaN NaN 7\\
NaN NaN NaN 7\\
NaN NaN NaN 7\\
NaN NaN NaN 7\\
6.6 8.6 0 7\\
6.6 9.4 0 7\\
NaN NaN NaN 7\\
NaN NaN NaN 7\\
NaN NaN NaN 7\\
NaN NaN NaN 7\\
6.6 9.6 0 7\\
6.6 10.4 0 7\\
NaN NaN NaN 7\\
NaN NaN NaN 7\\
NaN NaN NaN 7\\
NaN NaN NaN 7\\
6.6 10.6 0 7\\
6.6 11.4 0 7\\
NaN NaN NaN 7\\
NaN NaN NaN 7\\
NaN NaN NaN 7\\
6.6 0.6 0 7\\
6.6 0.6 0 7\\
6.6 1.4 0 7\\
6.6 1.4 0 7\\
6.6 0.6 0 7\\
NaN NaN NaN 7\\
6.6 1.6 0 7\\
6.6 1.6 2 7\\
6.6 2.4 2 7\\
6.6 2.4 0 7\\
6.6 1.6 0 7\\
NaN NaN NaN 7\\
6.6 2.6 0 7\\
6.6 2.6 2 7\\
6.6 3.4 2 7\\
6.6 3.4 0 7\\
6.6 2.6 0 7\\
NaN NaN NaN 7\\
6.6 3.6 0 7\\
6.6 3.6 3 7\\
6.6 4.4 3 7\\
6.6 4.4 0 7\\
6.6 3.6 0 7\\
NaN NaN NaN 7\\
6.6 4.6 0 7\\
6.6 4.6 3 7\\
6.6 5.4 3 7\\
6.6 5.4 0 7\\
6.6 4.6 0 7\\
NaN NaN NaN 7\\
6.6 5.6 0 7\\
6.6 5.6 3 7\\
6.6 6.4 3 7\\
6.6 6.4 0 7\\
6.6 5.6 0 7\\
NaN NaN NaN 7\\
6.6 6.6 0 7\\
6.6 6.6 3 7\\
6.6 7.4 3 7\\
6.6 7.4 0 7\\
6.6 6.6 0 7\\
NaN NaN NaN 7\\
6.6 7.6 0 7\\
6.6 7.6 3 7\\
6.6 8.4 3 7\\
6.6 8.4 0 7\\
6.6 7.6 0 7\\
NaN NaN NaN 7\\
6.6 8.6 0 7\\
6.6 8.6 4 7\\
6.6 9.4 4 7\\
6.6 9.4 0 7\\
6.6 8.6 0 7\\
NaN NaN NaN 7\\
6.6 9.6 0 7\\
6.6 9.6 4 7\\
6.6 10.4 4 7\\
6.6 10.4 0 7\\
6.6 9.6 0 7\\
NaN NaN NaN 7\\
6.6 10.6 0 7\\
6.6 10.6 4 7\\
6.6 11.4 4 7\\
6.6 11.4 0 7\\
6.6 10.6 0 7\\
NaN NaN NaN 7\\
7.4 0.6 0 7\\
7.4 0.6 0 7\\
7.4 1.4 0 7\\
7.4 1.4 0 7\\
7.4 0.6 0 7\\
NaN NaN NaN 7\\
7.4 1.6 0 7\\
7.4 1.6 2 7\\
7.4 2.4 2 7\\
7.4 2.4 0 7\\
7.4 1.6 0 7\\
NaN NaN NaN 7\\
7.4 2.6 0 7\\
7.4 2.6 2 7\\
7.4 3.4 2 7\\
7.4 3.4 0 7\\
7.4 2.6 0 7\\
NaN NaN NaN 7\\
7.4 3.6 0 7\\
7.4 3.6 3 7\\
7.4 4.4 3 7\\
7.4 4.4 0 7\\
7.4 3.6 0 7\\
NaN NaN NaN 7\\
7.4 4.6 0 7\\
7.4 4.6 3 7\\
7.4 5.4 3 7\\
7.4 5.4 0 7\\
7.4 4.6 0 7\\
NaN NaN NaN 7\\
7.4 5.6 0 7\\
7.4 5.6 3 7\\
7.4 6.4 3 7\\
7.4 6.4 0 7\\
7.4 5.6 0 7\\
NaN NaN NaN 7\\
7.4 6.6 0 7\\
7.4 6.6 3 7\\
7.4 7.4 3 7\\
7.4 7.4 0 7\\
7.4 6.6 0 7\\
NaN NaN NaN 7\\
7.4 7.6 0 7\\
7.4 7.6 3 7\\
7.4 8.4 3 7\\
7.4 8.4 0 7\\
7.4 7.6 0 7\\
NaN NaN NaN 7\\
7.4 8.6 0 7\\
7.4 8.6 4 7\\
7.4 9.4 4 7\\
7.4 9.4 0 7\\
7.4 8.6 0 7\\
NaN NaN NaN 7\\
7.4 9.6 0 7\\
7.4 9.6 4 7\\
7.4 10.4 4 7\\
7.4 10.4 0 7\\
7.4 9.6 0 7\\
NaN NaN NaN 7\\
7.4 10.6 0 7\\
7.4 10.6 4 7\\
7.4 11.4 4 7\\
7.4 11.4 0 7\\
7.4 10.6 0 7\\
NaN NaN NaN 7\\
NaN NaN NaN 7\\
7.4 0.6 0 7\\
7.4 1.4 0 7\\
NaN NaN NaN 7\\
NaN NaN NaN 7\\
NaN NaN NaN 7\\
NaN NaN NaN 7\\
7.4 1.6 0 7\\
7.4 2.4 0 7\\
NaN NaN NaN 7\\
NaN NaN NaN 7\\
NaN NaN NaN 7\\
NaN NaN NaN 7\\
7.4 2.6 0 7\\
7.4 3.4 0 7\\
NaN NaN NaN 7\\
NaN NaN NaN 7\\
NaN NaN NaN 7\\
NaN NaN NaN 7\\
7.4 3.6 0 7\\
7.4 4.4 0 7\\
NaN NaN NaN 7\\
NaN NaN NaN 7\\
NaN NaN NaN 7\\
NaN NaN NaN 7\\
7.4 4.6 0 7\\
7.4 5.4 0 7\\
NaN NaN NaN 7\\
NaN NaN NaN 7\\
NaN NaN NaN 7\\
NaN NaN NaN 7\\
7.4 5.6 0 7\\
7.4 6.4 0 7\\
NaN NaN NaN 7\\
NaN NaN NaN 7\\
NaN NaN NaN 7\\
NaN NaN NaN 7\\
7.4 6.6 0 7\\
7.4 7.4 0 7\\
NaN NaN NaN 7\\
NaN NaN NaN 7\\
NaN NaN NaN 7\\
NaN NaN NaN 7\\
7.4 7.6 0 7\\
7.4 8.4 0 7\\
NaN NaN NaN 7\\
NaN NaN NaN 7\\
NaN NaN NaN 7\\
NaN NaN NaN 7\\
7.4 8.6 0 7\\
7.4 9.4 0 7\\
NaN NaN NaN 7\\
NaN NaN NaN 7\\
NaN NaN NaN 7\\
NaN NaN NaN 7\\
7.4 9.6 0 7\\
7.4 10.4 0 7\\
NaN NaN NaN 7\\
NaN NaN NaN 7\\
NaN NaN NaN 7\\
NaN NaN NaN 7\\
7.4 10.6 0 7\\
7.4 11.4 0 7\\
NaN NaN NaN 7\\
NaN NaN NaN 7\\
NaN NaN NaN 7\\
};

\addplot3[%
surf,
shader=flat,
draw=black,
point meta=explicit,
mesh/rows=4]
table[row sep=crcr,header=false,meta index=3] {
NaN NaN NaN 6\\
5.6 0.6 0 6\\
5.6 1.4 0 6\\
NaN NaN NaN 6\\
NaN NaN NaN 6\\
NaN NaN NaN 6\\
NaN NaN NaN 6\\
5.6 1.6 0 6\\
5.6 2.4 0 6\\
NaN NaN NaN 6\\
NaN NaN NaN 6\\
NaN NaN NaN 6\\
NaN NaN NaN 6\\
5.6 2.6 0 6\\
5.6 3.4 0 6\\
NaN NaN NaN 6\\
NaN NaN NaN 6\\
NaN NaN NaN 6\\
NaN NaN NaN 6\\
5.6 3.6 0 6\\
5.6 4.4 0 6\\
NaN NaN NaN 6\\
NaN NaN NaN 6\\
NaN NaN NaN 6\\
NaN NaN NaN 6\\
5.6 4.6 0 6\\
5.6 5.4 0 6\\
NaN NaN NaN 6\\
NaN NaN NaN 6\\
NaN NaN NaN 6\\
NaN NaN NaN 6\\
5.6 5.6 0 6\\
5.6 6.4 0 6\\
NaN NaN NaN 6\\
NaN NaN NaN 6\\
NaN NaN NaN 6\\
NaN NaN NaN 6\\
5.6 6.6 0 6\\
5.6 7.4 0 6\\
NaN NaN NaN 6\\
NaN NaN NaN 6\\
NaN NaN NaN 6\\
NaN NaN NaN 6\\
5.6 7.6 0 6\\
5.6 8.4 0 6\\
NaN NaN NaN 6\\
NaN NaN NaN 6\\
NaN NaN NaN 6\\
NaN NaN NaN 6\\
5.6 8.6 0 6\\
5.6 9.4 0 6\\
NaN NaN NaN 6\\
NaN NaN NaN 6\\
NaN NaN NaN 6\\
NaN NaN NaN 6\\
5.6 9.6 0 6\\
5.6 10.4 0 6\\
NaN NaN NaN 6\\
NaN NaN NaN 6\\
NaN NaN NaN 6\\
NaN NaN NaN 6\\
5.6 10.6 0 6\\
5.6 11.4 0 6\\
NaN NaN NaN 6\\
NaN NaN NaN 6\\
NaN NaN NaN 6\\
5.6 0.6 0 6\\
5.6 0.6 0 6\\
5.6 1.4 0 6\\
5.6 1.4 0 6\\
5.6 0.6 0 6\\
NaN NaN NaN 6\\
5.6 1.6 0 6\\
5.6 1.6 2 6\\
5.6 2.4 2 6\\
5.6 2.4 0 6\\
5.6 1.6 0 6\\
NaN NaN NaN 6\\
5.6 2.6 0 6\\
5.6 2.6 2 6\\
5.6 3.4 2 6\\
5.6 3.4 0 6\\
5.6 2.6 0 6\\
NaN NaN NaN 6\\
5.6 3.6 0 6\\
5.6 3.6 3 6\\
5.6 4.4 3 6\\
5.6 4.4 0 6\\
5.6 3.6 0 6\\
NaN NaN NaN 6\\
5.6 4.6 0 6\\
5.6 4.6 3 6\\
5.6 5.4 3 6\\
5.6 5.4 0 6\\
5.6 4.6 0 6\\
NaN NaN NaN 6\\
5.6 5.6 0 6\\
5.6 5.6 3 6\\
5.6 6.4 3 6\\
5.6 6.4 0 6\\
5.6 5.6 0 6\\
NaN NaN NaN 6\\
5.6 6.6 0 6\\
5.6 6.6 3 6\\
5.6 7.4 3 6\\
5.6 7.4 0 6\\
5.6 6.6 0 6\\
NaN NaN NaN 6\\
5.6 7.6 0 6\\
5.6 7.6 3 6\\
5.6 8.4 3 6\\
5.6 8.4 0 6\\
5.6 7.6 0 6\\
NaN NaN NaN 6\\
5.6 8.6 0 6\\
5.6 8.6 4 6\\
5.6 9.4 4 6\\
5.6 9.4 0 6\\
5.6 8.6 0 6\\
NaN NaN NaN 6\\
5.6 9.6 0 6\\
5.6 9.6 4 6\\
5.6 10.4 4 6\\
5.6 10.4 0 6\\
5.6 9.6 0 6\\
NaN NaN NaN 6\\
5.6 10.6 0 6\\
5.6 10.6 4 6\\
5.6 11.4 4 6\\
5.6 11.4 0 6\\
5.6 10.6 0 6\\
NaN NaN NaN 6\\
6.4 0.6 0 6\\
6.4 0.6 0 6\\
6.4 1.4 0 6\\
6.4 1.4 0 6\\
6.4 0.6 0 6\\
NaN NaN NaN 6\\
6.4 1.6 0 6\\
6.4 1.6 2 6\\
6.4 2.4 2 6\\
6.4 2.4 0 6\\
6.4 1.6 0 6\\
NaN NaN NaN 6\\
6.4 2.6 0 6\\
6.4 2.6 2 6\\
6.4 3.4 2 6\\
6.4 3.4 0 6\\
6.4 2.6 0 6\\
NaN NaN NaN 6\\
6.4 3.6 0 6\\
6.4 3.6 3 6\\
6.4 4.4 3 6\\
6.4 4.4 0 6\\
6.4 3.6 0 6\\
NaN NaN NaN 6\\
6.4 4.6 0 6\\
6.4 4.6 3 6\\
6.4 5.4 3 6\\
6.4 5.4 0 6\\
6.4 4.6 0 6\\
NaN NaN NaN 6\\
6.4 5.6 0 6\\
6.4 5.6 3 6\\
6.4 6.4 3 6\\
6.4 6.4 0 6\\
6.4 5.6 0 6\\
NaN NaN NaN 6\\
6.4 6.6 0 6\\
6.4 6.6 3 6\\
6.4 7.4 3 6\\
6.4 7.4 0 6\\
6.4 6.6 0 6\\
NaN NaN NaN 6\\
6.4 7.6 0 6\\
6.4 7.6 3 6\\
6.4 8.4 3 6\\
6.4 8.4 0 6\\
6.4 7.6 0 6\\
NaN NaN NaN 6\\
6.4 8.6 0 6\\
6.4 8.6 4 6\\
6.4 9.4 4 6\\
6.4 9.4 0 6\\
6.4 8.6 0 6\\
NaN NaN NaN 6\\
6.4 9.6 0 6\\
6.4 9.6 4 6\\
6.4 10.4 4 6\\
6.4 10.4 0 6\\
6.4 9.6 0 6\\
NaN NaN NaN 6\\
6.4 10.6 0 6\\
6.4 10.6 4 6\\
6.4 11.4 4 6\\
6.4 11.4 0 6\\
6.4 10.6 0 6\\
NaN NaN NaN 6\\
NaN NaN NaN 6\\
6.4 0.6 0 6\\
6.4 1.4 0 6\\
NaN NaN NaN 6\\
NaN NaN NaN 6\\
NaN NaN NaN 6\\
NaN NaN NaN 6\\
6.4 1.6 0 6\\
6.4 2.4 0 6\\
NaN NaN NaN 6\\
NaN NaN NaN 6\\
NaN NaN NaN 6\\
NaN NaN NaN 6\\
6.4 2.6 0 6\\
6.4 3.4 0 6\\
NaN NaN NaN 6\\
NaN NaN NaN 6\\
NaN NaN NaN 6\\
NaN NaN NaN 6\\
6.4 3.6 0 6\\
6.4 4.4 0 6\\
NaN NaN NaN 6\\
NaN NaN NaN 6\\
NaN NaN NaN 6\\
NaN NaN NaN 6\\
6.4 4.6 0 6\\
6.4 5.4 0 6\\
NaN NaN NaN 6\\
NaN NaN NaN 6\\
NaN NaN NaN 6\\
NaN NaN NaN 6\\
6.4 5.6 0 6\\
6.4 6.4 0 6\\
NaN NaN NaN 6\\
NaN NaN NaN 6\\
NaN NaN NaN 6\\
NaN NaN NaN 6\\
6.4 6.6 0 6\\
6.4 7.4 0 6\\
NaN NaN NaN 6\\
NaN NaN NaN 6\\
NaN NaN NaN 6\\
NaN NaN NaN 6\\
6.4 7.6 0 6\\
6.4 8.4 0 6\\
NaN NaN NaN 6\\
NaN NaN NaN 6\\
NaN NaN NaN 6\\
NaN NaN NaN 6\\
6.4 8.6 0 6\\
6.4 9.4 0 6\\
NaN NaN NaN 6\\
NaN NaN NaN 6\\
NaN NaN NaN 6\\
NaN NaN NaN 6\\
6.4 9.6 0 6\\
6.4 10.4 0 6\\
NaN NaN NaN 6\\
NaN NaN NaN 6\\
NaN NaN NaN 6\\
NaN NaN NaN 6\\
6.4 10.6 0 6\\
6.4 11.4 0 6\\
NaN NaN NaN 6\\
NaN NaN NaN 6\\
NaN NaN NaN 6\\
};

\addplot3[%
surf,
shader=flat,
draw=black,
point meta=explicit,
mesh/rows=4]
table[row sep=crcr,header=false,meta index=3] {
NaN NaN NaN 5\\
4.6 0.6 0 5\\
4.6 1.4 0 5\\
NaN NaN NaN 5\\
NaN NaN NaN 5\\
NaN NaN NaN 5\\
NaN NaN NaN 5\\
4.6 1.6 0 5\\
4.6 2.4 0 5\\
NaN NaN NaN 5\\
NaN NaN NaN 5\\
NaN NaN NaN 5\\
NaN NaN NaN 5\\
4.6 2.6 0 5\\
4.6 3.4 0 5\\
NaN NaN NaN 5\\
NaN NaN NaN 5\\
NaN NaN NaN 5\\
NaN NaN NaN 5\\
4.6 3.6 0 5\\
4.6 4.4 0 5\\
NaN NaN NaN 5\\
NaN NaN NaN 5\\
NaN NaN NaN 5\\
NaN NaN NaN 5\\
4.6 4.6 0 5\\
4.6 5.4 0 5\\
NaN NaN NaN 5\\
NaN NaN NaN 5\\
NaN NaN NaN 5\\
NaN NaN NaN 5\\
4.6 5.6 0 5\\
4.6 6.4 0 5\\
NaN NaN NaN 5\\
NaN NaN NaN 5\\
NaN NaN NaN 5\\
NaN NaN NaN 5\\
4.6 6.6 0 5\\
4.6 7.4 0 5\\
NaN NaN NaN 5\\
NaN NaN NaN 5\\
NaN NaN NaN 5\\
NaN NaN NaN 5\\
4.6 7.6 0 5\\
4.6 8.4 0 5\\
NaN NaN NaN 5\\
NaN NaN NaN 5\\
NaN NaN NaN 5\\
NaN NaN NaN 5\\
4.6 8.6 0 5\\
4.6 9.4 0 5\\
NaN NaN NaN 5\\
NaN NaN NaN 5\\
NaN NaN NaN 5\\
NaN NaN NaN 5\\
4.6 9.6 0 5\\
4.6 10.4 0 5\\
NaN NaN NaN 5\\
NaN NaN NaN 5\\
NaN NaN NaN 5\\
NaN NaN NaN 5\\
4.6 10.6 0 5\\
4.6 11.4 0 5\\
NaN NaN NaN 5\\
NaN NaN NaN 5\\
NaN NaN NaN 5\\
4.6 0.6 0 5\\
4.6 0.6 0 5\\
4.6 1.4 0 5\\
4.6 1.4 0 5\\
4.6 0.6 0 5\\
NaN NaN NaN 5\\
4.6 1.6 0 5\\
4.6 1.6 2 5\\
4.6 2.4 2 5\\
4.6 2.4 0 5\\
4.6 1.6 0 5\\
NaN NaN NaN 5\\
4.6 2.6 0 5\\
4.6 2.6 2 5\\
4.6 3.4 2 5\\
4.6 3.4 0 5\\
4.6 2.6 0 5\\
NaN NaN NaN 5\\
4.6 3.6 0 5\\
4.6 3.6 3 5\\
4.6 4.4 3 5\\
4.6 4.4 0 5\\
4.6 3.6 0 5\\
NaN NaN NaN 5\\
4.6 4.6 0 5\\
4.6 4.6 3 5\\
4.6 5.4 3 5\\
4.6 5.4 0 5\\
4.6 4.6 0 5\\
NaN NaN NaN 5\\
4.6 5.6 0 5\\
4.6 5.6 3 5\\
4.6 6.4 3 5\\
4.6 6.4 0 5\\
4.6 5.6 0 5\\
NaN NaN NaN 5\\
4.6 6.6 0 5\\
4.6 6.6 3 5\\
4.6 7.4 3 5\\
4.6 7.4 0 5\\
4.6 6.6 0 5\\
NaN NaN NaN 5\\
4.6 7.6 0 5\\
4.6 7.6 3 5\\
4.6 8.4 3 5\\
4.6 8.4 0 5\\
4.6 7.6 0 5\\
NaN NaN NaN 5\\
4.6 8.6 0 5\\
4.6 8.6 4 5\\
4.6 9.4 4 5\\
4.6 9.4 0 5\\
4.6 8.6 0 5\\
NaN NaN NaN 5\\
4.6 9.6 0 5\\
4.6 9.6 4 5\\
4.6 10.4 4 5\\
4.6 10.4 0 5\\
4.6 9.6 0 5\\
NaN NaN NaN 5\\
4.6 10.6 0 5\\
4.6 10.6 4 5\\
4.6 11.4 4 5\\
4.6 11.4 0 5\\
4.6 10.6 0 5\\
NaN NaN NaN 5\\
5.4 0.6 0 5\\
5.4 0.6 0 5\\
5.4 1.4 0 5\\
5.4 1.4 0 5\\
5.4 0.6 0 5\\
NaN NaN NaN 5\\
5.4 1.6 0 5\\
5.4 1.6 2 5\\
5.4 2.4 2 5\\
5.4 2.4 0 5\\
5.4 1.6 0 5\\
NaN NaN NaN 5\\
5.4 2.6 0 5\\
5.4 2.6 2 5\\
5.4 3.4 2 5\\
5.4 3.4 0 5\\
5.4 2.6 0 5\\
NaN NaN NaN 5\\
5.4 3.6 0 5\\
5.4 3.6 3 5\\
5.4 4.4 3 5\\
5.4 4.4 0 5\\
5.4 3.6 0 5\\
NaN NaN NaN 5\\
5.4 4.6 0 5\\
5.4 4.6 3 5\\
5.4 5.4 3 5\\
5.4 5.4 0 5\\
5.4 4.6 0 5\\
NaN NaN NaN 5\\
5.4 5.6 0 5\\
5.4 5.6 3 5\\
5.4 6.4 3 5\\
5.4 6.4 0 5\\
5.4 5.6 0 5\\
NaN NaN NaN 5\\
5.4 6.6 0 5\\
5.4 6.6 3 5\\
5.4 7.4 3 5\\
5.4 7.4 0 5\\
5.4 6.6 0 5\\
NaN NaN NaN 5\\
5.4 7.6 0 5\\
5.4 7.6 3 5\\
5.4 8.4 3 5\\
5.4 8.4 0 5\\
5.4 7.6 0 5\\
NaN NaN NaN 5\\
5.4 8.6 0 5\\
5.4 8.6 4 5\\
5.4 9.4 4 5\\
5.4 9.4 0 5\\
5.4 8.6 0 5\\
NaN NaN NaN 5\\
5.4 9.6 0 5\\
5.4 9.6 4 5\\
5.4 10.4 4 5\\
5.4 10.4 0 5\\
5.4 9.6 0 5\\
NaN NaN NaN 5\\
5.4 10.6 0 5\\
5.4 10.6 4 5\\
5.4 11.4 4 5\\
5.4 11.4 0 5\\
5.4 10.6 0 5\\
NaN NaN NaN 5\\
NaN NaN NaN 5\\
5.4 0.6 0 5\\
5.4 1.4 0 5\\
NaN NaN NaN 5\\
NaN NaN NaN 5\\
NaN NaN NaN 5\\
NaN NaN NaN 5\\
5.4 1.6 0 5\\
5.4 2.4 0 5\\
NaN NaN NaN 5\\
NaN NaN NaN 5\\
NaN NaN NaN 5\\
NaN NaN NaN 5\\
5.4 2.6 0 5\\
5.4 3.4 0 5\\
NaN NaN NaN 5\\
NaN NaN NaN 5\\
NaN NaN NaN 5\\
NaN NaN NaN 5\\
5.4 3.6 0 5\\
5.4 4.4 0 5\\
NaN NaN NaN 5\\
NaN NaN NaN 5\\
NaN NaN NaN 5\\
NaN NaN NaN 5\\
5.4 4.6 0 5\\
5.4 5.4 0 5\\
NaN NaN NaN 5\\
NaN NaN NaN 5\\
NaN NaN NaN 5\\
NaN NaN NaN 5\\
5.4 5.6 0 5\\
5.4 6.4 0 5\\
NaN NaN NaN 5\\
NaN NaN NaN 5\\
NaN NaN NaN 5\\
NaN NaN NaN 5\\
5.4 6.6 0 5\\
5.4 7.4 0 5\\
NaN NaN NaN 5\\
NaN NaN NaN 5\\
NaN NaN NaN 5\\
NaN NaN NaN 5\\
5.4 7.6 0 5\\
5.4 8.4 0 5\\
NaN NaN NaN 5\\
NaN NaN NaN 5\\
NaN NaN NaN 5\\
NaN NaN NaN 5\\
5.4 8.6 0 5\\
5.4 9.4 0 5\\
NaN NaN NaN 5\\
NaN NaN NaN 5\\
NaN NaN NaN 5\\
NaN NaN NaN 5\\
5.4 9.6 0 5\\
5.4 10.4 0 5\\
NaN NaN NaN 5\\
NaN NaN NaN 5\\
NaN NaN NaN 5\\
NaN NaN NaN 5\\
5.4 10.6 0 5\\
5.4 11.4 0 5\\
NaN NaN NaN 5\\
NaN NaN NaN 5\\
NaN NaN NaN 5\\
};

\addplot3[%
surf,
shader=flat,
draw=black,
point meta=explicit,
mesh/rows=4]
table[row sep=crcr,header=false,meta index=3] {
NaN NaN NaN 4\\
3.6 0.6 0 4\\
3.6 1.4 0 4\\
NaN NaN NaN 4\\
NaN NaN NaN 4\\
NaN NaN NaN 4\\
NaN NaN NaN 4\\
3.6 1.6 0 4\\
3.6 2.4 0 4\\
NaN NaN NaN 4\\
NaN NaN NaN 4\\
NaN NaN NaN 4\\
NaN NaN NaN 4\\
3.6 2.6 0 4\\
3.6 3.4 0 4\\
NaN NaN NaN 4\\
NaN NaN NaN 4\\
NaN NaN NaN 4\\
NaN NaN NaN 4\\
3.6 3.6 0 4\\
3.6 4.4 0 4\\
NaN NaN NaN 4\\
NaN NaN NaN 4\\
NaN NaN NaN 4\\
NaN NaN NaN 4\\
3.6 4.6 0 4\\
3.6 5.4 0 4\\
NaN NaN NaN 4\\
NaN NaN NaN 4\\
NaN NaN NaN 4\\
NaN NaN NaN 4\\
3.6 5.6 0 4\\
3.6 6.4 0 4\\
NaN NaN NaN 4\\
NaN NaN NaN 4\\
NaN NaN NaN 4\\
NaN NaN NaN 4\\
3.6 6.6 0 4\\
3.6 7.4 0 4\\
NaN NaN NaN 4\\
NaN NaN NaN 4\\
NaN NaN NaN 4\\
NaN NaN NaN 4\\
3.6 7.6 0 4\\
3.6 8.4 0 4\\
NaN NaN NaN 4\\
NaN NaN NaN 4\\
NaN NaN NaN 4\\
NaN NaN NaN 4\\
3.6 8.6 0 4\\
3.6 9.4 0 4\\
NaN NaN NaN 4\\
NaN NaN NaN 4\\
NaN NaN NaN 4\\
NaN NaN NaN 4\\
3.6 9.6 0 4\\
3.6 10.4 0 4\\
NaN NaN NaN 4\\
NaN NaN NaN 4\\
NaN NaN NaN 4\\
NaN NaN NaN 4\\
3.6 10.6 0 4\\
3.6 11.4 0 4\\
NaN NaN NaN 4\\
NaN NaN NaN 4\\
NaN NaN NaN 4\\
3.6 0.6 0 4\\
3.6 0.6 0 4\\
3.6 1.4 0 4\\
3.6 1.4 0 4\\
3.6 0.6 0 4\\
NaN NaN NaN 4\\
3.6 1.6 0 4\\
3.6 1.6 2 4\\
3.6 2.4 2 4\\
3.6 2.4 0 4\\
3.6 1.6 0 4\\
NaN NaN NaN 4\\
3.6 2.6 0 4\\
3.6 2.6 2 4\\
3.6 3.4 2 4\\
3.6 3.4 0 4\\
3.6 2.6 0 4\\
NaN NaN NaN 4\\
3.6 3.6 0 4\\
3.6 3.6 3 4\\
3.6 4.4 3 4\\
3.6 4.4 0 4\\
3.6 3.6 0 4\\
NaN NaN NaN 4\\
3.6 4.6 0 4\\
3.6 4.6 3 4\\
3.6 5.4 3 4\\
3.6 5.4 0 4\\
3.6 4.6 0 4\\
NaN NaN NaN 4\\
3.6 5.6 0 4\\
3.6 5.6 3 4\\
3.6 6.4 3 4\\
3.6 6.4 0 4\\
3.6 5.6 0 4\\
NaN NaN NaN 4\\
3.6 6.6 0 4\\
3.6 6.6 3 4\\
3.6 7.4 3 4\\
3.6 7.4 0 4\\
3.6 6.6 0 4\\
NaN NaN NaN 4\\
3.6 7.6 0 4\\
3.6 7.6 3 4\\
3.6 8.4 3 4\\
3.6 8.4 0 4\\
3.6 7.6 0 4\\
NaN NaN NaN 4\\
3.6 8.6 0 4\\
3.6 8.6 4 4\\
3.6 9.4 4 4\\
3.6 9.4 0 4\\
3.6 8.6 0 4\\
NaN NaN NaN 4\\
3.6 9.6 0 4\\
3.6 9.6 4 4\\
3.6 10.4 4 4\\
3.6 10.4 0 4\\
3.6 9.6 0 4\\
NaN NaN NaN 4\\
3.6 10.6 0 4\\
3.6 10.6 4 4\\
3.6 11.4 4 4\\
3.6 11.4 0 4\\
3.6 10.6 0 4\\
NaN NaN NaN 4\\
4.4 0.6 0 4\\
4.4 0.6 0 4\\
4.4 1.4 0 4\\
4.4 1.4 0 4\\
4.4 0.6 0 4\\
NaN NaN NaN 4\\
4.4 1.6 0 4\\
4.4 1.6 2 4\\
4.4 2.4 2 4\\
4.4 2.4 0 4\\
4.4 1.6 0 4\\
NaN NaN NaN 4\\
4.4 2.6 0 4\\
4.4 2.6 2 4\\
4.4 3.4 2 4\\
4.4 3.4 0 4\\
4.4 2.6 0 4\\
NaN NaN NaN 4\\
4.4 3.6 0 4\\
4.4 3.6 3 4\\
4.4 4.4 3 4\\
4.4 4.4 0 4\\
4.4 3.6 0 4\\
NaN NaN NaN 4\\
4.4 4.6 0 4\\
4.4 4.6 3 4\\
4.4 5.4 3 4\\
4.4 5.4 0 4\\
4.4 4.6 0 4\\
NaN NaN NaN 4\\
4.4 5.6 0 4\\
4.4 5.6 3 4\\
4.4 6.4 3 4\\
4.4 6.4 0 4\\
4.4 5.6 0 4\\
NaN NaN NaN 4\\
4.4 6.6 0 4\\
4.4 6.6 3 4\\
4.4 7.4 3 4\\
4.4 7.4 0 4\\
4.4 6.6 0 4\\
NaN NaN NaN 4\\
4.4 7.6 0 4\\
4.4 7.6 3 4\\
4.4 8.4 3 4\\
4.4 8.4 0 4\\
4.4 7.6 0 4\\
NaN NaN NaN 4\\
4.4 8.6 0 4\\
4.4 8.6 4 4\\
4.4 9.4 4 4\\
4.4 9.4 0 4\\
4.4 8.6 0 4\\
NaN NaN NaN 4\\
4.4 9.6 0 4\\
4.4 9.6 4 4\\
4.4 10.4 4 4\\
4.4 10.4 0 4\\
4.4 9.6 0 4\\
NaN NaN NaN 4\\
4.4 10.6 0 4\\
4.4 10.6 4 4\\
4.4 11.4 4 4\\
4.4 11.4 0 4\\
4.4 10.6 0 4\\
NaN NaN NaN 4\\
NaN NaN NaN 4\\
4.4 0.6 0 4\\
4.4 1.4 0 4\\
NaN NaN NaN 4\\
NaN NaN NaN 4\\
NaN NaN NaN 4\\
NaN NaN NaN 4\\
4.4 1.6 0 4\\
4.4 2.4 0 4\\
NaN NaN NaN 4\\
NaN NaN NaN 4\\
NaN NaN NaN 4\\
NaN NaN NaN 4\\
4.4 2.6 0 4\\
4.4 3.4 0 4\\
NaN NaN NaN 4\\
NaN NaN NaN 4\\
NaN NaN NaN 4\\
NaN NaN NaN 4\\
4.4 3.6 0 4\\
4.4 4.4 0 4\\
NaN NaN NaN 4\\
NaN NaN NaN 4\\
NaN NaN NaN 4\\
NaN NaN NaN 4\\
4.4 4.6 0 4\\
4.4 5.4 0 4\\
NaN NaN NaN 4\\
NaN NaN NaN 4\\
NaN NaN NaN 4\\
NaN NaN NaN 4\\
4.4 5.6 0 4\\
4.4 6.4 0 4\\
NaN NaN NaN 4\\
NaN NaN NaN 4\\
NaN NaN NaN 4\\
NaN NaN NaN 4\\
4.4 6.6 0 4\\
4.4 7.4 0 4\\
NaN NaN NaN 4\\
NaN NaN NaN 4\\
NaN NaN NaN 4\\
NaN NaN NaN 4\\
4.4 7.6 0 4\\
4.4 8.4 0 4\\
NaN NaN NaN 4\\
NaN NaN NaN 4\\
NaN NaN NaN 4\\
NaN NaN NaN 4\\
4.4 8.6 0 4\\
4.4 9.4 0 4\\
NaN NaN NaN 4\\
NaN NaN NaN 4\\
NaN NaN NaN 4\\
NaN NaN NaN 4\\
4.4 9.6 0 4\\
4.4 10.4 0 4\\
NaN NaN NaN 4\\
NaN NaN NaN 4\\
NaN NaN NaN 4\\
NaN NaN NaN 4\\
4.4 10.6 0 4\\
4.4 11.4 0 4\\
NaN NaN NaN 4\\
NaN NaN NaN 4\\
NaN NaN NaN 4\\
};

\addplot3[%
surf,
shader=flat,
draw=black,
point meta=explicit,
mesh/rows=4]
table[row sep=crcr,header=false,meta index=3] {
NaN NaN NaN 3\\
2.6 0.6 0 3\\
2.6 1.4 0 3\\
NaN NaN NaN 3\\
NaN NaN NaN 3\\
NaN NaN NaN 3\\
NaN NaN NaN 3\\
2.6 1.6 0 3\\
2.6 2.4 0 3\\
NaN NaN NaN 3\\
NaN NaN NaN 3\\
NaN NaN NaN 3\\
NaN NaN NaN 3\\
2.6 2.6 0 3\\
2.6 3.4 0 3\\
NaN NaN NaN 3\\
NaN NaN NaN 3\\
NaN NaN NaN 3\\
NaN NaN NaN 3\\
2.6 3.6 0 3\\
2.6 4.4 0 3\\
NaN NaN NaN 3\\
NaN NaN NaN 3\\
NaN NaN NaN 3\\
NaN NaN NaN 3\\
2.6 4.6 0 3\\
2.6 5.4 0 3\\
NaN NaN NaN 3\\
NaN NaN NaN 3\\
NaN NaN NaN 3\\
NaN NaN NaN 3\\
2.6 5.6 0 3\\
2.6 6.4 0 3\\
NaN NaN NaN 3\\
NaN NaN NaN 3\\
NaN NaN NaN 3\\
NaN NaN NaN 3\\
2.6 6.6 0 3\\
2.6 7.4 0 3\\
NaN NaN NaN 3\\
NaN NaN NaN 3\\
NaN NaN NaN 3\\
NaN NaN NaN 3\\
2.6 7.6 0 3\\
2.6 8.4 0 3\\
NaN NaN NaN 3\\
NaN NaN NaN 3\\
NaN NaN NaN 3\\
NaN NaN NaN 3\\
2.6 8.6 0 3\\
2.6 9.4 0 3\\
NaN NaN NaN 3\\
NaN NaN NaN 3\\
NaN NaN NaN 3\\
NaN NaN NaN 3\\
2.6 9.6 0 3\\
2.6 10.4 0 3\\
NaN NaN NaN 3\\
NaN NaN NaN 3\\
NaN NaN NaN 3\\
NaN NaN NaN 3\\
2.6 10.6 0 3\\
2.6 11.4 0 3\\
NaN NaN NaN 3\\
NaN NaN NaN 3\\
NaN NaN NaN 3\\
2.6 0.6 0 3\\
2.6 0.6 0 3\\
2.6 1.4 0 3\\
2.6 1.4 0 3\\
2.6 0.6 0 3\\
NaN NaN NaN 3\\
2.6 1.6 0 3\\
2.6 1.6 2 3\\
2.6 2.4 2 3\\
2.6 2.4 0 3\\
2.6 1.6 0 3\\
NaN NaN NaN 3\\
2.6 2.6 0 3\\
2.6 2.6 2 3\\
2.6 3.4 2 3\\
2.6 3.4 0 3\\
2.6 2.6 0 3\\
NaN NaN NaN 3\\
2.6 3.6 0 3\\
2.6 3.6 3 3\\
2.6 4.4 3 3\\
2.6 4.4 0 3\\
2.6 3.6 0 3\\
NaN NaN NaN 3\\
2.6 4.6 0 3\\
2.6 4.6 3 3\\
2.6 5.4 3 3\\
2.6 5.4 0 3\\
2.6 4.6 0 3\\
NaN NaN NaN 3\\
2.6 5.6 0 3\\
2.6 5.6 3 3\\
2.6 6.4 3 3\\
2.6 6.4 0 3\\
2.6 5.6 0 3\\
NaN NaN NaN 3\\
2.6 6.6 0 3\\
2.6 6.6 3 3\\
2.6 7.4 3 3\\
2.6 7.4 0 3\\
2.6 6.6 0 3\\
NaN NaN NaN 3\\
2.6 7.6 0 3\\
2.6 7.6 3 3\\
2.6 8.4 3 3\\
2.6 8.4 0 3\\
2.6 7.6 0 3\\
NaN NaN NaN 3\\
2.6 8.6 0 3\\
2.6 8.6 3 3\\
2.6 9.4 3 3\\
2.6 9.4 0 3\\
2.6 8.6 0 3\\
NaN NaN NaN 3\\
2.6 9.6 0 3\\
2.6 9.6 3 3\\
2.6 10.4 3 3\\
2.6 10.4 0 3\\
2.6 9.6 0 3\\
NaN NaN NaN 3\\
2.6 10.6 0 3\\
2.6 10.6 4 3\\
2.6 11.4 4 3\\
2.6 11.4 0 3\\
2.6 10.6 0 3\\
NaN NaN NaN 3\\
3.4 0.6 0 3\\
3.4 0.6 0 3\\
3.4 1.4 0 3\\
3.4 1.4 0 3\\
3.4 0.6 0 3\\
NaN NaN NaN 3\\
3.4 1.6 0 3\\
3.4 1.6 2 3\\
3.4 2.4 2 3\\
3.4 2.4 0 3\\
3.4 1.6 0 3\\
NaN NaN NaN 3\\
3.4 2.6 0 3\\
3.4 2.6 2 3\\
3.4 3.4 2 3\\
3.4 3.4 0 3\\
3.4 2.6 0 3\\
NaN NaN NaN 3\\
3.4 3.6 0 3\\
3.4 3.6 3 3\\
3.4 4.4 3 3\\
3.4 4.4 0 3\\
3.4 3.6 0 3\\
NaN NaN NaN 3\\
3.4 4.6 0 3\\
3.4 4.6 3 3\\
3.4 5.4 3 3\\
3.4 5.4 0 3\\
3.4 4.6 0 3\\
NaN NaN NaN 3\\
3.4 5.6 0 3\\
3.4 5.6 3 3\\
3.4 6.4 3 3\\
3.4 6.4 0 3\\
3.4 5.6 0 3\\
NaN NaN NaN 3\\
3.4 6.6 0 3\\
3.4 6.6 3 3\\
3.4 7.4 3 3\\
3.4 7.4 0 3\\
3.4 6.6 0 3\\
NaN NaN NaN 3\\
3.4 7.6 0 3\\
3.4 7.6 3 3\\
3.4 8.4 3 3\\
3.4 8.4 0 3\\
3.4 7.6 0 3\\
NaN NaN NaN 3\\
3.4 8.6 0 3\\
3.4 8.6 3 3\\
3.4 9.4 3 3\\
3.4 9.4 0 3\\
3.4 8.6 0 3\\
NaN NaN NaN 3\\
3.4 9.6 0 3\\
3.4 9.6 3 3\\
3.4 10.4 3 3\\
3.4 10.4 0 3\\
3.4 9.6 0 3\\
NaN NaN NaN 3\\
3.4 10.6 0 3\\
3.4 10.6 4 3\\
3.4 11.4 4 3\\
3.4 11.4 0 3\\
3.4 10.6 0 3\\
NaN NaN NaN 3\\
NaN NaN NaN 3\\
3.4 0.6 0 3\\
3.4 1.4 0 3\\
NaN NaN NaN 3\\
NaN NaN NaN 3\\
NaN NaN NaN 3\\
NaN NaN NaN 3\\
3.4 1.6 0 3\\
3.4 2.4 0 3\\
NaN NaN NaN 3\\
NaN NaN NaN 3\\
NaN NaN NaN 3\\
NaN NaN NaN 3\\
3.4 2.6 0 3\\
3.4 3.4 0 3\\
NaN NaN NaN 3\\
NaN NaN NaN 3\\
NaN NaN NaN 3\\
NaN NaN NaN 3\\
3.4 3.6 0 3\\
3.4 4.4 0 3\\
NaN NaN NaN 3\\
NaN NaN NaN 3\\
NaN NaN NaN 3\\
NaN NaN NaN 3\\
3.4 4.6 0 3\\
3.4 5.4 0 3\\
NaN NaN NaN 3\\
NaN NaN NaN 3\\
NaN NaN NaN 3\\
NaN NaN NaN 3\\
3.4 5.6 0 3\\
3.4 6.4 0 3\\
NaN NaN NaN 3\\
NaN NaN NaN 3\\
NaN NaN NaN 3\\
NaN NaN NaN 3\\
3.4 6.6 0 3\\
3.4 7.4 0 3\\
NaN NaN NaN 3\\
NaN NaN NaN 3\\
NaN NaN NaN 3\\
NaN NaN NaN 3\\
3.4 7.6 0 3\\
3.4 8.4 0 3\\
NaN NaN NaN 3\\
NaN NaN NaN 3\\
NaN NaN NaN 3\\
NaN NaN NaN 3\\
3.4 8.6 0 3\\
3.4 9.4 0 3\\
NaN NaN NaN 3\\
NaN NaN NaN 3\\
NaN NaN NaN 3\\
NaN NaN NaN 3\\
3.4 9.6 0 3\\
3.4 10.4 0 3\\
NaN NaN NaN 3\\
NaN NaN NaN 3\\
NaN NaN NaN 3\\
NaN NaN NaN 3\\
3.4 10.6 0 3\\
3.4 11.4 0 3\\
NaN NaN NaN 3\\
NaN NaN NaN 3\\
NaN NaN NaN 3\\
};

\addplot3[%
surf,
shader=flat,
draw=black,
point meta=explicit,
mesh/rows=4]
table[row sep=crcr,header=false,meta index=3] {
NaN NaN NaN 2\\
1.6 0.6 0 2\\
1.6 1.4 0 2\\
NaN NaN NaN 2\\
NaN NaN NaN 2\\
NaN NaN NaN 2\\
NaN NaN NaN 2\\
1.6 1.6 0 2\\
1.6 2.4 0 2\\
NaN NaN NaN 2\\
NaN NaN NaN 2\\
NaN NaN NaN 2\\
NaN NaN NaN 2\\
1.6 2.6 0 2\\
1.6 3.4 0 2\\
NaN NaN NaN 2\\
NaN NaN NaN 2\\
NaN NaN NaN 2\\
NaN NaN NaN 2\\
1.6 3.6 0 2\\
1.6 4.4 0 2\\
NaN NaN NaN 2\\
NaN NaN NaN 2\\
NaN NaN NaN 2\\
NaN NaN NaN 2\\
1.6 4.6 0 2\\
1.6 5.4 0 2\\
NaN NaN NaN 2\\
NaN NaN NaN 2\\
NaN NaN NaN 2\\
NaN NaN NaN 2\\
1.6 5.6 0 2\\
1.6 6.4 0 2\\
NaN NaN NaN 2\\
NaN NaN NaN 2\\
NaN NaN NaN 2\\
NaN NaN NaN 2\\
1.6 6.6 0 2\\
1.6 7.4 0 2\\
NaN NaN NaN 2\\
NaN NaN NaN 2\\
NaN NaN NaN 2\\
NaN NaN NaN 2\\
1.6 7.6 0 2\\
1.6 8.4 0 2\\
NaN NaN NaN 2\\
NaN NaN NaN 2\\
NaN NaN NaN 2\\
NaN NaN NaN 2\\
1.6 8.6 0 2\\
1.6 9.4 0 2\\
NaN NaN NaN 2\\
NaN NaN NaN 2\\
NaN NaN NaN 2\\
NaN NaN NaN 2\\
1.6 9.6 0 2\\
1.6 10.4 0 2\\
NaN NaN NaN 2\\
NaN NaN NaN 2\\
NaN NaN NaN 2\\
NaN NaN NaN 2\\
1.6 10.6 0 2\\
1.6 11.4 0 2\\
NaN NaN NaN 2\\
NaN NaN NaN 2\\
NaN NaN NaN 2\\
1.6 0.6 0 2\\
1.6 0.6 0 2\\
1.6 1.4 0 2\\
1.6 1.4 0 2\\
1.6 0.6 0 2\\
NaN NaN NaN 2\\
1.6 1.6 0 2\\
1.6 1.6 2 2\\
1.6 2.4 2 2\\
1.6 2.4 0 2\\
1.6 1.6 0 2\\
NaN NaN NaN 2\\
1.6 2.6 0 2\\
1.6 2.6 2 2\\
1.6 3.4 2 2\\
1.6 3.4 0 2\\
1.6 2.6 0 2\\
NaN NaN NaN 2\\
1.6 3.6 0 2\\
1.6 3.6 3 2\\
1.6 4.4 3 2\\
1.6 4.4 0 2\\
1.6 3.6 0 2\\
NaN NaN NaN 2\\
1.6 4.6 0 2\\
1.6 4.6 3 2\\
1.6 5.4 3 2\\
1.6 5.4 0 2\\
1.6 4.6 0 2\\
NaN NaN NaN 2\\
1.6 5.6 0 2\\
1.6 5.6 3 2\\
1.6 6.4 3 2\\
1.6 6.4 0 2\\
1.6 5.6 0 2\\
NaN NaN NaN 2\\
1.6 6.6 0 2\\
1.6 6.6 3 2\\
1.6 7.4 3 2\\
1.6 7.4 0 2\\
1.6 6.6 0 2\\
NaN NaN NaN 2\\
1.6 7.6 0 2\\
1.6 7.6 3 2\\
1.6 8.4 3 2\\
1.6 8.4 0 2\\
1.6 7.6 0 2\\
NaN NaN NaN 2\\
1.6 8.6 0 2\\
1.6 8.6 3 2\\
1.6 9.4 3 2\\
1.6 9.4 0 2\\
1.6 8.6 0 2\\
NaN NaN NaN 2\\
1.6 9.6 0 2\\
1.6 9.6 3 2\\
1.6 10.4 3 2\\
1.6 10.4 0 2\\
1.6 9.6 0 2\\
NaN NaN NaN 2\\
1.6 10.6 0 2\\
1.6 10.6 4 2\\
1.6 11.4 4 2\\
1.6 11.4 0 2\\
1.6 10.6 0 2\\
NaN NaN NaN 2\\
2.4 0.6 0 2\\
2.4 0.6 0 2\\
2.4 1.4 0 2\\
2.4 1.4 0 2\\
2.4 0.6 0 2\\
NaN NaN NaN 2\\
2.4 1.6 0 2\\
2.4 1.6 2 2\\
2.4 2.4 2 2\\
2.4 2.4 0 2\\
2.4 1.6 0 2\\
NaN NaN NaN 2\\
2.4 2.6 0 2\\
2.4 2.6 2 2\\
2.4 3.4 2 2\\
2.4 3.4 0 2\\
2.4 2.6 0 2\\
NaN NaN NaN 2\\
2.4 3.6 0 2\\
2.4 3.6 3 2\\
2.4 4.4 3 2\\
2.4 4.4 0 2\\
2.4 3.6 0 2\\
NaN NaN NaN 2\\
2.4 4.6 0 2\\
2.4 4.6 3 2\\
2.4 5.4 3 2\\
2.4 5.4 0 2\\
2.4 4.6 0 2\\
NaN NaN NaN 2\\
2.4 5.6 0 2\\
2.4 5.6 3 2\\
2.4 6.4 3 2\\
2.4 6.4 0 2\\
2.4 5.6 0 2\\
NaN NaN NaN 2\\
2.4 6.6 0 2\\
2.4 6.6 3 2\\
2.4 7.4 3 2\\
2.4 7.4 0 2\\
2.4 6.6 0 2\\
NaN NaN NaN 2\\
2.4 7.6 0 2\\
2.4 7.6 3 2\\
2.4 8.4 3 2\\
2.4 8.4 0 2\\
2.4 7.6 0 2\\
NaN NaN NaN 2\\
2.4 8.6 0 2\\
2.4 8.6 3 2\\
2.4 9.4 3 2\\
2.4 9.4 0 2\\
2.4 8.6 0 2\\
NaN NaN NaN 2\\
2.4 9.6 0 2\\
2.4 9.6 3 2\\
2.4 10.4 3 2\\
2.4 10.4 0 2\\
2.4 9.6 0 2\\
NaN NaN NaN 2\\
2.4 10.6 0 2\\
2.4 10.6 4 2\\
2.4 11.4 4 2\\
2.4 11.4 0 2\\
2.4 10.6 0 2\\
NaN NaN NaN 2\\
NaN NaN NaN 2\\
2.4 0.6 0 2\\
2.4 1.4 0 2\\
NaN NaN NaN 2\\
NaN NaN NaN 2\\
NaN NaN NaN 2\\
NaN NaN NaN 2\\
2.4 1.6 0 2\\
2.4 2.4 0 2\\
NaN NaN NaN 2\\
NaN NaN NaN 2\\
NaN NaN NaN 2\\
NaN NaN NaN 2\\
2.4 2.6 0 2\\
2.4 3.4 0 2\\
NaN NaN NaN 2\\
NaN NaN NaN 2\\
NaN NaN NaN 2\\
NaN NaN NaN 2\\
2.4 3.6 0 2\\
2.4 4.4 0 2\\
NaN NaN NaN 2\\
NaN NaN NaN 2\\
NaN NaN NaN 2\\
NaN NaN NaN 2\\
2.4 4.6 0 2\\
2.4 5.4 0 2\\
NaN NaN NaN 2\\
NaN NaN NaN 2\\
NaN NaN NaN 2\\
NaN NaN NaN 2\\
2.4 5.6 0 2\\
2.4 6.4 0 2\\
NaN NaN NaN 2\\
NaN NaN NaN 2\\
NaN NaN NaN 2\\
NaN NaN NaN 2\\
2.4 6.6 0 2\\
2.4 7.4 0 2\\
NaN NaN NaN 2\\
NaN NaN NaN 2\\
NaN NaN NaN 2\\
NaN NaN NaN 2\\
2.4 7.6 0 2\\
2.4 8.4 0 2\\
NaN NaN NaN 2\\
NaN NaN NaN 2\\
NaN NaN NaN 2\\
NaN NaN NaN 2\\
2.4 8.6 0 2\\
2.4 9.4 0 2\\
NaN NaN NaN 2\\
NaN NaN NaN 2\\
NaN NaN NaN 2\\
NaN NaN NaN 2\\
2.4 9.6 0 2\\
2.4 10.4 0 2\\
NaN NaN NaN 2\\
NaN NaN NaN 2\\
NaN NaN NaN 2\\
NaN NaN NaN 2\\
2.4 10.6 0 2\\
2.4 11.4 0 2\\
NaN NaN NaN 2\\
NaN NaN NaN 2\\
NaN NaN NaN 2\\
};

\addplot3[%
surf,
shader=flat,
draw=black,
point meta=explicit,
mesh/rows=4]
table[row sep=crcr,header=false,meta index=3] {
NaN NaN NaN 1\\
0.6 0.6 0 1\\
0.6 1.4 0 1\\
NaN NaN NaN 1\\
NaN NaN NaN 1\\
NaN NaN NaN 1\\
NaN NaN NaN 1\\
0.6 1.6 0 1\\
0.6 2.4 0 1\\
NaN NaN NaN 1\\
NaN NaN NaN 1\\
NaN NaN NaN 1\\
NaN NaN NaN 1\\
0.6 2.6 0 1\\
0.6 3.4 0 1\\
NaN NaN NaN 1\\
NaN NaN NaN 1\\
NaN NaN NaN 1\\
NaN NaN NaN 1\\
0.6 3.6 0 1\\
0.6 4.4 0 1\\
NaN NaN NaN 1\\
NaN NaN NaN 1\\
NaN NaN NaN 1\\
NaN NaN NaN 1\\
0.6 4.6 0 1\\
0.6 5.4 0 1\\
NaN NaN NaN 1\\
NaN NaN NaN 1\\
NaN NaN NaN 1\\
NaN NaN NaN 1\\
0.6 5.6 0 1\\
0.6 6.4 0 1\\
NaN NaN NaN 1\\
NaN NaN NaN 1\\
NaN NaN NaN 1\\
NaN NaN NaN 1\\
0.6 6.6 0 1\\
0.6 7.4 0 1\\
NaN NaN NaN 1\\
NaN NaN NaN 1\\
NaN NaN NaN 1\\
NaN NaN NaN 1\\
0.6 7.6 0 1\\
0.6 8.4 0 1\\
NaN NaN NaN 1\\
NaN NaN NaN 1\\
NaN NaN NaN 1\\
NaN NaN NaN 1\\
0.6 8.6 0 1\\
0.6 9.4 0 1\\
NaN NaN NaN 1\\
NaN NaN NaN 1\\
NaN NaN NaN 1\\
NaN NaN NaN 1\\
0.6 9.6 0 1\\
0.6 10.4 0 1\\
NaN NaN NaN 1\\
NaN NaN NaN 1\\
NaN NaN NaN 1\\
NaN NaN NaN 1\\
0.6 10.6 0 1\\
0.6 11.4 0 1\\
NaN NaN NaN 1\\
NaN NaN NaN 1\\
NaN NaN NaN 1\\
0.6 0.6 0 1\\
0.6 0.6 0 1\\
0.6 1.4 0 1\\
0.6 1.4 0 1\\
0.6 0.6 0 1\\
NaN NaN NaN 1\\
0.6 1.6 0 1\\
0.6 1.6 1 1\\
0.6 2.4 1 1\\
0.6 2.4 0 1\\
0.6 1.6 0 1\\
NaN NaN NaN 1\\
0.6 2.6 0 1\\
0.6 2.6 1 1\\
0.6 3.4 1 1\\
0.6 3.4 0 1\\
0.6 2.6 0 1\\
NaN NaN NaN 1\\
0.6 3.6 0 1\\
0.6 3.6 2 1\\
0.6 4.4 2 1\\
0.6 4.4 0 1\\
0.6 3.6 0 1\\
NaN NaN NaN 1\\
0.6 4.6 0 1\\
0.6 4.6 2 1\\
0.6 5.4 2 1\\
0.6 5.4 0 1\\
0.6 4.6 0 1\\
NaN NaN NaN 1\\
0.6 5.6 0 1\\
0.6 5.6 2 1\\
0.6 6.4 2 1\\
0.6 6.4 0 1\\
0.6 5.6 0 1\\
NaN NaN NaN 1\\
0.6 6.6 0 1\\
0.6 6.6 2 1\\
0.6 7.4 2 1\\
0.6 7.4 0 1\\
0.6 6.6 0 1\\
NaN NaN NaN 1\\
0.6 7.6 0 1\\
0.6 7.6 2 1\\
0.6 8.4 2 1\\
0.6 8.4 0 1\\
0.6 7.6 0 1\\
NaN NaN NaN 1\\
0.6 8.6 0 1\\
0.6 8.6 2 1\\
0.6 9.4 2 1\\
0.6 9.4 0 1\\
0.6 8.6 0 1\\
NaN NaN NaN 1\\
0.6 9.6 0 1\\
0.6 9.6 3 1\\
0.6 10.4 3 1\\
0.6 10.4 0 1\\
0.6 9.6 0 1\\
NaN NaN NaN 1\\
0.6 10.6 0 1\\
0.6 10.6 3 1\\
0.6 11.4 3 1\\
0.6 11.4 0 1\\
0.6 10.6 0 1\\
NaN NaN NaN 1\\
1.4 0.6 0 1\\
1.4 0.6 0 1\\
1.4 1.4 0 1\\
1.4 1.4 0 1\\
1.4 0.6 0 1\\
NaN NaN NaN 1\\
1.4 1.6 0 1\\
1.4 1.6 1 1\\
1.4 2.4 1 1\\
1.4 2.4 0 1\\
1.4 1.6 0 1\\
NaN NaN NaN 1\\
1.4 2.6 0 1\\
1.4 2.6 1 1\\
1.4 3.4 1 1\\
1.4 3.4 0 1\\
1.4 2.6 0 1\\
NaN NaN NaN 1\\
1.4 3.6 0 1\\
1.4 3.6 2 1\\
1.4 4.4 2 1\\
1.4 4.4 0 1\\
1.4 3.6 0 1\\
NaN NaN NaN 1\\
1.4 4.6 0 1\\
1.4 4.6 2 1\\
1.4 5.4 2 1\\
1.4 5.4 0 1\\
1.4 4.6 0 1\\
NaN NaN NaN 1\\
1.4 5.6 0 1\\
1.4 5.6 2 1\\
1.4 6.4 2 1\\
1.4 6.4 0 1\\
1.4 5.6 0 1\\
NaN NaN NaN 1\\
1.4 6.6 0 1\\
1.4 6.6 2 1\\
1.4 7.4 2 1\\
1.4 7.4 0 1\\
1.4 6.6 0 1\\
NaN NaN NaN 1\\
1.4 7.6 0 1\\
1.4 7.6 2 1\\
1.4 8.4 2 1\\
1.4 8.4 0 1\\
1.4 7.6 0 1\\
NaN NaN NaN 1\\
1.4 8.6 0 1\\
1.4 8.6 2 1\\
1.4 9.4 2 1\\
1.4 9.4 0 1\\
1.4 8.6 0 1\\
NaN NaN NaN 1\\
1.4 9.6 0 1\\
1.4 9.6 3 1\\
1.4 10.4 3 1\\
1.4 10.4 0 1\\
1.4 9.6 0 1\\
NaN NaN NaN 1\\
1.4 10.6 0 1\\
1.4 10.6 3 1\\
1.4 11.4 3 1\\
1.4 11.4 0 1\\
1.4 10.6 0 1\\
NaN NaN NaN 1\\
NaN NaN NaN 1\\
1.4 0.6 0 1\\
1.4 1.4 0 1\\
NaN NaN NaN 1\\
NaN NaN NaN 1\\
NaN NaN NaN 1\\
NaN NaN NaN 1\\
1.4 1.6 0 1\\
1.4 2.4 0 1\\
NaN NaN NaN 1\\
NaN NaN NaN 1\\
NaN NaN NaN 1\\
NaN NaN NaN 1\\
1.4 2.6 0 1\\
1.4 3.4 0 1\\
NaN NaN NaN 1\\
NaN NaN NaN 1\\
NaN NaN NaN 1\\
NaN NaN NaN 1\\
1.4 3.6 0 1\\
1.4 4.4 0 1\\
NaN NaN NaN 1\\
NaN NaN NaN 1\\
NaN NaN NaN 1\\
NaN NaN NaN 1\\
1.4 4.6 0 1\\
1.4 5.4 0 1\\
NaN NaN NaN 1\\
NaN NaN NaN 1\\
NaN NaN NaN 1\\
NaN NaN NaN 1\\
1.4 5.6 0 1\\
1.4 6.4 0 1\\
NaN NaN NaN 1\\
NaN NaN NaN 1\\
NaN NaN NaN 1\\
NaN NaN NaN 1\\
1.4 6.6 0 1\\
1.4 7.4 0 1\\
NaN NaN NaN 1\\
NaN NaN NaN 1\\
NaN NaN NaN 1\\
NaN NaN NaN 1\\
1.4 7.6 0 1\\
1.4 8.4 0 1\\
NaN NaN NaN 1\\
NaN NaN NaN 1\\
NaN NaN NaN 1\\
NaN NaN NaN 1\\
1.4 8.6 0 1\\
1.4 9.4 0 1\\
NaN NaN NaN 1\\
NaN NaN NaN 1\\
NaN NaN NaN 1\\
NaN NaN NaN 1\\
1.4 9.6 0 1\\
1.4 10.4 0 1\\
NaN NaN NaN 1\\
NaN NaN NaN 1\\
NaN NaN NaN 1\\
NaN NaN NaN 1\\
1.4 10.6 0 1\\
1.4 11.4 0 1\\
NaN NaN NaN 1\\
NaN NaN NaN 1\\
NaN NaN NaN 1\\
};
\end{axis}
\end{tikzpicture}%

%% file: Policy/NonMono1.tex
%
%
\begin{tikzpicture}

\begin{axis}[%
width=2.5in,
height=2.2in,
unbounded coords=jump,
view={-130}{40},
scale only axis,
xmin=0.5,
xmax=11.5,
xtick={1,2,3,4,5,6,7,8,9,10,11},
xticklabels={0.1,1,2,3,4,5,6,7,8,9,10},
xlabel={$\gamma_2$},
xmajorgrids,
y dir=reverse,
ymin=0.5,
ymax=11.5,
ytick={1,2,3,4,5,6,7,8,9,10,11},
yticklabels={0.1,1,2,3,4,5,6,7,8,9,10},
ylabel={$\gamma_1$},
ymajorgrids,
zmin=0,
zmax=9,
zlabel={$\underline{\theta}_1^*$},
zmajorgrids,
name=plot1,
axis x line*=bottom,
axis y line*=left,
axis z line*=left
]

\addplot3[%
surf,
shader=flat,
draw=black,
point meta=explicit,
mesh/rows=4]
table[row sep=crcr,header=false,meta index=3] {
NaN NaN NaN 11\\
10.6 0.6 0 11\\
10.6 1.4 0 11\\
NaN NaN NaN 11\\
NaN NaN NaN 11\\
NaN NaN NaN 11\\
NaN NaN NaN 11\\
10.6 1.6 0 11\\
10.6 2.4 0 11\\
NaN NaN NaN 11\\
NaN NaN NaN 11\\
NaN NaN NaN 11\\
NaN NaN NaN 11\\
10.6 2.6 0 11\\
10.6 3.4 0 11\\
NaN NaN NaN 11\\
NaN NaN NaN 11\\
NaN NaN NaN 11\\
NaN NaN NaN 11\\
10.6 3.6 0 11\\
10.6 4.4 0 11\\
NaN NaN NaN 11\\
NaN NaN NaN 11\\
NaN NaN NaN 11\\
NaN NaN NaN 11\\
10.6 4.6 0 11\\
10.6 5.4 0 11\\
NaN NaN NaN 11\\
NaN NaN NaN 11\\
NaN NaN NaN 11\\
NaN NaN NaN 11\\
10.6 5.6 0 11\\
10.6 6.4 0 11\\
NaN NaN NaN 11\\
NaN NaN NaN 11\\
NaN NaN NaN 11\\
NaN NaN NaN 11\\
10.6 6.6 0 11\\
10.6 7.4 0 11\\
NaN NaN NaN 11\\
NaN NaN NaN 11\\
NaN NaN NaN 11\\
NaN NaN NaN 11\\
10.6 7.6 0 11\\
10.6 8.4 0 11\\
NaN NaN NaN 11\\
NaN NaN NaN 11\\
NaN NaN NaN 11\\
NaN NaN NaN 11\\
10.6 8.6 0 11\\
10.6 9.4 0 11\\
NaN NaN NaN 11\\
NaN NaN NaN 11\\
NaN NaN NaN 11\\
NaN NaN NaN 11\\
10.6 9.6 0 11\\
10.6 10.4 0 11\\
NaN NaN NaN 11\\
NaN NaN NaN 11\\
NaN NaN NaN 11\\
NaN NaN NaN 11\\
10.6 10.6 0 11\\
10.6 11.4 0 11\\
NaN NaN NaN 11\\
NaN NaN NaN 11\\
NaN NaN NaN 11\\
10.6 0.6 0 11\\
10.6 0.6 9 11\\
10.6 1.4 9 11\\
10.6 1.4 0 11\\
10.6 0.6 0 11\\
NaN NaN NaN 11\\
10.6 1.6 0 11\\
10.6 1.6 1 11\\
10.6 2.4 1 11\\
10.6 2.4 0 11\\
10.6 1.6 0 11\\
NaN NaN NaN 11\\
10.6 2.6 0 11\\
10.6 2.6 1 11\\
10.6 3.4 1 11\\
10.6 3.4 0 11\\
10.6 2.6 0 11\\
NaN NaN NaN 11\\
10.6 3.6 0 11\\
10.6 3.6 2 11\\
10.6 4.4 2 11\\
10.6 4.4 0 11\\
10.6 3.6 0 11\\
NaN NaN NaN 11\\
10.6 4.6 0 11\\
10.6 4.6 2 11\\
10.6 5.4 2 11\\
10.6 5.4 0 11\\
10.6 4.6 0 11\\
NaN NaN NaN 11\\
10.6 5.6 0 11\\
10.6 5.6 2 11\\
10.6 6.4 2 11\\
10.6 6.4 0 11\\
10.6 5.6 0 11\\
NaN NaN NaN 11\\
10.6 6.6 0 11\\
10.6 6.6 2 11\\
10.6 7.4 2 11\\
10.6 7.4 0 11\\
10.6 6.6 0 11\\
NaN NaN NaN 11\\
10.6 7.6 0 11\\
10.6 7.6 2 11\\
10.6 8.4 2 11\\
10.6 8.4 0 11\\
10.6 7.6 0 11\\
NaN NaN NaN 11\\
10.6 8.6 0 11\\
10.6 8.6 3 11\\
10.6 9.4 3 11\\
10.6 9.4 0 11\\
10.6 8.6 0 11\\
NaN NaN NaN 11\\
10.6 9.6 0 11\\
10.6 9.6 3 11\\
10.6 10.4 3 11\\
10.6 10.4 0 11\\
10.6 9.6 0 11\\
NaN NaN NaN 11\\
10.6 10.6 0 11\\
10.6 10.6 3 11\\
10.6 11.4 3 11\\
10.6 11.4 0 11\\
10.6 10.6 0 11\\
NaN NaN NaN 11\\
11.4 0.6 0 11\\
11.4 0.6 9 11\\
11.4 1.4 9 11\\
11.4 1.4 0 11\\
11.4 0.6 0 11\\
NaN NaN NaN 11\\
11.4 1.6 0 11\\
11.4 1.6 1 11\\
11.4 2.4 1 11\\
11.4 2.4 0 11\\
11.4 1.6 0 11\\
NaN NaN NaN 11\\
11.4 2.6 0 11\\
11.4 2.6 1 11\\
11.4 3.4 1 11\\
11.4 3.4 0 11\\
11.4 2.6 0 11\\
NaN NaN NaN 11\\
11.4 3.6 0 11\\
11.4 3.6 2 11\\
11.4 4.4 2 11\\
11.4 4.4 0 11\\
11.4 3.6 0 11\\
NaN NaN NaN 11\\
11.4 4.6 0 11\\
11.4 4.6 2 11\\
11.4 5.4 2 11\\
11.4 5.4 0 11\\
11.4 4.6 0 11\\
NaN NaN NaN 11\\
11.4 5.6 0 11\\
11.4 5.6 2 11\\
11.4 6.4 2 11\\
11.4 6.4 0 11\\
11.4 5.6 0 11\\
NaN NaN NaN 11\\
11.4 6.6 0 11\\
11.4 6.6 2 11\\
11.4 7.4 2 11\\
11.4 7.4 0 11\\
11.4 6.6 0 11\\
NaN NaN NaN 11\\
11.4 7.6 0 11\\
11.4 7.6 2 11\\
11.4 8.4 2 11\\
11.4 8.4 0 11\\
11.4 7.6 0 11\\
NaN NaN NaN 11\\
11.4 8.6 0 11\\
11.4 8.6 3 11\\
11.4 9.4 3 11\\
11.4 9.4 0 11\\
11.4 8.6 0 11\\
NaN NaN NaN 11\\
11.4 9.6 0 11\\
11.4 9.6 3 11\\
11.4 10.4 3 11\\
11.4 10.4 0 11\\
11.4 9.6 0 11\\
NaN NaN NaN 11\\
11.4 10.6 0 11\\
11.4 10.6 3 11\\
11.4 11.4 3 11\\
11.4 11.4 0 11\\
11.4 10.6 0 11\\
NaN NaN NaN 11\\
NaN NaN NaN 11\\
11.4 0.6 0 11\\
11.4 1.4 0 11\\
NaN NaN NaN 11\\
NaN NaN NaN 11\\
NaN NaN NaN 11\\
NaN NaN NaN 11\\
11.4 1.6 0 11\\
11.4 2.4 0 11\\
NaN NaN NaN 11\\
NaN NaN NaN 11\\
NaN NaN NaN 11\\
NaN NaN NaN 11\\
11.4 2.6 0 11\\
11.4 3.4 0 11\\
NaN NaN NaN 11\\
NaN NaN NaN 11\\
NaN NaN NaN 11\\
NaN NaN NaN 11\\
11.4 3.6 0 11\\
11.4 4.4 0 11\\
NaN NaN NaN 11\\
NaN NaN NaN 11\\
NaN NaN NaN 11\\
NaN NaN NaN 11\\
11.4 4.6 0 11\\
11.4 5.4 0 11\\
NaN NaN NaN 11\\
NaN NaN NaN 11\\
NaN NaN NaN 11\\
NaN NaN NaN 11\\
11.4 5.6 0 11\\
11.4 6.4 0 11\\
NaN NaN NaN 11\\
NaN NaN NaN 11\\
NaN NaN NaN 11\\
NaN NaN NaN 11\\
11.4 6.6 0 11\\
11.4 7.4 0 11\\
NaN NaN NaN 11\\
NaN NaN NaN 11\\
NaN NaN NaN 11\\
NaN NaN NaN 11\\
11.4 7.6 0 11\\
11.4 8.4 0 11\\
NaN NaN NaN 11\\
NaN NaN NaN 11\\
NaN NaN NaN 11\\
NaN NaN NaN 11\\
11.4 8.6 0 11\\
11.4 9.4 0 11\\
NaN NaN NaN 11\\
NaN NaN NaN 11\\
NaN NaN NaN 11\\
NaN NaN NaN 11\\
11.4 9.6 0 11\\
11.4 10.4 0 11\\
NaN NaN NaN 11\\
NaN NaN NaN 11\\
NaN NaN NaN 11\\
NaN NaN NaN 11\\
11.4 10.6 0 11\\
11.4 11.4 0 11\\
NaN NaN NaN 11\\
NaN NaN NaN 11\\
NaN NaN NaN 11\\
};

\addplot3[%
surf,
shader=flat,
draw=black,
point meta=explicit,
mesh/rows=4]
table[row sep=crcr,header=false,meta index=3] {
NaN NaN NaN 10\\
9.6 0.6 0 10\\
9.6 1.4 0 10\\
NaN NaN NaN 10\\
NaN NaN NaN 10\\
NaN NaN NaN 10\\
NaN NaN NaN 10\\
9.6 1.6 0 10\\
9.6 2.4 0 10\\
NaN NaN NaN 10\\
NaN NaN NaN 10\\
NaN NaN NaN 10\\
NaN NaN NaN 10\\
9.6 2.6 0 10\\
9.6 3.4 0 10\\
NaN NaN NaN 10\\
NaN NaN NaN 10\\
NaN NaN NaN 10\\
NaN NaN NaN 10\\
9.6 3.6 0 10\\
9.6 4.4 0 10\\
NaN NaN NaN 10\\
NaN NaN NaN 10\\
NaN NaN NaN 10\\
NaN NaN NaN 10\\
9.6 4.6 0 10\\
9.6 5.4 0 10\\
NaN NaN NaN 10\\
NaN NaN NaN 10\\
NaN NaN NaN 10\\
NaN NaN NaN 10\\
9.6 5.6 0 10\\
9.6 6.4 0 10\\
NaN NaN NaN 10\\
NaN NaN NaN 10\\
NaN NaN NaN 10\\
NaN NaN NaN 10\\
9.6 6.6 0 10\\
9.6 7.4 0 10\\
NaN NaN NaN 10\\
NaN NaN NaN 10\\
NaN NaN NaN 10\\
NaN NaN NaN 10\\
9.6 7.6 0 10\\
9.6 8.4 0 10\\
NaN NaN NaN 10\\
NaN NaN NaN 10\\
NaN NaN NaN 10\\
NaN NaN NaN 10\\
9.6 8.6 0 10\\
9.6 9.4 0 10\\
NaN NaN NaN 10\\
NaN NaN NaN 10\\
NaN NaN NaN 10\\
NaN NaN NaN 10\\
9.6 9.6 0 10\\
9.6 10.4 0 10\\
NaN NaN NaN 10\\
NaN NaN NaN 10\\
NaN NaN NaN 10\\
NaN NaN NaN 10\\
9.6 10.6 0 10\\
9.6 11.4 0 10\\
NaN NaN NaN 10\\
NaN NaN NaN 10\\
NaN NaN NaN 10\\
9.6 0.6 0 10\\
9.6 0.6 9 10\\
9.6 1.4 9 10\\
9.6 1.4 0 10\\
9.6 0.6 0 10\\
NaN NaN NaN 10\\
9.6 1.6 0 10\\
9.6 1.6 1 10\\
9.6 2.4 1 10\\
9.6 2.4 0 10\\
9.6 1.6 0 10\\
NaN NaN NaN 10\\
9.6 2.6 0 10\\
9.6 2.6 1 10\\
9.6 3.4 1 10\\
9.6 3.4 0 10\\
9.6 2.6 0 10\\
NaN NaN NaN 10\\
9.6 3.6 0 10\\
9.6 3.6 2 10\\
9.6 4.4 2 10\\
9.6 4.4 0 10\\
9.6 3.6 0 10\\
NaN NaN NaN 10\\
9.6 4.6 0 10\\
9.6 4.6 2 10\\
9.6 5.4 2 10\\
9.6 5.4 0 10\\
9.6 4.6 0 10\\
NaN NaN NaN 10\\
9.6 5.6 0 10\\
9.6 5.6 2 10\\
9.6 6.4 2 10\\
9.6 6.4 0 10\\
9.6 5.6 0 10\\
NaN NaN NaN 10\\
9.6 6.6 0 10\\
9.6 6.6 2 10\\
9.6 7.4 2 10\\
9.6 7.4 0 10\\
9.6 6.6 0 10\\
NaN NaN NaN 10\\
9.6 7.6 0 10\\
9.6 7.6 2 10\\
9.6 8.4 2 10\\
9.6 8.4 0 10\\
9.6 7.6 0 10\\
NaN NaN NaN 10\\
9.6 8.6 0 10\\
9.6 8.6 2 10\\
9.6 9.4 2 10\\
9.6 9.4 0 10\\
9.6 8.6 0 10\\
NaN NaN NaN 10\\
9.6 9.6 0 10\\
9.6 9.6 2 10\\
9.6 10.4 2 10\\
9.6 10.4 0 10\\
9.6 9.6 0 10\\
NaN NaN NaN 10\\
9.6 10.6 0 10\\
9.6 10.6 3 10\\
9.6 11.4 3 10\\
9.6 11.4 0 10\\
9.6 10.6 0 10\\
NaN NaN NaN 10\\
10.4 0.6 0 10\\
10.4 0.6 9 10\\
10.4 1.4 9 10\\
10.4 1.4 0 10\\
10.4 0.6 0 10\\
NaN NaN NaN 10\\
10.4 1.6 0 10\\
10.4 1.6 1 10\\
10.4 2.4 1 10\\
10.4 2.4 0 10\\
10.4 1.6 0 10\\
NaN NaN NaN 10\\
10.4 2.6 0 10\\
10.4 2.6 1 10\\
10.4 3.4 1 10\\
10.4 3.4 0 10\\
10.4 2.6 0 10\\
NaN NaN NaN 10\\
10.4 3.6 0 10\\
10.4 3.6 2 10\\
10.4 4.4 2 10\\
10.4 4.4 0 10\\
10.4 3.6 0 10\\
NaN NaN NaN 10\\
10.4 4.6 0 10\\
10.4 4.6 2 10\\
10.4 5.4 2 10\\
10.4 5.4 0 10\\
10.4 4.6 0 10\\
NaN NaN NaN 10\\
10.4 5.6 0 10\\
10.4 5.6 2 10\\
10.4 6.4 2 10\\
10.4 6.4 0 10\\
10.4 5.6 0 10\\
NaN NaN NaN 10\\
10.4 6.6 0 10\\
10.4 6.6 2 10\\
10.4 7.4 2 10\\
10.4 7.4 0 10\\
10.4 6.6 0 10\\
NaN NaN NaN 10\\
10.4 7.6 0 10\\
10.4 7.6 2 10\\
10.4 8.4 2 10\\
10.4 8.4 0 10\\
10.4 7.6 0 10\\
NaN NaN NaN 10\\
10.4 8.6 0 10\\
10.4 8.6 2 10\\
10.4 9.4 2 10\\
10.4 9.4 0 10\\
10.4 8.6 0 10\\
NaN NaN NaN 10\\
10.4 9.6 0 10\\
10.4 9.6 2 10\\
10.4 10.4 2 10\\
10.4 10.4 0 10\\
10.4 9.6 0 10\\
NaN NaN NaN 10\\
10.4 10.6 0 10\\
10.4 10.6 3 10\\
10.4 11.4 3 10\\
10.4 11.4 0 10\\
10.4 10.6 0 10\\
NaN NaN NaN 10\\
NaN NaN NaN 10\\
10.4 0.6 0 10\\
10.4 1.4 0 10\\
NaN NaN NaN 10\\
NaN NaN NaN 10\\
NaN NaN NaN 10\\
NaN NaN NaN 10\\
10.4 1.6 0 10\\
10.4 2.4 0 10\\
NaN NaN NaN 10\\
NaN NaN NaN 10\\
NaN NaN NaN 10\\
NaN NaN NaN 10\\
10.4 2.6 0 10\\
10.4 3.4 0 10\\
NaN NaN NaN 10\\
NaN NaN NaN 10\\
NaN NaN NaN 10\\
NaN NaN NaN 10\\
10.4 3.6 0 10\\
10.4 4.4 0 10\\
NaN NaN NaN 10\\
NaN NaN NaN 10\\
NaN NaN NaN 10\\
NaN NaN NaN 10\\
10.4 4.6 0 10\\
10.4 5.4 0 10\\
NaN NaN NaN 10\\
NaN NaN NaN 10\\
NaN NaN NaN 10\\
NaN NaN NaN 10\\
10.4 5.6 0 10\\
10.4 6.4 0 10\\
NaN NaN NaN 10\\
NaN NaN NaN 10\\
NaN NaN NaN 10\\
NaN NaN NaN 10\\
10.4 6.6 0 10\\
10.4 7.4 0 10\\
NaN NaN NaN 10\\
NaN NaN NaN 10\\
NaN NaN NaN 10\\
NaN NaN NaN 10\\
10.4 7.6 0 10\\
10.4 8.4 0 10\\
NaN NaN NaN 10\\
NaN NaN NaN 10\\
NaN NaN NaN 10\\
NaN NaN NaN 10\\
10.4 8.6 0 10\\
10.4 9.4 0 10\\
NaN NaN NaN 10\\
NaN NaN NaN 10\\
NaN NaN NaN 10\\
NaN NaN NaN 10\\
10.4 9.6 0 10\\
10.4 10.4 0 10\\
NaN NaN NaN 10\\
NaN NaN NaN 10\\
NaN NaN NaN 10\\
NaN NaN NaN 10\\
10.4 10.6 0 10\\
10.4 11.4 0 10\\
NaN NaN NaN 10\\
NaN NaN NaN 10\\
NaN NaN NaN 10\\
};

\addplot3[%
surf,
shader=flat,
draw=black,
point meta=explicit,
mesh/rows=4]
table[row sep=crcr,header=false,meta index=3] {
NaN NaN NaN 9\\
8.6 0.6 0 9\\
8.6 1.4 0 9\\
NaN NaN NaN 9\\
NaN NaN NaN 9\\
NaN NaN NaN 9\\
NaN NaN NaN 9\\
8.6 1.6 0 9\\
8.6 2.4 0 9\\
NaN NaN NaN 9\\
NaN NaN NaN 9\\
NaN NaN NaN 9\\
NaN NaN NaN 9\\
8.6 2.6 0 9\\
8.6 3.4 0 9\\
NaN NaN NaN 9\\
NaN NaN NaN 9\\
NaN NaN NaN 9\\
NaN NaN NaN 9\\
8.6 3.6 0 9\\
8.6 4.4 0 9\\
NaN NaN NaN 9\\
NaN NaN NaN 9\\
NaN NaN NaN 9\\
NaN NaN NaN 9\\
8.6 4.6 0 9\\
8.6 5.4 0 9\\
NaN NaN NaN 9\\
NaN NaN NaN 9\\
NaN NaN NaN 9\\
NaN NaN NaN 9\\
8.6 5.6 0 9\\
8.6 6.4 0 9\\
NaN NaN NaN 9\\
NaN NaN NaN 9\\
NaN NaN NaN 9\\
NaN NaN NaN 9\\
8.6 6.6 0 9\\
8.6 7.4 0 9\\
NaN NaN NaN 9\\
NaN NaN NaN 9\\
NaN NaN NaN 9\\
NaN NaN NaN 9\\
8.6 7.6 0 9\\
8.6 8.4 0 9\\
NaN NaN NaN 9\\
NaN NaN NaN 9\\
NaN NaN NaN 9\\
NaN NaN NaN 9\\
8.6 8.6 0 9\\
8.6 9.4 0 9\\
NaN NaN NaN 9\\
NaN NaN NaN 9\\
NaN NaN NaN 9\\
NaN NaN NaN 9\\
8.6 9.6 0 9\\
8.6 10.4 0 9\\
NaN NaN NaN 9\\
NaN NaN NaN 9\\
NaN NaN NaN 9\\
NaN NaN NaN 9\\
8.6 10.6 0 9\\
8.6 11.4 0 9\\
NaN NaN NaN 9\\
NaN NaN NaN 9\\
NaN NaN NaN 9\\
8.6 0.6 0 9\\
8.6 0.6 9 9\\
8.6 1.4 9 9\\
8.6 1.4 0 9\\
8.6 0.6 0 9\\
NaN NaN NaN 9\\
8.6 1.6 0 9\\
8.6 1.6 1 9\\
8.6 2.4 1 9\\
8.6 2.4 0 9\\
8.6 1.6 0 9\\
NaN NaN NaN 9\\
8.6 2.6 0 9\\
8.6 2.6 1 9\\
8.6 3.4 1 9\\
8.6 3.4 0 9\\
8.6 2.6 0 9\\
NaN NaN NaN 9\\
8.6 3.6 0 9\\
8.6 3.6 2 9\\
8.6 4.4 2 9\\
8.6 4.4 0 9\\
8.6 3.6 0 9\\
NaN NaN NaN 9\\
8.6 4.6 0 9\\
8.6 4.6 2 9\\
8.6 5.4 2 9\\
8.6 5.4 0 9\\
8.6 4.6 0 9\\
NaN NaN NaN 9\\
8.6 5.6 0 9\\
8.6 5.6 2 9\\
8.6 6.4 2 9\\
8.6 6.4 0 9\\
8.6 5.6 0 9\\
NaN NaN NaN 9\\
8.6 6.6 0 9\\
8.6 6.6 2 9\\
8.6 7.4 2 9\\
8.6 7.4 0 9\\
8.6 6.6 0 9\\
NaN NaN NaN 9\\
8.6 7.6 0 9\\
8.6 7.6 2 9\\
8.6 8.4 2 9\\
8.6 8.4 0 9\\
8.6 7.6 0 9\\
NaN NaN NaN 9\\
8.6 8.6 0 9\\
8.6 8.6 2 9\\
8.6 9.4 2 9\\
8.6 9.4 0 9\\
8.6 8.6 0 9\\
NaN NaN NaN 9\\
8.6 9.6 0 9\\
8.6 9.6 2 9\\
8.6 10.4 2 9\\
8.6 10.4 0 9\\
8.6 9.6 0 9\\
NaN NaN NaN 9\\
8.6 10.6 0 9\\
8.6 10.6 3 9\\
8.6 11.4 3 9\\
8.6 11.4 0 9\\
8.6 10.6 0 9\\
NaN NaN NaN 9\\
9.4 0.6 0 9\\
9.4 0.6 9 9\\
9.4 1.4 9 9\\
9.4 1.4 0 9\\
9.4 0.6 0 9\\
NaN NaN NaN 9\\
9.4 1.6 0 9\\
9.4 1.6 1 9\\
9.4 2.4 1 9\\
9.4 2.4 0 9\\
9.4 1.6 0 9\\
NaN NaN NaN 9\\
9.4 2.6 0 9\\
9.4 2.6 1 9\\
9.4 3.4 1 9\\
9.4 3.4 0 9\\
9.4 2.6 0 9\\
NaN NaN NaN 9\\
9.4 3.6 0 9\\
9.4 3.6 2 9\\
9.4 4.4 2 9\\
9.4 4.4 0 9\\
9.4 3.6 0 9\\
NaN NaN NaN 9\\
9.4 4.6 0 9\\
9.4 4.6 2 9\\
9.4 5.4 2 9\\
9.4 5.4 0 9\\
9.4 4.6 0 9\\
NaN NaN NaN 9\\
9.4 5.6 0 9\\
9.4 5.6 2 9\\
9.4 6.4 2 9\\
9.4 6.4 0 9\\
9.4 5.6 0 9\\
NaN NaN NaN 9\\
9.4 6.6 0 9\\
9.4 6.6 2 9\\
9.4 7.4 2 9\\
9.4 7.4 0 9\\
9.4 6.6 0 9\\
NaN NaN NaN 9\\
9.4 7.6 0 9\\
9.4 7.6 2 9\\
9.4 8.4 2 9\\
9.4 8.4 0 9\\
9.4 7.6 0 9\\
NaN NaN NaN 9\\
9.4 8.6 0 9\\
9.4 8.6 2 9\\
9.4 9.4 2 9\\
9.4 9.4 0 9\\
9.4 8.6 0 9\\
NaN NaN NaN 9\\
9.4 9.6 0 9\\
9.4 9.6 2 9\\
9.4 10.4 2 9\\
9.4 10.4 0 9\\
9.4 9.6 0 9\\
NaN NaN NaN 9\\
9.4 10.6 0 9\\
9.4 10.6 3 9\\
9.4 11.4 3 9\\
9.4 11.4 0 9\\
9.4 10.6 0 9\\
NaN NaN NaN 9\\
NaN NaN NaN 9\\
9.4 0.6 0 9\\
9.4 1.4 0 9\\
NaN NaN NaN 9\\
NaN NaN NaN 9\\
NaN NaN NaN 9\\
NaN NaN NaN 9\\
9.4 1.6 0 9\\
9.4 2.4 0 9\\
NaN NaN NaN 9\\
NaN NaN NaN 9\\
NaN NaN NaN 9\\
NaN NaN NaN 9\\
9.4 2.6 0 9\\
9.4 3.4 0 9\\
NaN NaN NaN 9\\
NaN NaN NaN 9\\
NaN NaN NaN 9\\
NaN NaN NaN 9\\
9.4 3.6 0 9\\
9.4 4.4 0 9\\
NaN NaN NaN 9\\
NaN NaN NaN 9\\
NaN NaN NaN 9\\
NaN NaN NaN 9\\
9.4 4.6 0 9\\
9.4 5.4 0 9\\
NaN NaN NaN 9\\
NaN NaN NaN 9\\
NaN NaN NaN 9\\
NaN NaN NaN 9\\
9.4 5.6 0 9\\
9.4 6.4 0 9\\
NaN NaN NaN 9\\
NaN NaN NaN 9\\
NaN NaN NaN 9\\
NaN NaN NaN 9\\
9.4 6.6 0 9\\
9.4 7.4 0 9\\
NaN NaN NaN 9\\
NaN NaN NaN 9\\
NaN NaN NaN 9\\
NaN NaN NaN 9\\
9.4 7.6 0 9\\
9.4 8.4 0 9\\
NaN NaN NaN 9\\
NaN NaN NaN 9\\
NaN NaN NaN 9\\
NaN NaN NaN 9\\
9.4 8.6 0 9\\
9.4 9.4 0 9\\
NaN NaN NaN 9\\
NaN NaN NaN 9\\
NaN NaN NaN 9\\
NaN NaN NaN 9\\
9.4 9.6 0 9\\
9.4 10.4 0 9\\
NaN NaN NaN 9\\
NaN NaN NaN 9\\
NaN NaN NaN 9\\
NaN NaN NaN 9\\
9.4 10.6 0 9\\
9.4 11.4 0 9\\
NaN NaN NaN 9\\
NaN NaN NaN 9\\
NaN NaN NaN 9\\
};

\addplot3[%
surf,
shader=flat,
draw=black,
point meta=explicit,
mesh/rows=4]
table[row sep=crcr,header=false,meta index=3] {
NaN NaN NaN 8\\
7.6 0.6 0 8\\
7.6 1.4 0 8\\
NaN NaN NaN 8\\
NaN NaN NaN 8\\
NaN NaN NaN 8\\
NaN NaN NaN 8\\
7.6 1.6 0 8\\
7.6 2.4 0 8\\
NaN NaN NaN 8\\
NaN NaN NaN 8\\
NaN NaN NaN 8\\
NaN NaN NaN 8\\
7.6 2.6 0 8\\
7.6 3.4 0 8\\
NaN NaN NaN 8\\
NaN NaN NaN 8\\
NaN NaN NaN 8\\
NaN NaN NaN 8\\
7.6 3.6 0 8\\
7.6 4.4 0 8\\
NaN NaN NaN 8\\
NaN NaN NaN 8\\
NaN NaN NaN 8\\
NaN NaN NaN 8\\
7.6 4.6 0 8\\
7.6 5.4 0 8\\
NaN NaN NaN 8\\
NaN NaN NaN 8\\
NaN NaN NaN 8\\
NaN NaN NaN 8\\
7.6 5.6 0 8\\
7.6 6.4 0 8\\
NaN NaN NaN 8\\
NaN NaN NaN 8\\
NaN NaN NaN 8\\
NaN NaN NaN 8\\
7.6 6.6 0 8\\
7.6 7.4 0 8\\
NaN NaN NaN 8\\
NaN NaN NaN 8\\
NaN NaN NaN 8\\
NaN NaN NaN 8\\
7.6 7.6 0 8\\
7.6 8.4 0 8\\
NaN NaN NaN 8\\
NaN NaN NaN 8\\
NaN NaN NaN 8\\
NaN NaN NaN 8\\
7.6 8.6 0 8\\
7.6 9.4 0 8\\
NaN NaN NaN 8\\
NaN NaN NaN 8\\
NaN NaN NaN 8\\
NaN NaN NaN 8\\
7.6 9.6 0 8\\
7.6 10.4 0 8\\
NaN NaN NaN 8\\
NaN NaN NaN 8\\
NaN NaN NaN 8\\
NaN NaN NaN 8\\
7.6 10.6 0 8\\
7.6 11.4 0 8\\
NaN NaN NaN 8\\
NaN NaN NaN 8\\
NaN NaN NaN 8\\
7.6 0.6 0 8\\
7.6 0.6 9 8\\
7.6 1.4 9 8\\
7.6 1.4 0 8\\
7.6 0.6 0 8\\
NaN NaN NaN 8\\
7.6 1.6 0 8\\
7.6 1.6 1 8\\
7.6 2.4 1 8\\
7.6 2.4 0 8\\
7.6 1.6 0 8\\
NaN NaN NaN 8\\
7.6 2.6 0 8\\
7.6 2.6 1 8\\
7.6 3.4 1 8\\
7.6 3.4 0 8\\
7.6 2.6 0 8\\
NaN NaN NaN 8\\
7.6 3.6 0 8\\
7.6 3.6 2 8\\
7.6 4.4 2 8\\
7.6 4.4 0 8\\
7.6 3.6 0 8\\
NaN NaN NaN 8\\
7.6 4.6 0 8\\
7.6 4.6 2 8\\
7.6 5.4 2 8\\
7.6 5.4 0 8\\
7.6 4.6 0 8\\
NaN NaN NaN 8\\
7.6 5.6 0 8\\
7.6 5.6 2 8\\
7.6 6.4 2 8\\
7.6 6.4 0 8\\
7.6 5.6 0 8\\
NaN NaN NaN 8\\
7.6 6.6 0 8\\
7.6 6.6 2 8\\
7.6 7.4 2 8\\
7.6 7.4 0 8\\
7.6 6.6 0 8\\
NaN NaN NaN 8\\
7.6 7.6 0 8\\
7.6 7.6 2 8\\
7.6 8.4 2 8\\
7.6 8.4 0 8\\
7.6 7.6 0 8\\
NaN NaN NaN 8\\
7.6 8.6 0 8\\
7.6 8.6 2 8\\
7.6 9.4 2 8\\
7.6 9.4 0 8\\
7.6 8.6 0 8\\
NaN NaN NaN 8\\
7.6 9.6 0 8\\
7.6 9.6 2 8\\
7.6 10.4 2 8\\
7.6 10.4 0 8\\
7.6 9.6 0 8\\
NaN NaN NaN 8\\
7.6 10.6 0 8\\
7.6 10.6 3 8\\
7.6 11.4 3 8\\
7.6 11.4 0 8\\
7.6 10.6 0 8\\
NaN NaN NaN 8\\
8.4 0.6 0 8\\
8.4 0.6 9 8\\
8.4 1.4 9 8\\
8.4 1.4 0 8\\
8.4 0.6 0 8\\
NaN NaN NaN 8\\
8.4 1.6 0 8\\
8.4 1.6 1 8\\
8.4 2.4 1 8\\
8.4 2.4 0 8\\
8.4 1.6 0 8\\
NaN NaN NaN 8\\
8.4 2.6 0 8\\
8.4 2.6 1 8\\
8.4 3.4 1 8\\
8.4 3.4 0 8\\
8.4 2.6 0 8\\
NaN NaN NaN 8\\
8.4 3.6 0 8\\
8.4 3.6 2 8\\
8.4 4.4 2 8\\
8.4 4.4 0 8\\
8.4 3.6 0 8\\
NaN NaN NaN 8\\
8.4 4.6 0 8\\
8.4 4.6 2 8\\
8.4 5.4 2 8\\
8.4 5.4 0 8\\
8.4 4.6 0 8\\
NaN NaN NaN 8\\
8.4 5.6 0 8\\
8.4 5.6 2 8\\
8.4 6.4 2 8\\
8.4 6.4 0 8\\
8.4 5.6 0 8\\
NaN NaN NaN 8\\
8.4 6.6 0 8\\
8.4 6.6 2 8\\
8.4 7.4 2 8\\
8.4 7.4 0 8\\
8.4 6.6 0 8\\
NaN NaN NaN 8\\
8.4 7.6 0 8\\
8.4 7.6 2 8\\
8.4 8.4 2 8\\
8.4 8.4 0 8\\
8.4 7.6 0 8\\
NaN NaN NaN 8\\
8.4 8.6 0 8\\
8.4 8.6 2 8\\
8.4 9.4 2 8\\
8.4 9.4 0 8\\
8.4 8.6 0 8\\
NaN NaN NaN 8\\
8.4 9.6 0 8\\
8.4 9.6 2 8\\
8.4 10.4 2 8\\
8.4 10.4 0 8\\
8.4 9.6 0 8\\
NaN NaN NaN 8\\
8.4 10.6 0 8\\
8.4 10.6 3 8\\
8.4 11.4 3 8\\
8.4 11.4 0 8\\
8.4 10.6 0 8\\
NaN NaN NaN 8\\
NaN NaN NaN 8\\
8.4 0.6 0 8\\
8.4 1.4 0 8\\
NaN NaN NaN 8\\
NaN NaN NaN 8\\
NaN NaN NaN 8\\
NaN NaN NaN 8\\
8.4 1.6 0 8\\
8.4 2.4 0 8\\
NaN NaN NaN 8\\
NaN NaN NaN 8\\
NaN NaN NaN 8\\
NaN NaN NaN 8\\
8.4 2.6 0 8\\
8.4 3.4 0 8\\
NaN NaN NaN 8\\
NaN NaN NaN 8\\
NaN NaN NaN 8\\
NaN NaN NaN 8\\
8.4 3.6 0 8\\
8.4 4.4 0 8\\
NaN NaN NaN 8\\
NaN NaN NaN 8\\
NaN NaN NaN 8\\
NaN NaN NaN 8\\
8.4 4.6 0 8\\
8.4 5.4 0 8\\
NaN NaN NaN 8\\
NaN NaN NaN 8\\
NaN NaN NaN 8\\
NaN NaN NaN 8\\
8.4 5.6 0 8\\
8.4 6.4 0 8\\
NaN NaN NaN 8\\
NaN NaN NaN 8\\
NaN NaN NaN 8\\
NaN NaN NaN 8\\
8.4 6.6 0 8\\
8.4 7.4 0 8\\
NaN NaN NaN 8\\
NaN NaN NaN 8\\
NaN NaN NaN 8\\
NaN NaN NaN 8\\
8.4 7.6 0 8\\
8.4 8.4 0 8\\
NaN NaN NaN 8\\
NaN NaN NaN 8\\
NaN NaN NaN 8\\
NaN NaN NaN 8\\
8.4 8.6 0 8\\
8.4 9.4 0 8\\
NaN NaN NaN 8\\
NaN NaN NaN 8\\
NaN NaN NaN 8\\
NaN NaN NaN 8\\
8.4 9.6 0 8\\
8.4 10.4 0 8\\
NaN NaN NaN 8\\
NaN NaN NaN 8\\
NaN NaN NaN 8\\
NaN NaN NaN 8\\
8.4 10.6 0 8\\
8.4 11.4 0 8\\
NaN NaN NaN 8\\
NaN NaN NaN 8\\
NaN NaN NaN 8\\
};

\addplot3[%
surf,
shader=flat,
draw=black,
point meta=explicit,
mesh/rows=4]
table[row sep=crcr,header=false,meta index=3] {
NaN NaN NaN 7\\
6.6 0.6 0 7\\
6.6 1.4 0 7\\
NaN NaN NaN 7\\
NaN NaN NaN 7\\
NaN NaN NaN 7\\
NaN NaN NaN 7\\
6.6 1.6 0 7\\
6.6 2.4 0 7\\
NaN NaN NaN 7\\
NaN NaN NaN 7\\
NaN NaN NaN 7\\
NaN NaN NaN 7\\
6.6 2.6 0 7\\
6.6 3.4 0 7\\
NaN NaN NaN 7\\
NaN NaN NaN 7\\
NaN NaN NaN 7\\
NaN NaN NaN 7\\
6.6 3.6 0 7\\
6.6 4.4 0 7\\
NaN NaN NaN 7\\
NaN NaN NaN 7\\
NaN NaN NaN 7\\
NaN NaN NaN 7\\
6.6 4.6 0 7\\
6.6 5.4 0 7\\
NaN NaN NaN 7\\
NaN NaN NaN 7\\
NaN NaN NaN 7\\
NaN NaN NaN 7\\
6.6 5.6 0 7\\
6.6 6.4 0 7\\
NaN NaN NaN 7\\
NaN NaN NaN 7\\
NaN NaN NaN 7\\
NaN NaN NaN 7\\
6.6 6.6 0 7\\
6.6 7.4 0 7\\
NaN NaN NaN 7\\
NaN NaN NaN 7\\
NaN NaN NaN 7\\
NaN NaN NaN 7\\
6.6 7.6 0 7\\
6.6 8.4 0 7\\
NaN NaN NaN 7\\
NaN NaN NaN 7\\
NaN NaN NaN 7\\
NaN NaN NaN 7\\
6.6 8.6 0 7\\
6.6 9.4 0 7\\
NaN NaN NaN 7\\
NaN NaN NaN 7\\
NaN NaN NaN 7\\
NaN NaN NaN 7\\
6.6 9.6 0 7\\
6.6 10.4 0 7\\
NaN NaN NaN 7\\
NaN NaN NaN 7\\
NaN NaN NaN 7\\
NaN NaN NaN 7\\
6.6 10.6 0 7\\
6.6 11.4 0 7\\
NaN NaN NaN 7\\
NaN NaN NaN 7\\
NaN NaN NaN 7\\
6.6 0.6 0 7\\
6.6 0.6 9 7\\
6.6 1.4 9 7\\
6.6 1.4 0 7\\
6.6 0.6 0 7\\
NaN NaN NaN 7\\
6.6 1.6 0 7\\
6.6 1.6 1 7\\
6.6 2.4 1 7\\
6.6 2.4 0 7\\
6.6 1.6 0 7\\
NaN NaN NaN 7\\
6.6 2.6 0 7\\
6.6 2.6 1 7\\
6.6 3.4 1 7\\
6.6 3.4 0 7\\
6.6 2.6 0 7\\
NaN NaN NaN 7\\
6.6 3.6 0 7\\
6.6 3.6 2 7\\
6.6 4.4 2 7\\
6.6 4.4 0 7\\
6.6 3.6 0 7\\
NaN NaN NaN 7\\
6.6 4.6 0 7\\
6.6 4.6 2 7\\
6.6 5.4 2 7\\
6.6 5.4 0 7\\
6.6 4.6 0 7\\
NaN NaN NaN 7\\
6.6 5.6 0 7\\
6.6 5.6 2 7\\
6.6 6.4 2 7\\
6.6 6.4 0 7\\
6.6 5.6 0 7\\
NaN NaN NaN 7\\
6.6 6.6 0 7\\
6.6 6.6 2 7\\
6.6 7.4 2 7\\
6.6 7.4 0 7\\
6.6 6.6 0 7\\
NaN NaN NaN 7\\
6.6 7.6 0 7\\
6.6 7.6 2 7\\
6.6 8.4 2 7\\
6.6 8.4 0 7\\
6.6 7.6 0 7\\
NaN NaN NaN 7\\
6.6 8.6 0 7\\
6.6 8.6 2 7\\
6.6 9.4 2 7\\
6.6 9.4 0 7\\
6.6 8.6 0 7\\
NaN NaN NaN 7\\
6.6 9.6 0 7\\
6.6 9.6 2 7\\
6.6 10.4 2 7\\
6.6 10.4 0 7\\
6.6 9.6 0 7\\
NaN NaN NaN 7\\
6.6 10.6 0 7\\
6.6 10.6 3 7\\
6.6 11.4 3 7\\
6.6 11.4 0 7\\
6.6 10.6 0 7\\
NaN NaN NaN 7\\
7.4 0.6 0 7\\
7.4 0.6 9 7\\
7.4 1.4 9 7\\
7.4 1.4 0 7\\
7.4 0.6 0 7\\
NaN NaN NaN 7\\
7.4 1.6 0 7\\
7.4 1.6 1 7\\
7.4 2.4 1 7\\
7.4 2.4 0 7\\
7.4 1.6 0 7\\
NaN NaN NaN 7\\
7.4 2.6 0 7\\
7.4 2.6 1 7\\
7.4 3.4 1 7\\
7.4 3.4 0 7\\
7.4 2.6 0 7\\
NaN NaN NaN 7\\
7.4 3.6 0 7\\
7.4 3.6 2 7\\
7.4 4.4 2 7\\
7.4 4.4 0 7\\
7.4 3.6 0 7\\
NaN NaN NaN 7\\
7.4 4.6 0 7\\
7.4 4.6 2 7\\
7.4 5.4 2 7\\
7.4 5.4 0 7\\
7.4 4.6 0 7\\
NaN NaN NaN 7\\
7.4 5.6 0 7\\
7.4 5.6 2 7\\
7.4 6.4 2 7\\
7.4 6.4 0 7\\
7.4 5.6 0 7\\
NaN NaN NaN 7\\
7.4 6.6 0 7\\
7.4 6.6 2 7\\
7.4 7.4 2 7\\
7.4 7.4 0 7\\
7.4 6.6 0 7\\
NaN NaN NaN 7\\
7.4 7.6 0 7\\
7.4 7.6 2 7\\
7.4 8.4 2 7\\
7.4 8.4 0 7\\
7.4 7.6 0 7\\
NaN NaN NaN 7\\
7.4 8.6 0 7\\
7.4 8.6 2 7\\
7.4 9.4 2 7\\
7.4 9.4 0 7\\
7.4 8.6 0 7\\
NaN NaN NaN 7\\
7.4 9.6 0 7\\
7.4 9.6 2 7\\
7.4 10.4 2 7\\
7.4 10.4 0 7\\
7.4 9.6 0 7\\
NaN NaN NaN 7\\
7.4 10.6 0 7\\
7.4 10.6 3 7\\
7.4 11.4 3 7\\
7.4 11.4 0 7\\
7.4 10.6 0 7\\
NaN NaN NaN 7\\
NaN NaN NaN 7\\
7.4 0.6 0 7\\
7.4 1.4 0 7\\
NaN NaN NaN 7\\
NaN NaN NaN 7\\
NaN NaN NaN 7\\
NaN NaN NaN 7\\
7.4 1.6 0 7\\
7.4 2.4 0 7\\
NaN NaN NaN 7\\
NaN NaN NaN 7\\
NaN NaN NaN 7\\
NaN NaN NaN 7\\
7.4 2.6 0 7\\
7.4 3.4 0 7\\
NaN NaN NaN 7\\
NaN NaN NaN 7\\
NaN NaN NaN 7\\
NaN NaN NaN 7\\
7.4 3.6 0 7\\
7.4 4.4 0 7\\
NaN NaN NaN 7\\
NaN NaN NaN 7\\
NaN NaN NaN 7\\
NaN NaN NaN 7\\
7.4 4.6 0 7\\
7.4 5.4 0 7\\
NaN NaN NaN 7\\
NaN NaN NaN 7\\
NaN NaN NaN 7\\
NaN NaN NaN 7\\
7.4 5.6 0 7\\
7.4 6.4 0 7\\
NaN NaN NaN 7\\
NaN NaN NaN 7\\
NaN NaN NaN 7\\
NaN NaN NaN 7\\
7.4 6.6 0 7\\
7.4 7.4 0 7\\
NaN NaN NaN 7\\
NaN NaN NaN 7\\
NaN NaN NaN 7\\
NaN NaN NaN 7\\
7.4 7.6 0 7\\
7.4 8.4 0 7\\
NaN NaN NaN 7\\
NaN NaN NaN 7\\
NaN NaN NaN 7\\
NaN NaN NaN 7\\
7.4 8.6 0 7\\
7.4 9.4 0 7\\
NaN NaN NaN 7\\
NaN NaN NaN 7\\
NaN NaN NaN 7\\
NaN NaN NaN 7\\
7.4 9.6 0 7\\
7.4 10.4 0 7\\
NaN NaN NaN 7\\
NaN NaN NaN 7\\
NaN NaN NaN 7\\
NaN NaN NaN 7\\
7.4 10.6 0 7\\
7.4 11.4 0 7\\
NaN NaN NaN 7\\
NaN NaN NaN 7\\
NaN NaN NaN 7\\
};

\addplot3[%
surf,
shader=flat,
draw=black,
point meta=explicit,
mesh/rows=4]
table[row sep=crcr,header=false,meta index=3] {
NaN NaN NaN 6\\
5.6 0.6 0 6\\
5.6 1.4 0 6\\
NaN NaN NaN 6\\
NaN NaN NaN 6\\
NaN NaN NaN 6\\
NaN NaN NaN 6\\
5.6 1.6 0 6\\
5.6 2.4 0 6\\
NaN NaN NaN 6\\
NaN NaN NaN 6\\
NaN NaN NaN 6\\
NaN NaN NaN 6\\
5.6 2.6 0 6\\
5.6 3.4 0 6\\
NaN NaN NaN 6\\
NaN NaN NaN 6\\
NaN NaN NaN 6\\
NaN NaN NaN 6\\
5.6 3.6 0 6\\
5.6 4.4 0 6\\
NaN NaN NaN 6\\
NaN NaN NaN 6\\
NaN NaN NaN 6\\
NaN NaN NaN 6\\
5.6 4.6 0 6\\
5.6 5.4 0 6\\
NaN NaN NaN 6\\
NaN NaN NaN 6\\
NaN NaN NaN 6\\
NaN NaN NaN 6\\
5.6 5.6 0 6\\
5.6 6.4 0 6\\
NaN NaN NaN 6\\
NaN NaN NaN 6\\
NaN NaN NaN 6\\
NaN NaN NaN 6\\
5.6 6.6 0 6\\
5.6 7.4 0 6\\
NaN NaN NaN 6\\
NaN NaN NaN 6\\
NaN NaN NaN 6\\
NaN NaN NaN 6\\
5.6 7.6 0 6\\
5.6 8.4 0 6\\
NaN NaN NaN 6\\
NaN NaN NaN 6\\
NaN NaN NaN 6\\
NaN NaN NaN 6\\
5.6 8.6 0 6\\
5.6 9.4 0 6\\
NaN NaN NaN 6\\
NaN NaN NaN 6\\
NaN NaN NaN 6\\
NaN NaN NaN 6\\
5.6 9.6 0 6\\
5.6 10.4 0 6\\
NaN NaN NaN 6\\
NaN NaN NaN 6\\
NaN NaN NaN 6\\
NaN NaN NaN 6\\
5.6 10.6 0 6\\
5.6 11.4 0 6\\
NaN NaN NaN 6\\
NaN NaN NaN 6\\
NaN NaN NaN 6\\
5.6 0.6 0 6\\
5.6 0.6 9 6\\
5.6 1.4 9 6\\
5.6 1.4 0 6\\
5.6 0.6 0 6\\
NaN NaN NaN 6\\
5.6 1.6 0 6\\
5.6 1.6 1 6\\
5.6 2.4 1 6\\
5.6 2.4 0 6\\
5.6 1.6 0 6\\
NaN NaN NaN 6\\
5.6 2.6 0 6\\
5.6 2.6 1 6\\
5.6 3.4 1 6\\
5.6 3.4 0 6\\
5.6 2.6 0 6\\
NaN NaN NaN 6\\
5.6 3.6 0 6\\
5.6 3.6 2 6\\
5.6 4.4 2 6\\
5.6 4.4 0 6\\
5.6 3.6 0 6\\
NaN NaN NaN 6\\
5.6 4.6 0 6\\
5.6 4.6 2 6\\
5.6 5.4 2 6\\
5.6 5.4 0 6\\
5.6 4.6 0 6\\
NaN NaN NaN 6\\
5.6 5.6 0 6\\
5.6 5.6 2 6\\
5.6 6.4 2 6\\
5.6 6.4 0 6\\
5.6 5.6 0 6\\
NaN NaN NaN 6\\
5.6 6.6 0 6\\
5.6 6.6 2 6\\
5.6 7.4 2 6\\
5.6 7.4 0 6\\
5.6 6.6 0 6\\
NaN NaN NaN 6\\
5.6 7.6 0 6\\
5.6 7.6 2 6\\
5.6 8.4 2 6\\
5.6 8.4 0 6\\
5.6 7.6 0 6\\
NaN NaN NaN 6\\
5.6 8.6 0 6\\
5.6 8.6 2 6\\
5.6 9.4 2 6\\
5.6 9.4 0 6\\
5.6 8.6 0 6\\
NaN NaN NaN 6\\
5.6 9.6 0 6\\
5.6 9.6 2 6\\
5.6 10.4 2 6\\
5.6 10.4 0 6\\
5.6 9.6 0 6\\
NaN NaN NaN 6\\
5.6 10.6 0 6\\
5.6 10.6 3 6\\
5.6 11.4 3 6\\
5.6 11.4 0 6\\
5.6 10.6 0 6\\
NaN NaN NaN 6\\
6.4 0.6 0 6\\
6.4 0.6 9 6\\
6.4 1.4 9 6\\
6.4 1.4 0 6\\
6.4 0.6 0 6\\
NaN NaN NaN 6\\
6.4 1.6 0 6\\
6.4 1.6 1 6\\
6.4 2.4 1 6\\
6.4 2.4 0 6\\
6.4 1.6 0 6\\
NaN NaN NaN 6\\
6.4 2.6 0 6\\
6.4 2.6 1 6\\
6.4 3.4 1 6\\
6.4 3.4 0 6\\
6.4 2.6 0 6\\
NaN NaN NaN 6\\
6.4 3.6 0 6\\
6.4 3.6 2 6\\
6.4 4.4 2 6\\
6.4 4.4 0 6\\
6.4 3.6 0 6\\
NaN NaN NaN 6\\
6.4 4.6 0 6\\
6.4 4.6 2 6\\
6.4 5.4 2 6\\
6.4 5.4 0 6\\
6.4 4.6 0 6\\
NaN NaN NaN 6\\
6.4 5.6 0 6\\
6.4 5.6 2 6\\
6.4 6.4 2 6\\
6.4 6.4 0 6\\
6.4 5.6 0 6\\
NaN NaN NaN 6\\
6.4 6.6 0 6\\
6.4 6.6 2 6\\
6.4 7.4 2 6\\
6.4 7.4 0 6\\
6.4 6.6 0 6\\
NaN NaN NaN 6\\
6.4 7.6 0 6\\
6.4 7.6 2 6\\
6.4 8.4 2 6\\
6.4 8.4 0 6\\
6.4 7.6 0 6\\
NaN NaN NaN 6\\
6.4 8.6 0 6\\
6.4 8.6 2 6\\
6.4 9.4 2 6\\
6.4 9.4 0 6\\
6.4 8.6 0 6\\
NaN NaN NaN 6\\
6.4 9.6 0 6\\
6.4 9.6 2 6\\
6.4 10.4 2 6\\
6.4 10.4 0 6\\
6.4 9.6 0 6\\
NaN NaN NaN 6\\
6.4 10.6 0 6\\
6.4 10.6 3 6\\
6.4 11.4 3 6\\
6.4 11.4 0 6\\
6.4 10.6 0 6\\
NaN NaN NaN 6\\
NaN NaN NaN 6\\
6.4 0.6 0 6\\
6.4 1.4 0 6\\
NaN NaN NaN 6\\
NaN NaN NaN 6\\
NaN NaN NaN 6\\
NaN NaN NaN 6\\
6.4 1.6 0 6\\
6.4 2.4 0 6\\
NaN NaN NaN 6\\
NaN NaN NaN 6\\
NaN NaN NaN 6\\
NaN NaN NaN 6\\
6.4 2.6 0 6\\
6.4 3.4 0 6\\
NaN NaN NaN 6\\
NaN NaN NaN 6\\
NaN NaN NaN 6\\
NaN NaN NaN 6\\
6.4 3.6 0 6\\
6.4 4.4 0 6\\
NaN NaN NaN 6\\
NaN NaN NaN 6\\
NaN NaN NaN 6\\
NaN NaN NaN 6\\
6.4 4.6 0 6\\
6.4 5.4 0 6\\
NaN NaN NaN 6\\
NaN NaN NaN 6\\
NaN NaN NaN 6\\
NaN NaN NaN 6\\
6.4 5.6 0 6\\
6.4 6.4 0 6\\
NaN NaN NaN 6\\
NaN NaN NaN 6\\
NaN NaN NaN 6\\
NaN NaN NaN 6\\
6.4 6.6 0 6\\
6.4 7.4 0 6\\
NaN NaN NaN 6\\
NaN NaN NaN 6\\
NaN NaN NaN 6\\
NaN NaN NaN 6\\
6.4 7.6 0 6\\
6.4 8.4 0 6\\
NaN NaN NaN 6\\
NaN NaN NaN 6\\
NaN NaN NaN 6\\
NaN NaN NaN 6\\
6.4 8.6 0 6\\
6.4 9.4 0 6\\
NaN NaN NaN 6\\
NaN NaN NaN 6\\
NaN NaN NaN 6\\
NaN NaN NaN 6\\
6.4 9.6 0 6\\
6.4 10.4 0 6\\
NaN NaN NaN 6\\
NaN NaN NaN 6\\
NaN NaN NaN 6\\
NaN NaN NaN 6\\
6.4 10.6 0 6\\
6.4 11.4 0 6\\
NaN NaN NaN 6\\
NaN NaN NaN 6\\
NaN NaN NaN 6\\
};

\addplot3[%
surf,
shader=flat,
draw=black,
point meta=explicit,
mesh/rows=4]
table[row sep=crcr,header=false,meta index=3] {
NaN NaN NaN 5\\
4.6 0.6 0 5\\
4.6 1.4 0 5\\
NaN NaN NaN 5\\
NaN NaN NaN 5\\
NaN NaN NaN 5\\
NaN NaN NaN 5\\
4.6 1.6 0 5\\
4.6 2.4 0 5\\
NaN NaN NaN 5\\
NaN NaN NaN 5\\
NaN NaN NaN 5\\
NaN NaN NaN 5\\
4.6 2.6 0 5\\
4.6 3.4 0 5\\
NaN NaN NaN 5\\
NaN NaN NaN 5\\
NaN NaN NaN 5\\
NaN NaN NaN 5\\
4.6 3.6 0 5\\
4.6 4.4 0 5\\
NaN NaN NaN 5\\
NaN NaN NaN 5\\
NaN NaN NaN 5\\
NaN NaN NaN 5\\
4.6 4.6 0 5\\
4.6 5.4 0 5\\
NaN NaN NaN 5\\
NaN NaN NaN 5\\
NaN NaN NaN 5\\
NaN NaN NaN 5\\
4.6 5.6 0 5\\
4.6 6.4 0 5\\
NaN NaN NaN 5\\
NaN NaN NaN 5\\
NaN NaN NaN 5\\
NaN NaN NaN 5\\
4.6 6.6 0 5\\
4.6 7.4 0 5\\
NaN NaN NaN 5\\
NaN NaN NaN 5\\
NaN NaN NaN 5\\
NaN NaN NaN 5\\
4.6 7.6 0 5\\
4.6 8.4 0 5\\
NaN NaN NaN 5\\
NaN NaN NaN 5\\
NaN NaN NaN 5\\
NaN NaN NaN 5\\
4.6 8.6 0 5\\
4.6 9.4 0 5\\
NaN NaN NaN 5\\
NaN NaN NaN 5\\
NaN NaN NaN 5\\
NaN NaN NaN 5\\
4.6 9.6 0 5\\
4.6 10.4 0 5\\
NaN NaN NaN 5\\
NaN NaN NaN 5\\
NaN NaN NaN 5\\
NaN NaN NaN 5\\
4.6 10.6 0 5\\
4.6 11.4 0 5\\
NaN NaN NaN 5\\
NaN NaN NaN 5\\
NaN NaN NaN 5\\
4.6 0.6 0 5\\
4.6 0.6 0 5\\
4.6 1.4 0 5\\
4.6 1.4 0 5\\
4.6 0.6 0 5\\
NaN NaN NaN 5\\
4.6 1.6 0 5\\
4.6 1.6 1 5\\
4.6 2.4 1 5\\
4.6 2.4 0 5\\
4.6 1.6 0 5\\
NaN NaN NaN 5\\
4.6 2.6 0 5\\
4.6 2.6 1 5\\
4.6 3.4 1 5\\
4.6 3.4 0 5\\
4.6 2.6 0 5\\
NaN NaN NaN 5\\
4.6 3.6 0 5\\
4.6 3.6 2 5\\
4.6 4.4 2 5\\
4.6 4.4 0 5\\
4.6 3.6 0 5\\
NaN NaN NaN 5\\
4.6 4.6 0 5\\
4.6 4.6 2 5\\
4.6 5.4 2 5\\
4.6 5.4 0 5\\
4.6 4.6 0 5\\
NaN NaN NaN 5\\
4.6 5.6 0 5\\
4.6 5.6 2 5\\
4.6 6.4 2 5\\
4.6 6.4 0 5\\
4.6 5.6 0 5\\
NaN NaN NaN 5\\
4.6 6.6 0 5\\
4.6 6.6 2 5\\
4.6 7.4 2 5\\
4.6 7.4 0 5\\
4.6 6.6 0 5\\
NaN NaN NaN 5\\
4.6 7.6 0 5\\
4.6 7.6 2 5\\
4.6 8.4 2 5\\
4.6 8.4 0 5\\
4.6 7.6 0 5\\
NaN NaN NaN 5\\
4.6 8.6 0 5\\
4.6 8.6 2 5\\
4.6 9.4 2 5\\
4.6 9.4 0 5\\
4.6 8.6 0 5\\
NaN NaN NaN 5\\
4.6 9.6 0 5\\
4.6 9.6 2 5\\
4.6 10.4 2 5\\
4.6 10.4 0 5\\
4.6 9.6 0 5\\
NaN NaN NaN 5\\
4.6 10.6 0 5\\
4.6 10.6 3 5\\
4.6 11.4 3 5\\
4.6 11.4 0 5\\
4.6 10.6 0 5\\
NaN NaN NaN 5\\
5.4 0.6 0 5\\
5.4 0.6 0 5\\
5.4 1.4 0 5\\
5.4 1.4 0 5\\
5.4 0.6 0 5\\
NaN NaN NaN 5\\
5.4 1.6 0 5\\
5.4 1.6 1 5\\
5.4 2.4 1 5\\
5.4 2.4 0 5\\
5.4 1.6 0 5\\
NaN NaN NaN 5\\
5.4 2.6 0 5\\
5.4 2.6 1 5\\
5.4 3.4 1 5\\
5.4 3.4 0 5\\
5.4 2.6 0 5\\
NaN NaN NaN 5\\
5.4 3.6 0 5\\
5.4 3.6 2 5\\
5.4 4.4 2 5\\
5.4 4.4 0 5\\
5.4 3.6 0 5\\
NaN NaN NaN 5\\
5.4 4.6 0 5\\
5.4 4.6 2 5\\
5.4 5.4 2 5\\
5.4 5.4 0 5\\
5.4 4.6 0 5\\
NaN NaN NaN 5\\
5.4 5.6 0 5\\
5.4 5.6 2 5\\
5.4 6.4 2 5\\
5.4 6.4 0 5\\
5.4 5.6 0 5\\
NaN NaN NaN 5\\
5.4 6.6 0 5\\
5.4 6.6 2 5\\
5.4 7.4 2 5\\
5.4 7.4 0 5\\
5.4 6.6 0 5\\
NaN NaN NaN 5\\
5.4 7.6 0 5\\
5.4 7.6 2 5\\
5.4 8.4 2 5\\
5.4 8.4 0 5\\
5.4 7.6 0 5\\
NaN NaN NaN 5\\
5.4 8.6 0 5\\
5.4 8.6 2 5\\
5.4 9.4 2 5\\
5.4 9.4 0 5\\
5.4 8.6 0 5\\
NaN NaN NaN 5\\
5.4 9.6 0 5\\
5.4 9.6 2 5\\
5.4 10.4 2 5\\
5.4 10.4 0 5\\
5.4 9.6 0 5\\
NaN NaN NaN 5\\
5.4 10.6 0 5\\
5.4 10.6 3 5\\
5.4 11.4 3 5\\
5.4 11.4 0 5\\
5.4 10.6 0 5\\
NaN NaN NaN 5\\
NaN NaN NaN 5\\
5.4 0.6 0 5\\
5.4 1.4 0 5\\
NaN NaN NaN 5\\
NaN NaN NaN 5\\
NaN NaN NaN 5\\
NaN NaN NaN 5\\
5.4 1.6 0 5\\
5.4 2.4 0 5\\
NaN NaN NaN 5\\
NaN NaN NaN 5\\
NaN NaN NaN 5\\
NaN NaN NaN 5\\
5.4 2.6 0 5\\
5.4 3.4 0 5\\
NaN NaN NaN 5\\
NaN NaN NaN 5\\
NaN NaN NaN 5\\
NaN NaN NaN 5\\
5.4 3.6 0 5\\
5.4 4.4 0 5\\
NaN NaN NaN 5\\
NaN NaN NaN 5\\
NaN NaN NaN 5\\
NaN NaN NaN 5\\
5.4 4.6 0 5\\
5.4 5.4 0 5\\
NaN NaN NaN 5\\
NaN NaN NaN 5\\
NaN NaN NaN 5\\
NaN NaN NaN 5\\
5.4 5.6 0 5\\
5.4 6.4 0 5\\
NaN NaN NaN 5\\
NaN NaN NaN 5\\
NaN NaN NaN 5\\
NaN NaN NaN 5\\
5.4 6.6 0 5\\
5.4 7.4 0 5\\
NaN NaN NaN 5\\
NaN NaN NaN 5\\
NaN NaN NaN 5\\
NaN NaN NaN 5\\
5.4 7.6 0 5\\
5.4 8.4 0 5\\
NaN NaN NaN 5\\
NaN NaN NaN 5\\
NaN NaN NaN 5\\
NaN NaN NaN 5\\
5.4 8.6 0 5\\
5.4 9.4 0 5\\
NaN NaN NaN 5\\
NaN NaN NaN 5\\
NaN NaN NaN 5\\
NaN NaN NaN 5\\
5.4 9.6 0 5\\
5.4 10.4 0 5\\
NaN NaN NaN 5\\
NaN NaN NaN 5\\
NaN NaN NaN 5\\
NaN NaN NaN 5\\
5.4 10.6 0 5\\
5.4 11.4 0 5\\
NaN NaN NaN 5\\
NaN NaN NaN 5\\
NaN NaN NaN 5\\
};

\addplot3[%
surf,
shader=flat,
draw=black,
point meta=explicit,
mesh/rows=4]
table[row sep=crcr,header=false,meta index=3] {
NaN NaN NaN 4\\
3.6 0.6 0 4\\
3.6 1.4 0 4\\
NaN NaN NaN 4\\
NaN NaN NaN 4\\
NaN NaN NaN 4\\
NaN NaN NaN 4\\
3.6 1.6 0 4\\
3.6 2.4 0 4\\
NaN NaN NaN 4\\
NaN NaN NaN 4\\
NaN NaN NaN 4\\
NaN NaN NaN 4\\
3.6 2.6 0 4\\
3.6 3.4 0 4\\
NaN NaN NaN 4\\
NaN NaN NaN 4\\
NaN NaN NaN 4\\
NaN NaN NaN 4\\
3.6 3.6 0 4\\
3.6 4.4 0 4\\
NaN NaN NaN 4\\
NaN NaN NaN 4\\
NaN NaN NaN 4\\
NaN NaN NaN 4\\
3.6 4.6 0 4\\
3.6 5.4 0 4\\
NaN NaN NaN 4\\
NaN NaN NaN 4\\
NaN NaN NaN 4\\
NaN NaN NaN 4\\
3.6 5.6 0 4\\
3.6 6.4 0 4\\
NaN NaN NaN 4\\
NaN NaN NaN 4\\
NaN NaN NaN 4\\
NaN NaN NaN 4\\
3.6 6.6 0 4\\
3.6 7.4 0 4\\
NaN NaN NaN 4\\
NaN NaN NaN 4\\
NaN NaN NaN 4\\
NaN NaN NaN 4\\
3.6 7.6 0 4\\
3.6 8.4 0 4\\
NaN NaN NaN 4\\
NaN NaN NaN 4\\
NaN NaN NaN 4\\
NaN NaN NaN 4\\
3.6 8.6 0 4\\
3.6 9.4 0 4\\
NaN NaN NaN 4\\
NaN NaN NaN 4\\
NaN NaN NaN 4\\
NaN NaN NaN 4\\
3.6 9.6 0 4\\
3.6 10.4 0 4\\
NaN NaN NaN 4\\
NaN NaN NaN 4\\
NaN NaN NaN 4\\
NaN NaN NaN 4\\
3.6 10.6 0 4\\
3.6 11.4 0 4\\
NaN NaN NaN 4\\
NaN NaN NaN 4\\
NaN NaN NaN 4\\
3.6 0.6 0 4\\
3.6 0.6 0 4\\
3.6 1.4 0 4\\
3.6 1.4 0 4\\
3.6 0.6 0 4\\
NaN NaN NaN 4\\
3.6 1.6 0 4\\
3.6 1.6 1 4\\
3.6 2.4 1 4\\
3.6 2.4 0 4\\
3.6 1.6 0 4\\
NaN NaN NaN 4\\
3.6 2.6 0 4\\
3.6 2.6 1 4\\
3.6 3.4 1 4\\
3.6 3.4 0 4\\
3.6 2.6 0 4\\
NaN NaN NaN 4\\
3.6 3.6 0 4\\
3.6 3.6 1 4\\
3.6 4.4 1 4\\
3.6 4.4 0 4\\
3.6 3.6 0 4\\
NaN NaN NaN 4\\
3.6 4.6 0 4\\
3.6 4.6 2 4\\
3.6 5.4 2 4\\
3.6 5.4 0 4\\
3.6 4.6 0 4\\
NaN NaN NaN 4\\
3.6 5.6 0 4\\
3.6 5.6 2 4\\
3.6 6.4 2 4\\
3.6 6.4 0 4\\
3.6 5.6 0 4\\
NaN NaN NaN 4\\
3.6 6.6 0 4\\
3.6 6.6 2 4\\
3.6 7.4 2 4\\
3.6 7.4 0 4\\
3.6 6.6 0 4\\
NaN NaN NaN 4\\
3.6 7.6 0 4\\
3.6 7.6 2 4\\
3.6 8.4 2 4\\
3.6 8.4 0 4\\
3.6 7.6 0 4\\
NaN NaN NaN 4\\
3.6 8.6 0 4\\
3.6 8.6 2 4\\
3.6 9.4 2 4\\
3.6 9.4 0 4\\
3.6 8.6 0 4\\
NaN NaN NaN 4\\
3.6 9.6 0 4\\
3.6 9.6 2 4\\
3.6 10.4 2 4\\
3.6 10.4 0 4\\
3.6 9.6 0 4\\
NaN NaN NaN 4\\
3.6 10.6 0 4\\
3.6 10.6 3 4\\
3.6 11.4 3 4\\
3.6 11.4 0 4\\
3.6 10.6 0 4\\
NaN NaN NaN 4\\
4.4 0.6 0 4\\
4.4 0.6 0 4\\
4.4 1.4 0 4\\
4.4 1.4 0 4\\
4.4 0.6 0 4\\
NaN NaN NaN 4\\
4.4 1.6 0 4\\
4.4 1.6 1 4\\
4.4 2.4 1 4\\
4.4 2.4 0 4\\
4.4 1.6 0 4\\
NaN NaN NaN 4\\
4.4 2.6 0 4\\
4.4 2.6 1 4\\
4.4 3.4 1 4\\
4.4 3.4 0 4\\
4.4 2.6 0 4\\
NaN NaN NaN 4\\
4.4 3.6 0 4\\
4.4 3.6 1 4\\
4.4 4.4 1 4\\
4.4 4.4 0 4\\
4.4 3.6 0 4\\
NaN NaN NaN 4\\
4.4 4.6 0 4\\
4.4 4.6 2 4\\
4.4 5.4 2 4\\
4.4 5.4 0 4\\
4.4 4.6 0 4\\
NaN NaN NaN 4\\
4.4 5.6 0 4\\
4.4 5.6 2 4\\
4.4 6.4 2 4\\
4.4 6.4 0 4\\
4.4 5.6 0 4\\
NaN NaN NaN 4\\
4.4 6.6 0 4\\
4.4 6.6 2 4\\
4.4 7.4 2 4\\
4.4 7.4 0 4\\
4.4 6.6 0 4\\
NaN NaN NaN 4\\
4.4 7.6 0 4\\
4.4 7.6 2 4\\
4.4 8.4 2 4\\
4.4 8.4 0 4\\
4.4 7.6 0 4\\
NaN NaN NaN 4\\
4.4 8.6 0 4\\
4.4 8.6 2 4\\
4.4 9.4 2 4\\
4.4 9.4 0 4\\
4.4 8.6 0 4\\
NaN NaN NaN 4\\
4.4 9.6 0 4\\
4.4 9.6 2 4\\
4.4 10.4 2 4\\
4.4 10.4 0 4\\
4.4 9.6 0 4\\
NaN NaN NaN 4\\
4.4 10.6 0 4\\
4.4 10.6 3 4\\
4.4 11.4 3 4\\
4.4 11.4 0 4\\
4.4 10.6 0 4\\
NaN NaN NaN 4\\
NaN NaN NaN 4\\
4.4 0.6 0 4\\
4.4 1.4 0 4\\
NaN NaN NaN 4\\
NaN NaN NaN 4\\
NaN NaN NaN 4\\
NaN NaN NaN 4\\
4.4 1.6 0 4\\
4.4 2.4 0 4\\
NaN NaN NaN 4\\
NaN NaN NaN 4\\
NaN NaN NaN 4\\
NaN NaN NaN 4\\
4.4 2.6 0 4\\
4.4 3.4 0 4\\
NaN NaN NaN 4\\
NaN NaN NaN 4\\
NaN NaN NaN 4\\
NaN NaN NaN 4\\
4.4 3.6 0 4\\
4.4 4.4 0 4\\
NaN NaN NaN 4\\
NaN NaN NaN 4\\
NaN NaN NaN 4\\
NaN NaN NaN 4\\
4.4 4.6 0 4\\
4.4 5.4 0 4\\
NaN NaN NaN 4\\
NaN NaN NaN 4\\
NaN NaN NaN 4\\
NaN NaN NaN 4\\
4.4 5.6 0 4\\
4.4 6.4 0 4\\
NaN NaN NaN 4\\
NaN NaN NaN 4\\
NaN NaN NaN 4\\
NaN NaN NaN 4\\
4.4 6.6 0 4\\
4.4 7.4 0 4\\
NaN NaN NaN 4\\
NaN NaN NaN 4\\
NaN NaN NaN 4\\
NaN NaN NaN 4\\
4.4 7.6 0 4\\
4.4 8.4 0 4\\
NaN NaN NaN 4\\
NaN NaN NaN 4\\
NaN NaN NaN 4\\
NaN NaN NaN 4\\
4.4 8.6 0 4\\
4.4 9.4 0 4\\
NaN NaN NaN 4\\
NaN NaN NaN 4\\
NaN NaN NaN 4\\
NaN NaN NaN 4\\
4.4 9.6 0 4\\
4.4 10.4 0 4\\
NaN NaN NaN 4\\
NaN NaN NaN 4\\
NaN NaN NaN 4\\
NaN NaN NaN 4\\
4.4 10.6 0 4\\
4.4 11.4 0 4\\
NaN NaN NaN 4\\
NaN NaN NaN 4\\
NaN NaN NaN 4\\
};

\addplot3[%
surf,
shader=flat,
draw=black,
point meta=explicit,
mesh/rows=4]
table[row sep=crcr,header=false,meta index=3] {
NaN NaN NaN 3\\
2.6 0.6 0 3\\
2.6 1.4 0 3\\
NaN NaN NaN 3\\
NaN NaN NaN 3\\
NaN NaN NaN 3\\
NaN NaN NaN 3\\
2.6 1.6 0 3\\
2.6 2.4 0 3\\
NaN NaN NaN 3\\
NaN NaN NaN 3\\
NaN NaN NaN 3\\
NaN NaN NaN 3\\
2.6 2.6 0 3\\
2.6 3.4 0 3\\
NaN NaN NaN 3\\
NaN NaN NaN 3\\
NaN NaN NaN 3\\
NaN NaN NaN 3\\
2.6 3.6 0 3\\
2.6 4.4 0 3\\
NaN NaN NaN 3\\
NaN NaN NaN 3\\
NaN NaN NaN 3\\
NaN NaN NaN 3\\
2.6 4.6 0 3\\
2.6 5.4 0 3\\
NaN NaN NaN 3\\
NaN NaN NaN 3\\
NaN NaN NaN 3\\
NaN NaN NaN 3\\
2.6 5.6 0 3\\
2.6 6.4 0 3\\
NaN NaN NaN 3\\
NaN NaN NaN 3\\
NaN NaN NaN 3\\
NaN NaN NaN 3\\
2.6 6.6 0 3\\
2.6 7.4 0 3\\
NaN NaN NaN 3\\
NaN NaN NaN 3\\
NaN NaN NaN 3\\
NaN NaN NaN 3\\
2.6 7.6 0 3\\
2.6 8.4 0 3\\
NaN NaN NaN 3\\
NaN NaN NaN 3\\
NaN NaN NaN 3\\
NaN NaN NaN 3\\
2.6 8.6 0 3\\
2.6 9.4 0 3\\
NaN NaN NaN 3\\
NaN NaN NaN 3\\
NaN NaN NaN 3\\
NaN NaN NaN 3\\
2.6 9.6 0 3\\
2.6 10.4 0 3\\
NaN NaN NaN 3\\
NaN NaN NaN 3\\
NaN NaN NaN 3\\
NaN NaN NaN 3\\
2.6 10.6 0 3\\
2.6 11.4 0 3\\
NaN NaN NaN 3\\
NaN NaN NaN 3\\
NaN NaN NaN 3\\
2.6 0.6 0 3\\
2.6 0.6 0 3\\
2.6 1.4 0 3\\
2.6 1.4 0 3\\
2.6 0.6 0 3\\
NaN NaN NaN 3\\
2.6 1.6 0 3\\
2.6 1.6 1 3\\
2.6 2.4 1 3\\
2.6 2.4 0 3\\
2.6 1.6 0 3\\
NaN NaN NaN 3\\
2.6 2.6 0 3\\
2.6 2.6 1 3\\
2.6 3.4 1 3\\
2.6 3.4 0 3\\
2.6 2.6 0 3\\
NaN NaN NaN 3\\
2.6 3.6 0 3\\
2.6 3.6 1 3\\
2.6 4.4 1 3\\
2.6 4.4 0 3\\
2.6 3.6 0 3\\
NaN NaN NaN 3\\
2.6 4.6 0 3\\
2.6 4.6 2 3\\
2.6 5.4 2 3\\
2.6 5.4 0 3\\
2.6 4.6 0 3\\
NaN NaN NaN 3\\
2.6 5.6 0 3\\
2.6 5.6 2 3\\
2.6 6.4 2 3\\
2.6 6.4 0 3\\
2.6 5.6 0 3\\
NaN NaN NaN 3\\
2.6 6.6 0 3\\
2.6 6.6 2 3\\
2.6 7.4 2 3\\
2.6 7.4 0 3\\
2.6 6.6 0 3\\
NaN NaN NaN 3\\
2.6 7.6 0 3\\
2.6 7.6 2 3\\
2.6 8.4 2 3\\
2.6 8.4 0 3\\
2.6 7.6 0 3\\
NaN NaN NaN 3\\
2.6 8.6 0 3\\
2.6 8.6 2 3\\
2.6 9.4 2 3\\
2.6 9.4 0 3\\
2.6 8.6 0 3\\
NaN NaN NaN 3\\
2.6 9.6 0 3\\
2.6 9.6 2 3\\
2.6 10.4 2 3\\
2.6 10.4 0 3\\
2.6 9.6 0 3\\
NaN NaN NaN 3\\
2.6 10.6 0 3\\
2.6 10.6 2 3\\
2.6 11.4 2 3\\
2.6 11.4 0 3\\
2.6 10.6 0 3\\
NaN NaN NaN 3\\
3.4 0.6 0 3\\
3.4 0.6 0 3\\
3.4 1.4 0 3\\
3.4 1.4 0 3\\
3.4 0.6 0 3\\
NaN NaN NaN 3\\
3.4 1.6 0 3\\
3.4 1.6 1 3\\
3.4 2.4 1 3\\
3.4 2.4 0 3\\
3.4 1.6 0 3\\
NaN NaN NaN 3\\
3.4 2.6 0 3\\
3.4 2.6 1 3\\
3.4 3.4 1 3\\
3.4 3.4 0 3\\
3.4 2.6 0 3\\
NaN NaN NaN 3\\
3.4 3.6 0 3\\
3.4 3.6 1 3\\
3.4 4.4 1 3\\
3.4 4.4 0 3\\
3.4 3.6 0 3\\
NaN NaN NaN 3\\
3.4 4.6 0 3\\
3.4 4.6 2 3\\
3.4 5.4 2 3\\
3.4 5.4 0 3\\
3.4 4.6 0 3\\
NaN NaN NaN 3\\
3.4 5.6 0 3\\
3.4 5.6 2 3\\
3.4 6.4 2 3\\
3.4 6.4 0 3\\
3.4 5.6 0 3\\
NaN NaN NaN 3\\
3.4 6.6 0 3\\
3.4 6.6 2 3\\
3.4 7.4 2 3\\
3.4 7.4 0 3\\
3.4 6.6 0 3\\
NaN NaN NaN 3\\
3.4 7.6 0 3\\
3.4 7.6 2 3\\
3.4 8.4 2 3\\
3.4 8.4 0 3\\
3.4 7.6 0 3\\
NaN NaN NaN 3\\
3.4 8.6 0 3\\
3.4 8.6 2 3\\
3.4 9.4 2 3\\
3.4 9.4 0 3\\
3.4 8.6 0 3\\
NaN NaN NaN 3\\
3.4 9.6 0 3\\
3.4 9.6 2 3\\
3.4 10.4 2 3\\
3.4 10.4 0 3\\
3.4 9.6 0 3\\
NaN NaN NaN 3\\
3.4 10.6 0 3\\
3.4 10.6 2 3\\
3.4 11.4 2 3\\
3.4 11.4 0 3\\
3.4 10.6 0 3\\
NaN NaN NaN 3\\
NaN NaN NaN 3\\
3.4 0.6 0 3\\
3.4 1.4 0 3\\
NaN NaN NaN 3\\
NaN NaN NaN 3\\
NaN NaN NaN 3\\
NaN NaN NaN 3\\
3.4 1.6 0 3\\
3.4 2.4 0 3\\
NaN NaN NaN 3\\
NaN NaN NaN 3\\
NaN NaN NaN 3\\
NaN NaN NaN 3\\
3.4 2.6 0 3\\
3.4 3.4 0 3\\
NaN NaN NaN 3\\
NaN NaN NaN 3\\
NaN NaN NaN 3\\
NaN NaN NaN 3\\
3.4 3.6 0 3\\
3.4 4.4 0 3\\
NaN NaN NaN 3\\
NaN NaN NaN 3\\
NaN NaN NaN 3\\
NaN NaN NaN 3\\
3.4 4.6 0 3\\
3.4 5.4 0 3\\
NaN NaN NaN 3\\
NaN NaN NaN 3\\
NaN NaN NaN 3\\
NaN NaN NaN 3\\
3.4 5.6 0 3\\
3.4 6.4 0 3\\
NaN NaN NaN 3\\
NaN NaN NaN 3\\
NaN NaN NaN 3\\
NaN NaN NaN 3\\
3.4 6.6 0 3\\
3.4 7.4 0 3\\
NaN NaN NaN 3\\
NaN NaN NaN 3\\
NaN NaN NaN 3\\
NaN NaN NaN 3\\
3.4 7.6 0 3\\
3.4 8.4 0 3\\
NaN NaN NaN 3\\
NaN NaN NaN 3\\
NaN NaN NaN 3\\
NaN NaN NaN 3\\
3.4 8.6 0 3\\
3.4 9.4 0 3\\
NaN NaN NaN 3\\
NaN NaN NaN 3\\
NaN NaN NaN 3\\
NaN NaN NaN 3\\
3.4 9.6 0 3\\
3.4 10.4 0 3\\
NaN NaN NaN 3\\
NaN NaN NaN 3\\
NaN NaN NaN 3\\
NaN NaN NaN 3\\
3.4 10.6 0 3\\
3.4 11.4 0 3\\
NaN NaN NaN 3\\
NaN NaN NaN 3\\
NaN NaN NaN 3\\
};

\addplot3[%
surf,
shader=flat,
draw=black,
point meta=explicit,
mesh/rows=4]
table[row sep=crcr,header=false,meta index=3] {
NaN NaN NaN 2\\
1.6 0.6 0 2\\
1.6 1.4 0 2\\
NaN NaN NaN 2\\
NaN NaN NaN 2\\
NaN NaN NaN 2\\
NaN NaN NaN 2\\
1.6 1.6 0 2\\
1.6 2.4 0 2\\
NaN NaN NaN 2\\
NaN NaN NaN 2\\
NaN NaN NaN 2\\
NaN NaN NaN 2\\
1.6 2.6 0 2\\
1.6 3.4 0 2\\
NaN NaN NaN 2\\
NaN NaN NaN 2\\
NaN NaN NaN 2\\
NaN NaN NaN 2\\
1.6 3.6 0 2\\
1.6 4.4 0 2\\
NaN NaN NaN 2\\
NaN NaN NaN 2\\
NaN NaN NaN 2\\
NaN NaN NaN 2\\
1.6 4.6 0 2\\
1.6 5.4 0 2\\
NaN NaN NaN 2\\
NaN NaN NaN 2\\
NaN NaN NaN 2\\
NaN NaN NaN 2\\
1.6 5.6 0 2\\
1.6 6.4 0 2\\
NaN NaN NaN 2\\
NaN NaN NaN 2\\
NaN NaN NaN 2\\
NaN NaN NaN 2\\
1.6 6.6 0 2\\
1.6 7.4 0 2\\
NaN NaN NaN 2\\
NaN NaN NaN 2\\
NaN NaN NaN 2\\
NaN NaN NaN 2\\
1.6 7.6 0 2\\
1.6 8.4 0 2\\
NaN NaN NaN 2\\
NaN NaN NaN 2\\
NaN NaN NaN 2\\
NaN NaN NaN 2\\
1.6 8.6 0 2\\
1.6 9.4 0 2\\
NaN NaN NaN 2\\
NaN NaN NaN 2\\
NaN NaN NaN 2\\
NaN NaN NaN 2\\
1.6 9.6 0 2\\
1.6 10.4 0 2\\
NaN NaN NaN 2\\
NaN NaN NaN 2\\
NaN NaN NaN 2\\
NaN NaN NaN 2\\
1.6 10.6 0 2\\
1.6 11.4 0 2\\
NaN NaN NaN 2\\
NaN NaN NaN 2\\
NaN NaN NaN 2\\
1.6 0.6 0 2\\
1.6 0.6 0 2\\
1.6 1.4 0 2\\
1.6 1.4 0 2\\
1.6 0.6 0 2\\
NaN NaN NaN 2\\
1.6 1.6 0 2\\
1.6 1.6 1 2\\
1.6 2.4 1 2\\
1.6 2.4 0 2\\
1.6 1.6 0 2\\
NaN NaN NaN 2\\
1.6 2.6 0 2\\
1.6 2.6 1 2\\
1.6 3.4 1 2\\
1.6 3.4 0 2\\
1.6 2.6 0 2\\
NaN NaN NaN 2\\
1.6 3.6 0 2\\
1.6 3.6 1 2\\
1.6 4.4 1 2\\
1.6 4.4 0 2\\
1.6 3.6 0 2\\
NaN NaN NaN 2\\
1.6 4.6 0 2\\
1.6 4.6 2 2\\
1.6 5.4 2 2\\
1.6 5.4 0 2\\
1.6 4.6 0 2\\
NaN NaN NaN 2\\
1.6 5.6 0 2\\
1.6 5.6 2 2\\
1.6 6.4 2 2\\
1.6 6.4 0 2\\
1.6 5.6 0 2\\
NaN NaN NaN 2\\
1.6 6.6 0 2\\
1.6 6.6 2 2\\
1.6 7.4 2 2\\
1.6 7.4 0 2\\
1.6 6.6 0 2\\
NaN NaN NaN 2\\
1.6 7.6 0 2\\
1.6 7.6 2 2\\
1.6 8.4 2 2\\
1.6 8.4 0 2\\
1.6 7.6 0 2\\
NaN NaN NaN 2\\
1.6 8.6 0 2\\
1.6 8.6 2 2\\
1.6 9.4 2 2\\
1.6 9.4 0 2\\
1.6 8.6 0 2\\
NaN NaN NaN 2\\
1.6 9.6 0 2\\
1.6 9.6 2 2\\
1.6 10.4 2 2\\
1.6 10.4 0 2\\
1.6 9.6 0 2\\
NaN NaN NaN 2\\
1.6 10.6 0 2\\
1.6 10.6 2 2\\
1.6 11.4 2 2\\
1.6 11.4 0 2\\
1.6 10.6 0 2\\
NaN NaN NaN 2\\
2.4 0.6 0 2\\
2.4 0.6 0 2\\
2.4 1.4 0 2\\
2.4 1.4 0 2\\
2.4 0.6 0 2\\
NaN NaN NaN 2\\
2.4 1.6 0 2\\
2.4 1.6 1 2\\
2.4 2.4 1 2\\
2.4 2.4 0 2\\
2.4 1.6 0 2\\
NaN NaN NaN 2\\
2.4 2.6 0 2\\
2.4 2.6 1 2\\
2.4 3.4 1 2\\
2.4 3.4 0 2\\
2.4 2.6 0 2\\
NaN NaN NaN 2\\
2.4 3.6 0 2\\
2.4 3.6 1 2\\
2.4 4.4 1 2\\
2.4 4.4 0 2\\
2.4 3.6 0 2\\
NaN NaN NaN 2\\
2.4 4.6 0 2\\
2.4 4.6 2 2\\
2.4 5.4 2 2\\
2.4 5.4 0 2\\
2.4 4.6 0 2\\
NaN NaN NaN 2\\
2.4 5.6 0 2\\
2.4 5.6 2 2\\
2.4 6.4 2 2\\
2.4 6.4 0 2\\
2.4 5.6 0 2\\
NaN NaN NaN 2\\
2.4 6.6 0 2\\
2.4 6.6 2 2\\
2.4 7.4 2 2\\
2.4 7.4 0 2\\
2.4 6.6 0 2\\
NaN NaN NaN 2\\
2.4 7.6 0 2\\
2.4 7.6 2 2\\
2.4 8.4 2 2\\
2.4 8.4 0 2\\
2.4 7.6 0 2\\
NaN NaN NaN 2\\
2.4 8.6 0 2\\
2.4 8.6 2 2\\
2.4 9.4 2 2\\
2.4 9.4 0 2\\
2.4 8.6 0 2\\
NaN NaN NaN 2\\
2.4 9.6 0 2\\
2.4 9.6 2 2\\
2.4 10.4 2 2\\
2.4 10.4 0 2\\
2.4 9.6 0 2\\
NaN NaN NaN 2\\
2.4 10.6 0 2\\
2.4 10.6 2 2\\
2.4 11.4 2 2\\
2.4 11.4 0 2\\
2.4 10.6 0 2\\
NaN NaN NaN 2\\
NaN NaN NaN 2\\
2.4 0.6 0 2\\
2.4 1.4 0 2\\
NaN NaN NaN 2\\
NaN NaN NaN 2\\
NaN NaN NaN 2\\
NaN NaN NaN 2\\
2.4 1.6 0 2\\
2.4 2.4 0 2\\
NaN NaN NaN 2\\
NaN NaN NaN 2\\
NaN NaN NaN 2\\
NaN NaN NaN 2\\
2.4 2.6 0 2\\
2.4 3.4 0 2\\
NaN NaN NaN 2\\
NaN NaN NaN 2\\
NaN NaN NaN 2\\
NaN NaN NaN 2\\
2.4 3.6 0 2\\
2.4 4.4 0 2\\
NaN NaN NaN 2\\
NaN NaN NaN 2\\
NaN NaN NaN 2\\
NaN NaN NaN 2\\
2.4 4.6 0 2\\
2.4 5.4 0 2\\
NaN NaN NaN 2\\
NaN NaN NaN 2\\
NaN NaN NaN 2\\
NaN NaN NaN 2\\
2.4 5.6 0 2\\
2.4 6.4 0 2\\
NaN NaN NaN 2\\
NaN NaN NaN 2\\
NaN NaN NaN 2\\
NaN NaN NaN 2\\
2.4 6.6 0 2\\
2.4 7.4 0 2\\
NaN NaN NaN 2\\
NaN NaN NaN 2\\
NaN NaN NaN 2\\
NaN NaN NaN 2\\
2.4 7.6 0 2\\
2.4 8.4 0 2\\
NaN NaN NaN 2\\
NaN NaN NaN 2\\
NaN NaN NaN 2\\
NaN NaN NaN 2\\
2.4 8.6 0 2\\
2.4 9.4 0 2\\
NaN NaN NaN 2\\
NaN NaN NaN 2\\
NaN NaN NaN 2\\
NaN NaN NaN 2\\
2.4 9.6 0 2\\
2.4 10.4 0 2\\
NaN NaN NaN 2\\
NaN NaN NaN 2\\
NaN NaN NaN 2\\
NaN NaN NaN 2\\
2.4 10.6 0 2\\
2.4 11.4 0 2\\
NaN NaN NaN 2\\
NaN NaN NaN 2\\
NaN NaN NaN 2\\
};

\addplot3[%
surf,
shader=flat,
draw=black,
point meta=explicit,
mesh/rows=4]
table[row sep=crcr,header=false,meta index=3] {
NaN NaN NaN 1\\
0.6 0.6 0 1\\
0.6 1.4 0 1\\
NaN NaN NaN 1\\
NaN NaN NaN 1\\
NaN NaN NaN 1\\
NaN NaN NaN 1\\
0.6 1.6 0 1\\
0.6 2.4 0 1\\
NaN NaN NaN 1\\
NaN NaN NaN 1\\
NaN NaN NaN 1\\
NaN NaN NaN 1\\
0.6 2.6 0 1\\
0.6 3.4 0 1\\
NaN NaN NaN 1\\
NaN NaN NaN 1\\
NaN NaN NaN 1\\
NaN NaN NaN 1\\
0.6 3.6 0 1\\
0.6 4.4 0 1\\
NaN NaN NaN 1\\
NaN NaN NaN 1\\
NaN NaN NaN 1\\
NaN NaN NaN 1\\
0.6 4.6 0 1\\
0.6 5.4 0 1\\
NaN NaN NaN 1\\
NaN NaN NaN 1\\
NaN NaN NaN 1\\
NaN NaN NaN 1\\
0.6 5.6 0 1\\
0.6 6.4 0 1\\
NaN NaN NaN 1\\
NaN NaN NaN 1\\
NaN NaN NaN 1\\
NaN NaN NaN 1\\
0.6 6.6 0 1\\
0.6 7.4 0 1\\
NaN NaN NaN 1\\
NaN NaN NaN 1\\
NaN NaN NaN 1\\
NaN NaN NaN 1\\
0.6 7.6 0 1\\
0.6 8.4 0 1\\
NaN NaN NaN 1\\
NaN NaN NaN 1\\
NaN NaN NaN 1\\
NaN NaN NaN 1\\
0.6 8.6 0 1\\
0.6 9.4 0 1\\
NaN NaN NaN 1\\
NaN NaN NaN 1\\
NaN NaN NaN 1\\
NaN NaN NaN 1\\
0.6 9.6 0 1\\
0.6 10.4 0 1\\
NaN NaN NaN 1\\
NaN NaN NaN 1\\
NaN NaN NaN 1\\
NaN NaN NaN 1\\
0.6 10.6 0 1\\
0.6 11.4 0 1\\
NaN NaN NaN 1\\
NaN NaN NaN 1\\
NaN NaN NaN 1\\
0.6 0.6 0 1\\
0.6 0.6 0 1\\
0.6 1.4 0 1\\
0.6 1.4 0 1\\
0.6 0.6 0 1\\
NaN NaN NaN 1\\
0.6 1.6 0 1\\
0.6 1.6 1 1\\
0.6 2.4 1 1\\
0.6 2.4 0 1\\
0.6 1.6 0 1\\
NaN NaN NaN 1\\
0.6 2.6 0 1\\
0.6 2.6 1 1\\
0.6 3.4 1 1\\
0.6 3.4 0 1\\
0.6 2.6 0 1\\
NaN NaN NaN 1\\
0.6 3.6 0 1\\
0.6 3.6 1 1\\
0.6 4.4 1 1\\
0.6 4.4 0 1\\
0.6 3.6 0 1\\
NaN NaN NaN 1\\
0.6 4.6 0 1\\
0.6 4.6 1 1\\
0.6 5.4 1 1\\
0.6 5.4 0 1\\
0.6 4.6 0 1\\
NaN NaN NaN 1\\
0.6 5.6 0 1\\
0.6 5.6 9 1\\
0.6 6.4 9 1\\
0.6 6.4 0 1\\
0.6 5.6 0 1\\
NaN NaN NaN 1\\
0.6 6.6 0 1\\
0.6 6.6 9 1\\
0.6 7.4 9 1\\
0.6 7.4 0 1\\
0.6 6.6 0 1\\
NaN NaN NaN 1\\
0.6 7.6 0 1\\
0.6 7.6 3 1\\
0.6 8.4 3 1\\
0.6 8.4 0 1\\
0.6 7.6 0 1\\
NaN NaN NaN 1\\
0.6 8.6 0 1\\
0.6 8.6 3 1\\
0.6 9.4 3 1\\
0.6 9.4 0 1\\
0.6 8.6 0 1\\
NaN NaN NaN 1\\
0.6 9.6 0 1\\
0.6 9.6 3 1\\
0.6 10.4 3 1\\
0.6 10.4 0 1\\
0.6 9.6 0 1\\
NaN NaN NaN 1\\
0.6 10.6 0 1\\
0.6 10.6 3 1\\
0.6 11.4 3 1\\
0.6 11.4 0 1\\
0.6 10.6 0 1\\
NaN NaN NaN 1\\
1.4 0.6 0 1\\
1.4 0.6 0 1\\
1.4 1.4 0 1\\
1.4 1.4 0 1\\
1.4 0.6 0 1\\
NaN NaN NaN 1\\
1.4 1.6 0 1\\
1.4 1.6 1 1\\
1.4 2.4 1 1\\
1.4 2.4 0 1\\
1.4 1.6 0 1\\
NaN NaN NaN 1\\
1.4 2.6 0 1\\
1.4 2.6 1 1\\
1.4 3.4 1 1\\
1.4 3.4 0 1\\
1.4 2.6 0 1\\
NaN NaN NaN 1\\
1.4 3.6 0 1\\
1.4 3.6 1 1\\
1.4 4.4 1 1\\
1.4 4.4 0 1\\
1.4 3.6 0 1\\
NaN NaN NaN 1\\
1.4 4.6 0 1\\
1.4 4.6 1 1\\
1.4 5.4 1 1\\
1.4 5.4 0 1\\
1.4 4.6 0 1\\
NaN NaN NaN 1\\
1.4 5.6 0 1\\
1.4 5.6 9 1\\
1.4 6.4 9 1\\
1.4 6.4 0 1\\
1.4 5.6 0 1\\
NaN NaN NaN 1\\
1.4 6.6 0 1\\
1.4 6.6 9 1\\
1.4 7.4 9 1\\
1.4 7.4 0 1\\
1.4 6.6 0 1\\
NaN NaN NaN 1\\
1.4 7.6 0 1\\
1.4 7.6 3 1\\
1.4 8.4 3 1\\
1.4 8.4 0 1\\
1.4 7.6 0 1\\
NaN NaN NaN 1\\
1.4 8.6 0 1\\
1.4 8.6 3 1\\
1.4 9.4 3 1\\
1.4 9.4 0 1\\
1.4 8.6 0 1\\
NaN NaN NaN 1\\
1.4 9.6 0 1\\
1.4 9.6 3 1\\
1.4 10.4 3 1\\
1.4 10.4 0 1\\
1.4 9.6 0 1\\
NaN NaN NaN 1\\
1.4 10.6 0 1\\
1.4 10.6 3 1\\
1.4 11.4 3 1\\
1.4 11.4 0 1\\
1.4 10.6 0 1\\
NaN NaN NaN 1\\
NaN NaN NaN 1\\
1.4 0.6 0 1\\
1.4 1.4 0 1\\
NaN NaN NaN 1\\
NaN NaN NaN 1\\
NaN NaN NaN 1\\
NaN NaN NaN 1\\
1.4 1.6 0 1\\
1.4 2.4 0 1\\
NaN NaN NaN 1\\
NaN NaN NaN 1\\
NaN NaN NaN 1\\
NaN NaN NaN 1\\
1.4 2.6 0 1\\
1.4 3.4 0 1\\
NaN NaN NaN 1\\
NaN NaN NaN 1\\
NaN NaN NaN 1\\
NaN NaN NaN 1\\
1.4 3.6 0 1\\
1.4 4.4 0 1\\
NaN NaN NaN 1\\
NaN NaN NaN 1\\
NaN NaN NaN 1\\
NaN NaN NaN 1\\
1.4 4.6 0 1\\
1.4 5.4 0 1\\
NaN NaN NaN 1\\
NaN NaN NaN 1\\
NaN NaN NaN 1\\
NaN NaN NaN 1\\
1.4 5.6 0 1\\
1.4 6.4 0 1\\
NaN NaN NaN 1\\
NaN NaN NaN 1\\
NaN NaN NaN 1\\
NaN NaN NaN 1\\
1.4 6.6 0 1\\
1.4 7.4 0 1\\
NaN NaN NaN 1\\
NaN NaN NaN 1\\
NaN NaN NaN 1\\
NaN NaN NaN 1\\
1.4 7.6 0 1\\
1.4 8.4 0 1\\
NaN NaN NaN 1\\
NaN NaN NaN 1\\
NaN NaN NaN 1\\
NaN NaN NaN 1\\
1.4 8.6 0 1\\
1.4 9.4 0 1\\
NaN NaN NaN 1\\
NaN NaN NaN 1\\
NaN NaN NaN 1\\
NaN NaN NaN 1\\
1.4 9.6 0 1\\
1.4 10.4 0 1\\
NaN NaN NaN 1\\
NaN NaN NaN 1\\
NaN NaN NaN 1\\
NaN NaN NaN 1\\
1.4 10.6 0 1\\
1.4 11.4 0 1\\
NaN NaN NaN 1\\
NaN NaN NaN 1\\
NaN NaN NaN 1\\
};

\end{axis}
\end{tikzpicture}%